%% file: main.tex
\documentclass[
11pt,
english,
singlespacing,
liststotoc,
headsepline,
table,
]{DoctoralThesis}

\usepackage[T1]{fontenc}
\usepackage{hyperref}
\usepackage{amsmath,amssymb,amsfonts}
\usepackage{algorithmic}
\usepackage[ruled,linesnumbered,noend]{algorithm2e}
\usepackage{xcolor}
\usepackage{graphicx}
\usepackage{textcomp}
\usepackage{xcolor}
\usepackage{nccmath}
\usepackage{environ}
\usepackage{tcolorbox}
\usepackage{soul}
\usepackage{booktabs}
\usepackage{multirow}
\usepackage{array}
\usepackage{float}
\usepackage{makecell}
\usepackage{flushend}
\usepackage{enumitem}
\usepackage{hhline}
\usepackage{comment}

\usepackage{blindtext, graphicx}
\usepackage{csquotes}
\usepackage{soul}
\usepackage{diagbox}
\usepackage{multirow}
\usepackage{listings}
\usepackage{longtable}
\usepackage{tcolorbox}
\usepackage{hhline}
\usepackage{tikz}
\usepackage{pdflscape}
\usepackage{comment}
\usepackage{afterpage}

\usepackage{lettrine}

\newcommand*\circled[1]{\tikz[baseline=(char.base)]{
            \node[shape=circle,fill,inner sep=2pt] (char) {\textcolor{white}{#1}};}}
            
\newlist{contribution}{enumerate}{10}
\setlist[contribution]{label*=\arabic*.}
\setlistdepth{10} 

\newif\iffull
\fullfalse

\SetKwFor{ForEach}{foreach}{do}{end~foreach}
\SetKwIF{If}{ElseIf}{Else}{if}{then}{else if}{else}{end~if}

\newcommand{\tabitem}{~~\llap{\textbullet}~~}

\usepackage{mathtools}
\usepackage{bbold}

\newcolumntype{P}[1]{>{\centering\arraybackslash}m{#1}}

\usepackage{makecell}

\usepackage{hyperref}
\usepackage{amsmath,amssymb,amsfonts}
\usepackage{algorithmic}
\usepackage[ruled,linesnumbered,noend]{algorithm2e}
\usepackage{xcolor}
\usepackage{graphicx}
\usepackage{textcomp}
\usepackage{xcolor}
\usepackage{nccmath}
\usepackage{environ}
\usepackage{tcolorbox}
\usepackage{soul}
\usepackage{booktabs}
\usepackage{multirow}
\usepackage{array}
\usepackage{float}
\usepackage{makecell}
\usepackage{flushend}
\usepackage{enumitem}
\usepackage{hhline}
\usepackage{comment}
\usepackage{epstopdf}
\usepackage{caption}
\usepackage{subcaption}

\usepackage{xcolor}
\usepackage{listings}

\usepackage{acronym}

\NewEnviron{myequation}{
\begin{equation}
\scalebox{1}{$\BODY$}
\end{equation}
}

\newcommand{\fig}{Fig.}
\newcommand{\tab}{Table}

\newcolumntype{P}[1]{>{\centering\arraybackslash}p{#1}}

\newcommand{\code}[1]{\textnormal{\texttt{#1}}}

\hypersetup{bookmarksnumbered}

\thesistitle{Towards an Improved Understanding of Software Vulnerability Assessment Using Data-Driven Approaches}
\supervisor{Professor \textsc{M. Ali Babar}}
\degree{Doctor of Philosophy}
\author{\textsc{Triet Huynh Minh Le}}
\addresses{}

\subject{Computer Science}
\university{\href{http://www.adelaide.edu.au}{The University of Adelaide}}
\department{\href{https://cs.adelaide.edu.au}{School of Computer Science}}
\group{\href{https://cs.adelaide.edu.au/research/cseducation/}{Computer Science Education Research}}
\faculty{\href{http://faculty.university.com}{Faculty Name}}

\AtBeginDocument{
\hypersetup{pdftitle=\ttitle}
\hypersetup{pdfauthor=\authorname}
\hypersetup{pdfkeywords=\keywordnames}
}

\begin{document}

\frontmatter

\pagestyle{plain}

\begin{titlepage}
\begin{center}

\includegraphics{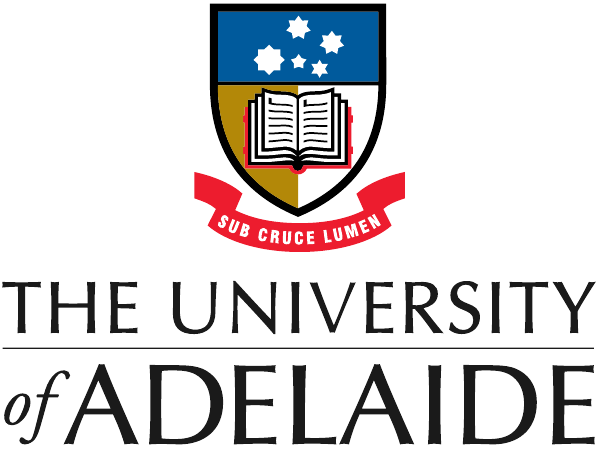}\\
\vspace{0.8cm}{\huge \bfseries \ttitle\par}

\vspace{1.5cm}
{Author: \textbf{\authorname}}\\
{Centre for Research on Engineering Software Technologies (CREST)\\
School of Computer Science\\
Faculty of Engineering, Computer and Mathematical Sciences\\
The University of Adelaide}

\vspace{0.7cm}
{Principal Supervisor: Professor Muhammad Ali Babar}\\
{Co-Supervisor: Professor Cheng-Chew Lim}

\vspace{1.5cm}
\large A thesis submitted for the degree of\\ DOCTOR OF PHILOSOPHY\\The University of Adelaide\\[0.3cm]
\vspace{0.7cm}

{\large March 21, 2022}\\[4cm]

\end{center}
\end{titlepage}

\hypersetup{
    colorlinks=true,
}

{
\hypersetup{linkcolor=[rgb]{0,0,0.52}}
\hypersetup{linktoc=all}

\tableofcontents

\listoffigures

\listoftables
}

\begin{abstract}
\addchaptertocentry{\abstractname}

Software Vulnerabilities (SVs) can expose software systems to cyber-attacks, potentially causing enormous financial and reputational damage for organizations.
There have been significant research efforts to detect these SVs so that developers can promptly fix them.
However, fixing SVs is complex and time-consuming in practice, and thus developers usually do not have sufficient time and resources to fix all SVs at once. As a result, developers often need SV information, such as exploitability, impact, and overall severity, to prioritize fixing more critical SVs. Such information required for fixing planning and prioritization is typically provided in the \textit{SV assessment} step of the SV lifecycle.
Recently, data-driven methods have been increasingly proposed to automate SV assessment tasks.
However, there are still numerous shortcomings with the existing studies on data-driven SV assessment that would hinder their application in practice.

This PhD thesis aims to contribute to the growing literature in data-driven SV assessment by investigating and addressing the constant changes in SV data as well as the lacking considerations of source code and developers' needs for SV assessment that impede the practical applicability of the field.
Particularly, we have made the following five contributions in this thesis.
(\textbf{1}) We systematize the knowledge of data-driven SV assessment to reveal the best practices of the field and the main challenges affecting its application in practice.
Subsequently, we propose various solutions to tackle these challenges to better support the real-world applications of data-driven SV assessment.
(\textbf{2}) We first demonstrate the existence of the concept drift (changing data) issue in descriptions of SV reports that current studies have mostly used for predicting the Common Vulnerability Scoring System (CVSS) metrics.
We augment report-level SV assessment models with subwords of terms extracted from SV descriptions to help the models more effectively capture the semantics of ever-increasing SVs.
(\textbf{3}) We also identify that SV reports are usually released after SV fixing. Thus, we propose using vulnerable code to enable earlier SV assessment without waiting for SV reports.
We are the first to use Machine Learning techniques to predict CVSS metrics on the function level leveraging vulnerable statements directly causing SVs and their context in code functions.
The performance of our function-level SV assessment models is promising, opening up research opportunities in this new direction.
(\textbf{4}) To facilitate continuous integration of software code nowadays, we present a novel deep multi-task learning model, DeepCVA, to simultaneously and efficiently predict multiple CVSS assessment metrics on the commit level, specifically using vulnerability-contributing commits.
DeepCVA is the first work that enables practitioners to perform SV assessment as soon as vulnerable changes are added to a codebase, supporting just-in-time prioritization of SV fixing.
(\textbf{5}) Besides code artifacts produced from a software project of interest, SV assessment tasks can also benefit from SV crowdsourcing information on developer Question and Answer (Q\&A) websites.
We automatically retrieve large-scale security/SV-related posts from these Q\&A websites. We then apply a topic modeling technique on these posts to distill developers' real-world SV concerns that can be used for data-driven SV assessment.
Overall, we believe that this thesis has provided evidence-based knowledge and useful guidelines for researchers and practitioners to automate SV assessment using data-driven approaches.

\end{abstract}

\begin{declaration}
\addchaptertocentry{\authorshipname}

I certify that this work contains no material which has been accepted for the award of any other degree or diploma in my name, in any university or other tertiary institution and, to the best of my knowledge and belief, contains no material previously published or written by another person, except where due reference has been made in the text. In addition, I certify that no part of this work will, in the future, be used in a submission in my name, for any other degree or diploma in any university or other tertiary institution without the prior approval of the University of Adelaide and where applicable, any partner institution responsible for the joint-award of this degree.

I acknowledge that copyright of published works contained within this thesis resides with the copyright holder(s) of those works.

I also give permission for the digital version of my thesis to be made available on the web, via the University's digital research repository, the Library Search and also through web search engines, unless permission has been granted by the University to restrict access for a period of time.

I acknowledge the support I have received for my research through the provision of University of Adelaide International Wildcard Scholarship.

\vspace{40mm}
\vspace{-1cm}
\begin{flushright}
Triet Huynh Minh Le
\end{flushright}
\begin{flushright}March 2022\end{flushright}
 
\end{declaration}

\cleardoublepage

\begin{acknowledgements}
\addchaptertocentry{\acknowledgementname}

This thesis would not have been possible without continuous support, guidance, and encouragement from many people and entities.
I would like to hereby acknowledge them.

Firstly, I express my deepest gratitude to my principal supervisor, Professor M. Ali Babar, for giving me a valuable opportunity to conduct PhD research under his supervision. His constructive feedback has motivated me to continuously reflect and improve myself to become a better researcher and a more-rounded person in life. With his kind patience and persistent guidance, I have also managed to navigate myself through the challenging COVID-19 pandemic and complete my PhD research to the best of my ability. Besides research, he has also given me great opportunities to engage in numerous teaching and supervision activities that have tremendously helped me to enhance my communication and interpersonal skills. All in all, working under his mentorship has profoundly transformed me and enabled me to go beyond my limits and better prepare myself for my future career.

Secondly, I sincerely thank my co-supervisor, Professor Cheng-Chew Lim for providing insightful comments on the research carried out in this thesis.

Thirdly, I am extremely grateful to many current/former members in the Centre for Research on Engineering Software Technologies (CREST) at the University of Adelaide. Special thanks to Faheem Ullah, Chadni Islam, Bushra Sabir, Aufeef Chauhan, Huaming Chen, Bakheet Aljedaani, Mansooreh Zahedi, Hao Chen, Roland Croft, David Hin, and Mubin Ul Haque for not only academic contributions and feedback on the research papers related to this thesis, but also for being wonderful colleagues from whom I have learned a lot. Specifically, I cannot give enough appreciation to Roland Croft and David Hin for their great technical insights and contributions to improve the quality of many research endeavors I have pursued during my PhD. In addition, I am happy to be accompanied by Faheem Ullah and Aufeef Chauhan during weekly Friday dinners, which has helped me to relax and recharge each week. I also appreciate Nguyen Khoi Tran for introducing me to the CREST family. Thank you all for making my PhD journey memorable.

Fourthly, I have also had the chance to collaborate with and learn from many world-class researchers outside of CREST such as Xuanyu Duan, Mengmeng Ge, Shang Gao, and Xuequan Lu.
I am also thankful for all the constructive feedback from the paper and thesis reviewers that helped significantly improve the research conducted in the thesis.

Fifthly, I fully acknowledge the University of Adelaide for providing me with the University of Adelaide International Wildcard Scholarship and world-class facilities that have supported me to pursue my doctoral research and activities.

Sixthly, I highly appreciate the Urbanest at the University of Adelaide for providing me with the best-conditioned accommodation that I can ever ask for so that I can enjoy my personal life and recharge after working hours during my PhD. I am also extremely fortunate that Urbanest has also given me sufficient facilities to work effectively from home during the pandemic. I also want to deeply thank my roommates, especially Zach Li, for cheering me up during my down days.

Seventhly, I am greatly appreciative of my ASUS laptop for always being my reliable companion and working restlessly to enable me to obtain experimental results as well as write research papers and this thesis in a timely manner.

Finally and most importantly, I am immensely and eternally indebted to my family, especially my grandmother, mother, and father, who always stand by my side during the ups and downs of my PhD. Without their constant support and unconditional caring, I would not have been able to pursue my dreams and be where I am now. I love all of you from the bottom of my heart.

\end{acknowledgements}

\newpage 
\thispagestyle{empty}
\ 
\newpage

\thispagestyle{plain}
\par\vspace*{.35\textheight}{\centering \textit{I would like to dedicate this thesis to my parents}.\par}
\addchaptertocentry{Dedication}

\mainmatter

\pagestyle{thesis}

{
\hypersetup{linkcolor=[rgb]{0,0,0.52}}
\hypersetup{urlcolor=[rgb]{0,0,0.52}}
\hypersetup{citecolor=[rgb]{0,0,0.52}}

\include{Chapters/Chapter_1_Introduction}

\include{Chapters/Chapter_2_LitReview}

\include{Chapters/Chapter_3_MSR2019}

\include{Chapters/Chapter_4_MSR2022}

\include{Chapters/Chapter_5_ASE2021}

\include{Chapters/Chapter_6_EASE2021}

\include{Chapters/Chapter_7_Conclusions}

\renewcommand{\bibname}{References}
\addcontentsline{toc}{chapter}{References}
\bibliographystyle{IEEEtran}
\bibliography{reference}
}

\end{document}

%% file: Chapters/Chapter_1_Introduction.tex
\chapter{Introduction}
\label{chap:thesis_introduction}

Software has become an integral part of the modern world~\cite{andreessen2011software}.
Software systems are rapidly increasing in size and complexity.
For example, the whole software ecosystems at Google, which host many popular applications/services such as Google Search, YouTube, and Google Maps, contain more than two million lines of code~\cite{google_codebase}. Quality assurance of such large systems is a focal point for both researchers and practitioners to minimize disruptions to millions of people around the world.

Software Vulnerabilities (SVs)\footnote{In this thesis, ``software vulnerability'' and ``security vulnerability'' are used interchangeably.} have been long-standing issues that negatively affect software quality~\cite{ghaffarian2017software}.
SVs are security bugs that are detrimental to the confidentiality, integrity and availability of software systems, potentially resulting in catastrophic cybersecurity attacks~\cite{nvd_vuln}.
The exploitation of these SVs such as the Heartbleed~\cite{heartbleed} or Log4j~\cite{log4j} attacks can damage the operations and reputation of millions of software systems and organizations globally. These cyber-attacks caused by SVs have led to huge financial losses as well. According to the Australian Cyber Security Centre, a loss of more than 30 billion dollars due to cyber-attacks have been reported worldwide from 2020 to 2021~\cite{cyber_loss_2021}. Therefore, it is important to remediate critical SVs as promptly as possible.

In practice, different types of SVs have varying levels of security threats to software-intensive systems~\cite{nayak2014some}. However, fixing all SVs at the same time is not always practical due to limited resources and time~\cite{khan2018review}. A common practice in this situation is to prioritize fixing SVs posing imminent and serious threats to a system of interest. Such fixing prioritization usually requires inputs from SV assessment~\cite{smyth2017software,le2021survey}.

\begin{figure}[t]
    \centering
    \includegraphics[width=\columnwidth,keepaspectratio]{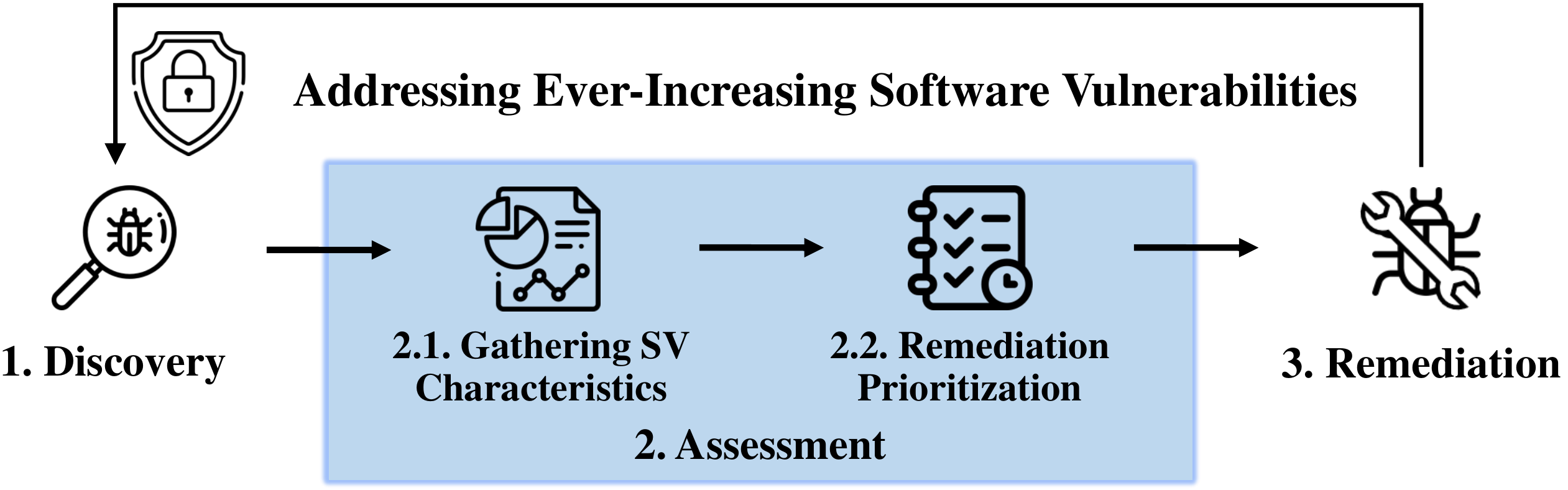}
    \caption[Phases in a software vulnerability lifecycle.]{Phases in an SV lifecycle. \textbf{Note}: The main focus of this thesis is SV assessment.}
    \label{fig:SV_lifecyle}
\end{figure}

The \textit{SV assessment} phase is between the SV \textit{discovery}/\textit{detection} and SV \textit{remediation}/\textit{mitigation}/\textit{fixing}/\textit{patching} phases in the SV management lifecycle~\cite{foreman2019vulnerability}, as shown in \fig~\ref{fig:SV_lifecyle}. The \textit{assessment} phase first unveils the characteristics of the SVs found in the \textit{discovery} phase to locate ``hot spots'' that contain many highly critical/severe SVs and require higher attention in a system.
Practitioners then use the assessment outputs to devise an optimal remediation plan, i.e., the order/priority of fixing each SV, based on available human and technological resources.
For example, an identified cross-site scripting (XSS) or SQL injection vulnerability in a web application will likely require an urgent remediation plan. These two types of SVs are well-known and can be easily exploited by attackers to gain unauthorized access and compromise sensitive data/information. On the other hand, an SV that requires admin access or happens only in a local network will probably have a lower priority since only a few people can initiate an attack.
According to the plan devised in the assessment phase, SVs would be prioritized for fixing in the \textit{remediation} phase.
In practice, the tasks in the SV assessment phase for ever-increasing SVs are repetitive and time-consuming, and thus require automation to save time and effort for practitioners.

Given the increasing size and complexity of software systems nowadays, automation of SV assessment tasks has attracted significant attention in the Software Engineering community. Traditionally, static analysis tools have been the de-facto approach to SV assessment~\cite{kritikos2019survey}. These tools rely on pre-defined rules to determine SV characteristics. However, these pre-defined rules require significant expertise and effort to define and may be easily error-prone~\cite{bell1985expert}. In addition, these rules need manual modifications and extensions to adapt to ever-changing patterns of new SVs~\cite{han2011data}. Such manual changes do not scale well with the fast-paced growth of SVs, potentially leading to delays in SV assessment and in turn untimely SV mitigation. Hence, there is an apparent need for automated techniques that can perform SV assessment without using manually-defined rules.

In the last decade, data-driven techniques have emerged as a promising alternative to static analysis counterparts for SV assessment, as indicated in our extensive review~\cite{le2021survey} (to be discussed in-depth in Chapter~\ref{chapter:lit_review}).
The emergence of data-driven SV assessment is mainly because of the rapid increase in size of SV data in the wild (e.g., more than 170,000 SVs were reported on National Vulnerability Database (NVD)~\cite{nvd} from 2002 to 2021~\cite{vuln_stat}).
These approaches are underpinned by Machine Learning (ML), Deep Learning (DL) and Natural Language Processing (NLP) models that are capable of automatically extracting complex patterns and rules from large-scale SV data, reducing the reliance on experts' knowledge. Overall, these data-driven models have opened up new opportunities for the field of automated SV assessment.

\section{Problem Statement and Research Objectives} \label{sec:thesis_problem_statement}

Data-driven models have many promising applications for SV assessment; however, according to our review on data-driven SV assessment~\cite{le2021survey}, there are still several unaddressed yet important challenges/gaps that affect the practical application of this field. Specifically, this PhD thesis focuses on the three key practical challenges of data-driven SV assessment.
The first challenge is the missing treatment for changing SV data over time that leads to degraded robustness and performance of assessment models.
The second challenge is the lack of using rich and relevant knowledge of (vulnerable) source code that can result in untimely SV assessment.
The third challenge is the negligence of incorporating developers' needs into data-driven models that limits customized SV assessment.
These three challenges are captured in the problem statement of this thesis, which is stated as follows.

\begin{tcolorbox}
\textbf{Problem statement}:
Practical applicability of SV assessment using data-driven approaches is negatively affected by the changing data of SVs along with the missing considerations/utilization of source code and developers' real-world needs.
It is important to investigate, understand, and address these challenges to make data-driven SV assessment more timely and applicable in practice.
\end{tcolorbox}

\textbf{\textit{The objectives of this thesis are to present and evaluate solutions to address the lack of (1) treatment for changing SV data, (2) usage of SV-related code, and (3) consideration of developers' real-world SV concerns to improve the practicality of data-driven SV assessment}}.

\begin{figure}[t]
    \centering
    \includegraphics[width=\columnwidth,keepaspectratio]{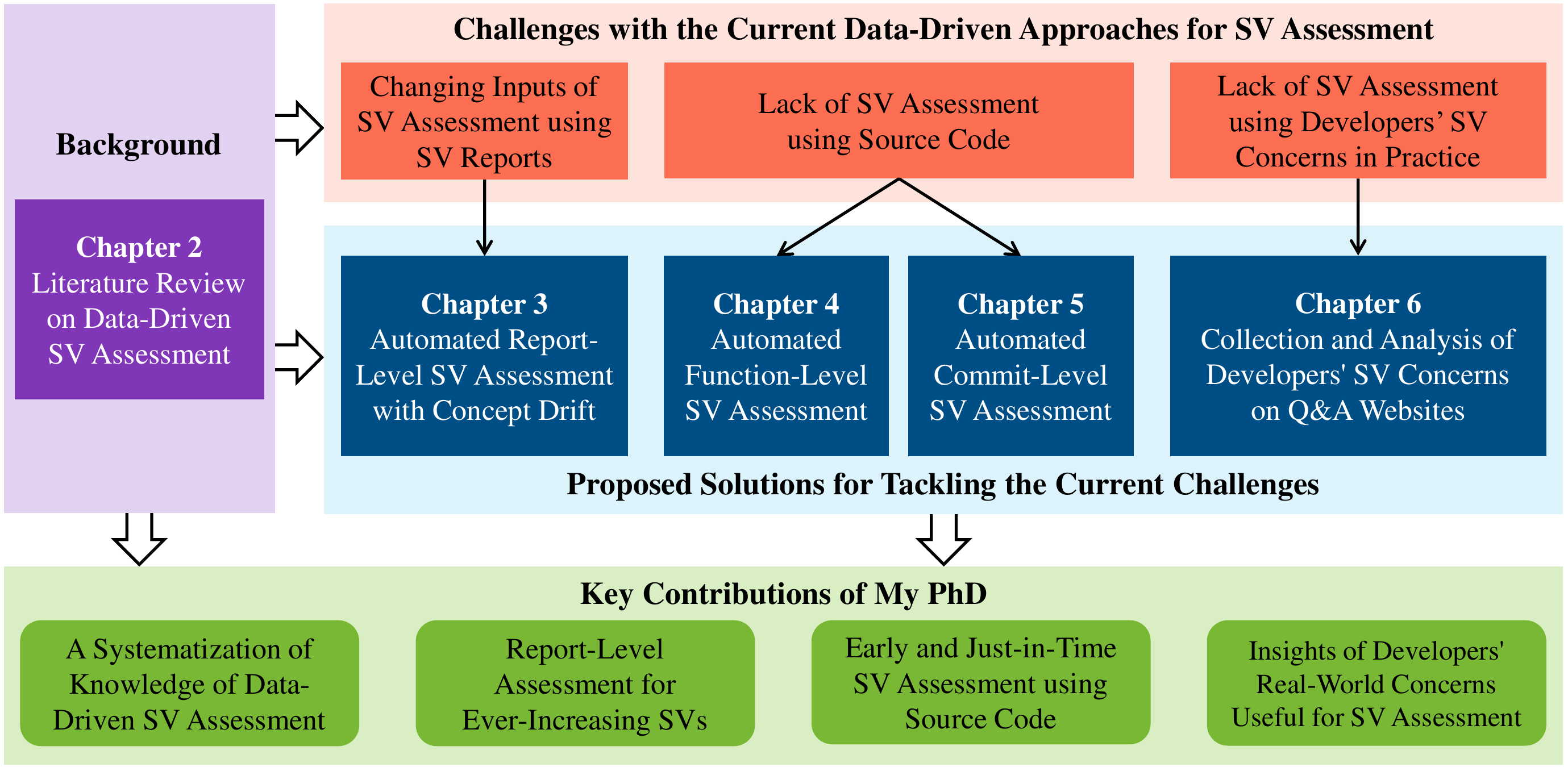}
    \caption[Overview of the thesis.]{Overview of the thesis. \textbf{Note}: SV stands for Software Vulnerability.}
    \label{fig:thesis_overview}
\end{figure}

\section{Thesis Overview and Contributions}
\label{sec:thesis_overview}

A brief overview of the thesis is given in \fig~\ref{fig:thesis_overview}. The research objectives have been realized in six chapters, which are described hereafter. While I (the author of this thesis) am mainly responsible for research activities done in this thesis, most of the work was conducted in collaboration with other researchers. Thus, the pronoun ``\textbf{we}'' was used in this thesis to reflect the collaborative research efforts.

\bigskip

\noindent \textbf{Chapter 2: Literature Review on Data-Driven SV Assessment}

Chapter~\ref{chapter:lit_review} (the purple box in \fig~\ref{fig:thesis_overview}) provides a background on the topic of data-driven SV assessment for this thesis. We conduct a literature review and use thematic analysis~\mbox{\cite{cruzes2011research}} to devise a taxonomy of the key SV assessment tasks that have been automated using data-driven approaches. For each theme of tasks, we identify the key practices that have been followed by the primary studies and highlighted the challenges with the current approaches to pave the way for future research. Particularly, we pinpoint three main challenges (the red boxes in \fig~\ref{fig:thesis_overview}) that potentially hinder the practical application of current data-driven approaches for SV assessment. These challenges lead to the four proposed solutions (the blue boxes in \fig~\ref{fig:thesis_overview}) in this thesis described in Chapters~\ref{chap:msr19},~\ref{chap:msr22},~\ref{chap:ase21}, and~\ref{chap:ease21} below.

\bigskip

\noindent \textbf{Chapter 3: Automated Report-Level SV Assessment with Concept Drift}

Chapter~\ref{chap:msr19} addresses the first challenge of changing data in the context of report-level SV assessment.
Our review in Chapter~\ref{chapter:lit_review} has pointed out that data-driven SV assessment tasks have been most commonly carried out using the information provided in SV reports. Each of these reports contains an expert-verified summary of an SV, e.g., a brief description of the type, the associated attack vectors and potential impacts if exploited.
However, we observe that the content of these reports continuously changes in practice as experts usually need to use different/new terms to describe newly introduced SVs. Such issue is referred to as \textit{concept drift}~\cite{gama2014survey} that can degrade the performance of SV assessment models over time. However, most of the existing report-level SV assessment models have not accounted for this concept drift issue, potentially affecting their performance when deployed in the wild.
Using more than 100,000 SV reports from NVD, Chapter~\ref{chap:msr19} performs a large-scale investigation of the prevalence and impacts of the concept drift issue on report-level SV assessment models. Moreover, we present a novel SV assessment model that combines characters and words extracted from SV descriptions to better capture the semantics of new terms, increasing the model robustness against concept drift.

\bigskip

\noindent \textbf{Chapter 4: Automated Function-Level SV Assessment}

Chapter~\ref{chap:msr22} addresses the second challenge, i.e., the need for SV assessment using source code.
Real-world SVs are usually rooted in source code, but Chapter~\ref{chapter:lit_review} has found that the current data-driven SV assessment efforts have been mainly done on the report level. While report-level models have shown promising performance for various SV assessment tasks~\cite{spanos2017assessment,han2017learning,spanos2018multi,le2019automated}, these models are still heavily dependent on SV reports that require expertise and manual effort to generate. Our analysis in Chapter~\ref{chap:msr22} has also found that most (97\%) of the reports used for SV assessment in the previous studies were not available at the fixing time of respective SVs. Such unavailability of SV reports affects the timeliness of report-level SV assessment in practice.
A promising alternative is to directly utilize vulnerable code available for SV fixing to enable earlier SV assessment.
Thus, in Chapter~\ref{chap:msr22}, we propose to use (vulnerable) code functions instead of SV reports for SV assessment. Using 1,782 vulnerable functions curated from 200 open-source projects, we particularly investigate the use of different code parts (i.e., (non-)vulnerable statements) in these functions for building effective function-level SV assessment models. To the best of our knowledge, we are the first to distill practices of performing data-driven SV assessment using source code. Such an approach can enable earlier fixing prioritization of SVs than the report-level SV assessment counterpart.

\bigskip

\noindent \textbf{Chapter 5: Automated Commit-Level SV Assessment}

Chapter~\ref{chap:ase21} extends the work in Chapter~\ref{chap:msr22} by addressing another practical scenario of code-based SV assessment.
In real-world software development, developers have increasingly adopted DevOps for continuous integration, in which incremental changes are made to codebases via code commits to implement new features or fix bugs/SVs~\cite{rajapakse2022challenges}. Meneely et al.~\cite{meneely2013patch} showed that such code commits/changes can introduce SVs. However, it is wastage of resources to use function-level SV assessment for these cases because functions are often not entirely changed/added in code commits~\cite{kamei2012large}. Instead, performing SV assessment directly on these changes enables \textit{just-in-time}, i.e., as soon as SVs are introduced, provision of SV characteristics for SV fixing. To the best of our knowledge, just-in-time SV assessment using code commits has never been explored.
Therefore, in Chapter~\ref{chap:ase21}, we propose DeepCVA, the first model that automates commit-level SV assessment. This model leverages the multi-task DL paradigm~\mbox{\cite{zhang2017survey}} to automate various SV assessment tasks simultaneously in a unified model. We evaluate the effectiveness and efficiency of DeepCVA on 1,229 vulnerability-contributing commits in 246 open-source projects. DeepCVA is expected to increase the efficiency in model (re-)training and maintenance for continuous integration using DevOps in practice compared to conventional task-wise models.

\bigskip

\noindent \textbf{Chapter 6: Collection and Analysis of Developers' SV Concerns on Question and Answer Websites}

Chapter~\ref{chap:ease21} tackles the third challenge of lacking considerations of developers' real-world SV concerns for SV assessment.
Most of the current data-driven models, including the ones in Chapter~\ref{chap:msr19},~\ref{chap:msr22}, and~\ref{chap:ase21}, have automated the SV assessment tasks that are based on the expert-defined SV taxonomies/standards such as Common Weakness Enumeration (CWE)~\cite{cwe} or Common Vulnerability Scoring System (CVSS)~\cite{cvss}.
While these taxonomies are designed to be as general as possible, they may not well represent the SV-related concerns that developers regularly have. For instance, developers often encounter only a small subset of SVs/SV types rather than all available ones; these SVs should be given a higher priority during SV assessment as they are of more use/interest to developers.
Chapter~\ref{chap:ease21} conducts the first empirical study on developers' real-world SV concerns using more than 70,000 SV-related posts curated from Question and Answer (Q\&A) websites.
We use the Latent Dirichlet Allocation topic modeling technique~\cite{blei2003latent} to identify the commonly encountered issues. We then characterize these posts in terms of their popularity, difficulty, provided expertise, and available solutions. We also provide implications on leveraging such characteristics for making (data-driven) SV assessment more practical.

\bigskip

The key \textbf{contributions} of this thesis (the green boxes in \fig~\ref{fig:thesis_overview}) from the six aforementioned chapters are summarized as follows.

\begin{contribution}

\item \textbf{A systematization of knowledge of data-driven SV assessment (Chapter~\ref{chapter:lit_review})}: (\textit{i}) A taxonomy of five SV assessment tasks. (\textit{ii}) Detailed analysis of the pros and cons of frequent data sources, features, prediction models, evaluation techniques and evaluation metrics used for developing data-driven SV assessment models. (\textit{iii}) Three key challenges limiting the practical application of data-driven SV assessment.

\item \textbf{Report-level assessment for ever-increasing SVs (Chapter~\ref{chap:msr19})}. (\textit{i}) Impact analysis of changing data (concept drift) of SV reports on the development and evaluation of SV assessment models. (\textit{ii}) Concept-drift-aware models for automating report-level SV assessment.

\item \textbf{Early and just-in-time SV assessment using source code (Chapters~\ref{chap:msr22} and~\ref{chap:ase21})}. (\textit{i}) Practices of developing effective data-driven models using vulnerable code statements and context in functions for early SV assessment without delays caused by missing SV reports. (\textit{ii}) Just-in-time and efficient SV assessment using code commits (where SVs are first added) with deep multi-task learning.

\item \textbf{Insights of developers' real-world concerns that are useful for SV assessment (Chapter~\ref{chap:ease21})}. (\textit{i}) A taxonomy of 13 key SV concerns commonly encountered by developers in practice. (\textit{ii}) Analysis of popularity, difficulty, expertise level, solutions provided for these real-world concerns. (\textit{iii}) Implications of these concerns and their characteristics for practical data-driven SV assessment.

\end{contribution}

\section{Related Publications}
\label{sec:thesis_contributions}

All of the core chapters and contributions of this thesis have been published during my PhD candidature. The list of directly related publications with respect to each chapter is given below.

\bigskip

\noindent \circled{1} \textbf{Triet Huynh Minh Le}, Huaming Chen, and Muhammad Ali Babar, ``A Survey on Data-Driven Software Vulnerability Assessment and Prioritization,'' ACM Computing Surveys (CSUR), 2021. [CORE ranking: \textbf{rank A*}, Impact factor (2020): 10.282, SJR rating: Q1] (Chapter~\ref{chapter:lit_review})

\bigskip

\noindent \circled{2} \textbf{Triet Huynh Minh Le} and Muhammad Ali Babar, ``On the Use of Fine-Grained Vulnerable Code Statements for Software Vulnerability Assessment Models,'' in Proceedings of the 19\textsuperscript{th} International Conference on Mining Software Repositories (MSR). ACM, 2022. [CORE ranking: \textbf{rank A}, Acceptance rate: 34\%] (Chapter~\ref{chap:msr19})

\bigskip

\noindent \circled{3} \textbf{Triet Huynh Minh Le}, Bushra Sabir, and Muhammad Ali Babar, ``Automated Software Vulnerability Assessment with Concept Drift,'' in Proceedings of the 16\textsuperscript{th} International Conference on Mining Software Repositories (MSR). IEEE, 2019, pp. 371–382. [CORE ranking: \textbf{rank A}, Acceptance rate: 25\%] (Chapter~\ref{chap:msr22})

\bigskip

\noindent \circled{4} \textbf{Triet Huynh Minh Le}, David Hin, Roland Croft, and Muhammad Ali Babar, ``DeepCVA: Automated Commit-Level Vulnerability Assessment with Deep Multi-Task Learning,'' in Proceedings of the 36\textsuperscript{th} IEEE/ACM International Conference on Automated Software Engineering (ASE). IEEE, 2021, pp. 717–729. [CORE ranking: \textbf{rank A*}, Acceptance rate: 19\%] (Chapter~\ref{chap:ase21})

\bigskip

\noindent \circled{5} \textbf{Triet Huynh Minh Le}, David Hin, Roland Croft, and Muhammad Ali Babar, ``PUMiner: Mining Security Posts from Developer Question and Answer Websites with PU Learning,'' in Proceedings of the 17\textsuperscript{th} International Conference on Mining Software Repositories (MSR). ACM, 2020, pp. 350–361. [CORE ranking: \textbf{rank A}, Acceptance rate: 25.7\%] (Chapter~\ref{chap:ease21})

\bigskip

\noindent \circled{6} \textbf{Triet Huynh Minh Le}, Roland Croft, David Hin, and Muhammad Ali Babar, ``A Large-Scale Study of Security Vulnerability Support on Developer Q\&A Websites,'' in Proceedings of the 25\textsuperscript{th} Evaluation and Assessment in Software Engineering (EASE). ACM, 2021, pp. 109–118. [CORE ranking: \textbf{rank A}, Acceptance rate: 27\%, \textbf{Nominated for the Best Paper Award}] (Chapter~\ref{chap:ease21})
\bigskip

In addition to the six aforementioned publications, I contributed as the first author or coauthor of the following two publications during my PhD candidature, which are not directly related to the materials in this thesis.

\bigskip

\noindent \circled{7} \textbf{Triet Huynh Minh Le}, Hao Chen, and Muhammad Ali Babar, ``Deep Learning for Source Code Modeling and Generation: Models, Applications, and Challenges,'' ACM Computing Surveys (CSUR), vol. 53, no. 3, pp. 1–38, 2020. [CORE ranking: \textbf{rank A*}, Impact factor (2020): 10.282, SJR rating: Q1, \textbf{High-impact research work} selected by Faculty of Engineering, Computer \& Mathematical Sciences at the University of Adelaide.]

\bigskip

\noindent \circled{8} Xuanyu Duan, Mengmeng Ge, \textbf{Triet Huynh Minh Le}, Faheem Ullah, Shang Gao, Xuequan Lu, and Muhammad Ali Babar, ``Automated Security Assessment for the Internet of Things,'' in 2021 IEEE 26th Pacific Rim International Symposium on Dependable Computing (PRDC). IEEE, 2021, pp. 47–56. [CORE ranking: \textbf{rank B}]

\section{Thesis Organization}
\label{sec:thesis_organization}

The remainder of the thesis is organized as follows. Chapter~\ref{chapter:lit_review} reports a literature review on the tasks, practices and challenges of data-driven SV assessment.
Chapter~\ref{chap:msr19} investigates the impacts of the concept drift issue on report-level SV assessment models and proposes an automated technique to address this issue.
Chapter~\ref{chap:msr22} explores the practices of building function-level SV assessment models. Chapter~\ref{chap:ase21} describes DeepCVA, a novel multi-task DL model for automated commit-level SV assessment.
Chapter~\ref{chap:ease21} presents an empirical study of the common SV concerns encountered by developers on Q\&A websites and distills respective implications for data-driven SV assessment. Finally, Chapter~\ref{chap:conclusions_future_work} summarizes the main contributions and findings of the thesis and suggests potential research avenues for future work in the area of data-driven SV assessment.

%% file: Chapters/Chapter_2_LitReview.tex
\chapter{Literature Review on Data-Driven Software Vulnerability Assessment}
\label{chapter:lit_review}

\begin{tcolorbox}
\textbf{Related publication}: This chapter is based on our paper titled ``\textit{A Survey on Data-Driven Software Vulnerability Assessment and Prioritization}'' published in the ACM Computing Surveys journal (CORE A*)~\cite{le2021survey}.
\end{tcolorbox}
\bigskip

As mentioned in Chapter~\ref{chap:thesis_introduction}, Software Vulnerabilities (SVs) are increasing in complexity and scale, posing great security risks to many software systems. Given the limited resources in practice, SV assessment\footnote{In the original review paper~\cite{le2021survey}, we used the term \textit{SV assessment and prioritization} instead of \textit{SV assessment}. However, in Chapter 2, the term \textit{SV assessment} is used to ensure consistency with the other parts of the thesis, and it is important to note that \textit{SV assessment} and \textit{SV assessment and prioritization} can be used interchangeably as most SV assessment tasks can be used for prioritizing SV fixing.} help practitioners devise optimal SV mitigation plans based on various SV characteristics. The recent surges in SV data sources and data-driven techniques such as Machine Learning and Deep Learning have taken SV assessment to the next level. Chapter~\ref{chapter:lit_review} provides a taxonomy of the key tasks performed by the past research efforts in the area. We also highlight the best practices, in terms of data sources, features, prediction models, evaluation techniques and evaluation metrics, for data-driven SV assessment. At the end of the review, we discuss some of the current and important challenges in the field that set the stage for the key contributions of this thesis in Chapters~\ref{chap:msr19},~\ref{chap:msr22},~\ref{chap:ase21} and~\ref{chap:ease21}.

\newpage

\lstset{language=Java,keywordstyle={\bfseries}}

\section{Introduction}\label{sec:introduction_csur22}

As discussed in Chapter~\ref{chap:thesis_introduction}, Software Vulnerability (SV) assessment is required to prioritize the remediation of critical SVs (i.e., the ones that can lead to devastating cyber-attacks)~\cite{smyth2017software}.
SV assessment includes tasks that determine various characteristics such as the types, exploitability, impact and severity levels of SVs~\cite{kamongi2013vulcan}.
Such characteristics help understand and select high-priority SVs to resolve early given the limited effort and resources.
For example, SVs with simple exploitation and severe impacts likely require high fixing priority.

There has been an active research area to assess and prioritize SVs using increasingly large data from multiple sources.
Many studies in this area have proposed different Natural Language Processing (NLP), Machine Learning (ML) and Deep Learning (DL) techniques to leverage such data to automate various tasks such as predicting the Common Vulnerability Scoring System~\cite{cvss} (CVSS) metrics (e.g.,~\cite{le2019automated,spanos2018multi,han2017learning}) or public exploits (e.g.,~\cite{bozorgi2010beyond,sabottke2015vulnerability,bullough2017predicting}).
These prediction models can learn the patterns automatically from vast SV data, which would be otherwise impossible to do manually. Such patterns are utilized to speed up the assessment processes of ever-increasing and more complex SVs, significantly reducing practitioners' effort.
Despite the rising research interest in data-driven SV assessment, to the best of our knowledge, there has been no comprehensive review on the state-of-the-art methods and existing challenges in this area. To bridge this gap, we are the first to review in-depth the research studies that automate \textit{data-driven SV assessment} tasks leveraging SV data and NLP/ML/DL techniques.

The \textbf{contributions} of our review are summarized as follows:

\begin{contribution}
\item We categorize and describe the key tasks of data-driven SV assessment performed in relevant primary studies.
\item We synthesize and discuss the pros and cons of data, features, models, evaluation methods and metrics commonly used in the reviewed studies.
\item We highlight some key challenges with the current practices.
\end{contribution}

\noindent We believe that our findings can provide useful guidelines for researchers and practitioners to effectively utilize data to perform SV assessment.

\noindent \textbf{Related Work}.
There have been several existing surveys/reviews on SV analysis and prediction, but they are fundamentally different from ours (see Table~\ref{tab:survey_comparison_csur22}).
Ghaffarian et al.~\cite{ghaffarian2017software} conducted a seminal survey on ML-based SV analysis and discovery.
Subsequently, several studies~\cite{zeng2020software,singh2020applying,semasaba2020literature,lin2020software} reviewed DL techniques for detecting vulnerable code.
However, these prior reviews did not describe how ML/DL techniques can be used to assess and prioritize the detected SVs.
There have been other relevant reviews on using Open Source Intelligence (OSINT) (e.g., phishing or malicious emails/URLs/IPs) to make informed security decisions~\cite{pastor2020not,evangelista2020systematic,sun2018data}. However, these OSINT reviews did not explicitly discuss the use of SV data and how such data can be leveraged to automate the assessment processes. Moreover, most of the reviews on SV assessment have focused on either static analysis tools~\cite{kritikos2019survey} or rule-based approaches (e.g., expert systems or ontologies)~\cite{khan2018review}.
These methods rely on pre-defined patterns and struggle to work with new types and different data sources of SVs compared to contemporary ML or DL approaches presented in this chapter~\cite{bell1985expert,han2011data,goodfellow2016deep}.
Recently, Dissanayake et al.~\cite{dissanayake2020software} reviewed the socio-technical challenges and solutions for security patch management that involves SV assessment after SV patches are identified.
Unlike~\cite{dissanayake2020software}, we focus on the challenges, solutions and practices of automating various SV assessment tasks with data-driven techniques. We also consider all types of SV assessment regardless of the patch availability.

\begin{table}[!t]
\fontsize{7}{8}\selectfont
\caption{Comparison of contributions between our review and the existing related surveys/reviews.}

\label{tab:survey_comparison_csur22}
\centering
\begin{tabular}{|l|P{3.2cm}|P{2.9cm}|P{3.2cm}|}
\hline
\multicolumn{1}{|l|}{\diagbox[height=1.4cm, width=3.8cm]{\raisebox{2\height}{\enspace \textbf{Study}}}{\raisebox{-1\height}{\enspace \textbf{Contribution}}}} &
\multicolumn{1}{P{3.2cm}|}{\centering\textbf{Focus on SV assessment}} & \multicolumn{1}{P{2.9cm}|}{\centering\textbf{Analysis of SV\\data sources}} &
\multicolumn{1}{P{3.2cm}|}{\centering\textbf{Analysis of data-\\driven approaches}}
\\
\hline
\makecell[l]{Ghaffarian et al. 2017~\cite{ghaffarian2017software}} & \multicolumn{1}{c|}{--} & \multicolumn{1}{c|}{--} & \multicolumn{1}{c|}{\checkmark (Mostly ML)} \\
\hline
\makecell[l]{Lin et al. 2020~\cite{lin2020software}\\ Semasaba et al. 2020~\cite{semasaba2020literature}\\ Singh et al. 2020~\cite{singh2020applying}\\ Zeng et al. 2020~\cite{zeng2020software}} & \multicolumn{1}{c|}{--} & \multicolumn{1}{c|}{--} & \multicolumn{1}{c|}{\checkmark (Mostly DL)} \\
\hline
\multicolumn{1}{|p{2.8cm}|}{Pastor et al. 2020~\cite{pastor2020not}} & \multicolumn{1}{c|}{--} & \multicolumn{1}{c|}{\checkmark (OSINT)} & \multicolumn{1}{c|}{--} \\
\hline
\makecell[l]{Sun et al. 2018~\cite{sun2018data}\\ Evangelista et al. 2020~\cite{evangelista2020systematic}} & \multicolumn{1}{c|}{--} & \multicolumn{1}{c|}{\checkmark (OSINT)} & \multicolumn{1}{c|}{\checkmark} \\
\hline
\makecell[l]{Khan et al. 2018~\cite{khan2018review}} & \multicolumn{1}{c|}{\checkmark (Rule-based methods)} & \multicolumn{1}{c|}{--} & \multicolumn{1}{c|}{--} \\
\hline
\multicolumn{1}{|p{2.8cm}|}{Kritikos et al. 2019~\cite{kritikos2019survey}} & \multicolumn{1}{c|}{\checkmark (Static analysis)} & \multicolumn{1}{c|}{\checkmark} & \multicolumn{1}{c|}{--} \\
\hline
\multicolumn{1}{|p{3.2cm}|}{Dissanayake et al. 2020~\cite{dissanayake2020software}} & \multicolumn{1}{c|}{\checkmark (Socio-technical aspects)} & \multicolumn{1}{c|}{--} & \multicolumn{1}{c|}{--} \\
\hline\hline
\rowcolor{lightgray}
\multicolumn{1}{|l|}{\textbf{Our review}} & \multicolumn{1}{c|}{\textbf{\checkmark}} & \multicolumn{1}{c|}{\textbf{\checkmark}} & \multicolumn{1}{c|}{\textbf{\checkmark}} \\
\hline
\end{tabular}
\end{table}

\noindent \textbf{Chapter Organization}. The remainder of the chapter is organized as follows. Section~\ref{sec:survey_overview} presents the scope, methodology and taxonomy covered in this chapter. Sections~\ref{sec:exploit_prediction},~\ref{sec:impact_prediction},~\ref{sec:severity_prediction},~\ref{sec:type_prediction} and~\ref{sec:miscellaneous} review the studies in each theme of the taxonomy and discuss the limitations/gaps and open opportunities at the end of each theme.
Section~\ref{sec:elements_analysis} identifies and discusses the common practices and respective implications for data-driven SV assessment.
Finally, section~\ref{sec:conclusions_chap2} concludes the review and discusses the open challenges of this research area.

\section{Overview of the Literature Review}
\label{sec:survey_overview}

\subsection{Scope}\label{subsec:scope_csur22}

This review's focus is on \textit{data-driven SV assessment}.
Unlike the existing surveys on rule-based or experience-based SV assessment~\cite{kritikos2019survey,khan2018review,dissanayake2020software} that hardly utilize the potential of SV data in the wild, this chapter aims to review research papers that have leveraged such data to automate tasks in this area using data-driven models.
To keep our focus, we do not consider papers that only perform manual analyses or descriptive statistics (e.g., taking mean/median/variation of data) without using any data-driven models as these techniques cannot automatically assess or prioritize new SVs.
We also do not directly compare the absolute performance of all the related studies as they did not use exactly the same experimental setup (e.g., data sources and model configurations). While it is theoretically possible to perform a comparative evaluation of the identified techniques by establishing and using a common setup, this type of evaluation is out of the scope of this chapter.
However, we still cover the key directions/techniques of the studies in sections~\mbox{\ref{sec:exploit_prediction}},~\mbox{\ref{sec:impact_prediction}},~\mbox{\ref{sec:severity_prediction}},~\mbox{\ref{sec:type_prediction}} and~\mbox{\ref{sec:miscellaneous}}.
We also provide in-depth discussion on the common practices of these studies in section~\mbox{\ref{sec:elements_analysis}} and identify some current challenges with the field in section~\ref{sec:conclusions_chap2}.

\subsection{Methodology}
\label{subsec:methodology_csur22}
\noindent \textbf{Study selection}. Our study selection was inspired by the Systematic Literature Review guidelines~\mbox{\cite{keele2007guidelines}}.
We first designed the search string: ``\textit{`software' AND vulner* AND (learn* OR data* OR predict*) AND (priority* OR assess* OR impact* OR exploit* OR severity*) AND NOT (fuzz* OR dynamic* OR intrusion OR adversari* OR malware* OR `vulnerability detection' OR `vulnerability discovery' OR `vulnerability identification' OR `vulnerability prediction')}''. This search string covered the key papers (i.e., with more than 50 citations) in the area and excluded many papers on general security and SV detection. We then adapted this search string\footnote{This search string was customized for each database and the database-wise search strings can be found at \url{https://figshare.com/s/da4d238ecdf9123dc0b8}.} to retrieve an initial list of 1,765 papers up to April 2021\footnote{Given the ever-growing nature of this field, we maintain an up-to-date list of papers and resources on data-driven SV assessment at \url{https://github.com/lhmtriet/awesome-vulnerability-assessment}. More details are also given in Appendix~\ref{sec:appendix_lit_review}.} from various commonly used databases such as IEEE Xplore, ACM Digital Library, Scopus, SpringerLink and Wiley. We also defined the inclusion/exclusion criteria (see~\tab~\ref{tab:inclusion_exclusion_csur22}) to filter out irrelevant/low-quality studies with respect to our scope in section~\mbox{\ref{subsec:scope_csur22}}. Based on these criteria and the titles and abstracts and keywords of 1,765 initial papers, we removed 1,550 papers. After reading the full-text and applying the criteria on the remaining 215 papers, we obtained 70 papers directly related to data-driven SV assessment. To further increase the coverage of studies, we performed backward and forward snowballing~\mbox{\cite{wohlin2014guidelines}} on these 70 papers (using the above sources and Google Scholar) and identified 14 more papers that satisfied the inclusion/exclusion criteria. In total, we included 84 studies in our review. We do not claim that we have collected all the papers in this area, but we believe that our selection covered most of the key studies to unveil the practices of data-driven SV assessment.

\begin{table}[t]
    \centering
    \caption[Inclusion and exclusion criteria for study selection.]{Inclusion and exclusion criteria for study selection. \textbf{Notes}: We did not limit the selection to only peer-reviewed papers as this is an emerging field with many (high-quality) papers on Arxiv and most of them are under-submission at the time of writing.
    However, to ensure the quality of the papers on Arxiv, we only selected the ones with at least one citation.}
    \begin{tabular}{l}
        \hline
         \textbf{Inclusion criteria}\\
         \hline
         \makecell[l]{\tabitem \textit{I1}. Studies that focused on SVs rather than hardware or human vulnerabilities\\
         \tabitem \textit{I2}. Studies that focused on assessment task(s) of SVs\\
         \tabitem \textit{I3}. Studies that used data-driven approaches (e.g., ML/DL/NLP techniques) for SV\\assessment}\\
         \hline
         \textbf{Exclusion criteria}\\
         \hline
         \makecell[l]{\tabitem \textit{E1}. Studies that were not written in English\\
         \tabitem \textit{E2}. Studies that we could not retrieve their full texts\\
         \tabitem \textit{E3}. Studies that were not related to Computer Science\\
         \tabitem \textit{E4}. Studies that were literature review or survey\\
         \tabitem \textit{E5}. Studies that only performed statistical analysis of SV assessment metrics\\
         \tabitem \textit{E6}. Studies that only focused on (automated) collection of SV data \\}\\
         \hline
         
    \end{tabular}
    \label{tab:inclusion_exclusion_csur22}
\end{table}

\noindent \textbf{Data extraction and synthesis of the selected studies}. We followed the steps of thematic analysis~\mbox{\cite{cruzes2011research}} to identify the taxonomy of data-driven SV assessment tasks in sections~\mbox{\ref{sec:exploit_prediction}},~\mbox{\ref{sec:impact_prediction}},~\mbox{\ref{sec:severity_prediction}},~\mbox{\ref{sec:type_prediction}} and~\mbox{\ref{sec:miscellaneous}} as well as the key practices of data-driven model building for automating these tasks in section~\mbox{\ref{sec:elements_analysis}}. We first conducted a pilot study of 20 papers to familiarize ourselves with data to be extracted from the primary studies. After that, we generated initial codes and then merged them iteratively in several rounds to create themes. Two of the authors performed the analysis independently, in which each author analyzed half of the selected papers and then reviewed the analysis output of the other author. Any disagreements were resolved through discussions.

\begin{figure*}[t]
  \centering
  \includegraphics[width=0.99\linewidth,keepaspectratio]{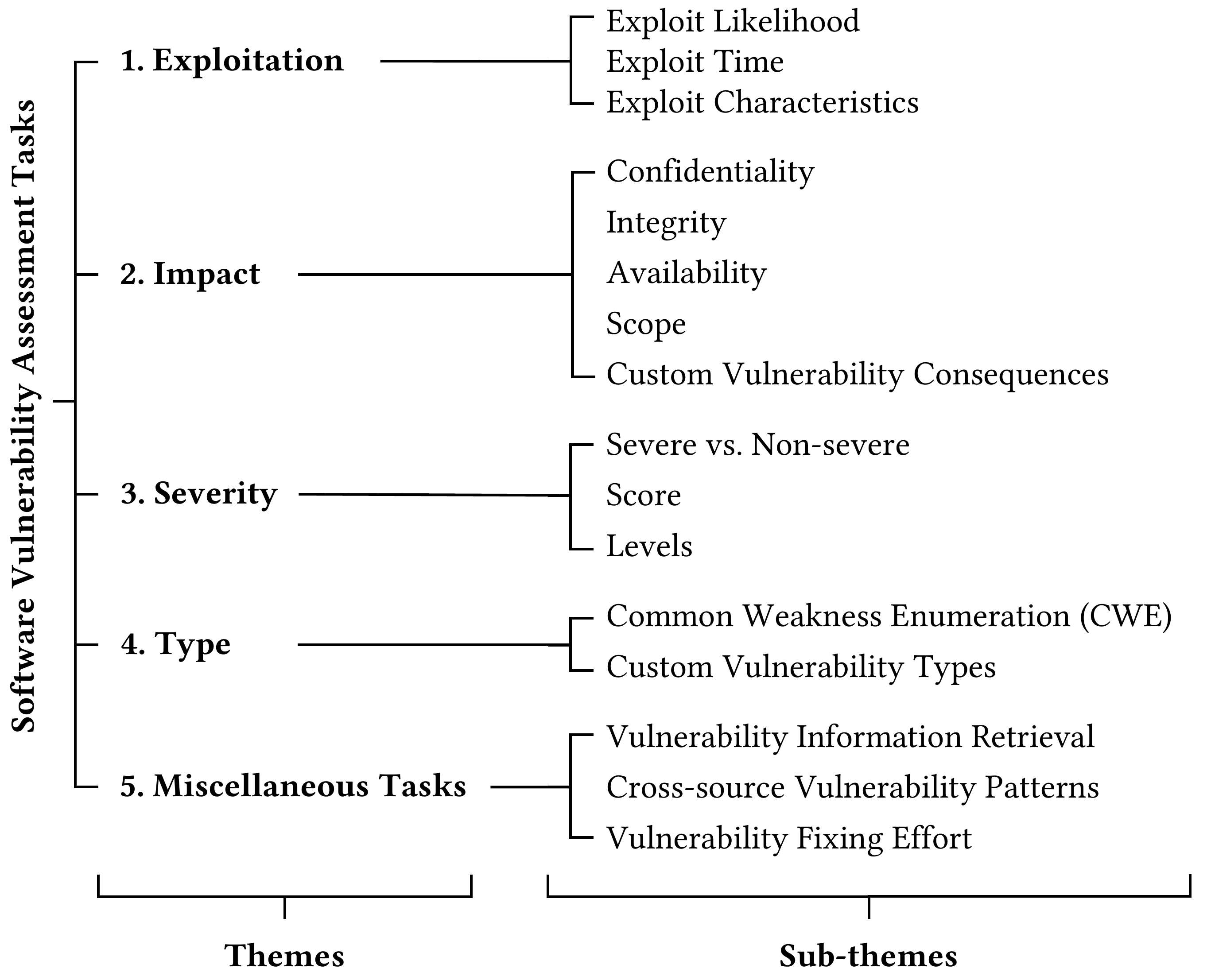}

  \caption[Taxonomy of studies on data-driven software vulnerability assessment.]{Taxonomy of studies on data-driven SV assessment.}
  \label{fig:taxonomy_csur22}
\end{figure*}

\subsection{Taxonomy of Data-Driven Software Vulnerability Assessment}\label{subsec:taxonomy_csur22}

Based on the scope in section~\ref{subsec:scope_csur22} and the methodology in section~\ref{subsec:methodology_csur22}, we identified five main themes of the relevant studies in the area of data-driven SV assessment (see Figure~\ref{fig:taxonomy_csur22}). Specifically, we extracted the themes by grouping related SV assessment tasks that the reviewed studies aim to automate/predict using data-driven models. Note that a paper is categorized into more than one theme if that paper develops models for multiple cross-theme tasks.

We acknowledge that there can be other ways to categorize the studies. However, we assert the reliability of our taxonomy as all of our themes (except theme 5) align with the security standards used in practice. For example, Common Vulnerability Scoring System (CVSS)~\cite{cvss} provides a framework to characterize exploitability, impact and severity of SVs (themes 1-3), while Common Weakness Enumeration (CWE)~\cite{cwe} includes many vulnerability types (theme 4). Hence, we believe our taxonomy can help identify and bridge the knowledge gap between the academic literature and industrial practices, making it relevant and potentially beneficial for both researchers and practitioners. Details of each theme in our taxonomy are covered in subsequent sections.

\begin{table}
\fontsize{8}{9}\selectfont
\centering
\caption[List of the reviewed papers in the \textit{Exploit Likelihood} sub-theme of the \textit{Exploitation} theme.]{List of the reviewed papers in the \textit{Exploit Likelihood} sub-theme of the \textit{Exploitation} theme.
\textbf{Note}: The nature of task of this sub-theme is binary classification of existence/possibility of proof-of-concept and/or real-world exploits.
}
\label{tab:exploit_studies_likelihood}

\begin{tabular}[!t]{|p{1.5cm}|p{6.2cm}|p{5cm}|}

  \hline \textbf{Study} & \textbf{Data source} & \textbf{Data-driven technique}\\\hline

  Bozorgi et al. 2010~\cite{bozorgi2010beyond} & CVE, Open Source Vulnerability Database (OSVDB) & Linear Support Vector Machine (SVM)\\\hline
  Sabottke et al. 2015~\cite{sabottke2015vulnerability} & NVD, Twitter, OSVDB, ExploitDB, Symantec security advisories, private Microsoft security advisories & Linear SVM\\\hline
  Edkrantz et al. 2015~\cite{edkrantz2015predictingthesis,edkrantz2015predicting} & NVD, Recorded Future security advisories, ExploitDB & Na\"ive Bayes, Linear SVM, Random forest\\\hline
  Bullough et al. 2017~\cite{bullough2017predicting} & NVD, Twitter, ExploitDB & Linear SVM\\\hline
      
  Almukaynizi et al.~\cite{almukaynizi2017proactive,almukaynizi2019patch} & NVD, ExploitDB, Zero Day Initiative security advisories \& Darkweb forums/markets & SVM, Random forest, Na\"ive Bayes, Bayesian network, Decision tree, Logistic regression\\\hline
  Xiao et al. 2018~\cite{xiao2018patching} & NVD, SecurityFocus security advisories, Symantec \newline Spam/malicious activities based on daily blacklists from abuseat.org, spamhaus.org, spamcop.net, uceprotect.net, wpbl.info \& list of unpatched SVs in hosts & Identification of malicious activity groups with community detection algorithms + Random forest for exploit prediction\\\hline
  Tavabi et al. 2018~\cite{tavabi2018darkembed} & NVD, 200 sites on Darkweb, ExploitDB, Symantec, Metasploit & Paragraph embedding + Radial basis function kernel SVM\\\hline
  de Sousa et al. 2020~\cite{de2020evaluating} & NVD, Twitter, ExploitDB, Symantec\newline Avast, ESET, Trend Micro security advisories & Linear SVM, Logistic regression, XGBoost, Light Gradient Boosting Machine (LGBM)\\\hline
  Fang et al. 2020~\cite{fang2020fastembed} & NVD, ExploitDB, SecurityFocus, Symantec & fastText + LGBM\\\hline
  Huang et al. 2020~\cite{huang2020poster} & NVD, CVE Details, Twitter, ExploitDB, Symantec security advisories & Random forest\\\hline
  Jacobs et al. 2020~\cite{jacobs2020improving} & NVD, Kenna Security \newline Exploit sources: Exploit DB, Metasploit, FortiGuard Labs, SANS Internet Storm Center, Securewords CTU, Alienvault OSSIM, Canvas/D2 Security's Elliot Exploitation Frameworks, Contagio, Reversing Labs
  & XGBoost\\\hline
  Yin et al. 2020~\cite{yin2020apply} & NVD, ExploitDB, General text: Book Corpus \& Wikipedia for pretraining BERT models & Fine-tuning BERT models pretrained on general text\\\hline
  Bhatt et al. 2021~\cite{bhatt2021exploitability} & NVD, ExploitDB & Features augmented by SV types + Decision tree, Random forest, Na\"ive Bayes, Logistic regression, SVM\\\hline
  Suciu et al. 2021~\cite{suciu2021expected} & NVD, Vulners database, Twitter, Symantec, SecurityFocus, IBM X-Force Threat Intelligence \newline Exploit sources: ExploitDB, Metasploit, Canvas, D2 Security's Elliot, Tenable, Skybox, AlienVault, Contagio & Multi-layer perceptron\\\hline\hline
  
  Younis et al. 2014~\cite{younis2014using} & Vulnerable functions from NVD (Apache HTTP Server project), ExploitDB, OSVDB & SVM\\\hline
  Yan et al. 2017~\cite{yan2017exploitmeter} & Executables (binary code) of 100 Linux applications & Combining ML (Decision tree) output \& fuzzing with a Bayesian network\\\hline
  Tripathi et al. 2017~\cite{tripathi2017exniffer} & Program crashes from VDiscovery~\cite{cha2012unleashing,grieco2016toward} \& LAVA~\cite{dolan2016lava} datasets & Static/Dynamic analysis features + Linear/Radial basis function kernel SVM\\\hline
  Zhang et al. 2018~\cite{zhang2018assisting} & Program crashes from VDiscovery~\cite{cha2012unleashing,grieco2016toward} dataset & $n$-grams of system calls from execution traces + Online passive-aggressive classifier\\\hline
  
\end{tabular}
\end{table}

\section{Exploitation Prediction}
\label{sec:exploit_prediction}

This section covers the \textit{Exploitation} theme that automates the detection and understanding of both Proof-of-Concept (PoC) and real-world exploits\footnote{An \textit{exploit} is a piece of code used to compromise vulnerable software~\cite{sabottke2015vulnerability}. Real-world exploits are harmful \& used in real host/network-based attacks. PoC exploits are unharmful \& used to show the potential threats of SVs in penetration tests.} targeting identified SVs.
This theme outputs the origin of SVs and how/when attackers would take advantage of such SVs to compromise a system of interest, assisting practitioners to quickly react to the more easily exploitable or already exploited SVs.
The papers in this theme can be categorized into three groups/sub-themes: (\textit{i}) Exploit likelihood, (\textit{ii}) Exploit time, (\textit{iii}) Exploit characteristics, as given in Tables~\ref{tab:exploit_studies_likelihood},~\ref{tab:exploit_studies_time} and~\ref{tab:exploit_studies_prop}, respectively.

\subsection{Summary of Primary Studies}
\subsubsection{Exploit Likelihood}
\label{subsubsec:exploit_likelihood}

The first sub-theme is \textit{exploit likelihood} that predicts whether SVs would be exploited in the wild or PoC exploits would be released publicly (see Table~\ref{tab:exploit_studies_likelihood}). In 2010, Bozorgi et al.~\cite{bozorgi2010beyond} were the first to use SV descriptions on Common Vulnerabilities and Exposures (CVE)~\cite{cve} and Open Source Vulnerability Database (OSVDB)\footnote{\url{http://osvdb.org}. Note that this database has been discontinued since 2016.} to predict exploit existence based on the labels on OSVDB. In 2015, Sabottke et al.~\cite{sabottke2015vulnerability} conducted a seminal study that used Linear SVM and SV information on Twitter to predict PoC exploits on ExploitDB~\cite{exploitdb} as well as real-world exploits on OSVDB, Symantec's attack signatures~\cite{symantec} and private Microsoft's security advisories~\cite{ms_security}. These authors urged to explicitly consider real-world exploits as \textit{not} all PoC exploits would result in exploitation in practice.
They also showed SV-related information on Twitter\footnote{\url{https://twitter.com}} can enable earlier detection of exploits than using expert-verified SV sources (e.g., NVD).

Built upon these two foundational studies~\cite{bozorgi2010beyond,sabottke2015vulnerability}, the literature has mainly aimed to improve the performance and applicability of exploit prediction models by leveraging more exploit sources and/or better data-driven techniques/practices.
Many researchers~\cite{edkrantz2015predicting,edkrantz2015predictingthesis,almukaynizi2017proactive,tavabi2018darkembed,almukaynizi2019patch,jacobs2020improving} increased the amount of ground-truth exploits using extensive sources other than ExploitDB and Symantec in~\cite{bozorgi2010beyond,sabottke2015vulnerability}. The sources were security advisories such as Zero Day Initiative~\cite{zeroday_initiative}, Metasploit~\cite{metasploit}, SecurityFocus~\cite{securityfocus}, Recorded Future~\cite{recorded_future}, Kenna Security~\cite{kenna_security}, Avast\footnote{\url{https://avast.com/exploit-protection.php}. This link was provided by de Sousa et al.~\cite{de2020evaluating}, but it is no longer available.}, ESET~\cite{eset}, Trend Micro~\cite{trend_micro}, malicious activities in hosts based on traffic of spam/malicious IP addresses~\cite{xiao2018patching} and Darkweb sites/forums/markets~\cite{nunes2016darknet}. In addition to enriching exploit sources, better data-driven models and practices for exploit prediction were also studied. Ensemble models (e.g., Random forest, eXtreme Gradient Boosting (XGBoost)~\cite{chen2016xgboost}, Light Gradient Boosting Machine (LGBM)~\cite{ke2017lightgbm}) were shown to outperform single-model baselines (e.g., Na\"ive Bayes, SVM, Logistic regression and Decision tree) for exploit prediction~\cite{fang2020fastembed,jacobs2020improving,de2020evaluating,huang2020poster}.
Additionally, Bullough et al.~\cite{bullough2017predicting} identified and addressed several issues with exploit prediction models, e.g., time sensitivity of SV data, already-exploited SVs before disclosure and training data imbalance, helping to improve the practical application of such models. Recently, Yin et al.~\cite{yin2020apply} demonstrated that transfer learning is an alternative solution for improving the performance of exploit prediction with scarcely labeled exploits. Specifically, these authors pre-trained a DL model, BERT~\cite{devlin2018bert}, on massive non-SV sources (e.g., text on Book Corpus~\cite{zhu2015aligning} and Wikipedia~\cite{wikipedia}) and then fine-tuned this pre-trained model on SV data using additional pooling and dense layers. Bhatt et al.~\cite{bhatt2021exploitability} also suggested that incorporating the types of SVs (e.g., SQL injection) into ML models can further enhance the predictive effectiveness. Suciu et al.~\cite{suciu2021expected} empirically showed that unifying SV-related sources used in prior work (e.g., SV databases~\cite{bozorgi2010beyond}, social media~\cite{sabottke2015vulnerability}, SV-related discussions~\cite{tavabi2018darkembed} and PoC code in ExploitDB~\cite{jacobs2020improving}) supports more effective and timely prediction of \textit{functional} exploits~\cite{cvss_v31}.

Besides using SV descriptions as input for exploit prediction, several studies in this sub-theme have also predicted exploits on the code level.
Younis et al.~\cite{younis2014using} predicted the exploitability of vulnerable functions in the Apache HTTP Server project. Specifically, these authors used an SVM model with features extracted from the dangerous system calls~\cite{bernaschi2002remus} in entry points/functions~\cite{manadhata2010attack} and the reachability from any of these entry points to vulnerable functions~\cite{horwitz1990interprocedural}.
Moving from high-level to binary code, Yan et al.~\cite{yan2017exploitmeter} first used a Decision tree to obtain prior beliefs about SV types in 100 Linux applications using static features (e.g., \textit{hexdump}) extracted from executables. Subsequently, they applied various fuzzing tools (i.e., Basic Fuzzing Framework~\cite{basic_fuzzing_framework} and OFuzz~\cite{ofuzz}) to detect SVs with the ML-predicted types. They finally updated the posterior beliefs about the exploitability based on the outputs of the ML model and fuzzers using a Bayesian network. The proposed method outperformed \textit{!exploitable},\footnote{\url{https://microsoft.com/security/blog/2013/06/13/exploitable-crash-analyzer-version-1-6}} a static crash analyzer provided by Microsoft.
Tripathi et al.~\cite{tripathi2017exniffer} also predicted SV exploitability from crashes (i.e., VDiscovery~\cite{cha2012unleashing,grieco2016toward} and LAVA~\cite{dolan2016lava} datasets) using an SVM model and static features from core dumps and dynamic features generated by the Last Branch Record hardware debugging utility.
Zhang et al.~\cite{zhang2018assisting} proposed two improvements to Tripathi et al.~\cite{tripathi2017exniffer}'s approach. These authors first replaced the hardware utility in~\cite{tripathi2017exniffer} that may not be available for resource-constrained devices (e.g., IoT) with sequence/$n$-grams of system calls extracted from execution traces. They also used an online passive-aggressive classifier~\cite{crammer2006online} to enable online/incremental learning of exploitability for new crash batches on-the-fly.

\begin{table}
\fontsize{8}{9}\selectfont
\centering
\caption{List of the reviewed papers in the \textit{Exploit Time} sub-theme of the \textit{Exploitation} theme.}
\label{tab:exploit_studies_time}

\begin{tabular}[!t]{|p{1.5cm}|p{4cm}|p{3.5cm}|p{3cm}|}

  \hline \textbf{Study} & \textbf{Nature of task} & \textbf{Data source} & \textbf{Data-driven technique}\\\hline

  Bozorgi et al. 2010~\cite{bozorgi2010beyond} & \textit{Binary classification}: Likelihood that SVs would be exploited within 2 to 30 days after disclosure & CVE, OSVDB & Linear SVM\\\hline
  
  Edkrantz 2015~\cite{edkrantz2015predictingthesis} & \textit{Binary classification}: Likelihood of SV exploits within 12 months after disclosure & NVD, ExploitDB, Recorded Future security advisories & SVM, K-Nearest Neighbors (KNN), Na\"ive Bayes, Random forest\\\hhline{-~*{2}{-}}
  Jacobs et al. 2019~\cite{jacobs2019exploit,jacobs2021exploit} & & NVD, Kenna Security \newline Exploit sources: Exploit DB, Metasploit, D2 Security's Elliot \& Canvas Exploitation Frameworks, Fortinet, Proofpoint, AlienVault \& GreyNoise & Logistic regression\\\hline
  
  Chen et al. 2019~\cite{chen2019using,chen2019vest} & \textit{Binary classification}: Likelihood that SVs would be exploited within 1/3/6/9/12 months after disclosure \newline \textit{Regression}: number of days until SV exploits after disclosure & CVE, Twitter, ExploitDB, Symantec security advisories & Graph neural network embedding + Linear regression, Bayes, Random forest, XGBoost, Lasso/Ridge regression\\\hline

\end{tabular}
\end{table}

\subsubsection{Exploit Time}
\label{subsubsec:exploit_time}

After predicting the likelihood of SV exploits in the previous sub-theme, this sub-theme provides more fine-grained information about \textit{exploit time} (see Table~\ref{tab:exploit_studies_time}).
Besides performing binary classification of exploits, Bozorgi et al.~\cite{bozorgi2010beyond} and Edkrantz~\cite{edkrantz2015predictingthesis} also predicted the time frame (2-30 days in~\cite{bozorgi2010beyond} and 12 months in~\cite{edkrantz2015predictingthesis}) within which exploits would happen after the disclosure of SVs. Jacobs et al.~\cite{jacobs2019exploit,jacobs2021exploit} then leveraged multiple sources containing both PoC and real-world exploits, as given in Table~\ref{tab:exploit_studies_time}, to improve the number of labeled exploits, enhancing the prediction of exploit appearance within 12 months. Chen et al.~\cite{chen2019using} predicted whether SVs would be exploited within 1-12 months and the exploit time (number of days) after SV disclosure using Twitter data. The authors proposed a novel regression model whose feature embedding was a multi-layer graph neural network~\cite{kivela2014multilayer} capturing the content and relationships among tweets, respective tweets' authors and SVs. The proposed model outperformed many baselines and was integrated into the VEST system~\cite{chen2019vest} to provide timely SV assessment information for practitioners. To the best of our knowledge, at the time of writing, Chen et al.~\cite{chen2019using,chen2019vest} have been the only ones pinpointing the exact exploit time of SVs rather than large/uncertain time-frames (e.g., months) in other studies, helping practitioners to devise much more fine-grained remediation plans.

\begin{table}
\fontsize{8}{9}\selectfont
\centering
\caption{List of the reviewed papers in the \textit{Exploit Characteristics} sub-theme of the \textit{Exploitation} theme.}
\label{tab:exploit_studies_prop}

\begin{tabular}[!t]{|p{1.5cm}|p{4.5cm}|p{3.2cm}|p{3cm}|}

  \hline \textbf{Study} & \textbf{Nature of task} & \textbf{Data source} & \textbf{Data-driven technique}\\\hline

  Yamamoto et al. 2015~\cite{yamamoto2015text} & \multirowcell{3}[0ex][l]{\textit{Multi-class classification}: CVSS v2\\(Access Vector \& Access Complex-\\ity metrics)\\\\ \textit{Binary classification}: CVSS v2\\(Authentication metric)} & NVD & Supervised Latent Dirichlet Allocation (LDA)\\\hhline{-~*{2}{-}}
  Wen et al. 2015~\cite{wen2015novel} & & NVD, OSVDB, SecurityFocus, IBM X-Force & Radial basis function kernel SVM\\\hhline{-~*{2}{-}}
  Le et al. 2019~\cite{le2019automated}
  & & NVD & Concept-drift-aware models with Na\"ive Bayes, KNN, Linear SVM, Random forest, XGBoost, LGBM\\\hline
 
  Toloudis et al. 2016~\cite{toloudis2016associating} & \textit{Correlation analysis}: CVSS v2 & NVD & Principal component analysis \& Spearman correlation coefficient\\\hline
 
  Ognawala et al. 2018~\cite{ognawala2018automatically} & \multirowcell{2}[0ex][l]{\textit{Multi-class classification}: CVSS v3\\(Attack Vector, Attack Complexity\\ \& Privileges Required metrics)\\\\ \textit{Binary classification}: CVSS v3\\ (User Interaction metric)} & NVD (buffer overflow SVs) \& Source code of vulnerable software/components & Combining static analysis tool (Macke~\cite{ognawala2016macke}) \& ML classifiers (Na\"ive Bayes \& Random forest)\\\hhline{-~*{2}{-}}
  Chen et al. 2019~\cite{chen2019vest} & & CVE, NVD, Twitter & Graph convolutional network\\\hline
 
  Elbaz et al. 2020~\cite{elbaz2020fighting} & \textit{Multi-class/Binary classification}: CVSS v2/v3 & NVD & Mapping outputs of Linear regression to CVSS metrics with closest values\\\hhline{-~*{2}{-}}
  Jiang et al. 2020~\cite{jiang2020approach} & & NVD, ICS Cert, Vendor websites (Resolve inconsistencies with a majority vote) & Logistic regression\\\hline
 
  Gawron et al. 2017~\cite{gawron2017automatic} & \textit{Multi-target classification}: CVSS v2 & NVD & Na\"ive Bayes, Multi-layer Perceptron (MLP)\\\hhline{-~*{2}{-}}
  Spanos et al. 2018~\cite{spanos2018multi} & & NVD & Random forest, boosting model, Decision tree\\\hline
 
  Gong et al. 2019~\cite{gong2019joint} & \textit{Multi-task classification}: CVSS v2 & NVD & Bi-LSTM with attention mechanism\\\hline\hline
 
  Chen et al. 2010~\cite{chen2010categorization} & \textit{Multi-class classification}: Platform-specific vulnerability locations (Local, Remote, Local area network) \& vulnerability causes (e.g., Access/Input/Origin validation error) & NVD, Secunia vulnerability database, SecurityFocus, IBM X-Force & Linear SVM\\\hline

  Ruohonen et al. 2017~\cite{ruohonen2017classifying} & \textit{Binary classification}: Web-related exploits or not & ExploitDB & LDA + Random forest\\\hline
 
  Aksu et al. 2018~\cite{aksu2018automated} & \textit{Multi-class classification}: author-defined pre-/post-condition privileges (None, OS (Admin/User), App (Admin/User)) & NVD & RBF network, Linear SVM, NEAT~\cite{stanley2002evolving}, MLP\\\hhline{-~*{2}{-}}
  Liu et al. 2019~\cite{liu2019automated} & & NVD & Information gain + Convolutional neural network\\\hline
  
  Kanakogi et al. 2021~\cite{kanakogi2021tracing} & \textit{Multi-class classification}: Common Attack Pattern Enumeration and Classification (CAPEC) & NVD, CAPEC & Doc2vec/tf-idf with cosine similarity\\\hline
  
\end{tabular}
\end{table}

\subsubsection{Exploit Characteristics}\label{subsubsec:exploit_prop}

\textit{Exploit characteristics} is the final sub-theme that reveals various requirements/means of exploits (see Table~\ref{tab:exploit_studies_prop}), informing the potential scale of SVs; e.g., remote exploits likely affect more systems than local ones.
The commonly used outputs are the Exploitability metrics provided by versions 2~\cite{cvss_v2} and 3~\cite{cvss_v3,cvss_v31} of Common Vulnerability Scoring System (CVSS).

Many studies have focused on predicting and analyzing version 2 of CVSS exploitability metrics (i.e., Access Vector, Access Complexity and Authentication). Yamamoto et al.~\cite{yamamoto2015text} were the first one to leverage descriptions of SVs on NVD together with a supervised Latent Dirichlet Allocation topic model~\cite{blei2007supervised} to predict these CVSS metrics.
Subsequently, Wen et al.~\cite{wen2015novel} used  Radial Basis Function (RBF)-kernel SVM and various SV databases/advisories other than NVD (e.g., SecurityFocus, OSVDB and IBM X-Force~\cite{xforce}) to predict the metrics.
Le et al.~\cite{le2019automated}\footnote{This study is presented in Chapter~\ref{chap:msr19}.} later showed that the prediction of CVSS metrics suffered from the \textit{concept drift} issue; i.e., descriptions of new SVs may contain Out-of-Vocabulary terms for prediction models.
They proposed to combine sub-word features with traditional Bag-of-Word (BoW) features to infer the semantics of novel terms/words from existing ones, helping assessment models be more robust against concept drift. Besides prediction, Toloudis et al.~\cite{toloudis2016associating} used principal component analysis~\cite{wold1987principal} and Spearman's $\rho$ correlation coefficient to reveal the predictive contribution of each word in SV descriptions to each CVSS metric. However, this technique does not directly produce the value of each metric.

Recently, several studies have started to predict CVSS version 3 exploitability metrics including the new Privileges and User Interactions.
Ognawala et al.~\cite{ognawala2018automatically} fed the features generated by a static analysis tool, Macke~\cite{ognawala2016macke}, to a Random forest model to predict these CVSS version 3 metrics for vulnerable software/components.
Later, Chen et al.~\cite{chen2019vest} found that many SVs were disclosed on Twitter before on NVD. Therefore, these authors developed a system built on top of a Graph Convolutional Network~\cite{kipf2016semi} capturing the content and relationships of related Twitter posts about SVs to enable more timely prediction of the CVSS version 3 metrics.
Elbaz et al.~\cite{elbaz2020fighting} developed a linear regression model to predict the numerical output of each metric and then obtained the respective categorical value with the numerical value closest to the predicted value. For example, a predicted value of 0.8 for Attack Vector CVSS v3 is mapped to \textit{Network} (0.85)~\cite{cvss_v3}.
To prepare a clean dataset to predict these CVSS metrics, Jiang et al.~\cite{jiang2020approach} replaced inconsistent CVSS values in various SV sources (i.e., NVD, ICS CERT and vendor websites) with the most frequent value.

Instead of building a separate model for each CVSS metric, there has been another family of approaches predicting these metrics using a single model to increase efficiency. Gawron et al.~\cite{gawron2017automatic} and Spanos et al.~\cite{spanos2018multi} predicted multiple CVSS metrics as a unique string instead of individual values. The output of each metric is then extracted from the concatenated string.
Later, Gong et al.~\cite{gong2019joint} adopted the idea of a unified model from the DL perspective by using the multi-task learning paradigm~\cite{zhang2021survey} to predict CVSS metrics simultaneously. The model has a feature extraction module (based on a Bi-LSTM model with attention mechanism~\cite{bahdanau2014neural}) shared among all the CVSS metrics/tasks, yet specific prediction head/layer for each metric/task. This model outperformed single-task counterparts while requiring much less time to (re-)train.

Although CVSS exploitability metrics were most commonly used, several studies used other schemes for characterizing exploitation. Chen et al.~\cite{chen2010categorization} used Linear SVM and SV descriptions to predict multiple SV characteristics, including three \textit{SV locations} (i.e., Local, LAN and Remote) on SecurityFocus~\cite{securityfocus} and Secunia~\cite{secunia} databases as well as 11 \textit{SV causes}\footnote{Access/Input/Origin validation error, Atomicity/Configuration/Design/Environment/Serialization error, Boundary condition error, Failure on exceptions, Race condition error} on SecurityFocus.
Regarding the exploit types, Rouhonen et al.~\cite{ruohonen2017classifying} used LDA~\cite{blei2003latent} and Random forest to classify whether an exploit would affect a web application. This study can help find relevant exploits in components/sub-systems of a large system.
For privileges, Aksu et al.~\cite{aksu2018automated} extended the Privileges Required metric of CVSS by incorporating the context (i.e., Operating system or Application) to which privileges are applied (see Table~\ref{tab:exploit_studies_prop}).
They found MLP~\cite{hastie2009elements} to be the best-performing model for obtaining these privileges from SV descriptions. They also utilized the predicted privileges to generate attack graphs (sequence of attacks from source to sink nodes).
Liu et al.~\cite{liu2019automated} advanced this task by combining information gain for feature selection and Convolutional Neural Network (CNN)~\cite{kim2014convolutional} for feature extraction.
Regarding attack patterns, Kanakogi et al.~\cite{kanakogi2021tracing} found Doc2vec~\cite{le2014distributed} to be more effective than term-frequency inverse document frequency (tf-idf) when combined with cosine similarity to find the most relevant Common Attack Pattern Enumeration and Classification (CAPEC)~\cite{capec} for a given SV on NVD. Such attack patterns can manifest how identified SVs can be exploited by adversaries, assisting the selection of suitable countermeasures.

\subsection{Theme Discussion}
In the \textit{Exploitation} theme, the primary tasks are binary classification of whether Proof-of-Concept (PoC)/real-world exploits of SVs would appear and multi-classification of exploit characteristics based on CVSS. PoC exploits mostly come from ExploitDB~\mbox{\cite{exploitdb}}; whereas, real-world exploits, despite coming from multiple sources, are still much scarcer than PoC counterparts. Consequently, the models predicting real-world exploits have generally performed worse than those for PoC exploits. Similarly, the performance of the models determining CVSS v3 exploitability metrics has been mostly lower than that of the CVSS v2 based models. However, real exploits and CVSS v3 are usually of more interest to the community. The former can lead to real cyber-attacks and the latter is the current standard in practice. To improve the performance of these practical tasks, future work can consider adapting the patterns learned from PoC exploits and old CVSS versions to real exploits and newer CVSS versions, respectively, e.g., using transfer learning~\mbox{\cite{pan2009survey}}.

Besides the above tasks, there are other under-explored tasks targeting fine-grained prediction of exploits. In fact, mitigation of exploits in practice usually requires more information besides simply determining whether an SV would be exploited. Information gathered from predicting \textit{when} and \textit{how} the exploits would happen is also needed to devise better SV fixing prioritization and mitigation plans. VEST~\mbox{\cite{chen2019vest}} is one of the first and few systems aiming to provide such all-in-one information about SV exploitation. However, this system currently only uses data from NVD/CVE and Twitter, which can be extended to incorporate more (exploit-related) sources and more sophisticated data-driven techniques in the future.

Most of the current studies have used SV descriptions on NVD and other security advisories to predict the exploitation-related metrics. This is surprising as SV descriptions do not contain root causes of SVs. Instead, SVs are rooted in source code, yet there is little work on code-based exploit prediction. So far, Younis et al.~\mbox{\cite{younis2014using}} have been the only ones using source code for exploit prediction, but their approach still requires manual identification of dangerous function calls in C/C++. More work is required to employ data-driven approaches to alleviate the need for manually defined rules to improve the effectiveness and generalizability of code-based exploit prediction.

\vspace{-1pt}
\section{Impact Prediction}
\label{sec:impact_prediction}

This section describes the \textit{Impact} theme that determines the (negative) effects that SVs have on a system of interest if such SVs are exploited. There are five key tasks that the papers in this theme have automated/predicted: (\textit{i}) Confidentiality impact, (\textit{ii}) Integrity impact, (\textit{iii}) Availability impact, (\textit{iv}) Scope and (\textit{v}) Custom vulnerability consequences (see Table~\ref{tab:impact_studies}). 

\begin{table}[!t]
\fontsize{8}{9}\selectfont
\caption[List of the reviewed papers in the \textit{Impact} theme.]{List of the reviewed papers in the \textit{Impact} theme. \textbf{Note}: We grouped the first four sub-themes as they were mostly predicted together.}
\label{tab:impact_studies}
\centering
\begin{tabular}{|p{1.5cm}|p{3.8cm}|p{3.2cm}|p{3.8cm}|}

 \hline \multicolumn{1}{|c|}{\textbf{Study}} & \multicolumn{1}{c|}{\textbf{Nature of task}} & \multicolumn{1}{c|}{\textbf{Data source}} & \multicolumn{1}{c|}{\textbf{Data-driven technique}}\\\hline
 
 \multicolumn{4}{|c|}{\cellcolor[HTML]{C0C0C0}}\\
 \multicolumn{4}{|c|}{\multirow{-2}{*}{\cellcolor[HTML]{C0C0C0} \textbf{Sub-themes: 1. Confidentiality, 2. Integrity, 3. Availability \& 4. Scope (only in CVSS v3)}}}\\\hline

 Yamamoto et al. 2015~\cite{yamamoto2015text} & \textit{Multi-class classification}: CVSS v2 & NVD & Supervised Latent Dirichlet Allocation\\\hhline{-~*{2}{-}}
 Wen et al. 2015~\cite{wen2015novel} & & NVD, OSVDB, SecurityFocus, IBM X-Force & Radial basis function kernel SVM\\\hhline{-~*{2}{-}}
 Le et al. 2019~\cite{le2019automated} & & NVD & Concept-drift-aware models with Na\"ive Bayes, KNN, Linear SVM, Random forest, XGBoost, LGBM\\\hline
 
  Toloudis et al. 2016~\cite{toloudis2016associating} & \textit{Correlation analysis}: CVSS v2 & NVD & Principal component analysis \& Spearman correlation coefficient\\\hline
 
  Ognawala et al. 2018~\cite{ognawala2018automatically} & \multirowcell{2}[0ex][l]{\textit{Multi-class classification}:\\ CVSS v3\\\\ \textit{Binary classification}: Scope\\ in CVSS v3} & NVD (buffer overflow SVs) \& Source code of vulnerable software/components & Combining static analysis tool (Macke~\cite{ognawala2016macke}) \& ML classifiers (Na\"ive Bayes \& Random forest)\\\hhline{-~*{2}{-}}
  Chen et al. 2019~\cite{chen2019vest} & & CVE, NVD, Twitter & Graph convolutional network\\\hline
 
 Elbaz et al. 2020~\cite{elbaz2020fighting} & \multirowcell{2}[0ex][l]{\textit{Multi-class classification}:\\ CVSS v2/v3\\\\ \textit{Binary classification}: Scope\\ in CVSS v3} & NVD & Mapping outputs of Linear regression outputs to CVSS metrics with closest values\\\hhline{-~*{2}{-}}
 Jiang et al. 2020~\cite{jiang2020approach} & & NVD, ICS Cert, Vendor websites (Resolve inconsistencies with a majority vote) & Logistic regression\\\hline
 
 Gawron et al. 2017~\cite{gawron2017automatic} & \textit{Multi-target classification}: CVSS v2 & NVD & Na\"ive Bayes, MLP\\\hhline{-~*{2}{-}}
 Spanos et al. 2018~\cite{spanos2018multi} & & NVD & Random forest, boosting model, Decision tree\\\hline
 
 Gong et al. 2019~\cite{gong2019joint} & \textit{Multi-task classification}: CVSS v2 & NVD & Bi-LSTM with attention mechanism\\\hline
 
 \multicolumn{4}{|c|}{\cellcolor[HTML]{C0C0C0}}\\
 \multicolumn{4}{|c|}{\multirow{-2}{*}{\cellcolor[HTML]{C0C0C0} \textbf{Sub-theme: 5. Custom Vulnerability Consequences}}}\\\arrayrulecolor{black}\hline
 
 Chen et al. 2010~\cite{chen2010categorization} & \textit{Multi-label classification}: Platform-specific impacts (e.g., Gain system access) & NVD, Secunia vulnerability database, SecurityFocus, IBM X-Force & Linear SVM\\\hline

\end{tabular}
\vspace{-3pt}
\end{table}

\subsection{Summary of Primary Studies}
\subsubsection{Confidentiality, Integrity, Availability, and Scope}\label{subsubsec:cvss_impact}

A majority of the papers have focused on the impact metrics provided by CVSS, including versions 2~\cite{cvss_v2} and 3~\cite{cvss_v3,cvss_v31}.
Versions 2 and 3 share three impact metrics \textit{Confidentiality}, \textit{Integrity} and \textit{Availability}. Version 3 also has a new metric, \textit{Scope}, that specifies whether an exploited SV would affect only the system that contains the SV. For example, \textit{Scope} changes when an SV occurring in a virtual machine affects the whole host machine, in turn increasing individual impacts.

The studies that predicted the CVSS impact metrics are mostly the same as the ones predicting the CVSS exploitability metrics in section~\ref{sec:exploit_prediction}.
Given the overlap, we hereby only describe the main directions and techniques of the \textit{Impact}-related tasks rather than iterating the details of each study.
Overall, a majority of the work has focused on classifying CVSS impact metrics (versions 2 and 3) using three main learning paradigms: single-task~\cite{yamamoto2015text,wen2015novel,le2019automated,ognawala2018automatically,chen2019vest,elbaz2020fighting,jiang2020approach}, multi-target~\cite{gawron2017automatic,spanos2018multi} and multi-task~\cite{gong2019joint} learning. Instead of developing a separate prediction model for each metric like the single-task approach, multi-target and multi-task approaches only need a single model for all tasks. Multi-target learning predicts concatenated output; whereas, multi-task learning uses shared feature extraction for all tasks and task-specific softmax layers to determine the output of each task. These three learning paradigms were powered by applying and/or customizing a wide range of data-driven methods. The first method was to use single ML classifiers like supervised Latent Dirichlet Allocation~\cite{yamamoto2015text}, Principal component analysis~\cite{toloudis2016associating}, Na\"ive Bayes~\cite{le2019automated,ognawala2018automatically,gawron2017automatic}, Logistic regression~\cite{jiang2020approach}, Kernel-based SVM~\cite{wen2015novel}, Linear SVM~\cite{le2019automated}, KNN~\cite{le2019automated} and Decision tree~\cite{spanos2018multi}. Other studies employed ensemble models combining the strength of multiple single models such as Random forest~\cite{le2019automated,ognawala2018automatically}, boosting model~\cite{spanos2018multi} and XGBoost/LGBM~\cite{le2019automated}. Recently, more studies moved towards more sophisticated DL architectures such as MLP~\cite{gawron2017automatic}, attention-based (Bi-)LSTM~\cite{gong2019joint} and graph neural network~\cite{chen2019vest}. Ensemble and DL models usually beat the single ones, but there is a lack of direct comparisons between these two emerging model types.

\subsubsection{Custom Vulnerability Consequences}\label{subsubsec:other_consequences}

To devise effective remediation strategies for a system of interest in practice, practitioners may want to know \textit{custom vulnerability consequences} which are more interpretable than the levels of impact provided by CVSS. Chen et al.~\cite{chen2010categorization} curated a list of 11 vulnerability consequences\footnote{Gain system access, Bypass security, Configuration manipulation, Data/file manipulation, Denial of Service, Privilege escalation, Information leakage, Session hijacking, Cross-site scripting (XSS), Source spoofing, Brute-force proneness.} from X-Force~\cite{xforce} and Secunia~\cite{secunia} vulnerability databases.
They then used a Linear SVM model to perform multi-label classification of these consequences for SVs, meaning that an SV can lead to more than one consequence. To the best of our knowledge, this is the only study that has pursued this research direction so far.

\subsection{Theme Discussion}
In the \textit{Impact} theme, the common task is to predict the impact base metrics provided by CVSS versions 2 and 3. Similar to the Exploitation theme, the models for CVSS v3 still require more attention and effort from the community to reach the same performance level as the models for CVSS v2. These impact metrics are also usually predicted together with the exploitability metrics given their similar nature (multi-class classification) using either task-wise models or a unified (multi-target or multi-task) model. Multi-target and multi-task learning are promising as they can reduce the time for continuous (re)training and maintenance when deployed in production.

Besides CVSS impact metrics, other fine-grained SV consequences have also been explored~\mbox{\cite{chen2010categorization}}, but there is still no widely accepted taxonomy for such consequences. Thus, these consequences have seen less adoption in practice than CVSS metrics, despite being potentially useful by providing more concrete information about what assets/components in a system that an SV can compromise.
We recommend that future work investigate SV-related issues that practitioners commonly encounter in practice to potentially create a systematic taxonomy of custom SV consequences.

\section{Severity Prediction}
\label{sec:severity_prediction}

This section discusses the work in the \textit{Severity} theme.
Severity is often a function/combination of Exploitation (section~\ref{sec:exploit_prediction}) and Impact (section~\ref{sec:impact_prediction}). SVs with higher severity usually require more urgent remediation.
There are three main prediction tasks in this theme: (\textit{i}) Severe vs. Non-severe, (\textit{ii}) Severity levels and (\textit{iii}) Severity score, shown in Tables~\ref{tab:severity_binary_studies},~\ref{tab:severity_level_studies} and~\ref{tab:severity_score_studies}, respectively.

Similar to the \textit{Exploitation} and \textit{Impact} themes, many studies in the \textit{Severity} theme have used CVSS versions 2 and 3. According to both CVSS versions, the severity score shares the same range from 0 to 10, with an increment of 0.1.
Based on the score, the existing studies have either defined a threshold to decide whether an SV is severe (requiring high attention), or predicted levels/groups of severity score that require a similar amount of attention or determined the raw score value.

\subsection{Summary of Primary Studies}
\subsubsection{Severe vs. Non-Severe}\label{subsubsec:severe_or_not}

\begin{table}
\fontsize{8}{9}\selectfont
\centering
\caption[List of the reviewed papers in the \textit{Severe vs. Non-Severe} sub-theme of the \textit{Severity} theme.]{List of the reviewed papers in the \textit{Severe vs. Non-Severe} sub-theme of the \textit{Severity} theme.
\textbf{Note}: The nature of task here is binary classification of severe SVs with High/Critical CVSS v2/v3 severity levels.
}
\label{tab:severity_binary_studies}

\begin{tabular}{|p{1.5cm}|p{6cm}|p{5.3cm}|}
 \hline \textbf{Study} & \textbf{Data source (software project)} & \textbf{Data-driven technique}\\\hline

 Kudjo et al. 2019~\cite{kudjo2019improving} & NVD (Mozilla Firefox, Google Chrome, Internet Explorer, Microsoft Edge, Sea Monkey, Linux Kernel, Windows 7, Windows 10, Mac OS, Chrome OS) & Term frequency \& inverse gravity moment weighting + KNN, Decision tree, Random forest\\\hline
 Chen et al. 2020~\cite{chen2020automatic} & NVD (Adobe Flash Player, Enterprise Linux, Linux Kernel, Foxit Reader, Safari, Windows 10, Microsoft Office, Oracle Business Suites, Chrome, QuickTime) & Term frequency \& inverse gravity moment weighting + KNN, Decision tree, Na\"ive Bayes, SVM, Random forest\\\hline
 Kudjo et al. 2020~\cite{kudjo2020effect} & NVD (Google Chrome, Mozilla Firefox, Internet Explorer and Linux Kernel) & Find the best smallest training dataset using KNN, Logistic regression, MLP, Random forest\\\hline
 Malhotra et al. 2021~\cite{malhotra2021severity} & NVD (Apache Tomcat) & Chi-square/Information gain + bagging technique, Random forest, Na\"ive Bayes, SVM\\\hline
     
\end{tabular}
\end{table}

The first group of studies have classified whether an SV is \textit{severe or non-severe}, making it a binary classification problem (see Table~\ref{tab:severity_binary_studies}). These studies have typically selected severe SVs as the ones with at least High severity level (i.e., CVSS severity score $\geq$ 7.0). Kudjo et al.~\cite{kudjo2019improving} showed that using term frequency (BoW) with inverse gravity moment weighting~\cite{chen2016turning} to extract features from SV descriptions can enhance the performance of ML models (i.e., KNN, Decision tree and Random forest) in predicting the severity of SVs. Later, Chen et al.~\cite{chen2020automatic} confirmed that this feature extraction method was also effective for more projects and classifiers (e.g., Na\"ive Bayes and SVM). Besides investigating feature extraction, Kudjo et al.~\cite{kudjo2020effect} also highlighted the possibility of finding Bellwether, i.e., the smallest set of data that can be used to train an optimal prediction model, for classifying severity. Recently, Malhotra et al.~\cite{malhotra2021severity} revisited this task by showing that Chi-square and information gain can be effective dimensionality reduction techniques for multiple classifiers, i.e., bagging technique, Random forest, Na\"ive Bayes and SVM.

\subsubsection{Severity Levels}\label{subsec:severity_levels}

\begin{table}
\fontsize{8}{9}\selectfont
\centering
\caption{List of the reviewed papers in the \textit{Severity Levels} sub-theme of the \textit{Severity} theme.}
\label{tab:severity_level_studies}

\begin{tabular}{|p{1.5cm}|p{4cm}|p{3cm}|p{3.8cm}|}
 \hline  \textbf{Study} & \textbf{Nature of task} & \textbf{Data source} & \textbf{Data-driven technique}\\\hline
 
 Spanos et al. 2017~\cite{spanos2017assessment} & \textit{Multi-class classification}: NVD severity levels based on CVSS v2 \& WIVSS (High, Medium, Low) & NVD & Decision tree, SVM, MLP\\\hline

 Wang et al. 2019~\cite{wang2019intelligent} & \multirowcell{2}[0ex][l]{\textit{Multi-class classification}: NVD\\severity levels based on CVSS\\ v2(High, Medium, Low)} & NVD (XSS attacks) & XGBoost, Logistic regression, SVM, Random forest\\\hhline{-~*{2}{-}}
 Le et al. 2019~\cite{le2019automated} & & NVD & Concept-drift-aware models with Na\"ive Bayes, KNN, Linear SVM, Random forest, XGBoost, LGBM\\\hhline{-~*{2}{-}}
 Liu et al. 2019~\cite{liu2019vulnerability} & & NVD, China National Vulnerability Database (XSS attacks) & Recurrent Convolutional Neural Network (RCNN), Convolutional Neural Network (CNN), Long-Short Term Memory (LSTM)\\\hhline{-~*{2}{-}}
 Sharma et al. 2020~\cite{sharma2021software} & & CVE Details & CNN\\\hline

 Han et al. 2017~\cite{han2017learning} & \multirowcell{2}[0ex][l]{\textit{Multi-class classification}:\\ Atlassian categories of CVSS\\ severity score (Critical, High,\\ Medium, Low)} & CVE Details & 1-layer CNN, 2-layer CNN, CNN-LSTM, Linear SVM\\\hhline{-~*{2}{-}}
 Sahin et al. 2019~\cite{sahin2019conceptual} & & NVD & 1-layer CNN, LSTM, XGBoost, Linear SVM\\\hhline{-~*{2}{-}}
 Nakagawa et al. 2019~\cite{nakagawa2019character} & & CVE Details & Character-level CNN vs. Word-based CNN + Linear SVM\\\hline
 
 Gong et al. 2019~\cite{gong2019joint} & \textit{Multi-task classification}: Atlassian categories of CVSS severity score (Critical, High, Medium, Low) & CVE Details & Bi-LSTM with attention mechanism\\\hline
 
 Chen et al. 2010~\cite{chen2010categorization} & \textit{Multi-class classification}: severity levels of Secunia (Extremely/highly/ moderately/less/non- critical) & CVE, Secunia vulnerability database, SecurityFocus, IBM X-Force & Linear SVM\\\hline
 
 Zhang et al. 2020~\cite{zhang2020general} & \textit{Multi-class classification}: Platform-specific levels (High/Medium/Low) & China National Vulnerability Database & Logistic regression, Linear discriminant analysis, KNN, CART, SVM, bagging/boosting models\\\hline

 Khazaei et al. 2016~\cite{khazaei2016automatic} & \textit{Multi-class classification}: 10 severity score bins (one unit/bin) & CVE \& OSVDB & Linear SVM, Random forest, Fuzzy system\\\hline

\end{tabular}
\end{table}

Rather than just performing binary classification of whether an SV is severe, several studies have identified one among multiple \textit{severity levels} that an SV belongs to (see Table~\ref{tab:severity_level_studies}). This setting can be considered as multi-class classification. Spanos et al.~\cite{spanos2017assessment} were to first one to show the applicability of ML to classify SVs into one of the three severity levels using SV descriptions. These three levels are provided by NVD and based on the severity score of CVSS version 2~\cite{cvss_v2} and WIVSS~\cite{spanos2013wivss}, i.e., Low (0.0 -- 3.9), Medium (4.0 -- 6.9), High (7.0 -- 10.0). Note that WIVSS assigns different weights for the Confidentiality, Integrity and Availability impact metrics of CVSS, enhancing the ability to capture varied contributions of these impacts to the final severity score. Later, Wang et al.~\cite{wang2019intelligent} showed that XGBoost~\cite{chen2016xgboost} performed the best among the investigated ML classifiers for predicting these three NVD-based severity levels. Le et al.~\cite{le2019automated} also confirmed that ensemble methods (e.g., XGBoost~\cite{chen2016xgboost}, LGBM~\cite{ke2017lightgbm} and Random forest) outperformed single models (e.g., Na\"ive Bayes, KNN and SVM) for this task. Predicting severity levels has also been tackled with DL techniques~\cite{liu2019vulnerability,sharma2021software} such as Recurrent Convolutional Neural Network (RCNN)~\cite{lai2015recurrent}, Convolutional Neural Network (CNN)~\cite{kim2014convolutional}, Long-Short Term Memory (LSTM)~\cite{hochreiter1997long}. These studies showed potential performance gain of DL models compared to traditional ML counterparts. Han et al.~\cite{han2017learning} showed that DL techniques (i.e., 1-layer CNN) also achieved promising results for predicting a different severity categorization, namely Atlassian's levels.\footnote{\url{https://www.atlassian.com/trust/security/security-severity-levels}} Such findings were successfully replicated by Sahin et al.~\cite{sahin2019conceptual}. Nakagawa et al.~\cite{nakagawa2019character} further enhanced the DL model performance for the same task by incorporating the character-level features into a CNN model~\cite{zhang2015character}. Complementary to performance enhancement, Gong et al.~\cite{gong2019joint} proposed to predict these severity levels concurrently with other CVSS metrics in a single model using multi-task learning~\cite{zhang2021survey} powered by an attention-based Bi-LSTM shared feature extraction model. The unified model was demonstrated to increase both the prediction effectiveness and efficiency.
Besides Atlassian's categories, several studies applied ML models to predict severity levels on other platforms such as Secunia~\cite{chen2010categorization} and China National Vulnerability Database\footnote{\url{https://www.cnvd.org.cn}}~\cite{zhang2020general}. Instead of using textual categories, Khazaei et al.~\cite{khazaei2016automatic} divided the CVSS severity score into 10 bins with 10 increments each (e.g., values of 0 -- 0.9 are in one bin) and obtained decent results (86-88\% Accuracy) using Linear SVM, Random forest and Fuzzy system.

\subsubsection{Severity Score}\label{subsec:severity_score}

\begin{table}
\fontsize{8}{9}\selectfont
\centering
\caption[List of the reviewed papers in the \textit{Severity Score} sub-theme of the \textit{Severity} theme.]{List of the reviewed papers in the \textit{Severity Score} sub-theme of the \textit{Severity} theme. \textbf{Notes}: \textsuperscript{\textdagger}denotes that the severity score is computed from ML-predicted base metrics using the formula provided by an assessment framework (CVSS and/or WIVSS).}
\label{tab:severity_score_studies}

\begin{tabular}{|p{1.5cm}|p{3cm}|p{3.2cm}|p{4.5cm}|}
 \hline \textbf{Study} & \textbf{Nature of task} & \textbf{Data source} & \textbf{Data-driven technique}\\\hline

 Sahin et al. 2019~\cite{sahin2019conceptual} & \textit{Regression}: CVSS v2 (0-10) & NVD & 1-layer CNN, LSTM, XGBoost regressor, Linear regression\\\hhline{-~*{2}{-}}
 Wen et al. 2015~\cite{wen2015novel} & & OSVDB, SecurityFocus, IBM X-Force & Radial basis function kernel SVM\textsuperscript{\textdagger}\\\hline
 
 Ognawala et al. 2018~\cite{ognawala2018automatically} & \textit{Regression}: CVSS v3 (0-10) & NVD (buffer overflow SVs) & Combining a static analysis tool (Macke~\cite{ognawala2016macke}) \& ML classifiers (Na\"ive Bayes \& Random forest)\textsuperscript{\textdagger}\\\hhline{-~*{2}{-}}
 Chen et al. 2019~\cite{chen2019vest,chen2019vase} & & CVE, NVD, Twitter & Graph convolutional network\\\hhline{-~*{2}{-}}
 Anwar et al. 2020~\cite{anwar2020cleaning,anwar2021cleaning} & & NVD & Linear regression, Support vector regression, CNN, MLP\\\hline
 
 Elbaz et al. 2020~\cite{elbaz2020fighting} & \textit{Regression}: CVSS v2/v3 (0-10) & NVD & Mapping outputs of Linear regression to CVSS metrics with closest values\textsuperscript{\textdagger}\\\hhline{-~*{2}{-}}
 Jiang et al. 2020~\cite{jiang2020approach} & & NVD, ICS Cert, Vendor websites (Resolve inconsistencies with a majority vote) & Logistic regression\textsuperscript{\textdagger}\\\hline
 
 Spanos et al. 2018~\cite{spanos2018multi} & \textit{Regression}: CVSS v2 \& WIVSS (0-10) & NVD & Random forest, boosting model, Decision tree\textsuperscript{\textdagger}\\\hline
 
 Toloudis et al. 2016~\cite{toloudis2016associating} & \textit{Correlation analysis}: CVSS v2 \& WIVSS (0-10) & NVD & Principal component analysis \& Spearman correlation coefficient\\\hline
     
\end{tabular}
\end{table}

To provide even more fine-grained severity value than the categories, the last sub-theme has predicted the \textit{severity score} (see Table~\ref{tab:severity_score_studies}).
Using SV descriptions on NVD, Sahin et al.~\cite{sahin2019conceptual} compared the performance of ML-based regressors (e.g., XGBoost~\cite{chen2016xgboost} and Linear regression) and DL-based ones (e.g., CNN~\cite{kim2014convolutional} and LSTM~\cite{hochreiter1997long}) for predicting the severity score of CVSS version 2~\cite{cvss_v2}. These authors showed that DL-based approaches generally outperformed the ML-based counterparts. For CVSS version 3~\cite{cvss_v3,cvss_v31}, Chen et al.~\cite{chen2019vest,chen2019vase} and Anwar et al.~\cite{anwar2020cleaning,anwar2021cleaning} also reported the strong performance of DL-based models (e.g., CNN and graph convolutional neural network~\cite{kipf2016semi}). Some other studies did not directly predict severity score from SV descriptions, instead they aggregated the predicted values of the CVSS Exploitability (see section~\ref{sec:exploit_prediction}) and Impact metrics (see section~\ref{sec:impact_prediction}) using the formulas of CVSS version 2~\cite{wen2015novel,spanos2018multi,elbaz2020fighting,jiang2020approach}, version 3~\cite{ognawala2018automatically,elbaz2020fighting,jiang2020approach} and WIVSS~\cite{spanos2018multi}.
We noticed the papers predicting both versions (e.g., CVSS versions 2 vs. 3 or CVSS version 2 vs. WIVSS) usually obtained better performance for version 3 and WIVSS than version 2~\cite{elbaz2020fighting,jiang2020approach}. These findings may suggest that the improvements made by experts in version 3 and WIVSS compared to version 2 help make the patterns in severity score clearer and easier for ML models to capture.
In addition to predicting severity score, Toloudis et al.~\cite{toloudis2016associating} examined the correlation between words in descriptions of SVs and the severity values of such SVs, aiming to shed light on words that increase or decrease the severity score of SVs.

\subsection{Theme Discussion}
In the \textit{Severity} theme, predicting the severity levels is the most prevalent task, followed by severity score prediction and then binary classification of the severity.
In practice, severity score gives more fine-grained information (fewer SVs per value) for practitioners to rank/prioritize SVs than categorical/binary levels. However, predicting continuous score values is usually challenging and requires more robust models as this task involves higher uncertainty to learn inherent patterns from data than classifying fixed/discrete levels. We observed that DL models such as graph neural networks~\mbox{\cite{chen2019vest,chen2019vase}}, LSTM~\mbox{\cite{sahin2019conceptual}} and CNN~\mbox{\cite{anwar2020cleaning,anwar2021cleaning}} have been shown to be better than traditional ML models for predicting severity score. However, most of these studies did not evaluate their models in a continuous deployment setting to investigate how the models will cope with changing patterns of new SVs over time. This issue particularly manifests and requires remediation in the context of report-level SV assessment (see section~\ref{sec:conclusions_chap2}) where SV descriptions, the main input for SV assessment models, contain changing/new terms to describe ever-increasing SVs.

\section{Type Prediction}
\label{sec:type_prediction}

\begin{table}
\fontsize{8}{9}\selectfont
\centering
\caption{List of the reviewed papers in the \textit{Type} theme.}\label{tab:type_studies}
\vspace{-2pt}

\begin{tabular}{|p{1.5cm}|p{3.9cm}|p{2.3cm}|p{4.5cm}|}

 \hline \textbf{Study} & \textbf{Nature of task} & \textbf{Data source} & \textbf{Data-driven technique}\\\hline
 
    \multicolumn{4}{|c|}{\cellcolor[HTML]{C0C0C0}}\\
    \multicolumn{4}{|c|}{\multirow{-2}{*}{\cellcolor[HTML]{C0C0C0} \textbf{Sub-theme: 1. Common Weakness Enumeration (CWE)}}}\\\hline

    Wang et al. 2010~\cite{wang2010vulnerability} & \textit{Multi-class classification}: CWE classes & NVD, CVSS & Na\"ive Bayes\\\hhline{-~*{2}{-}}
    Shuai et al. 2013~\cite{shuai2013automatic} & & NVD & SVM\\\hhline{-~*{2}{-}}
    Na et al. 2016 ~\cite{na2016study} & & NVD & Na\"ive Bayes\\\hhline{-~*{2}{-}}
    Ruohonen et al. 2018~\cite{ruohonen2018toward} & & NVD, CWE, Snyk & tf-idf with 1/2/3-grams and cosine similarity\\\hhline{-~*{2}{-}}
    Huang et al. 2019~\cite{huang2019automatic} & & NVD, CWE & MLP, Linear SVM, Na\"ive Bayes, KNN\\\hhline{-~*{2}{-}}
    Aota et al. 2020~\cite{aota2020automation} & & NVD & Random forest, Linear SVM, Logistic regression, Decision tree, Extremely randomized trees, LGBM\\\hhline{-~*{2}{-}}
    Aghaei et al. 2020~\cite{aghaei2020threatzoom} & & NVD, CVE & Adaptive fully-connected neural network with one hidden layer\\\hhline{-~*{2}{-}}
    Das et al. 2021~\cite{das2021v2w} & & NVD, CWE & BERT, Deep Siamese network\\\hhline{-~*{2}{-}}
    \noalign{\vskip\doublerulesep\vskip-\arrayrulewidth}\hhline{-~*{2}{-}}
    
    Zou et al. 2019~\cite{zou2019muvuldeepecker} & & NVD \& Software Assurance Reference Dataset (SARD) & Three Bi-LSTM models for extracting and combining global and local features from code functions\\\hline\hline
      
    Murtaza et al. 2016~\cite{murtaza2016mining} & \textit{Unsupervised learning}: sequence mining of SV types (over time) & NVD (CWE \& CPE) & 2/3/4/5-grams of CWEs\\\hline
      
    Lin et al. 2017~\cite{lin2017machine} & \textit{Unsupervised learning}: association rule mining of CWE-related aspects (prog. language, time of introduction \& consequence scope) & CWE & FP-growth association rule mining algorithm\\\hline
      
     Han et al. 2018~\cite{han2018deepweak} & \textit{Binary/Multi-class classification}: CWE relationships (CWE links, link types \& CWE consequences) & CWE & Deep knowledge graph embedding of CWE entities\\\hline
    
    \multicolumn{4}{|c|}{\cellcolor[HTML]{C0C0C0}}\\
    \multicolumn{4}{|c|}{\multirow{-2}{*}{\cellcolor[HTML]{C0C0C0} \textbf{Sub-theme: 2. Custom Vulnerability Types}}}\\\hline
      
    Venter et al. 2008~\cite{venter2008standardising}  & \textit{Unsupervised learning}: clustering & CVE & Self-organizing map\\\hline
  
    Neuhaus et al. 2010~\cite{neuhaus2010security}  & \textit{Unsupervised learning}: topic modeling & CVE & Latent Dirichlet Allocation (LDA)\\\hhline{-~*{2}{-}}
    Mounika et al.~\cite{mounika2019analyzing,vanamala2020topic} & & CVE, Open Web Application Security Project (OWASP) & LDA\\\hhline{-~*{2}{-}}
    Aljedaani et al. 2020~\cite{aljedaani2020lda} & & SV reports (Chromium project) & LDA\\\hline\hline
      
    Williams et al.~\cite{williams2018analyzing,williams2020vulnerability} & \multirowcell{2}[0ex][l]{\textit{Multi-class classification}:\\ manually coded SV types} & NVD & Supervised Topical Evolution Model \& Diffusion-based storytelling technique\\\hhline{-~*{2}{-}}
    Russo et al. 2019~\cite{russo2019summarizing} & & NVD & Bayesian network, J48 tree, Logistic regression, Na\"ive Bayes, Random forest\\\hhline{-~*{2}{-}}
    Yan et al. 2017~\cite{yan2017exploitmeter} & & Executables of 100 Linux applications & Decision tree\\\hline
    
    Zhang et al. 2020~\cite{zhang2020general} & \textit{Multi-class classification}: platform-specific vulnerability types & China National Vulnerability Database & Logistic regression, Linear discriminant analysis, KNN, CART, SVM, bagging/boosting models\\\hline
\end{tabular}
\end{table}
\vspace{-3pt}

This section reports the work done in the \textit{Type} theme. Type groups SVs with similar characteristics, e.g., causes, attack patterns and impacts, and thus facilitating the reuse of known prioritization and remediation strategies employed for prior SVs of the same types.
Two key prediction outputs are: (\textit{i}) Common Weakness Enumeration (CWE) and (\textit{ii}) Custom vulnerability types (see Table~\ref{tab:type_studies}).

\subsection{Summary of Primary Studies}
\subsubsection{Common Weakness Enumeration (CWE)}\label{subsubsec:cwe}

The first sub-theme determines and analyzes the patterns of the SV types provided by \textit{CWE}~\cite{cwe}.
CWE is currently the standard for SV types with more than 900 entries.
The first group of studies has focused on multi-class classification of these CWEs. Wang et al.~\cite{wang2010vulnerability} were the first to tackle this problem with a Na\"ive Bayes model using the CVSS metrics (version 2)~\cite{cvss_v2} and product names. Later, Shuai et al.~\cite{shuai2013automatic} used LDA~\cite{blei2003latent} with a location-aware weighting to extract important features from SV descriptions for building an effective SVM-based CWE classifier. Na et al.~\cite{na2016study} also showed that features extracted from SV descriptions can improve the Na\"ive Bayes model in~\cite{wang2010vulnerability}.
Ruohonen et al.~\cite{ruohonen2018toward} studied an information retrieval method, i.e., term-frequency inverse document frequency (tf-idf) and cosine similarity, to detect the CWE-ID with a description most similar to that of a given SV collected from NVD and Snyk.\footnote{\url{https://snyk.io/vuln}} This method performed well for CWEs without clear patterns/keywords in SV descriptions.
Aota et al.~\cite{aota2020automation} utilized the Boruta feature selection algorithm~\cite{kursa2010boruta} and Random forest to improve the performance of \textit{base} CWE classification.
Base CWEs give more fine-grained information for SV remediation than categorical CWEs used in~\cite{na2016study}.

There has been a recent rise in using neural network/DL based models for CWE classification. Huang et al.~\cite{huang2019automatic} implemented a deep neural network with tf-idf and information gain for the task and obtained better performance than SVM, Na\"ive Bayes and KNN. Aghaei et al.~\cite{aghaei2020threatzoom} improved upon~\cite{aota2020automation} for both categorical (coarse-grained) and base (fine-grained) CWE classification with an adaptive hierarchical neural network to determine sequences of less to more fine-grained CWEs. To capture the hierarchical structure and rare classes of CWEs, Das et al.~\cite{das2021v2w} matched SV and CWE descriptions instead of predicting CWEs directly.
They presented a deep Siamese network with a BERT-based~\cite{devlin2018bert} shared feature extractor that outperformed many baselines even for rare/unseen CWE classes. Recently, Zou et al.~\cite{zou2019muvuldeepecker} pioneered the multi-class classification of CWE in vulnerable functions curated from Software Assurance Reference Dataset (SARD)~\cite{sard} and NVD. They achieved high performance ($\sim$95\% F1-Score) with DL (Bi-LSTM) models. The strength of their model came from combining global (semantically related statements) and local (variables/statements affecting function calls) features. Note that this model currently only works for functions in C/C++ and 40 selected classes of CWE.

Another group of studies has considered unsupervised learning methods to extract CWE sequences, patterns and relationships.
Sequences of SV types over time were identified by Murtaza et al.~\cite{murtaza2016mining} using an $n$-gram model.
This model sheds light on both co-occurring and upcoming CWEs (grams), raising awareness of potential cascading attacks. Lin et al.~\cite{lin2017machine} applied an association rule mining algorithm, FP-growth~\cite{han2000mining}, to extract the rules/patterns of various CWEs aspects including types, programming language, time of introduction and consequence scope. For example, buffer overflow (CWE type) usually appears during the implementation phase (time of introduction) in C/C++ (programming language) and affects the availability (consequence scope). Lately, Han et al.~\cite{han2018deepweak} developed a deep knowledge graph embedding technique to mine the relationships among CWE types, assisting in finding relevant SV types with similar properties.

\subsubsection{Custom Vulnerability Types}\label{subsubsec:custom_types}

The second sub-theme is about \textit{custom vulnerability types} other than CWE.
Venter et al.~\cite{venter2008standardising} used Self-organizing map~\cite{kohonen1990self}, an unsupervised clustering algorithm, to group SVs with similar descriptions on CVE. This was one of the earliest studies that automated SV type classification.
Topic modeling is another popular unsupervised learning model~\cite{neuhaus2010security,mounika2019analyzing,vanamala2020topic,aljedaani2020lda} to categorize SVs without an existing taxonomy. Neuhaus et al.~\cite{neuhaus2010security} applied LDA~\cite{blei2003latent} on SV descriptions to identify 28 prevalent SV types and then analyzed the trends of such types over time. The identified SV topics/types had considerable overlaps (up to 98\% precision and 95\% recall) with CWEs. Mounika et al.~\cite{mounika2019analyzing,vanamala2020topic} extended~\cite{neuhaus2010security} to map the LDA topics with the top-10 OWASP~\cite{owasp_website}. However, the LDA topics/keywords did not agree well ($<$ 40\%) with the OWASP descriptions, probably because 10 topics did not cover all the underlying patterns of SV descriptions.
Aljedaani et al.~\cite{aljedaani2020lda} again used LDA to identify 10 types of SVs reported in the bug tracking system of Chromium\footnote{\url{https://bugs.chromium.org/p/chromium/issues/list}} and found memory-related issues were the most prevalent topics.

Another group of studies has classified manually defined/selected SV types rather than CWE as some SV types are encountered more often in practice and require more attention. Williams et al.~\cite{williams2018analyzing,williams2020vulnerability} applied a supervised topical evolution model~\cite{naim2017scalable} to identify the features that best described the 10 pre-defined SV types\footnote{1. Buffer errors, 2. Cross-site scripting, 3. Path traversal, 4. Permissions and Privileges, 5. Input validation, 6. SQL injection, 7. Information disclosure, 8. Resources Error, 9. Cryptographic issues, 10. Code injection.} prevalent in the wild.
These authors then used a diffusion-based storytelling technique~\cite{barranco2019analyzing} to show the evolution of a particular topic of SVs over time; e.g., increasing API-related SVs requires hardening the APIs used in a product. To support user-friendly SV assessment using ever-increasing unstructured SV data, Russo et al.~\cite{russo2019summarizing} used Bayesian network to predict 10 pre-defined SV types.\footnote{1. Authentication bypass or Improper Authorization, 2. Cross-site scripting or HTML injection, 3. Denial of service, 4. Directory Traversal, 5. Local/Remote file include and Arbitrary file upload, 6. Information disclosure and/or Arbitrary file read, 7. Buffer/stack/heap/integer overflow, 8. Remote code execution, 9. SQL injection, 10. Unspecified vulnerability}
Besides predicting manually defined SV types using SV natural language descriptions, Yan et al.~\cite{yan2017exploitmeter} used a decision tree to predict 22 SV types prevalent in the executables of Linux applications. The predicted type was then combined with fuzzers' outputs to predict SV exploitability (see section~\ref{subsubsec:exploit_likelihood}).
Besides author-defined types, custom SV types also come from specific SV platforms.
Zhang et al.~\cite{zhang2020general} designed an ML-based framework to predict the SV types collected from China National Vulnerability Database. Ensemble models (bagging and boosting models) achieved, on average, the highest performance for this task.

\subsection{Theme Discussion}
In the \textit{Type} theme, detecting and characterizing coarse-grained and fine-grained CWE-based SV types are the frequent tasks. The large number and hierarchical structure of classes are the main challenges with CWE classification/analysis. In terms of solutions, deep Siamese networks~\mbox{\cite{das2021v2w}} are more robust to the class imbalance issue (due to many CWE classes), while graph-based neural networks~\mbox{\cite{han2018deepweak}} can effectively capture the hierarchical structure of CWEs. Future work can investigate the combination of these two types of DL architectures to solve both issues simultaneously.
Besides model-level solutions, author-selected or platform-specific SV types have been considered to reduce the complexity of CWE. However, similar to custom SV consequences in section~\mbox{\ref{subsubsec:other_consequences}}, there is not yet a universally accepted taxonomy for these custom SV types. To reduce the subjectivity in selecting SV types for prediction, we suggest that future work should focus on the types that are commonly encountered and discussed by developers in the wild.

\vspace{-3pt}

\section{Miscellaneous Tasks}
\label{sec:miscellaneous}

The last theme is \textit{Miscellaneous Tasks} covering the studies that are representative yet do not fit into the four previous themes.
This theme has three main sub-themes/tasks: (\textit{i}) Vulnerability information retrieval, (\textit{ii}) Cross-source vulnerability patterns and (\textit{iii}) Vulnerability fixing effort (see Table ~\ref{tab:misc_studies}).

\vspace{-3pt}

\begin{table}
\fontsize{8}{9}\selectfont
\centering

\caption{List of the reviewed papers in the \textit{Miscellaneous Tasks} theme.}
\label{tab:misc_studies}

\begin{tabular}{|p{1.7cm}|p{4.7cm}|p{3cm}|p{3cm}|}

    \hline \textbf{Study} & \textbf{Nature of task} & \textbf{Data source} & \textbf{Data-driven technique}\\\hline
    
    \multicolumn{4}{|c|}{\cellcolor[HTML]{C0C0C0}}\\
    \multicolumn{4}{|c|}{\multirow{-2}{*}{\cellcolor[HTML]{C0C0C0} \textbf{Sub-theme: 1. Vulnerability Information Retrieval}}}\\\hline
    
    Weeraward-hana et al. 2014~\cite{weerawardhana2014automated} & \textit{Multi-class classification}: Extraction of entities (software name/version, impact, attacker/user actions) from SV descriptions & NVD (210 randomly selected and manually labeled SVs) & Stanford Named Entity Recognizer implementing a CRF classifier\\\hline
 
    Dong et al. 2019~\cite{dong2019towards} & \textit{Multi-class classification}: Vulnerable software names/versions & CVE Details, NVD, ExploitDB, SecurityFocus, SecurityFocus Forum, SecurityTracker, Openwall & Word-level and character-level Bi-LSTM with attention mechanism\\\hline
    
    Gonzalez et al. 2019~\cite{gonzalez2019automated} & \textit{Multi-class classification}: Extraction of 19 Vulnerability Description Ontology~\cite{vdo} classes from SV descriptions & NVD & Na\"ive Bayes, Decision tree, SVM, Random forest, Majority voting model\\\hline

    Binyamini et al. 2020~\cite{binyamini2020automated,binyamini2021framework} & Multi-class classification: Extraction of entities (attack vector/means/technique, privilege, impact, vulnerable platform/version/OS, network protocol/port) from SV descriptions to generate MulVal~\cite{ou2005mulval} interaction rules & NVD & Bi-LSTM with various feature extractors: word2vec, ELMo, BERT (pre-trained or trained from scratch)\\\hline

    Guo et al. 2020~\cite{guo2020predicting,guo2021detecting} & \textit{Multi-class classification}: Extraction of entities (SV type, root cause, attack type, attack vector) from SV descriptions & NVD, SecurityFocus & CNN, Bi-LSTM (with or without attention mechanism)\\\hline
 
    Waareus et al. 2020~\cite{waareus2020automated} & \textit{Multi-class classification}: Common Product Enumeration (CPE) & NVD & Word-level and character-level Bi-LSTM\\\hline
 
    Yitagesu et al. 2021~\cite{yitagesu2021vulnpos} & \textit{Multi-class classification}: Part-of-speech tagging of SV descriptions & NVD, CVE, CWE, CAPEC, CPE, Twitter, PTB corpus~\cite{marcus1993building} & Bi-LSTM\\\hline
 
    Sun et al. 2021~\cite{sun2021generating} & \textit{Multi-class classification}: Extraction of entities (vulnerable product/version/component, type, attack type, root cause, attack vector, impact) from ExploitDB to generate SV descriptions & NVD, ExploitDB & BERT models\\\hline
    
    \multicolumn{4}{|c|}{\cellcolor[HTML]{C0C0C0}}\\
    \multicolumn{4}{|c|}{\multirow{-2}{*}{\cellcolor[HTML]{C0C0C0} \textbf{Sub-theme: 2. Cross-source Vulnerability Patterns}}}\\\hline

    Horawalavith-ana et al. 2019~\cite{horawalavithana2019mentions} & \textit{Regression}: Number of software development activities on GitHub after disclosure of SVs & Twitter, Reddit, GitHub & MLP, LSTM\\\hline
    
    Xiao et al. 2019~\cite{xiao2019embedding} & \textit{Knowledge-graph reasoning}: modeling the relationships among SVs, its types and attack patterns & CVE, CWE, CAPEC (Linux project) & Translation-based knowledge-graph embedding\\\hline
    
    \multicolumn{4}{|c|}{\cellcolor[HTML]{C0C0C0}}\\
    \multicolumn{4}{|c|}{\multirow{-2}{*}{\cellcolor[HTML]{C0C0C0} \textbf{Sub-theme: 3. Vulnerability Fixing Effort}}}\\\hline

    Othmane et al. 2017~\cite{othmane2017time} & \textit{Regression}: time (days) to fix SVs & Proprietary SV data collected at the SAP company & Linear/Tree-based/Neural network regression\\\hline

\end{tabular}
\end{table}

\subsection{Summary of Primary Studies}
\subsubsection{Vulnerability Information Retrieval}\label{subsec:information_retrieval}

The first and major sub-theme is \textit{vulnerability information retrieval} that studies data-driven methods to extract different SV-related entities (e.g., affected products/versions) and their relationships from SV data. The current sub-theme extracts assessment information appearing explicitly in SV data (e.g., SV descriptions on NVD) rather than predicting implicit properties as done in prior sub-themes. For instance, CWE-119, i.e., ``Improper Restriction of Write Operations within the Bounds of a Memory Buffer'', can be retrieved directly from CVE-2020-28022,\footnote{\url{https://nvd.nist.gov/vuln/detail/CVE-2020-28022}} but not from CVE-2021-2122.\footnote{\url{https://nvd.nist.gov/vuln/detail/CVE-2021-21220}}
The latter case requires techniques from section~\ref{subsubsec:cwe}.

Most of the retrieval methods in this sub-theme have been formulated under the multi-class classification setting. One of the earliest works was conducted by Weerawardhana et al.~\cite{weerawardhana2014automated}. This study extracted software names/versions, impacts and attacker's/user's action from SV descriptions on NVD using Stanford Named Entity Recognition (NER) technique, a.k.a. CRF classifier~\cite{finkel2005incorporating}. Later, Dong et al.~\cite{dong2019towards} proposed to use a word/character-level Bi-LSTM to improve the performance of extracting vulnerable software names and versions from SV descriptions available on NVD and other SV databases/advisories (e.g., CVE Details~\cite{cve_details}, ExploitDB~\cite{exploitdb}, SecurityFocus~\cite{securityfocus}, SecurityTracker~\cite{securitytracker} and Openwall~\cite{openwall}). Based on the extracted entities, these authors also highlighted the inconsistencies in vulnerable software names and versions across different SV sources. Besides version products/names of SVs, Gonzalez et al.~\cite{gonzalez2019automated} used a majority vote of different ML models (e.g., SVM and Random forest) to extract the 19 entities of Vulnerability Description Ontology (VDO)~\cite{vdo} from SV descriptions to check the consistency of these descriptions based on the guidelines of VDO.
Since 2020, there has been a trend in using DL models (e.g., Bi-LSTM, CNNs or BERT~\cite{devlin2018bert}/ELMo~\cite{peters2018deep}) to extract different information from SV descriptions including required elements for generating MulVal~\cite{ou2005mulval} attack rules~\cite{binyamini2020automated,binyamini2021framework} or SV types/root cause, attack type/vector~\cite{guo2020predicting,guo2021detecting}, Common Product Enumeration (CPE)~\cite{cpe} for standardizing names of vulnerable vendors/products/versions~\cite{waareus2020automated}, part-of-speech~\cite{yitagesu2021vulnpos} and relevant entities (e.g., vulnerable products, attack type, root cause) from ExploitDB to generate SV descriptions~\cite{sun2021generating}. BERT models~\cite{devlin2018bert}, pre-trained on general text (e.g., Wikipedia pages~\cite{wikipedia} or PTB corpus~\cite{marcus1993building}) and fine-tuned on SV text, have also been increasingly used to address the data scarcity/imbalance for the retrieval tasks.

\subsubsection{Cross-Source Vulnerability Patterns}\label{subsubsec:cross_source_patterns}

The second sub-theme, \textit{cross-source vulnerability patterns}, finds commonality and/or discovers latent relationships among SV sources to enrich information for SV assessment. Horawalavithana et al.~\cite{horawalavithana2019mentions} found a positive correlation between development activities (e.g., push/pull requests and issues) on GitHub and SV mentions on Reddit\footnote{\url{https://reddit.com}} and Twitter. These authors then used DL models (i.e., MLP~\cite{hastie2009elements} and LSTM~\cite{hochreiter1997long}) to predict the appearance and sequence of development activities when SVs were mentioned on the two social media platforms. Xiao et al.~\cite{xiao2019embedding} applied a translation-based graph embedding method to encode and predict the relationships among different SVs and the respective attack patterns and types. This work~\cite{xiao2019embedding} was based on the DeepWeak model of Han et al.~\cite{han2018deepweak}, but it still belongs to this sub-theme as they provided a multi-dimensional view of SVs using three different sources (NVD~\cite{nvd}, CWE~\cite{cwe} and CAPEC~\cite{capec}).
Xiao et al.~\cite{xiao2019embedding} envisioned that their knowledge graph can be extended to incorporate the source code introducing/fixing SVs.

\subsubsection{Vulnerability Fixing Effort}\label{subsubsec:fixing_effort}

The last sub-theme is \textit{vulnerability fixing effort} that focuses on estimating SV fixing effort through proxies such as the SV fixing time, usually in days.
Othmane and the co-authors were among the first to approach this problem.
These authors first conducted a large-scale qualitative study at the SAP company and identified 65 important code-based, process-based and developer-based factors contributing to the SV fixing effort~\cite{ben2015factors}. Later, the same group of authors~\cite{othmane2017time} leveraged the identified factors in their prior qualitative study to develop various regression models such as linear regression, tree-based regression and neural network regression models, to predict time-to-fix SVs using the data collected at SAP. These authors found that code components containing detected SVs are more important for the prediction than SV types.

\subsection{Theme Discussion}
In the \textit{Miscellaneous Tasks} theme, the key focus is on retrieving SV-related entities and characteristics from SV descriptions. The retrieval tasks are usually formulated as Named Entity Recognition from SV descriptions. However, we observed that NVD descriptions do not follow a consistent template~\mbox{\cite{anwar2020cleaning,anwar2021cleaning}}, posing significant challenges in labeling the entities for retrieval. The affected versions and vendor/product names of SVs also contain inconsistencies~\mbox{\cite{dong2019towards,anwar2020cleaning,anwar2021cleaning}}, making the retrieval tasks difficult. We recommend that data normalization and cleaning should be performed before labeling entities and building respective retrieval models to ensure the reliability of results.

Besides information retrieval, other tasks such as linking multi-sources, extracting cross-source patterns or estimating fixing effort are also useful to obtain richer SV information for assessment, yet these tasks are still in early stages. Linking multiple sources and their patterns is the first step towards building an SV knowledge graph to answer different queries regarding a particular SV (e.g., what systems are affected, exploitation status, how to fix, or what SVs are similar). In the future, such a knowledge graph can be extended to capture artifacts of SVs in emerging software types like AI-based systems~\cite{kumar2020legal}.
Moreover, to advance SV fixing effort prediction, future work can consider adapting/customizing the existing practices/techniques used to predict fixing effort for general bugs~\mbox{\cite{zhang2013predicting,akbarinasaji2018predicting}}.

\section{Analysis of Data-Driven Approaches for Software Vulnerability Assessment}\label{sec:elements_analysis}

\begin{table}
\fontsize{5}{5.5}\selectfont
\centering

\caption[The frequent data sources, features, models, evaluation techniques and evaluation metrics used for the five identified SV assessment themes.]{The frequent data sources, features, models, evaluation techniques and evaluation metrics used for the five identified SV assessment themes. \textbf{Notes}: The values are organized based on their overall frequency across the five themes. For the Prediction Model and Evaluation Metric elements, the values are first organized by categories (ML then DL for Prediction Model and classification then regression for Evaluation Metric) and then by frequencies. k-CV stands for k-fold cross-validation.
The full list of values and their appearance frequencies for the five elements in the five themes can be found at \url{https://figshare.com/s/da4d238ecdf9123dc0b8}.}
\label{tab:data_driven_elements}

\begin{tabular}{|p{3.3cm}|p{6cm}|p{6.3cm}|}

    \hline \textbf{Source/Technique/Metric} & \textbf{Strengths} & \textbf{Weaknesses}\\\hline
    
    \multicolumn{3}{|c|}{\cellcolor[HTML]{C0C0C0}}\\
    \multicolumn{3}{|c|}{\multirow{-2}{*}{\cellcolor[HTML]{C0C0C0} \textbf{Element: Data Source}}}\\\hline
    
    \makecell[l]{NVD/CVE/CVE Details\\(deprecated  OSVDB)} &
    \makecell[l]{\tabitem Report expert-verified information (with CVE-ID)\\ \tabitem Contain CWE and CVSS entries for each SV\\ \tabitem Link to external sources (e.g., official fixes)} & \makecell[l]{\tabitem Missing/incomplete links to vulnerable code/fixes\\ \tabitem Inconsistencies due to human errors\\ \tabitem Delayed SV reporting and assignment of CVSS metrics}\\\hline
    
    ExploitDB & 
    \makecell[l]{\tabitem Report PoC exploits of SVs (with links to CVE-ID)} & \makecell[l]{\tabitem May not lead to real exploits in the wild}\\\hline
    
    \makecell[l]{Other security advisories (e.g.,\\ SecurityFocus, Symantec or\\ X-Force)} & 
    \makecell[l]{\tabitem Report real-world exploits of SVs\\ \tabitem Cover diverse SVs (including ones w/o CVE-ID)} & \makecell[l]{\tabitem Some exploits may not have links to CVE entries for\\ mapping with other assessment metrics}\\\hline
    
    \makecell[l]{Informal sources (e.g., Twitter\\ and darkweb)} & 
    \makecell[l]{\tabitem Early reporting of SVs (maybe earlier than NVD)\\ \tabitem Contain non-technical SV information (e.g., financial\\ damage or socio-technical challenges in addressing SVs)} & \makecell[l]{\tabitem Contain non-verified and even misleading information\\ \tabitem May cause adversarial attacks to assessment models}\\\hline
    
    \multicolumn{3}{|c|}{\cellcolor[HTML]{C0C0C0}}\\
    \multicolumn{3}{|c|}{\multirow{-2}{*}{\cellcolor[HTML]{C0C0C0} \textbf{Element: Model Feature}}}\\\hline
    
    BoW/tf-idf/n-grams & 
    \makecell[l]{\tabitem Simple to implement\\ \tabitem Strong baseline for text-based inputs (e.g., SV\\ descriptions in security databases/advisories)} & \makecell[l]{\tabitem May suffer from vocabulary explosion (e.g., many new\\ description words for new SVs)\\ \tabitem No consideration of word context/order (maybe needed\\ for code-based SV analysis)\\ \tabitem Cannot handle Out-of-Vocabulary (OoV) words\\ (can resolve with subwords~\cite{le2019automated})}\\\hline
    
    Word2vec & 
    \makecell[l]{\tabitem Capture nearby context of each word\\ \tabitem Can reuse existing pre-trained model(s)} & \makecell[l]{\tabitem Cannot handle OoV words (can resolve with\\ fastText~\cite{bojanowski2017enriching})\\ \tabitem No consideration of word order}\\\hline
    
    \makecell[l]{DL model end-to-end trainable\\ features} & 
    \makecell[l]{\tabitem Produce SV task-specific features} & \makecell[l]{\tabitem May not produce high-quality representation for tasks\\ with limited data (e.g., real-world exploit prediction)}\\\hline
    
    \makecell[l]{Bidirectional Encoder Repre-\\sentations from Transformers\\(BERT)} & 
    \makecell[l]{\tabitem Capture contextual representation of text (i.e., the\\ feature vector of a word is specific to each input)\\ \tabitem Capture word order in an input\\ \tabitem Can handle OoV words} & \makecell[l]{\tabitem May require GPU to speed up feature inference\\ \tabitem May be too computationally expensive and require too\\ much data to train a strong model from scratch\\ \tabitem May require fine-tuning to work well for a source task}\\\hline
    
    \makecell[l]{Source/expert-defined meta-\\data features} & 
    \makecell[l]{\tabitem Lightweight\\ \tabitem Human interpretable for a task of interest} & \makecell[l]{\tabitem Require SV expertise to define relevant features\\ \tabitem Hard to generalize to new tasks}\\\hline
    
    \multicolumn{3}{|c|}{\cellcolor[HTML]{C0C0C0}}\\
    \multicolumn{3}{|c|}{\multirow{-2}{*}{\cellcolor[HTML]{C0C0C0} \textbf{Element: Prediction Model}}}\\\hline
    
    \makecell[l]{Single ML models (e.g., Linear\\ SVM, Logistic regression,\\Naïve Bayes)} & 
    \makecell[l]{\tabitem Simple to implement\\ \tabitem Efficient to (re-)train on large data (e.g., using the\\ entire NVD database)} & \makecell[l]{\tabitem May be prone to overfitting\\ \tabitem Usually do not perform as well as ensemble/DL models}\\\hline
    
    \makecell[l]{Ensemble ML models (e.g.,\\Random forest, XGBoost,\\LGBM)} & 
    \makecell[l]{\tabitem Strong baseline (usually stronger than single models)\\ \tabitem Less prone to overfitting} & \makecell[l]{\tabitem Take longer to train than single models}\\\hline
    
    \makecell[l]{Latent Dirichlet Allocation\\(LDA -- topic modeling)} & 
    \makecell[l]{\tabitem Require no labeled data for training\\ \tabitem Can provide features for supervised learning models} & \makecell[l]{\tabitem Require SV expertise to manually label generated topics\\ \tabitem May generate human non-interpretable topics}\\\hline
    
    \makecell[l]{Deep Multi-Layer Perceptron\\ (MLP)} & 
    \makecell[l]{\tabitem Work readily with tabular data (e.g., manually defined\\ features or BoW/tf-idf/n-grams)} & \makecell[l]{\tabitem Perform comparably yet are more costly compared to\\ ensemble ML models\\ \tabitem Less effective for unstructured data (e.g., SV descrip-\\ tions)}\\\hline
    
    \makecell[l]{Deep Convolutional Neural\\ Networks (CNN)} & 
    \makecell[l]{\tabitem Capture local and hierarchical patterns of inputs\\ \tabitem Usually perform better than MLP for text-based data} & \makecell[l]{\tabitem Cannot effectively capture sequential order of inputs\\ (maybe needed for code-based SV analysis)}\\\hline
    
    \makecell[l]{Deep recurrent neural networks\\ (e.g., LSTM or Bi-LSTM)} & 
    \makecell[l]{\tabitem Capture short-/long-term dependencies from inputs\\ \tabitem Usually perform better than MLP for text-based data} & \makecell[l]{\tabitem May suffer from the information bottleneck issue\\ (can resolve with attention mechanism~\cite{bahdanau2014neural})\\ \tabitem Usually take longer to train than CNNs}\\\hline
    
    \makecell[l]{Deep graph neural networks\\ (e.g., Graph convolutional\\ network)} & 
    \makecell[l]{\tabitem Capture directed relationships among multiple SV\\ entities and sources} & \makecell[l]{\tabitem Require graph-structured inputs to work\\ \tabitem More computationally expensive than other DL models}\\\hline
    
    \makecell[l]{Deep transfer learning with\\ fine-tuning (e.g., BERT with\\task-specific classification\\layer(s))} & 
    \makecell[l]{\tabitem Can improve the performance for tasks with small\\ data (e.g., real-world exploit prediction)} & \makecell[l]{\tabitem Require target task to have similar nature as source task}\\\hline
    
    \makecell[l]{Deep constrastive learning\\ (e.g., Siamese neural networks)} & 
    \makecell[l]{\tabitem Can improve performance for tasks with small data\\ \tabitem Robust to class imbalance (e.g., CWE classes)} & \makecell[l]{\tabitem Computationally expensive (two inputs instead of one)\\ \tabitem Do not directly produce class-wise probabilities}\\\hline
    
    \makecell[l]{Deep multi-task learning} & 
    \makecell[l]{\tabitem Can share features for predicting multiple tasks (e.g.,\\ CVSS metrics) simultaneously\\ \tabitem Reduce training/maintenance cost} & \makecell[l]{\tabitem Require predicted tasks to be related\\ \tabitem Hard to tune the performance of individual tasks}\\\hline
    
    \multicolumn{3}{|c|}{\cellcolor[HTML]{C0C0C0}}\\
    \multicolumn{3}{|c|}{\multirow{-2}{*}{\cellcolor[HTML]{C0C0C0} \textbf{Element: Evaluation Technique}}}\\\hline
    
    \makecell[l]{Single k-CV without test} & 
    \makecell[l]{\tabitem Easy to implement\\ \tabitem Reduce the randomness of results with multiple folds} & \makecell[l]{\tabitem No separate test set for validating optimized models\\ (can resolve with separate test set(s))\\ \tabitem Maybe infeasible for expensive DL models\\ \tabitem Use future data/SVs for training, may bias results}\\\hline
    
    \makecell[l]{Single/multiple random train/\\test with/without val (using\\val to tune hyperparameters)} & 
    \makecell[l]{\tabitem Easy to implement\\ \tabitem Reduce the randomness of results (the multiple\\ version)} & \makecell[l]{\tabitem May produce unstable results (the single version)\\ \tabitem Maybe infeasible for expensive DL models (the multiple\\ version)\\ \tabitem Use future data/SVs for training, may bias results}\\\hline
    
    \makecell[l]{Single/multiple time-based\\train/test with/without val\\ (using val to tune\\ hyper-\\parameters)} & 
    \makecell[l]{\tabitem Consider the temporal properties of SVs, simulating\\ the realistic evaluation of ever-increasing SVs in practice\\ \tabitem Reduce the randomness of results (the multiple\\ version)} & \makecell[l]{\tabitem Similar drawbacks for the single \& multiple versions as\\ the random counterparts\\ \tabitem May result in uneven/small splits (e.g., many SVs in a\\ year)}\\\hline
    
    \multicolumn{3}{|c|}{\cellcolor[HTML]{C0C0C0}}\\
    \multicolumn{3}{|c|}{\multirow{-2}{*}{\cellcolor[HTML]{C0C0C0} \textbf{Element: Evaluation Metric}}}\\\hline
    
    \makecell[l]{F1-Score/Precision/Recall\\ (classification)} & 
    \makecell[l]{\tabitem Suitable for imbalanced data (common in SV assess-\\ment tasks)} & \makecell[l]{\tabitem Do not consider True Negatives in a confusion matrix\\ (can resolve with Matthews Correlation Coefficient (MCC))}\\\hline
    
    \makecell[l]{Accuracy (classification)} & 
    \makecell[l]{\tabitem Consider all the cells in a confusion matrix} & \makecell[l]{\tabitem Unsuitable for imbalanced data (can resolve with MCC)}\\\hline
    
    \makecell[l]{Area Under the Curve (AUC)\\ (classification)} & 
    \makecell[l]{\tabitem Independent of prediction thresholds} & \makecell[l]{\tabitem May not represent real-world settings (i.e., as models\\ in practice\\ mostly use fixed classification thresholds)\\ \tabitem ROC-AUC may not be suitable for imbalanced data\\ (can resolve with Precision-Recall-AUC)}\\\hline
    
    \makecell[l]{Mean absolute (percentage)\\error/Root mean squared error\\ (regression)} & 
    \makecell[l]{\tabitem Show absolute performance of models} & \makecell[l]{\tabitem Maybe hard to interpret a value on its own without\\ domain knowledge (i.e., whether an error of $x$ is sufficiently\\ effective)}\\\hline
    
    \makecell[l]{Correlation coefficient ($r$)/\\ Coef. of determination ($R^2$)\\ (regression)} & 
    \makecell[l]{\tabitem Show relative performance of models (0 -- 1), where 0\\ is worst \& 1 is best} & \makecell[l]{\tabitem $R^2$ always increases when adding any new feature\\ (can resolve with adjusted $R^2$)}\\\hline

\end{tabular}
\end{table}

We extract and analyze the five key elements for data-driven SV assessment: (\textit{i}) Data sources, (\textit{ii}) Model features, (\textit{iii}) Prediction models, (\textit{iv}) Evaluation techniques and (\textit{v}) Evaluation metrics.
These elements correspond to the four main steps in building data-driven models: data collection (data sources), feature engineering (model features), model training (prediction models) and model evaluation (evaluation techniques/metrics)~\cite{han2011data,sabir2021machine}.
We present the most common practices for each element in Table~\mbox{\ref{tab:data_driven_elements}}.

\subsection{Data Sources}\label{subsec:data_source}

Identifying and collecting rich and reliable SV-related data are the first tasks to build data-driven models for automating SV assessment tasks. As shown in Table~\ref{tab:data_driven_elements}, a wide variety of data sources have been considered to accomplish the five identified themes.

Across the five themes, NVD~\cite{nvd} and CVE~\cite{cve} have been the most prevalently used data sources.
The popularity of NVD/CVE is mainly because they publish expert-verified SV information that can be used to develop prediction models.
Firstly, many studies have considered SV descriptions on NVD/CVE as model inputs. Secondly, the SV characteristics on NVD have been heavily used as assessment outputs in all the themes, e.g., CVSS Exploitability metrics for \textit{Exploitation}, CVSS Impact/Scope metrics for \textit{Impact}, CVSS severity score/levels for \textit{Severity}, CWE for \textit{Type}, CWE/CPE for \textit{Miscellaneous tasks}. Thirdly, external sources on NVD/CVE have enabled many studies to obtain richer SV information (e.g., exploitation availability/time~\cite{chen2019vest} or vulnerable code/crashes~\cite{yan2017exploitmeter,tripathi2017exniffer}) and extract relationships among multiple SV sources to develop a knowledge graph of SVs (e.g.,~\cite{han2018deepweak,xiao2019embedding}).
However, NVD/CVE still suffer from information inconsistencies~\cite{dong2019towards,anwar2020cleaning,anwar2021cleaning} and missing relevant external sources (e.g., SV fixing code)~\cite{hommersom2021automated}.
Such issues motivate future work to validate/clean NVD data and utilize more sources for code-based SV assessment (see section~\ref{sec:conclusions_chap2}).

To enrich the SV information on NVD/CVE, many other security advisories and SV databases have been commonly leveraged by the reviewed studies, notably ExploitDB~\cite{exploitdb}, Symantec~\cite{symantec,symantec_threat}, SecurityFocus~\cite{securityfocus}, CVE Details~\cite{cve_details} and OSVDB. Most of these sources disclose PoC (ExploitDB and OSVDB) and/or real-world (Symantec and Security Focus) exploits. However, real-world exploits are much rarer and different compared to PoC ones~\cite{sabottke2015vulnerability,jacobs2020improving}.
It is recommended that future work should explore more data sources (other than the ones in Table~\ref{tab:exploit_studies_likelihood}) and better methods to retrieve real-world exploits, e.g., using semi-supervised learning~\cite{van2020survey} to maximize the data efficiency for exploit retrieval and/or few-shot learning for tackling the extreme exploit data imbalance issue~\cite{wang2020generalizing}.
Additionally, CVE Details and OSVDB are SV databases like NVD yet with a few key differences. CVE Details explicitly monitors Exploit-DB entries that may be missed on NVD and provides a more user-friendly interface to view/search SVs. OSVDB also reports SVs that do not appear on NVD (without CVE-ID), but this site was discontinued in 2016.

Besides official/expert-verified data sources, we have seen an increasing interest in mining SV information from informal sources that also contain non-expert generated content such as social media (e.g., Twitter) and darkweb. Especially, Twitter has been widely used for predicting exploits as this platform has been shown to contain many SV disclosures even before official databases like NVD~\cite{sabottke2015vulnerability,chen2019vase}. Recently, darkweb forums/sites/markets have also gained traction as SV mentions on these sites have a strong correlation with their exploits in the wild~\cite{almukaynizi2017proactive,almukaynizi2019patch}. However, SV-related data on these informal sources are much noisier because they neither follow any pre-defined structure nor have any verification and they are even prone to fake news~\cite{sabottke2015vulnerability}. Thus, the data integrity of these sources should be checked,
potentially by checking the reputation of posters, to avoid inputting unreliable data to prediction models and potentially producing misleading findings.

\subsection{Model Features}\label{subsec:model_feature}
Collected raw data need to be represented by suitable features for training prediction models.
There are three key types of feature representation methods in this area: term frequency (e.g., BoW, tf-idf and n-grams), DL learned features (e.g., BERT and word2vec) and source/expert-defined metadata (e.g., CVSS metrics and CPE on NVD or tweet properties on Twitter), as summarized in Table~\ref{tab:data_driven_elements}.

Regarding the term-frequency based methods, BoW has been the most popular one. Its popularity is probably because it is one of the simplest ways to extract features from natural language descriptions of SVs and directly compatible with popular ML models (e.g., Linear SVM, Logistic regression and Random forest) in section~\ref{subsec:prediction_model}. Besides plain term count/frequency, other studies have also considered different weighting mechanisms such as inverse document frequency weighting (tf-idf) or tf-igm~\cite{chen2016turning} inverse gravity moment weighting (tf-igm). Tf-igm has been shown to work better than BoW and tf-idf at classifying severity~\cite{kudjo2019improving,chen2020automatic}.
Future work is still needed to evaluate the applicability and generalizability of tf-igm for other SV assessment tasks.

Recently, Neural Network (NN) or DL based features such as word2vec~\cite{mikolov2013distributed} and BERT~\cite{devlin2018bert} have been increasingly used to improve the performance of predicting CVSS exploitation/impact/severity metrics~\cite{han2017learning,gong2019joint}, CWE types~\cite{das2021v2w} and SV information retrieval~\cite{guo2020predicting,guo2021detecting,waareus2020automated}. Compared to BoW and its variants, NN and DL can extract more efficient and context-aware features from vast SV data~\cite{le2020deep}. NN/DL techniques rely on distributed representation to encode SV-related words using fixed-length vectors much smaller than a vocabulary size. Moreover, these techniques capture the sequential order and context (nearby words) to enable better SV-related text comprehension (e.g., SV vs. general \textit{exploit}). Importantly, these NN/DL learned features can be first trained in a non-SV domain with abundant data (e.g., Wikipedia pages~\cite{wikipedia}) and then transferred/fine-tuned in the SV domain to address limited/imbalanced SV data~\cite{yin2020apply}. The main concern with these sophisticated NN/DL features is their limited interpretability, which is an exciting future research area~\cite{zhang2020survey}.

The metadata about SVs can also complement the missing information in descriptions or code for SV assessment. For example, prediction of exploits and their characteristics have been enhanced using CVSS metrics~\cite{almukaynizi2019patch}, CPE~\cite{aksu2018automated} and SV types~\cite{bhatt2021exploitability} on NVD. Additionally, Twitter-related statistics (e.g., number of followers, likes and retweets) have been shown to increase the performance of predicting SV exploitation, impact and severity~\cite{sabottke2015vulnerability,chen2019vest}. Recently, alongside features extracted from vulnerable code, the information about a software development process and involved developers have also been extracted to predict SV fixing effort~\cite{othmane2017time}. Currently, metadata-based and text-based features have been mainly integrated by concatenating their respective feature vectors (e.g.,~\mbox{\cite{chen2019vase,chen2019using,almukaynizi2017proactive,almukaynizi2019patch}}). An alternative yet unexplored way is to build separate models for each feature type and then combine these models using meta-learning (e.g., model stacking~\mbox{\cite{dzeroski2002combining}}).

\subsection{Prediction Models}\label{subsec:prediction_model}

The extracted features enter a wide variety of ML/DL-based prediction models shown in Table~\ref{tab:data_driven_elements} to automate various SV assessment tasks.
Classification techniques have the largest proportion, while regression and unsupervised techniques are less common.

Linear SVM~\cite{cortes1995support} has been the most frequently used classifier, especially in the Exploitation, Impact and Severity themes.
This popularity is reasonable as Linear SVM works well with the commonly used features, i.e., BoW and tf-idf, as mentioned in section~\ref{subsec:model_feature}.
Besides Linear SVM, Random forest, Na\"ive Bayes and Logistic regression have also been common classification models.
In recent years, advanced boosting models (e.g., XGBoost~\cite{chen2016xgboost} and LGBM~\cite{ke2017lightgbm}), and more lately, DL techniques (e.g., CNN~\cite{kim2014convolutional} and (Bi-)LSTM with attention~\cite{bahdanau2014neural}) have been increasingly utilized and shown better results than simple ML models like Linear SVM or Logistic regression.
In this area, some DL models are essential for certain tasks, e.g., building SV knowledge graph from multiple sources with graph neural networks~\mbox{\cite{kipf2016semi}}. DL models also offer solutions to data-related issues such as addressing class imbalance (e.g., deep Siamese network~\mbox{\cite{reimers2019sentence}}) or improving data efficiency (e.g., deep multi-task learning~\mbox{\cite{zhang2021survey}}).
Whenever applicable, it is recommended that future work should still consider simple baselines alongside sophisticated ones as simple methods can perform on par with advanced ones~\mbox{\cite{mazuera2021shallow}}.

Besides classification, various prediction models have also been investigated for regression (e.g., predicting exploit time, severity score and fixing time).
Linear SVM has again been the most commonly used regressor as SV descriptions have usually been the regression input.
Notably, many studies in the Severity theme did not build regression models to directly obtain the severity score (e.g.,~\cite{wen2015novel,ognawala2018automatically,elbaz2020fighting,jiang2020approach,spanos2018multi}). Instead, they used the formulas defined by assessment frameworks (e.g., CVSS versions 2/3~\cite{cvss_v2,cvss_v3} or WIVSS~\cite{spanos2013wivss}) to compute the severity score from the base metrics predicted by respective classification models.
We argue that more effort should be invested in determining the severity score directly from SV data as these severity formulas can be subjective~\mbox{\cite{spring2021time}}. We also observe that there is still limited use of DL models for regression compared to classification.

In addition to supervised (classification/regression) techniques, unsupervised learning has also been considered for extracting underlying patterns of SV data, especially in the Type theme. Latent Dirichlet Allocation (LDA)~\cite{blei2003latent} has been the most commonly used topic model to identify latent topics/types of SVs without relying on a labeled taxonomy. The identified topics were mapped to the existing SV taxonomies such as CWE~\cite{neuhaus2010security} and OWASP~\cite{mounika2019analyzing,vanamala2020topic}.
The topics generated by topic models like LDA can also be used as features for classification/regression models~\cite{ruohonen2017classifying} or building topic-wise models to capture local SV patterns~\cite{menzies2018500+}.
However, definite interpretations for unsupervised outputs are challenging to obtain as they usually rely on human judgement~\mbox{\cite{palacio2019evaluation}}.

\subsection{Evaluation Techniques}\label{subsec:evaluation_technique}

It is important to evaluate a trained model to ensure the model meets certain requirements (e.g., advancing the state-of-the-art).
The evaluation generally needs to be conducted on a different set of data other than the training set to avoid overfitting and objectively estimate model generalizability~\cite{hastie2009elements}. The commonly used evaluation techniques are summarized in Table~\ref{tab:data_driven_elements}.

The reviewed studies have mostly used one or multiple validation and/or test sets\footnote{Validation set(s) helps optimize/tune a model (finding the best task/data-specific hyperparameters), and test set(s) evaluates the optimized/tuned model. Using only validation set(s) means evaluating a model with default/pre-defined hyperparameters.} to evaluate their models, in which each validation/test set has been either randomly or time-based selected. Specifically, k-fold cross-validation has been one of the most commonly used techniques. The number of folds has usually been 5 or 10, but less standard values like 4~\cite{yan2017exploitmeter} have also been used. However, k-fold cross-validation uses all parts of data at least once for training; thus, there is no hidden test set to evaluate the optimal model with the highest (cross-)validation performance.

To address the lack of hidden test set(s), a common practice in the studied papers has been to split a dataset into single training and test sets, sometimes with an additional validation set for tuning hyperparameters to obtain an optimal model. Recently, data has been increasingly split based on the published time of SVs to better reflect the changing nature of ever-increasing SVs~\cite{le2019automated}.\footnote{This study is presented in Chapter~\ref{chap:msr19}.}
However, the results reported using single validation/test sets may be unstable (i.e., unreproducible results using different set(s))~\cite{raschka2018model}.

To ensure both the time order and reduce the result randomness, we recommend using multiple splits of training and test sets in combination with time-based validation in each training set. Statistical analyses (e.g., hypothesis testing and effect size) should also be conducted to confirm the reliability of findings with respect to the randomization of models/data in multiple runs~\cite{de2019evolution}.

\subsection{Evaluation Metrics}\label{subsec:evaluation_metric}

Evaluating different aspects of a model requires respective proper metrics.
The popular metrics for evaluating the tasks in each theme are given in Table~\ref{tab:data_driven_elements}.

Across the five themes, Accuracy, Precision, Recall and F1-Score~\cite{jiao2016performance} have been the most commonly used metrics because of a large number of classification tasks in the five themes.
However, Accuracy is not a suitable measure for SV assessment tasks with imbalanced data (e.g., SVs with real-world exploits vs. non-exploited SVs).
The sample size of one class is much smaller than the others, and thus the overall Accuracy would be dominated by the majority classes.
Besides these four commonly used metrics, AUC based on the ROC curve (ROC-AUC)~\cite{jiao2016performance} has also been considered as it is threshold-independent. However, we suggest that ROC-AUC should be used with caution in practice as most deployed models would have a fixed decision threshold (e.g., 0.5). Instead of ROC-AUC, we suggest Matthews Correlation Coefficient~\cite{jiao2016performance} (MCC) as a more meaningful evaluation metric to be considered as it explicitly captures all values in a confusion matrix, and thus has less bias in results.

For regression tasks, various metrics have been used such as Mean absolute error, Mean absolute percentage error, Root mean squared error~\cite{spanos2018multi} as well as Correlation coefficient ($r$) and Coefficient of determination ($R^2$)~\cite{othmane2017time}. Note that \textit{adjusted} $R^2$ should be preferred over $R^2$ as $R^2$ would always increase when adding a new (even irrelevant) feature.

A model can have a higher value of one metric yet lower values of others.\footnote{\url{https://stackoverflow.com/questions/34698161}} Therefore, we suggest using a combination of suitable metrics for a task of interest to avoid result bias towards a specific metric. Currently, most studies have focused on evaluating model effectiveness, i.e., how well the predicted outputs match the ground-truth values. Besides effectiveness, other aspects (e.g., efficiency in training/deployment and robustness to input changes) of models should also be evaluated to provide a complete picture of model applicability in practice.

\section{Chapter Summary}
\label{sec:conclusions_chap2}

SV assessment is crucial to optimize resource utilization in addressing SVs at scale. This phase has witnessed radical transformations following the increasing availability of SV data from multiple sources and advances in data-driven techniques. We presented a taxonomy to summarize the five main directions of the research work so far in this area. We also identified and analyzed the key practices to develop data-driven models for SV assessment in the reviewed studies.

Despite the great potential of data-driven approaches for SV assessment, we highlighted the following three open challenges limiting the practical application of the field:\footnote{The presented challenges are the ones that have been addressed in this thesis. More challenges related to data-driven SV assessment can be found in Chapter~\ref{chap:conclusions_future_work}.}

\begin{contribution} 
\item \textbf{Unrealistic evaluation settings for report-level SV assessment models}. Most of the reviewed studies have evaluated their prediction models without capturing many factors encountered during the deployment of such models to production.
Specifically, the models used in practice would require to handle new SV data over time.
In the context of report-level SV assessment, Out-of-Vocabulary (OoV) words in SV descriptions of new SVs need to be properly accommodated to avoid performance degradation of prediction models.
The impact of using time-based splits rather than random splits (e.g., k-fold cross-validation) for these models to avoid leaking unseen (future) patterns to the model training also requires further investigation.
We address the first challenge in Chapter~\ref{chap:msr19}.

\item \textbf{Untimely SV assessment models}. Although SV descriptions have been commonly used as model inputs (see section~\ref{subsec:data_source}), these descriptions are usually published long after SVs introduced/discovered~\cite{meneely2013patch} and even after SV are fixed~\cite{li2017large,piantadosi2019fixing} in codebases. One potential solution to this issue is to directly perform SV assessment using (vulnerable) source code. The key benefit of using vulnerable code for SV assessment is that such code is always available/required before SV fixing. Thus, performing code-based SV assessment can be done even when SV reports are not (yet) available. Version control systems like GitHub\footnote{\url{https://github.com}} can provide such vulnerable code for SV assessment.
Leveraging data from these version control systems, we address the second challenge in Chapters~\ref{chap:msr22} and~\ref{chap:ase21}, in which data-driven SV assessment is investigated on the code function and code commit levels, respectively.

\item \textbf{Lack of utilization of developers' real-world SV concerns}. The current SV data hardly contain specific developers' concerns and practices when addressing real-world SVs. \textit{Developer Question \& Answer (Q\&A) platforms} like Stack Overflow\footnote{\url{https://stackoverflow.com}} and Security StackExchange\footnote{\url{https://security.stackexchange.com}} contain millions of posts about different challenges and solutions shared by millions of developers when tackling different software-related issues in real-world scenarios. The rich data on Q\&A sites can be collected and analyzed to determine the key concerns that practitioners are facing while addressing SV in practice. Such SV concerns can be incorporated into other technical metrics like CVSS metrics for more thoroughly assessing and prioritizing SVs. For example, the fixing effort of SVs may depend on the technical difficulty of implementing the respective mitigation strategies in a language or system of interest.
We address the third challenge in Chapter~\ref{chap:ease21}.
\end{contribution}

\section{Appendix - Ever-Growing Literature on Data-Driven SV Assessment}
\label{sec:appendix_lit_review}
Given that data-driven SV assessment is an emerging field, there have been many new contributions in this area since this literature review was conducted in April 2021. In this appendix, we would like to briefly outline the recent trends in data-driven SV assessment from May 1, 2021 to February 28, 2022 (at the time of writing this thesis). We have also maintained a website (\url{https://github.com/lhmtriet/awesome-vulnerability-assessment}) to keep track of the latest contributions of data-driven SV assessment for researchers and practitioners in the field.

To obtain new papers from May 2021 to February 2022, we followed the same methodology of study selection, as described in section~\ref{subsec:methodology_csur22}. We obtained 228 new studies returned from the search on the online databases. We then applied the inclusion/exclusion criteria, as given in \tab~\ref{tab:inclusion_exclusion_csur22}, on the titles, abstracts and full-texts of the curated papers to obtain the list of relevant papers. After the filtering steps, we selected 18 papers relevant to data-driven SV assessment from May 2021 to February 2022.\footnote{We do not claim that this list of updated papers is complete, yet we believe that we have covered the representative ones. It should also be noted that we did not include our own papers as these papers would be presented in subsequent chapters of this thesis.} The categorization of these selected papers into the five themes, as described in section~\ref{subsec:taxonomy_csur22}, is given in \tab~\ref{tab:updated_papers_csur22}.

\begin{table}[t]
    \centering
    \caption{The mapping between the themes/tasks and the respective studies collected from May 2021 to February 2022.}
    \begin{tabular}{lll}
        \hline
         \textbf{Theme/Task} & \textbf{Studies}
         \\
         \hline
         Exploitation Prediction & \cite{ampel2021linking}, \cite{kuehn2021ovana}, \cite{kekul2021multiclass}, \cite{shahid2021cvss}, \cite{lyu2021character}, \cite{babalau2021severity}, \cite{charmanas2021predicting}, \cite{yin2022vulnerability}, \cite{bulut2021nl2vul}
         \\
         \hline
         Impact Prediction & \cite{kuehn2021ovana}, \cite{kekul2021multiclass}, \cite{shahid2021cvss}, \cite{babalau2021severity}, \cite{bulut2021nl2vul} \\
         \hline
         Severity Prediction & \cite{kuehn2021ovana}, \cite{kekul2021multiclass}, \cite{shahid2021cvss}, \cite{babalau2021severity}, \cite{bulut2021nl2vul} \\
         \hline
         Type Prediction & \cite{ramesh2021automatic}, \cite{aivatoglou2021tree}, \cite{vishnu2022deep}, \cite{wang2022automatic}, \cite{yosifova2021predicting}
         \\
         \hline
         Miscellaneous Tasks & \cite{kuehn2021ovana}, \cite{yang2021few}, \cite{yitagesu2021unsupervised}, \cite{yuan2021predicting}, \cite{guo2021key}
         \\
         \hline
         
    \end{tabular}
    \label{tab:updated_papers_csur22}
\end{table}

The key patterns and practices of these 18 papers are summarized as follows:

\begin{itemize}
    \item All the tasks tackled by the studies have aligned with the ones presented in this literature review, reinforcing the robustness of our taxonomy presented in \fig~\ref{fig:taxonomy_csur22}.
    \item Among the \textit{themes}, Exploitation has attracted the most contributions from these new studies, similar to that of the reviewed papers prior to May 2021. In addition, prediction of the CVSS exploitability, impact, and severity metrics is still the most common task. However, it is encouraging to see that more studies have worked on CVSS version 3.1, which is closer to the current industry standard.
    \item Regarding \textit{data sources}, NVD/CVE is still most prevalently used. Besides CAPEC~\cite{capec}, MITRE ATT\&CK Framework\footnote{\url{https://attack.mitre.org/}} is a new source of attack patterns being utilized~\cite{ampel2021linking}.
    \item Regarding \textit{model features}, BERT~\cite{devlin2018bert} has been commonly used to extract contextual feature embeddings from SV descriptions for various tasks.
    \item Regarding \textit{prediction models}, DL techniques, which are mainly based on CNN and/or (Bi-)LSTM with attention, have been increasingly used to improve the performance of the tasks.
    \item Regarding \textit{evaluation techniques} and \textit{evaluation metrics}, since most of the tasks are the same as before, evaluation practices have largely stayed the same as the ones presented in sections~\ref{subsec:evaluation_technique} and~\ref{subsec:evaluation_metric}.
    
\end{itemize}

Overall, the three key practical challenges presented in section~\ref{sec:conclusions_chap2} are still not addressed by the new studies. Thus, we believe that our contributions/solutions to address these challenges in this thesis would set the foundation for future research to further improve the practical applicability of SV assessment using data-driven approaches.

%% file: Chapters/Chapter_3_MSR2019.tex
\chapter{Automated Report-Level Software Vulnerability Assessment with Concept Drift}
\label{chap:msr19}

\begin{tcolorbox}
\textbf{Related publication}: This chapter is based on our paper titled ``\textit{Automated Software Vulnerability Assessment with Concept Drift}'' published in the 16\textsuperscript{th} International Conference on Mining Software Repositories (MSR), 2019 (CORE A)~\cite{le2019automated}.
\end{tcolorbox}
\bigskip

In our literature review in Chapter~\ref{chapter:lit_review}, Software Engineering researchers are increasingly using Natural Language Processing (NLP) techniques to automate Software Vulnerability (SV) assessment using SV descriptions in public repositories. However, the existing NLP-based approaches suffer from \emph{concept drift}. This problem is caused by a lack of proper treatment of new (out-of-vocabulary) terms for evaluating unseen SVs over time. To perform automated SV assessment with \emph{concept drift} using SV descriptions present in SV reports, in Chapter~\ref{chap:msr19}, we propose a systematic approach that combines both character and word features. The proposed approach is used to predict seven Vulnerability Characteristics (VCs). The optimal model of each VC is selected using our customized time-based cross-validation method from a list of eight NLP representations and six well-known Machine Learning models. We use the proposed approach to conduct large-scale experiments on more than 100,000 SVs in National Vulnerability Database (NVD). The results show that our approach can effectively tackle the \emph{concept drift} issue of the SVs' descriptions reported from 2000 to 2018 in NVD even without retraining the model. In addition, our approach performs competitively compared to the existing word-only method. We also investigate how to build compact \emph{concept-drift}-aware models with much fewer features and give some recommendations on the choice of classifiers and NLP representations for report-level SV assessment.
\newpage

\section{Introduction}
\label{sec:introduction_msr19}

Software Vulnerability (SV) reports in public repositories, such as National Vulnerability Database (NVD)~\cite{nvd}, have been widely leveraged to automatically predict SV characteristics using data-driven approaches (see Chapter~\ref{chapter:lit_review}). However, data in these SV reports have the temporal property since many new terms appear in the descriptions of SVs. Such terms are a result of the release of new technologies/products or the discovery of a zero-day attack or SV; for example, NVD received more than 13,000 new SVs in 2017~\cite{nvd}. The appearance of new concepts makes SV data and patterns change over time~\cite{williams2018analyzing,murtaza2016mining,neuhaus2010security}, which is known as \emph{concept drift}~\cite{bullough2017predicting}. For example, the keyword Android has only started appearing in NVD since 2008, the year when Google released Android. We assert that such new SV terms can cause problems for building report-level SV assessment models.

Previous studies~\cite{spanos2018multi,almukaynizi2019patch,bozorgi2010beyond} have suffered from \emph{concept drift} as they have usually mixed new and old SVs in the model validation step. Such approach accidentally merges new SV information with existing one, which can lead to biased results. Moreover, the previous work of SV analysis~\cite{spanos2018multi,spanos2017assessment,almukaynizi2019patch,bozorgi2010beyond,toloudis2016associating,yamamoto2015text,edkrantz2015predicting} used predictive models with only word features without reporting how to handle novel or extended concepts (e.g., new versions of the same software) in new SVs' descriptions. Research on machine translation~\cite{luong2014addressing,huang2011using,liu2018context,razmara2013graph} has shown that unseen (Out-of-Vocabulary (OoV)) terms can make existing word-only models less robust to future prediction due to their missing information. For SV prediction, Han et al.~\cite{han2017learning} did use random embedding vectors to represent the OoV words, which still discards the relationship between new and old concepts. Such observations motivated us to tackle the research problem ``\textbf{How to handle the \emph{concept drift} issue of the SV descriptions in public repositories to improve the robustness of automated report-level SV assessment?}'' It appears to us that it is important to address the issue of \emph{concept drift} to enable the practical applicability of automated SV assessment tools. To the best of our knowledge, there has been no existing work to systematically address the \emph{concept drift} issue in report-level SV assessment.

To perform report-level SV assessment with \emph{concept drift} using SV descriptions in public repositories, we present a Machine Learning (ML) model that utilizes both character-level and word-level features. We also propose a customized time-based version of cross-validation method for model selection and validation. Our cross-validation method splits the data by year to embrace the temporal relationship of SVs. We evaluate the proposed model on the prediction of seven Vulnerability Characteristics (VCs), i.e., Confidentiality, Integrity, Availability, Access Vector, Access Complexity, Authentication, and Severity. Our key \textbf{contributions} are:

\begin{contribution}
  \item We demonstrate the \emph{concept drift} issue of SVs using concrete examples from NVD.
  \item We investigate a customized time-based cross-validation method to select the optimal ML models for SV assessment. Our method can help prevent future SV information from being leaked into the past in model selection and validation steps.
  \item We propose and extensively evaluate an effective Character-Word Model (CWM) to assess SVs using the descriptions with \emph{concept drift}. We also investigate the performance of low-dimensional CWM models. We provide our models and associated source code for future research at \url{https://github.com/lhmtriet/MSR2019}.
\end{contribution}

\textbf{Chapter organization}. Section~\ref{sec:background_msr19} introduces SV descriptions and VCs. Section~\ref{sec:methodology_msr19} describes our proposed approach. Section~\ref{sec:expt_setup_msr19} presents the experimental design of this work. Section~\ref{sec:results_msr19} analyzes the experimental results and discusses the findings. Section~\ref{sec:discussion_msr19} identifies the threats to validity. Section~\ref{sec:related_work_msr19} covers the related works. Section~\ref{sec:conclusions_msr19} concludes and suggests some future directions.

\section{Background}
\label{sec:background_msr19}

\begin{figure}[t]
    \centering
    \includegraphics[width=\linewidth,keepaspectratio]{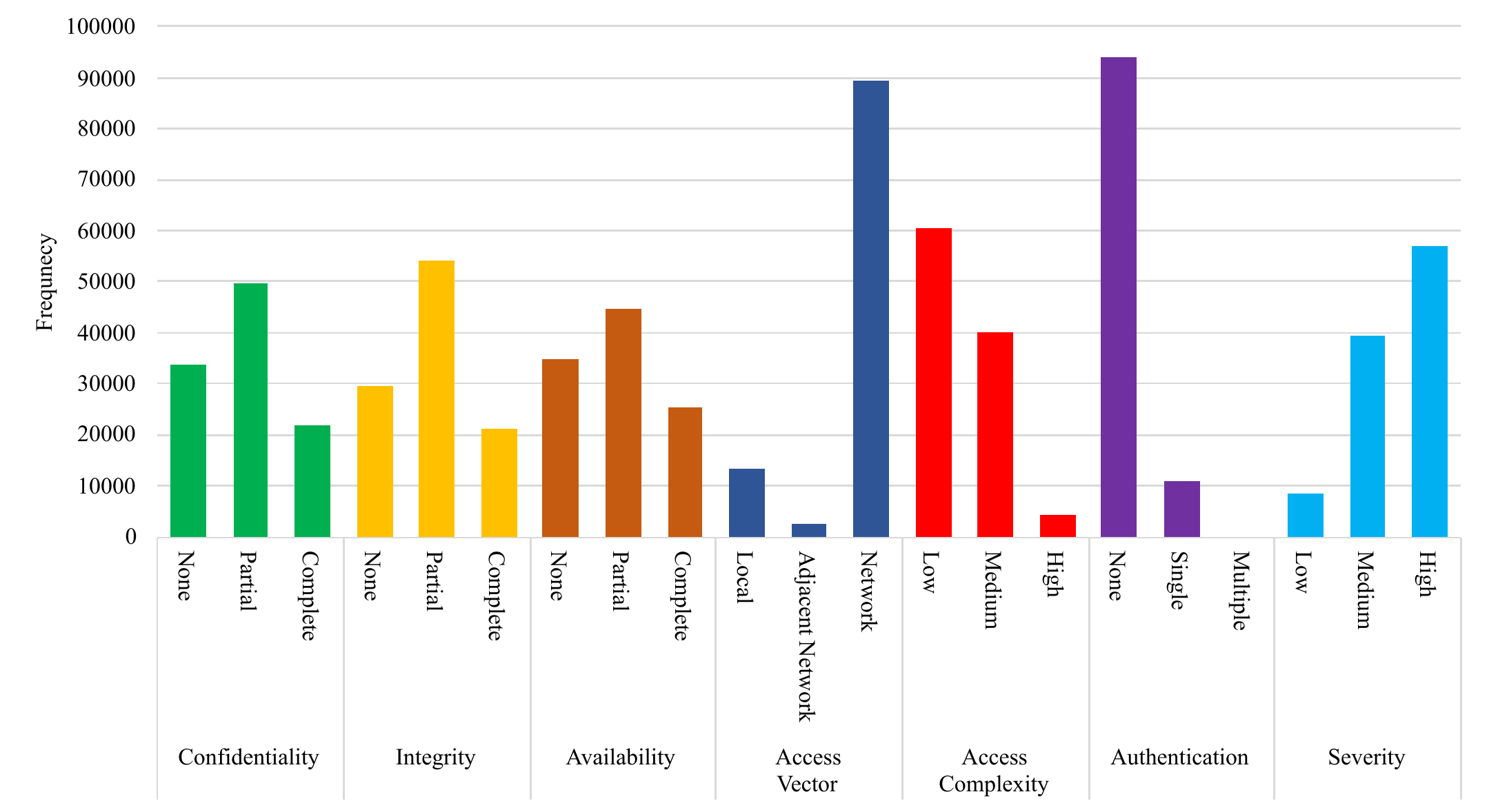}
    \caption[Frequencies of each class of the seven vulnerability characteristics.]{Frequencies of each class of the seven VCs.}
    \label{fig:vc_freqs_msr19}
\end{figure}

Software Vulnerability (SV) assessment is an important step in the SV lifecycle, which determines various characteristics of detected SVs~\cite{smyth2017software}. Such characteristics support developers to understand the nature of SVs, which can inform prioritization and remediation strategies. For example, if an SV can severely damage the confidentiality of a system, e.g., allowing attackers to access/steal sensitive information, this SV should have a high fixing priority. A fixing protocol to ensure confidentiality can then be followed, e.g., checking/enforcing privileges to access the affected component/data.

National Vulnerability Database~\cite{nvd} (NVD) is one of the most popular and trustworthy sources for SV assessment. NVD is maintained by governmental bodies (National Cyber Security and Division of the United States Department of Homeland Security). This site inherits unique SV identifiers and descriptions from Common Vulnerabilities and Exposures (CVE)~\cite{cve}. For SV assessment, NVD provides expert-verified assessment metrics, namely Common Vulnerability Scoring System (CVSS)~\cite{cvss}, for each reported SV.

CVSS is one of the most commonly used frameworks by both researchers and practitioners to perform SV assessment. There are two main versions of CVSS, namely versions 2 and 3, in which version 3 only came into effect in 2015. CVSS version 2 is still widely used as many SVs prior to 2015 can yet pose threats to contemporary systems. For instance, the SV with CVE-2004-0113 first found in 2004 was exploited in 2018~\mbox{\cite{old_sv_exploit}}. Hence, we adopt the assessment metrics of CVSS version 2 as the outputs for the SV assessment models in this study.

CVSS version 2 provides metrics to quantify the three main aspects of SVs, namely exploitability, impact, and severity.
We focus on the \textit{base} metrics because the temporal metrics (e.g., exploit availability in the wild) and environmental metrics (e.g., potential impact outside of a system) are unlikely obtainable from project artifacts (e.g., SV code/reports) alone.
Specifically, the base \textit{Exploitability} metrics examine the technique (Access Vector) and complexity to initiate an exploit (Access Complexity) as well as the authentication requirement (Authentication). The base \textit{Impact} metrics of CVSS focus on the system Confidentiality, Integrity, and Availability. The Exploitation and Impact metrics are used to compute the \textit{Severity} of SVs. Severity approximates the criticality of an SV. Nevertheless, relying solely on Severity may be insufficient because an SV with medium severity may still have high impacts as it is considerably complex to be exploited. It is important to assign a high fixing priority to such an SV as an affected system would face tremendous risks in case of a successful cyber-attack. Therefore, in this study, we consider all the base metrics of CVSS version 2 (i.e., Confidentiality, Integrity, Availability, Access Vector, Access Complexity, Authentication and Severity), as shown in Fig.~\ref{fig:vc_freqs_msr19}, for developing SV assessment models. From the perspective of ML, predicting CVSS metrics is a classification problem, which can be solved readily using ML algorithms. It is noted that Access Vector, Access Complexity and Authentication characteristics suffer the most from the issue of imbalanced data, in which the number of elements in the minority class is much smaller compared to those of the other classes.

\section{The Proposed Approach}
\label{sec:methodology_msr19}

\subsection{Approach Overview}
\label{subsec:method_overview_msr19}

\begin{figure*}[t]
    \centering
    \includegraphics[width=\linewidth,keepaspectratio]{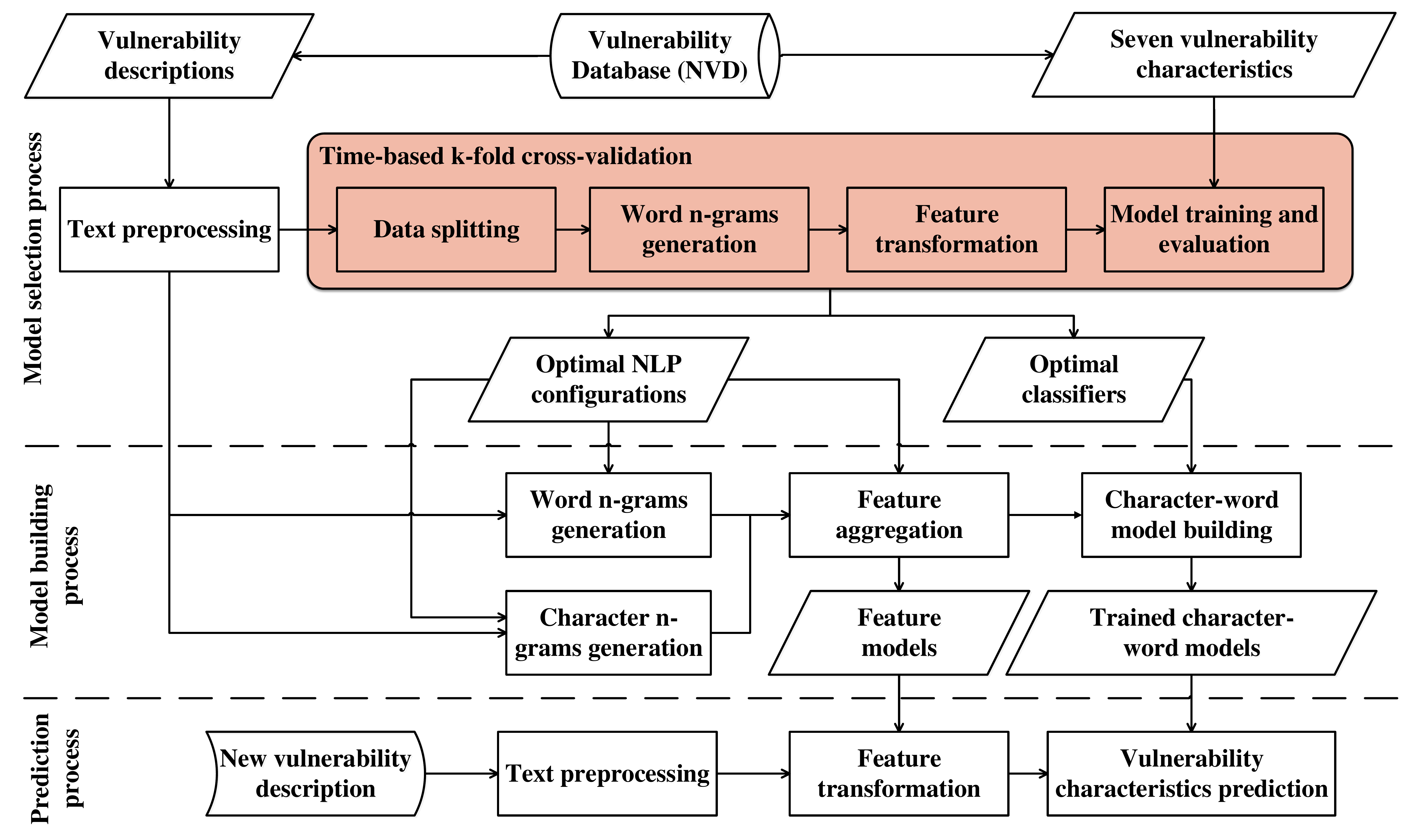}
    \caption{Workflow of our proposed model for report-level software vulnerability assessment with \emph{concept drift}.}
    \label{fig:workflow_msr19}
\end{figure*}

The overall workflow of our proposed approach is given in Fig.~\ref{fig:workflow_msr19}. Our approach consists of three processes: model selection, model building and prediction. The first two processes work on the training set, while the prediction process performs on either a separate testing set or new SV descriptions. The first model selection has two steps: Text preprocessing and Time-based k-fold cross-validation. The text preprocessing step (see section~\ref{subsec:preprocessing_msr19}) is necessary to reduce noise in text to build a better assessment model. Next, the preprocessed text enters the time-based k-fold cross-validation step to select the optimal classifier and Natural Language Processing (NLP) representations for each VC. It should be noted that this step only tunes the word-level models instead of the combined models of both word and character features. One reason is that the search space of the combined model is much larger than that of the word-only model since we at least have to consider different NLP representations for character-level features. The computational resource to extract character-level n-grams is also more than that of word-level counterparts. Section~\ref{subsec:time_cross-validation_msr19} provides more details about the time-based k-fold cross-validation method.

Next comes the model building process with four main steps: (\emph{i}) word n-grams generation, (\emph{ii}) character n-grams generation, (\emph{iii}) feature aggregation and (\emph{iv}) character-word model building. Steps (\emph{i}) and (\emph{ii}) use the preprocessed text in the previous process to generate word and character n-grams based on the identified optimal NLP representations of each VC. The word n-grams generation step (\emph{i}) here is the same as the one in the time-based k-fold cross-validation of the previous process. An example of the word and character n-grams in our approach is given in Table~\ref{tab:feature_examples_msr19}. Such character n-grams increase the probability of capturing parts of OoV terms due to \emph{concept drift} in SV descriptions. Subsequently, both levels of the n-grams and the optimal NLP representations are input into the feature aggregation step (\emph{iii}) to extract features from the preprocessed text using our proposed algorithm in section~\ref{subsec:feature_aggregation_msr19}. This step also combines the aggregated character and word vocabularies with the optimal NLP representations of each VC to create the feature models. We save such models to transform data of future prediction. In the last step (\emph{iv}), the extracted features are trained with the optimal classifiers found in the model selection process to build the complete character-word models for each VC to perform automated report-level SV assessment with \emph{concept drift}.

In the prediction process, a new SV description is first preprocessed using the same text preprocessing step. Then, the preprocessed text is transformed to create features by the feature models saved in the model building process. Finally, the trained character-word models use such features to determine each VC.

\begin{table}[t]
  \centering
  \caption{Word and character n-grams extracted from the sentence ``Hello World''. `\_' represents a space.}
    \begin{tabular}{cll}
    \hline
    \textbf{n-grams} & \multicolumn{1}{c}{\textbf{Words}} & \multicolumn{1}{c}{\textbf{Characters}} \\
    \hline
    1 & Hello, World & H, e, l, l, o, W, o, r, l, d \\
    \hline
    2 & Hello World & He, el, ll, lo, o\_, \_W, Wo, or, rl, ld \\
    \hline
    \end{tabular}
  \label{tab:feature_examples_msr19}
\end{table}

\subsection{Text Preprocessing of SV Descriptions}
\label{subsec:preprocessing_msr19}

The text preprocessing is an important step for any NLP task~\cite{kao2007natural}. We use the following text preprocessing techniques: (\emph{i}) removal of stop words and punctuations, (\emph{ii}) conversion to lowercase and (\emph{iii}) stemming. The stop words are combined from the default lists of the \emph{scikit-learn}~\cite{pedregosa2011scikit} and \emph{nltk}~\cite{loper2002nltk} libraries. We only remove the punctuations followed by at least one space or the ones at the end of a sentence. This punctuation removal method keeps important words in the software and security contexts such as ``\emph{input.c}'', ``\emph{man-in-the-middle}'', ``\emph{cross-site}''.

Subsequently, the stemming step is done using the Porter Stemmer algorithm~\cite{porter1980algorithm} in the \emph{nltk} library. Stemming is needed to avoid two or more words with the same meaning but in different forms (e.g., ``\emph{allow}'' vs. ``\emph{allows}''). The main goal of stemming is to retrieve consistent features (words), thus any algorithm that can return each word's root should work. Researchers may use lemmatization, which is relatively slower as it also considers the surrounding context.

\subsection{Model Selection with Time-based k-Fold Cross-Validation}
\label{subsec:time_cross-validation_msr19}

\begin{figure}[t]
    \centering
    \includegraphics[width=\linewidth,keepaspectratio]{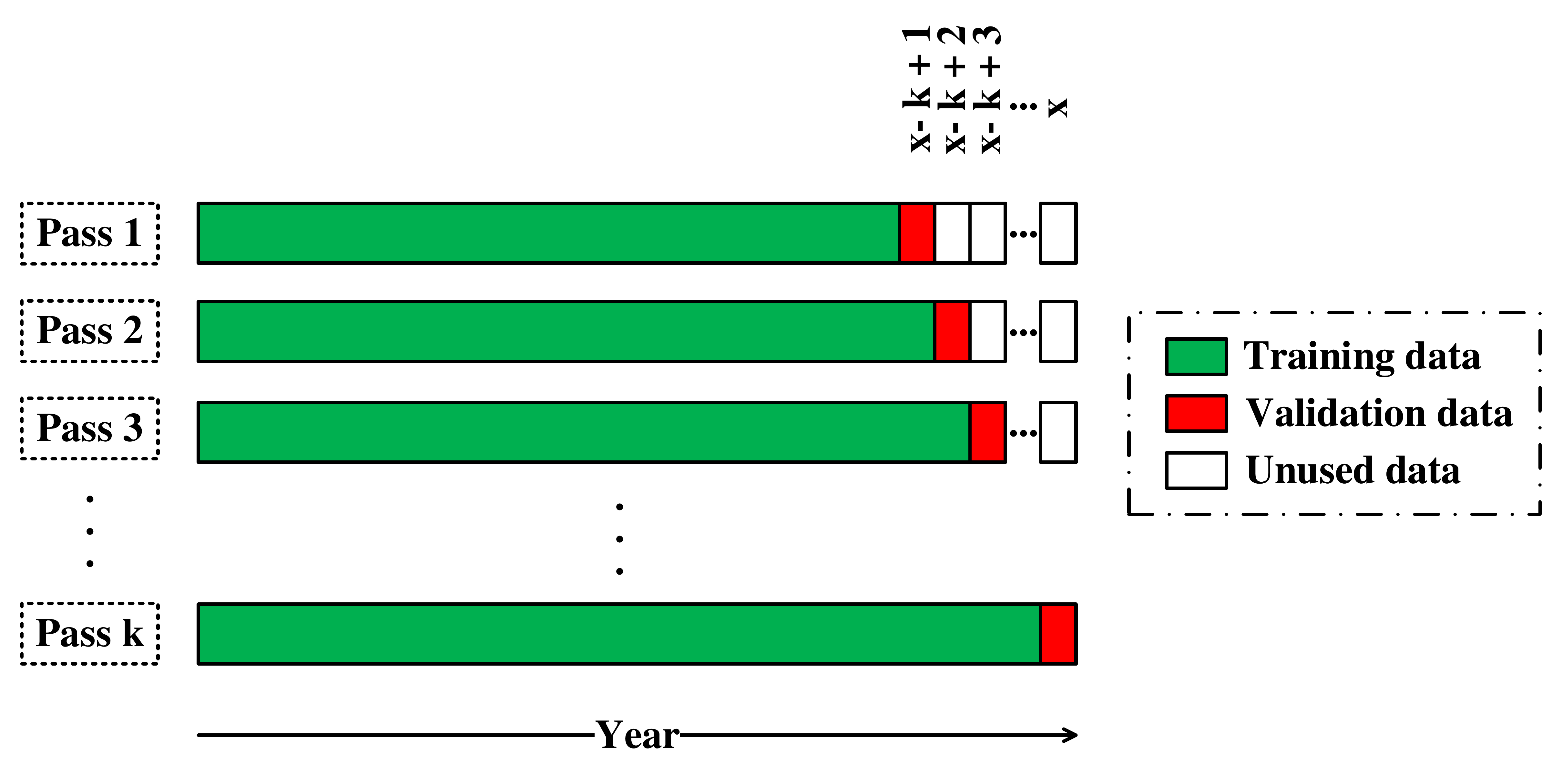}
    \caption[Our proposed time-based cross-validation method.]{Our proposed time-based cross-validation method. \textbf{Note}: x is the final year in the original training set and k is the number of cross-validation folds.}
    \label{fig:time_validation_msr19}
\end{figure}

We propose a time-based cross-validation method (see Fig.~\ref{fig:time_validation_msr19}) to select the best classifier and NLP representation for each VC. The idea has been inspired by the time-series domain~\cite{bergmeir2012use}. As shown in Fig.~\ref{fig:workflow_msr19}, our method has four steps: (\emph{i}) data splitting, (\emph{ii}) word n-grams generation, (\emph{iii}) feature transformation, and (\emph{iv}) model training and evaluation. Data splitting explicitly considers the time order of SVs to ensure that in each pass/fold, the new information of the validation set does not exist in the training set, which maintains the temporal property of SVs. New terms can appear at different time during a year; thus, the preprocessed text in each fold is split by year explicitly, not by equal sample size as in traditional time-series splits;\footnote{\url{https://scikit-learn.org/stable/modules/generated/sklearn.model_selection.TimeSeriesSplit.html}} e.g., SVs from 1999 to 2010 are for training and those in 2011 are for validation in a pass/fold.

\begin{table}[t]
  \centering
  \caption[The eight configurations of Natural Language Processing representations used for model selection.]{The eight configurations of NLP representations used for model selection. \textbf{Note}: `\checkmark' is selected, `-' is non-selected.}
    \begin{tabular}{ccc}
    \hline
    \textbf{Configuration} & \textbf{Word n-grams} & \textbf{tf-idf} \\
    \hline
    1     & 1     & - \\
    \hline
    2     & 1     & \checkmark \\
    \hline
    3     & 1-2   & - \\
    \hline
    4     & 1-3   & - \\
    \hline
    5     & 1-4   & - \\
    \hline
    6     & 1-2   & \checkmark \\
    \hline
    7     & 1-3   & \checkmark \\
    \hline
    8     & 1-4   & \checkmark \\
    \hline
    \end{tabular}
  \label{tab:model_configurations_msr19}
\end{table}

After data splitting in each fold, we use the training set to generate the word n-grams. Subsequently, with each of the eight NLP configurations in Table~\ref{tab:model_configurations_msr19}, the feature transformation step uses the word n-grams as the vocabulary to transform the preprocessed text of both training and validation sets into the features for building a model. We create the NLP configurations from various values of n-grams combined with either term frequency or tf-idf measure. Uni-gram with term frequency is also called Bag-of-Words (BoW). These NLP representations have been selected since they are popular and have performed well for SV analysis~\cite{spanos2018multi,almukaynizi2019patch,yamamoto2015text}. For each NLP configuration, the model training and evaluation step trains six classifiers (see section~\ref{subsec:ml_models_msr19}) on the training set and then evaluates the models on the validation set using different evaluation metrics (see section~\ref{subsec:evaluation_metrics_msr19}). The model with the highest average cross-validated performance is selected for a VC. The process is repeated for every VC, then the optimal classifiers and NLP representations are returned for all seven VCs.

\subsection{Feature Aggregation Algorithm}
\label{subsec:feature_aggregation_msr19}

We propose Algorithm~\ref{algo:feature_aggregation} to combine word and character n-grams in the model building process to create features for our character-word model. Six inputs of the algorithm are (\emph{i}) input descriptions, (\emph{ii}) word n-grams, (\emph{iii}) character n-grams, (\emph{iv}) the minimum, (\emph{v}) the maximum number of character n-grams, and (\emph{vi}) the optimal NLP configuration of a VC. The main output is a feature matrix containing the term weights of the documents transformed by the aggregated character and word vocabularies to build the character-word models. We also output the character and word feature models for future prediction of VCs.

\begin{algorithm}[t]
    \caption{Feature aggregation algorithm to transform the documents with the aggregated word and character-level features.}
    \label{algo:feature_aggregation}
    \DontPrintSemicolon
    \SetAlgoNoLine

    \KwIn{List of SV descriptions: ${D}_{in}$\\
    Set of word-level n-grams: ${F}_{w}=\{{{f}_{1w}},{{f}_{2w}},...,{{f}_{nw}}\}$\\
    Set of character-level n-grams: ${F}_{c}=\{{{f}_{1c}},{{f}_{2c}},...,{{f}_{mc}}\}$\\
    The minimum and maximum character n-grams: ${min}_{n-gram}$ and ${max}_{n-gram}$\\
    The optimal NLP configuration of the current VC: \emph{config}}
    \KwOut{The aggregated data matrix: ${\mathbf{X}}_{agg}$\\
    The word and character feature models: ${\text{model}}_{w}, {\text{model}}_{c}$}

    $slt\_chars \leftarrow \emptyset$\;

    \ForEach{${f}_{i}\in {F}_{c}$}{
        $tokens \leftarrow {f}_{i}$ trimmed and split by space\;

        \If{(size of $tokens$ = 1) {\bf and} ((length of the first element in $tokens$) $>$ 1)}{
            $slt\_chars \leftarrow slt\_chars + \{tokens\}$\;
        }
    }

    $diff\_words \leftarrow {F}_{w} - slt\_chars$\;

    ${\text{model}}_{w} \leftarrow \text{Feature\_transformation}(diff\_words, config)$\;

    ${\text{model}}_{c} \leftarrow \text{Feature\_transformation}(slt\_chars, {min}_{n-gram} - 1, {max}_{n-gram}, config)$\;

    ${\mathbf{X}}_{word} \leftarrow {D}_{in}$ transformed with ${\text{model}}_{w}$\;
    ${\mathbf{X}}_{char} \leftarrow {D}_{in}$ transformed with ${\text{model}}_{c}$\;
    ${\mathbf{X}}_{agg} \leftarrow \text{horizontal\_append}({\mathbf{X}}_{word}, {\mathbf{X}}_{char})$\;

    \Return ${\mathbf{X}}_{agg}$, ${\text{model}}_{w}$, ${\text{model}}_{c}$

\end{algorithm}

Steps 2-7 of the algorithm filter the character features. More specifically, step 3 removes (trims) spaces from both ends of each feature. Then, we split such feature by space(s) to determine how many words to which the character(s) belongs. Subsequently, steps 4-6 retain only the character features that are parts of single words (size of $tokens$ = 1), except the single characters such as $x$, $y$, $z$ ((length of the first element in $tokens$) $>$ 1). The n-gram characters with space(s) in between represent more than one word, which can make a classifier more prone to overfitting. Similarly, single characters are too short and they can belong to too many words, which is likely to make a model hardly generalizable. In fact, `\emph{a}' is a meaningful single character, but it has been already removed as a stop word. The characters can even represent a whole word (e.g., ``\emph{attack}'' token with ${max}_{n-gram}\ge \,6$). In such cases, step 8 removes duplicated word-level features (${F}_{w} - slt\_chars$). Based on the assumption that unseen or misspelled terms can share common characters with existing words, such choice can enhance the probability of a model capturing the OoV words in new descriptions. Retaining only the character features also helps reduce the number of features and model overfitting. After that, steps 9-10 define the feature models ${\text{model}}_{w}$ and ${\text{model}}_{c}$ using the word ($diff\_words$) and character ($slt\_chars$) vocabularies, respectively, along with the NLP configurations to transform the input documents into feature matrices for building an assessment model. Steps 11-12 then use the two defined word and character models to actually transform the input documents into the feature matrices ${\mathbf{X}}_{word}$ and ${\mathbf{X}}_{word}$, respectively. Step 13 concatenates the two feature matrices by columns. Finally, step 14 returns the final aggregated feature matrix ${\mathbf{X}}_{agg}$ along with the word and character feature models ${\text{model}}_{w}$ and ${\text{model}}_{c}$.

\section{Experimental Design and Setup}
\label{sec:expt_setup_msr19}

All the classifiers and NLP representations (n-grams, term frequency and tf-idf) in this work were implemented in the \emph{scikit-learn}~\cite{pedregosa2011scikit} and \emph{nltk}~\cite{loper2002nltk} libraries in Python. Our code ran on a fourth-generation Intel Core i7-4200HQ CPU (four cores) running at 2.6 GHz with 16 GB of RAM.

\subsection{Research Questions}
\label{subsec:rqs_msr19}

Our research aims at addressing the \emph{concept drift} issue in SVs' descriptions to improve the robustness of both model selection and prediction steps of report-level SV assessment. Specifically, we evaluate our two-phase character-word models. The first phase selects the optimal word-only models for each VC. The second phase incorporates character features to build character-word models. We raise and answer four Research Questions (RQs):

\begin{itemize}
  \item \textbf{RQ1}: \emph{Is our time-based cross-validation more effective than a non-temporal method to handle concept drift in the model selection step for report-level SV assessment?} To answer RQ1, we first identify the new terms in SV descriptions. We associate such terms with their release or discovery years. We then use qualitative examples to demonstrate information leakage in the non-temporal model selection step. We also quantitatively compare the effectiveness of the proposed time-based cross-validation method with a traditional non-temporal one for addressing the temporal relationship in the context of report-level SV assessment.
  \item \textbf{RQ2}: \emph{Which are the optimal models for multi-classification of each SV characteristic?} To answer RQ2, we present the optimal models (i.e., classifiers and NLP representations) using word features for each VC selected by a five-fold time-based cross-validation method (see section~\ref{subsec:time_cross-validation_msr19}). We also compare the performance of different classes of models (single vs. ensemble) and NLP representations to give recommendations for future use.
  \item \textbf{RQ3}: \emph{How effective is our character-word model to perform automated report-level SV assessment with concept drift?} For RQ3, we first demonstrate how the OoV phrases identified in RQ1 can affect the performance of the existing word-only models. We then highlight the ability of the character features to handle the \emph{concept drift} issue of SVs. We also compare the performance of our character-word model with those of the word-only (without handling \emph{concept drift}) and character-only models.
  \item \textbf{RQ4}: \emph{To what extent can low-dimensional model retain the original performance?} The features of our proposed model in RQ3 are high-dimensional and sparse. Hence, we evaluate a dimensionality reduction technique (i.e., Latent Semantic Analysis~\cite{deerwester1990indexing}) and the sub-word embeddings (i.e., fastText~\cite{bojanowski2017enriching,joulin2016bag}) to show how much information of the original model is approximated in lower dimensions. RQ4 findings can facilitate the building of more efficient \emph{concept-drift}-aware predictive models.
\end{itemize}

\subsection{Dataset}
\label{subsec:dataset_msr19}

We retrieved 113,292 SVs from NVD in JSON format. The dataset contains the SVs from 1988 to 2018. We discarded 5,926 SVs that contain ``** REJECT **'' in their descriptions since they had been confirmed duplicated or incorrect by experts. Seven VCs of CVSS 2 (see section~\ref{sec:background_msr19}) were used as our SV assessment metrics. It turned out that there are 2,242 SVs without any value of CVSS 2. Therefore, we also removed such SVs from our dataset. Finally, we obtained a dataset containing \textbf{105,124 SVs} along with their descriptions and the values of seven VCs indicated previously. For evaluation purposes, we followed the work in~\cite{spanos2018multi} to use the year of 2016 to divide our dataset into training and testing sets with the sizes of 76,241 and 28,883, respectively. The primary reason for splitting the dataset based on the time order is to consider the temporal relationship of SVs.

\subsection{Machine Learning Models for Report-Level SV Assessment}
\label{subsec:ml_models_msr19}

To solve our multi-class classification (report-level SV assessment) problem, we used six well-known ML models. These classifiers have achieved great results in recent data science competitions such as Kaggle~\cite{kaggle_website}. We provide brief descriptions and the hyperparameters of each classifier below.
\begin{itemize}
  \item Na\"ive Bayes (NB)~\cite{russell2002artificial} is a simple probabilistic model that is based on Bayes' theorem. This model assumes that all the features are conditionally independent with respect to each other. In this study, NB had no tuning hyperparameter during the validation step.
  \item Logistic Regression (LR)~\cite{walker1967estimation} is a linear classifier in which the logistic function is used to convert a linear output into a respective probability. The one-vs-rest scheme was applied to split the multi-class problem into multiple binary classification problems. In this work, we selected the optimal value of the regularization parameter for LR from the list of values: 0.01, 0.1, 1, 10, 100.
  \item Support Vector Machine (SVM)~\cite{cortes1995support} is a classification model in which a maximum margin is determined to separate the classes. For NLP, the linear kernel is preferred because of its more scalable computation and sparsity handling~\cite{basu2003support}. The tuning regularization values of SVM are the same as LR.
  \item Random Forest (RF)~\cite{ho1995random} is a bagging model in which multiple decision trees are combined to reduce the variance and sensitivity to noise. The complexity of RF is mainly controlled by (\emph{i}) the number of trees, (\emph{ii}) maximum depth, and (\emph{iii}) maximum number of leaves. (\emph{i}) tuning values were: 100, 300, 500. We set (\emph{ii}) to \emph{unlimited}, which makes the model the highest degree of flexibility and easier to adapt to new data. For (\emph{iii}), the tuning values were 100, 200, 300 and \emph{unlimited}.
  \item XGBoost - Extreme Gradient Boosting (XGB)~\cite{chen2016xgboost} is a variant Gradient Boosting Tree Model (GBTM) in which multiple weak tree-based classifiers are combined and regularized to enhance the robustness of the overall model. Three hyperparameters of XGB that require tuning were the same as RF. It should be noted that the \emph{unlimited} value of the maximum number of leaves is not applicable to XGB.
  \item Light Gradient Boosting Machine (LGBM)~\cite{ke2017lightgbm} is a light-weight version of GBTM. Its main advantage is the scalability since the sub-trees are grown in a leaf-wise manner rather than depth-wise of other GBT algorithms. Three hyperparameters of LGBM that require tuning were the same as XGB.
\end{itemize}

In this work, we considered NB, LR and SVM as single models, while RF, XGB and LGBM as ensemble models.

\subsection{Evaluation Metrics}
\label{subsec:evaluation_metrics_msr19}

Our multi-class classification problem can be decomposed into multiple binary classification problems. To define the standard evaluation metrics for a binary problem~\cite{spanos2018multi,spanos2017assessment,han2017learning}, we first describe four possibilities as follows.
\begin{itemize}
  \item True positive (\emph{TP}): The classifier correctly predicts that an SV has a particular characteristic.
  \item False positive (\emph{FP}): The classifier incorrectly predicts that an SV has a particular characteristic.
  \item True negative (\emph{TN}): The classifier correctly predicts that an SV does not have a particular characteristic.
  \item False negative (\emph{FN}): The classifier incorrectly predicts that an SV does not have a particular characteristic.
\end{itemize}

Based on \emph{TP}, \emph{FP}, \emph{TN}, \emph{FN}, \emph{Accuracy}, \emph{Precision}, \emph{Recall} and \emph{F1-Score} can be defined accordingly in~\eqref{eq:accuracy_msr19},~\eqref{eq:precision_msr19},~\eqref{eq:recall_msr19},~\eqref{eq:f_score_msr19}.

\begin{equation}\label{eq:accuracy_msr19}
  Accuracy\,\,=\,\,\frac{TP\,+\,TN}{TP+\,FP\,+\,TN\,+\,FN}
\end{equation}

\begin{equation}\label{eq:precision_msr19}
  Precision\,=\,\frac{TP}{TP\,+\,FP}
\end{equation}

\begin{equation}\label{eq:recall_msr19}
  Recall\,=\,\frac{TP}{TP\,+\,FN}
\end{equation}

\begin{equation}\label{eq:f_score_msr19}
  F1-Score\,=\,\frac{2\,\times \,Precision\,\times \,Recall}{Precision\,+\,Recall}
\end{equation}

Whilst \emph{Accuracy} measures the global performance of all classes, \emph{F1-Score} (a harmonic mean of \emph{Precision} and \emph{Recall}) evaluates each class separately. Such local estimate like F1-Score is more favorable to the imbalanced VCs such as Access Vector, Access Complexity, Authentication, Severity (see Fig.~\ref{fig:vc_freqs_msr19}). In fact, there are several variants of \emph{F1-Score} for a multi-class classification problem, namely \emph{Micro}, \emph{Macro} and Weighted\emph{ F1-Scores}. In the case of multi-class classification, \emph{Micro F1-Score} is actually the same as \emph{Accuracy}. For \emph{Macro} and \emph{Weighted F1-Scores}, the former does not consider class distribution (the number of elements in each class) for computing \emph{F1-Score} of each class; whereas, the latter does. To account for the balanced and imbalanced VCs globally and locally, we used \emph{Accuracy}, \emph{Macro}, and \emph{Weighted F1-Scores} to evaluate our models. For model selection, if there was a performance tie among models regarding \emph{Accuracy} and/or \emph{Macro F1-Score}, \emph{Weighted F1-Score} would be chosen as the discriminant criterion. The reason is that \emph{Weighted F1-Score} can be considered a compromise between \emph{Macro F1-Score} and \emph{Accuracy}. If the tie still existed, the less complex model with the smaller number of hyperparameters would be selected as per the Occam's razor principle~\cite{blumer1987occam}. In the last tie scenario, the model with shorter training time would be chosen.

\begin{figure}[t]
    \centering
    \includegraphics[width=\linewidth,keepaspectratio]{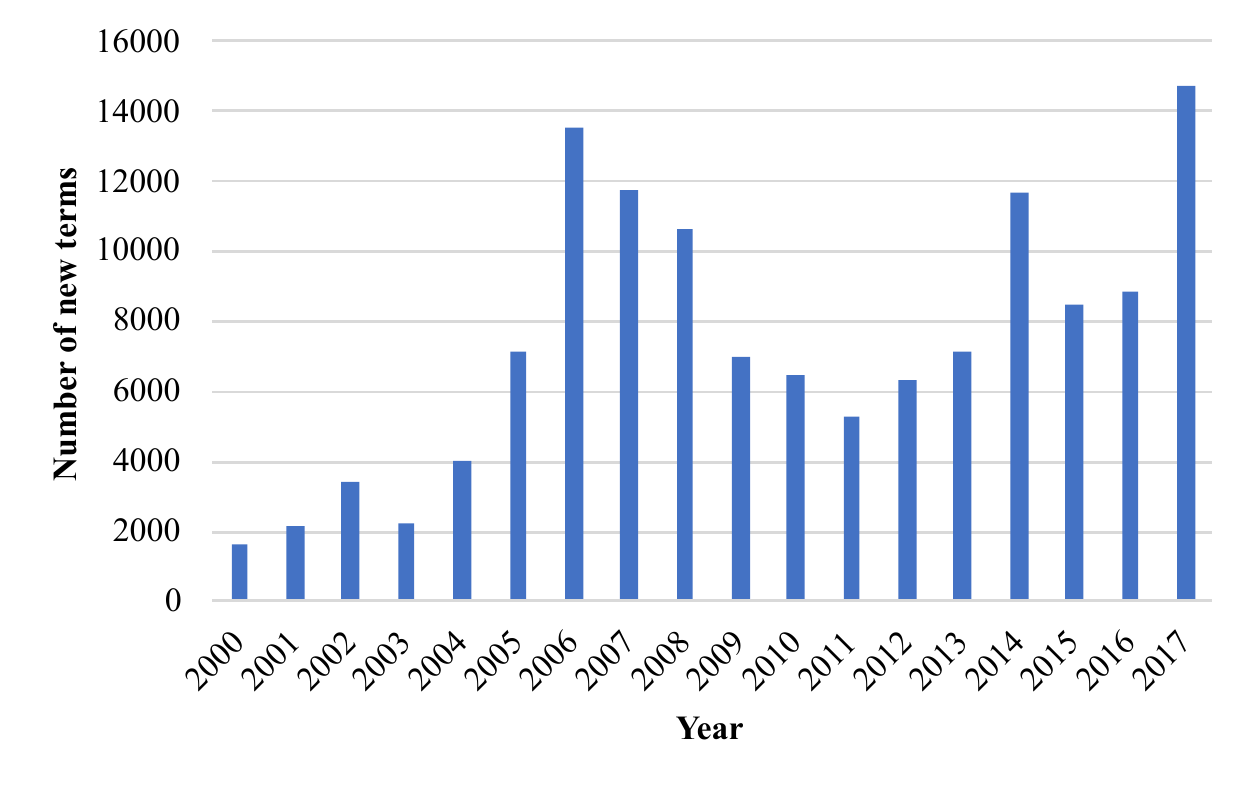}
    \caption[The number of new terms from 2000 to 2017 of SV descriptions in National Vulnerability Database.]{The number of new terms from 2000 to 2017 of SV descriptions in NVD.}
    \label{fig:new_terms_msr19}
\end{figure}

\section{Experimental Results and Discussion}
\label{sec:results_msr19}

\subsection{\textbf{RQ1}: Is Our Time-Based Cross-Validation More Effective Than a Non-Temporal Method to Handle Concept Drift in The Model Selection Step for Report-Level SV Assessment?}
\label{subsec:rq1_results_msr19}

\begin{figure}[t]
    \centering
    \includegraphics[width=\linewidth,keepaspectratio]{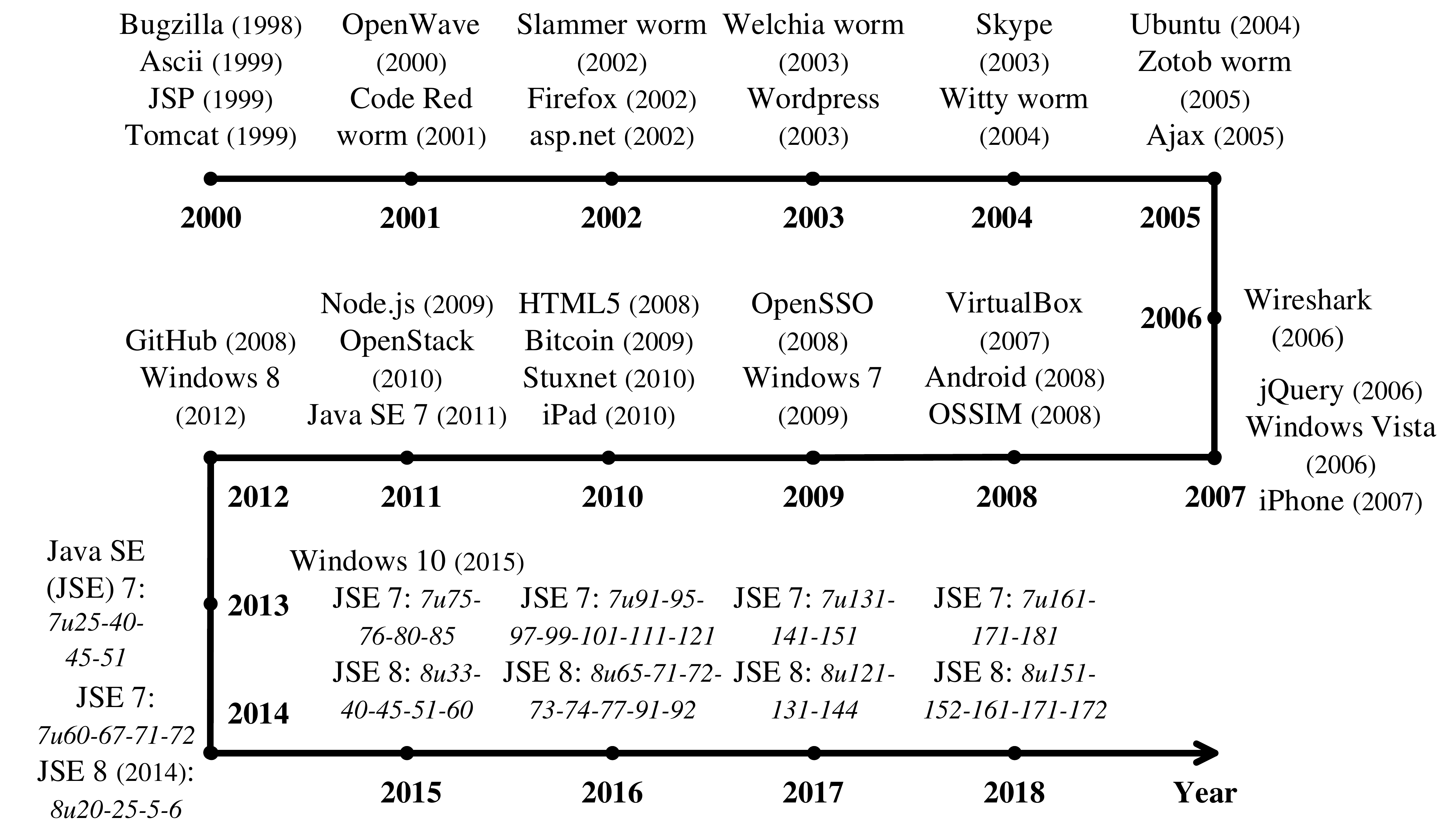}
    \caption[Examples of new terms in National Vulnerability Database corresponding to new products, software, cyber-attacks from 2000 to 2018.]{Examples of new terms in NVD corresponding to new products, software, cyber-attacks from 2000 to 2018. \textbf{Note}: The year of release/discovery is put in parentheses.}
    \label{fig:sv_timeline_msr19}
\end{figure}

We performed both qualitative and quantitative analyses to demonstrate the relationship between \emph{concept drift} and the model selection step of report-level SV assessment. Firstly, it is intuitive that data of SVs intrinsically change over time because of new products, software and attack vectors. The number of new terms appearing in the NVD descriptions each year during the period from 2000 to 2017 is given in Fig.~\ref{fig:new_terms_msr19}. On average each year, there were 7345 new terms added to the vocabulary. Moreover, from 2015 to 2017, the number of new terms had been consistently increasing and achieved an all-time high value of 14684 in 2017. We also highlight some concrete examples about the terms appearing in the database after a particular technology, product or attack was released in Fig.~\ref{fig:sv_timeline_msr19}. There seems to be a strong correlation between the time of appearance of some new terms in the descriptions and their years of release or discovery. Such unseen terms contained many concepts about new products (e.g., \emph{Firefox}, \emph{Skype}, and \emph{iPhone}), operating systems (e.g., \emph{Android}, \emph{Windows Vista/7/8/10}), technologies (e.g., \emph{Ajax}, \emph{jQuery}, and \emph{Node.js}), attacks (e.g., \emph{Code Red}, \emph{Slammer}, and \emph{Stuxnet} worms). There were also the extended forms of existing ones such as the updated versions of Java Standard Edition (Java SE) each year. These qualitative results depict that if the time property of SVs is not considered in the model selection step, then future terms can be mixed with past ones. Such information leakage can result in a discrepancy in the real-world model performance.

\begin{figure}[t]
    \centering
    \includegraphics[width=\linewidth,keepaspectratio]{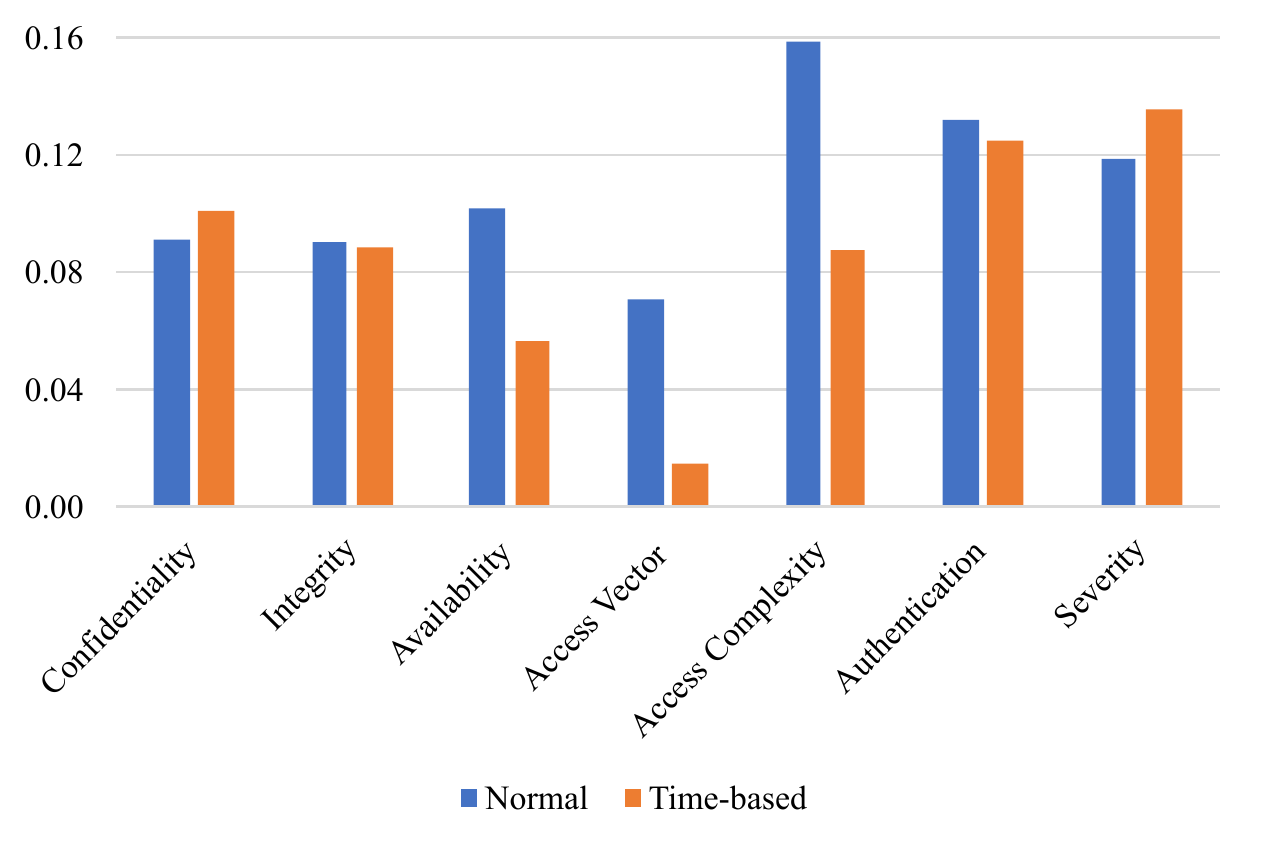}
    \caption{Performance differences between the validated and testing \emph{Weighted F1-Scores} of our time-based validation and a normal cross-validation methods.}
    \label{fig:time-norm_overfitting_msr19}
\end{figure}

In fact, the main goal of the validation step is to select the optimal models that can exhibit similar behavior on unseen data. Next, our approach quantitatively compared the degree of model overfitting between our time-based cross-validation method and a stratified non-temporal one used in~\cite{spanos2018multi,spanos2017assessment}. For each method, we computed the \emph{Weighted F1-Scores} difference between the cross-validated and testing results of the optimal models found in the validation step (see Fig.~\ref{fig:time-norm_overfitting_msr19}). The model selection and selection criteria procedures of the normal cross-validation method were the same as our temporal one. Fig.~\ref{fig:time-norm_overfitting_msr19} shows that traditional non-temporal cross-validation was overfitted in four out of seven cases (i.e., Availability, Access Vector, Access Complexity, and Authentication). Especially, the degrees of overfitting of non-temporal validation method were 1.8, 4.7 and 1.8 times higher than those of the time-based version for Availability, Access Vector, and Access Complexity, respectively. For the other three VCs, both methods were similar, in which the differences were within 0.02. Moreover, on average, the \emph{Weighted F1-Scores} on the testing set of the non-temporal cross-validation method were only 0.002 higher than our approach. This value is negligible compared to the difference of 0.02 (ten times more) in the validation step. It is worth noting that a similar comparison also held for non-stratified non-temporal cross-validation. Overall, both qualitative and quantitative findings suggest that the time-based cross-validation method should be preferred to lower the performance overestimation and mis-selection of report-level SV assessment models due to the effect of \emph{concept drift} in the model selection step.

\begin{tcolorbox}
\textbf{The summary answer to RQ1}: The qualitative results show that many new terms are regularly added to NVD, after the release or discovery of the corresponding software products or cyber-attacks. Normal random-based evaluation methods mixing these new terms can inflate the cross-validated model performance. Quantitatively, the optimal models found by our time-based cross-validation are also less overfitted, especially two to five times for Availability, Access Vector and Access Complexity. It is recommended that the time-based cross-validation should be adopted in the model selection step for report-level SV assessment.
\end{tcolorbox}

\subsection{\textbf{RQ2}: Which are the Optimal Models for Multi-Classification of Each SV Characteristic?}
\label{subsec:rq2_results_msr19}

The answer to RQ1 has shown that the temporal cross-validation should be used for selecting the optimal models in the context of report-level SV assessment. RQ2 presents the detailed results of the first phase of our model. Specifically, we used our five-fold time-based cross-validation to select the optimal word-only model for each of the seven VCs from six classifiers (see section~\ref{subsec:ml_models_msr19}) and eight NLP representations (see Table~\ref{tab:model_configurations_msr19}). We followed the guidelines of the previous work~\cite{spanos2018multi} to extract only the words appearing in more than 0.1\% of all descriptions as features for RQ2.

\begin{table}[t]
  \centering
  \caption{Optimal hyperparameters found for each classifier.}
    \begin{tabular}{ll}
    \hline
    \textbf{Classifier} & \multicolumn{1}{c}{\textbf{Hyperparameters}} \\
    \hline
    \textbf{NB} & None \\
    \hline
    \multirow{3}{*}{\textbf{LR}} & Regularization value: \\
     & + 0.1 for term frequency \\
     & + 10 for tf-idf \\
    \hline
    \multirow{2}{*}{\textbf{SVM}} & Kernel: linear \\
     & Regularization value: 0.1 \\
    \hline
    \multirow{3}{*}{\textbf{RF}} & Number of trees: 100 \\
     & Max. depth: unlimited \\
     & Max. number of leaf nodes: unlimited \\
    \hline
    \multirow{3}{*}{\textbf{XGB}} & Number of trees: 100 \\
     & Max. depth: unlimited \\
     & Max. number of leaf nodes: 100 \\
    \hline
    \multirow{3}{*}{\textbf{LGBM}} & Number of trees: 100 \\
     & Max. depth: unlimited \\
     & Max. number of leaf nodes: 100 \\
    \hline
    \end{tabular}
  \label{tab:optimal_hyperparameters_msr19}
\end{table}

Firstly, each classifier was tuned using random VCs to select its optimal set of hyperparameters. Such selected hyperparameters are reported in Table~\ref{tab:optimal_hyperparameters_msr19}. It is worth noting that we utilized local optimization as a filter to reduce the search space. We found that 0.1 was a consistently good value of regularization coefficient for SVM. Unlike SVM, for LR, 0.1 was suitable for term frequency representation; whereas, 10 performed better for the case of tf-idf. One possible explanation is that LR provides a decision boundary that is more sensitive to hyperparameter. Additionally, although tf-idf with l2-normalization helps a model converge faster, it usually requires more regularization to avoid overfitting~\cite{le2018identification}. For ensemble models, more hyperparameters need tuning, as mentioned in section~\ref{subsec:ml_models_msr19}. Regarding the maximum number of leaves, the optimal value for RF was \emph{unlimited}, which is expected since it would give more flexibility to the model.

\begin{table}[t]
  \centering
  \caption[Optimal models and results after the validation step.]{Optimal models and results after the validation step. \textbf{Note}: The NLP configuration number is put in parentheses.}
    \begin{tabular}{llccc}
    \hline
    \makecell{\textbf{SV}\\ \textbf{characteristic}} & \multicolumn{1}{c}{\makecell{\textbf{Classifier}\\ \textbf{(config)}}} & \textbf{\emph{Accuracy}} & \makecell{\textbf{\emph{Macro}}\\ \textbf{\emph{F1-Score}}} & \makecell{\textbf{\emph{Weighted}}\\ \textbf{\emph{F1-Score}}} \\
    \hline
    \textbf{Confidentiality} & LGBM (1) & 0.839 & 0.831 & 0.840 \\
    \hline
    \textbf{Integrity} & XGB (4) & 0.861 & 0.853 & 0.861 \\
    \hline
    \textbf{Availability} & LGBM (1) & 0.785 & 0.783 & 0.782 \\
    \hline
    \textbf{Access Vector} & XGB (7) & 0.936 & 0.643 & 0.914 \\
    \hline
    \textbf{Access Complexity} & LGBM (1) & 0.771 & 0.553 & 0.758 \\
    \hline
    \textbf{Authentication} & LR (3) & 0.973 & 0.626 & 0.972 \\
    \hline
    \textbf{Severity} & LGBM (5) & 0.814 & 0.763 & 0.811 \\
    \hline
    \end{tabular}
  \label{tab:optimal_models_msr19}
\end{table}

However, for XGB and LGBM, the \emph{unlimited} value was not available. In fact, the higher value did not improve the performance, but significantly increased the computational time. As a result, we chose 100 to be the number of leaves for XGB and LGBM. Similarly, we obtained 100 as a good value for the number of trees of each ensemble model. We noticed that the maximum depth of ensemble methods was the hyperparameter that affected the validation result the most; the others did not change the performance dramatically. Finally, we got a search space of size of 336 in the cross-validation step ((six classifiers) $\times$ (eight NLP configurations) $\times$ (seven characteristics)). The optimal validation results after using our five-fold time-based cross-validation method in section~\ref{subsec:time_cross-validation_msr19} are given in Table~\ref{tab:optimal_models_msr19}. 

\begin{table}[t]
  \centering
  \caption{Average cross-validated \emph{Weighted F-scores} of term frequency vs. tf-idf grouped by six classifiers.}
    \begin{tabular}{lcccccc}
    \cline{2-7}
    \multicolumn{1}{l}{}      & \multicolumn{6}{c}{\textbf{Classifier}} \\
    \cline{2-7}
    \multicolumn{1}{l}{} & \textbf{NB} & \textbf{LR} & \textbf{SVM} & \textbf{RF} & \textbf{XGB} & \textbf{LGBM} \\
    \hline
    \makecell[l]{\textbf{Term}\\ \textbf{frequency}} & 0.781 & \textbf{0.833} & \textbf{0.835} & \textbf{0.843} & \textbf{0.846} & \textbf{0.846} \\
    \hline
    \textbf{tf-idf} & \textbf{0.786} & 0.832 & 0.831 & 0.836 & 0.843 & 0.844 \\
    \hline
    \end{tabular}
  \label{tab:feature_clf_msr19}
\end{table}

\begin{table}[t]
  \centering
  \caption{Average cross-validated \emph{Weighted F-scores} of uni-gram vs. n-grams (2 $\leq$ n $\leq$ 4) grouped by six classifiers.}
    \begin{tabular}{lcccc}
    \hline
    \multicolumn{1}{c}{\textbf{Classifier}} & \textbf{1-gram} & \textbf{2-grams} & \textbf{3-grams} & \textbf{4-grams} \\
    \hline
    \textbf{NB} & 0.756 & 0.778 & 0.784 & \textbf{0.785} \\
    \hline
    \textbf{LR} & 0.821 & 0.835 & \textbf{0.836} & \textbf{0.836} \\
    \hline
    \textbf{SVM} & 0.823 & 0.835 & 0.836 & \textbf{0.837} \\
    \hline
    \textbf{RF} & 0.838 & \textbf{0.840} & 0.838 & 0.838 \\
    \hline
    \textbf{XGB} & 0.844 & 0.845 & \textbf{0.846} & \textbf{0.846} \\
    \hline
    \textbf{LGBM} & \textbf{0.845} & \textbf{0.845} & \textbf{0.845} & \textbf{0.845} \\
    \hline
    \end{tabular}
  \label{tab:ngrams_clf_msr19}
\end{table}

\begin{figure}[t]
    \centering
    \includegraphics[width=\linewidth,keepaspectratio]{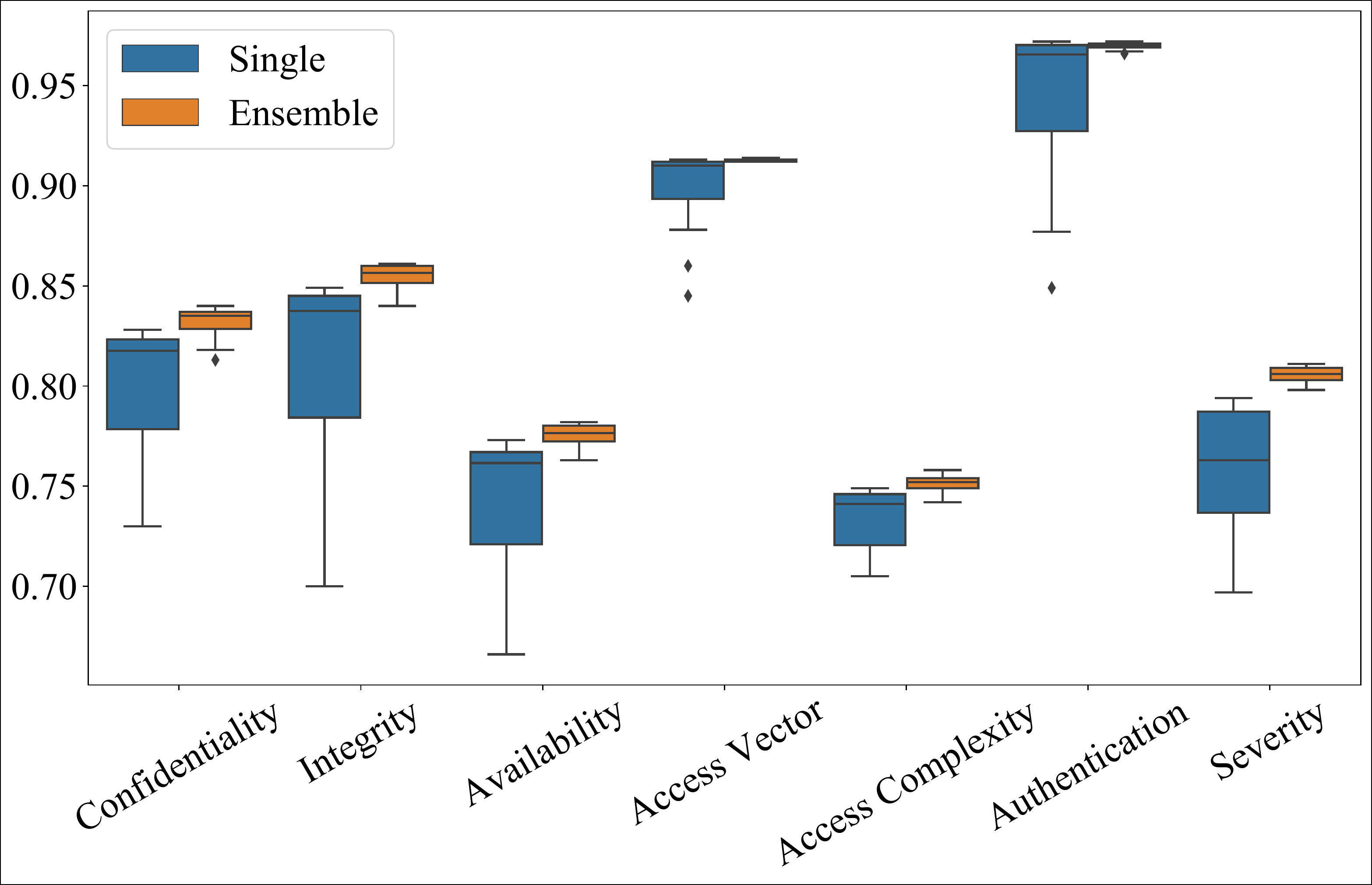}
    \caption[Average cross-validated \emph{Weighted F1-Score}s comparison between ensemble and single models for each vulnerability characteristic.]{Average cross-validated \emph{Weighted F1-Score}s comparison between ensemble and single models for each VC.}
    \label{fig:clf_types_msr19}
\end{figure}

Besides each output, we also examined the validated results across different types of classifiers (single vs. ensemble models) and NLP representations (n-grams and tf-idf vs. term frequency). Since the NLP representations mostly affect the classifiers, their validated results are grouped by six classifiers in Tables~\ref{tab:feature_clf_msr19} and~\ref{tab:ngrams_clf_msr19}. The result shows that tf-idf did not outperform term frequency for five out of six classifiers. This result agrees with the existing work~\cite{spanos2018multi,spanos2017assessment}. It seemed that n-grams with n $>$ 1 improved the result. We used a one-sided non-parametric Wilcoxon signed rank test~\cite{wilcoxon1992individual} to check the significance of such improvement of n-grams (n $>$ 1). The $p$-value was 0.169, which was larger than 0.01 (the significance level). Thus, we were unable to accept the improvement of n-grams over uni-gram. Furthermore, there was no performance improvement after increasing the number of n-grams. The above-reported three observations implied that the more complex NLP representations did not provide a statistically significant improvement over the simplest BoW (configuration 1 in Table~\ref{tab:model_configurations_msr19}). This argument helped explain why three out of seven optimal models in Table~\ref{tab:optimal_models_msr19} were BoW.

\begin{table}[t]
    \centering
    \caption[$P$-values of H\textsubscript{0}: Ensemble models $\leq$ Single models for each vulnerability characteristic.]{$P$-values of H\textsubscript{0}: Ensemble models $\leq$ Single models for each VC.}
    \begin{tabular}{lc}
        \hline
        \textbf{SV characteristic} & \textbf{$p$-value}       \\ \hline
        \textbf{Confidentiality} & $3.261 \times 10^{-5}$ \\ \hline
        \textbf{Integrity} & $9.719 \times 10^{-5}$  \\ \hline
        \textbf{Availability} & $3.855 \times 10^{-5}$  \\ \hline
        \textbf{Access Vector} & $2.320 \times 10^{-3}$  \\ \hline
        \textbf{Access Complexity} & $1.430 \times 10^{-5}$  \\ \hline
        \textbf{Authentication} & $1.670 \times 10^{-3}$  \\ \hline
        \textbf{Severity} & $1.060 \times 10^{-7}$  \\ \hline
    \end{tabular}
    \label{tab:clf_test_msr19}
\end{table}

Along with the NLP representations, we also investigated the performance difference between single (NB, LR, and SVM) and ensemble (RF, XGB, and LGBM) models. The average \emph{Weighted F1-Scores} grouped by VCs for single and ensemble models are illustrated in Fig.~\ref{fig:clf_types_msr19}. The ensemble models seemed to consistently demonstrate the superior performance compared to the single counterparts. We also observed that the ensemble methods produced mostly consistent results (i.e., small variances) for Access Vector and Authentication characteristics. We performed the one-sided non-parametric Wilcoxon signed rank tests~\cite{wilcoxon1992individual} to check the significance of the better performance of the ensemble over the single models. Table~\ref{tab:clf_test_msr19} reports the $p$-values of the results from the hypothesis testing. The tests confirmed that the superiority of the ensemble methods was significant since all $p$-values were smaller than the significance level of 0.01. The validated results in Table~\ref{tab:optimal_models_msr19} also affirmed that six out of seven optimal classifiers were ensemble (i.e., LGBM and XGB). It is noted that the XGB model usually took more time to train than the LGBM model, especially for tf-idf representation. Our findings suggest that LGBM, XGB and BoW should be considered as baseline classifiers and NLP representations for report-level SV assessment.

\begin{tcolorbox}
\textbf{The summary answer to RQ2}: LGBM and BoW are the most frequent optimal classifiers and NLP representations. Overall, the more complex NLP representations such as n-grams, tf-idf do not provide a statistically significant performance improvement than BoW. The ensemble models perform statistically better than single ones. It is recommended that the ensemble classifiers (e.g., XGB and LGBM) and BoW should be used as baseline models for report-level SV assessment.
\end{tcolorbox}

\subsection{\textbf{RQ3}: How Effective is Our Character-Word Model to Perform Automated Report-Level SV Assessment with Concept Drift?}
\label{subsec:rq3_results_msr19}

The OoV terms presented in RQ1 actually directly have an impact on the word-only models. Such missing features can make a model unable to produce reliable results. Especially when no existing term is found (i.e., all features are zero), a model would have the same output regardless of the context. To answer RQ3, we first tried to identify such all-zero cases in the SV descriptions from 2000 to 2018. For each year from 2000 to 2018, we split the dataset into (\emph{i}) training set (data from the previous year backward) for building the vocabulary, and (\emph{ii}) testing set (data from the current year to 2018) for checking the vocabulary existence. We found 64 cases from 2000 to 2018 in the testing data, in which all the features were missing (see Appendix~\ref{sec:Appendix_msr19}). We used the terms appearing at least 0.1\% in all descriptions. It should be noted that the number of all-zero cases may be reduced using a larger vocabulary with the trade-off for larger computational time. We also investigated the descriptions of these vulnerabilities and found several interesting patterns. The average length of these abnormal descriptions was only 7.98 words compared to 39.17 of all descriptions. It turned out that the information about the threats and sources of such SVs was limited. Most of them just included the assets and attack/SV types. For example, the vulnerabilities with ID of CVE-2016-10001xx had nearly the same format ``Reflected XSS in WordPress plugin'' with the only differences were the name and version of the plugin. This format made it hard for a model to evaluate the impact of each SV separately. Another issue was due to specialized or abbreviated terms such as $/redirect?url= XSS, SEGV, CSRF$ without proper explanation. The above issues suggest that SV descriptions should be written with sufficient information to enhance the comprehensibility of SVs.

For RQ3, the solution to the issue of the word-only models using character-level features is evaluated. We considered the non-stop-words with high frequency (i.e., appearing in more than 10\% of all descriptions) to generate the character features. Using the same 0.1\% value as RQ2 increased the dimensions more than 30 times, but the performance only changed within 0.02. According to Algorithm~\ref{algo:feature_aggregation}, the output minimum number of character n-grams was chosen to be two. We first tested the robustness of the character-only models by setting the maximum number of characters to only three. For each year \textit{y} from 1999 to 2017, we used such character model to generate the characters from the data of the considering year \textit{y} backward. We then verified the existence of such features using the descriptions of the other part of the data (i.e., from year \emph{y} + 1 towards 2018). Surprisingly, the model using only two-to-three-character n-grams could produce at least one non-zero feature for all the descriptions even using only training data in 1999 (i.e., the first year in our dataset based on CVE-ID). Such finding shows that our approach is stable to SV data changes (\emph{concept drift}) in testing data from 2000 to 2018 even with the limited amount of data and without retraining.

\begin{figure}[t]
    \centering
    \includegraphics[width=0.9\linewidth,keepaspectratio]{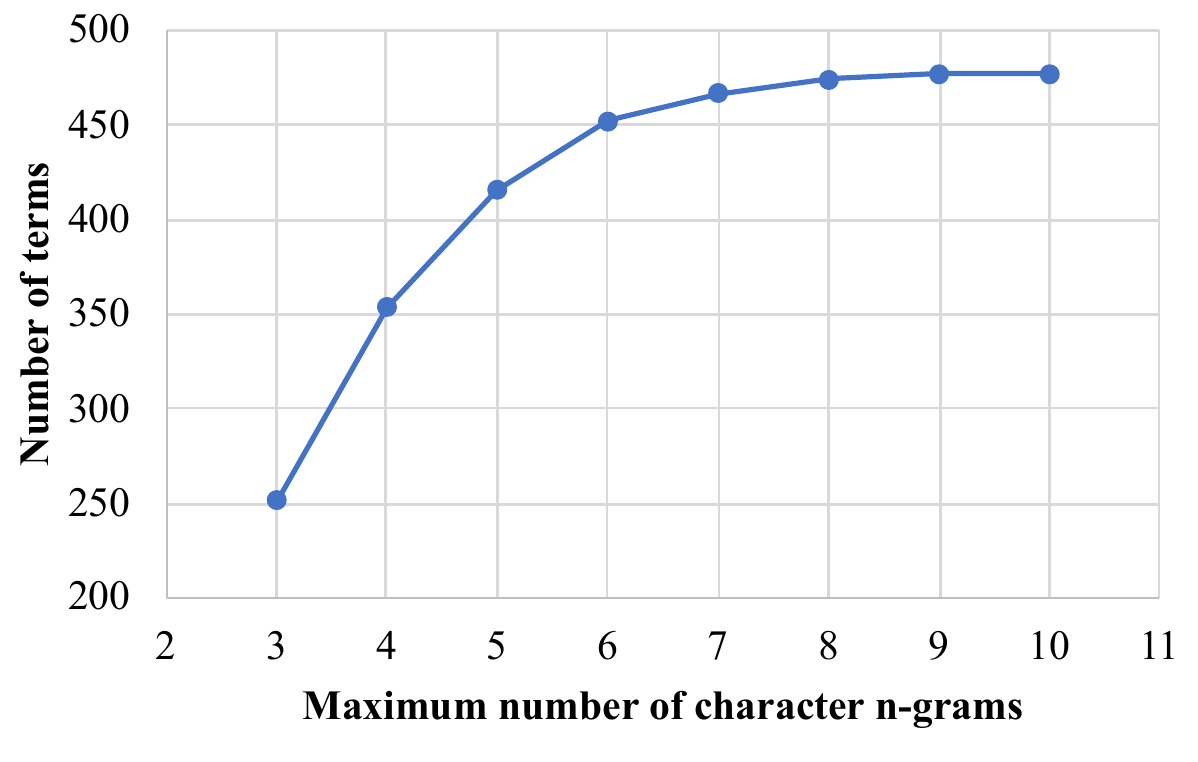}
    \caption{The relationship between the size of vocabulary and the maximum number of character n-grams.}
    \label{fig:char_ngrams_msr19}
\end{figure}

Next, to increase the generalizability of our approach, values of 3-10 were considered for selecting the maximum number of character n-grams based on their corresponding vocabulary sizes (see Fig.~\ref{fig:char_ngrams_msr19}). Using the elbow method in cluster analysis~\cite{marutho2018determination}, six was selected since the vocabulary size did not increase dramatically after this point. The selected minimum and maximum values of character n-grams matched the minimum and average word lengths of all NVD descriptions in our dataset, respectively.

We then used the feature aggregation algorithm (see section~\ref{subsec:feature_aggregation_msr19}) to create the aggregated features from the character n-grams (2 $\leq$ n $\leq$ 6) and word n-grams to build the final model set and compared it with two baselines: Word-only Model (WoM) and Character-only Model (CoM).

\begin{landscape}
\begin{table}
  \centering
  \caption{Performance (\emph{Accuracy}, \emph{Macro F1-Score}, \emph{Weighted F1-Score}) of our character-word vs. word-only and character-only models.}
    \begin{tabular}{|l|c|c|c|c|c|c|c|c|c|}
    \hline
    \multicolumn{1}{|c|}{\multirow{2}{*}{\textbf{SV characteristic}}} & \multicolumn{3}{c|}{\textbf{Our optimal model (CWM)}} & \multicolumn{3}{c|}{\textbf{Word-only model (WoM)}} & \multicolumn{3}{c|}{\textbf{Character-only model (CoM)}} \\
    \cline{2-10}
     & \textbf{\emph{Accuracy}} & \makecell{\textbf{\emph{Macro}}\\ \textbf{\emph{F1-Score}}} & \makecell{\textbf{\emph{Weighted}}\\ \textbf{\emph{F1-Score}}} & \textbf{\emph{Accuracy}} & \makecell{\textbf{\emph{Macro}}\\ \textbf{\emph{F1-Score}}} & \makecell{\textbf{\emph{Weighted}}\\ \textbf{\emph{F1-Score}}} & \textbf{\emph{Accuracy}} & \makecell{\textbf{\emph{Macro}}\\ \textbf{\emph{F1-Score}}} & \makecell{\textbf{\emph{Weighted}}\\ \textbf{\emph{F1-Score}}} \\
    \hline
    \textbf{Confidentiality} & \textbf{0.727} & \textbf{0.717} & \textbf{0.728} & 0.722 & 0.708 & 0.723 & 0.694 & 0.683 & 0.698 \\
    \hline
    \textbf{Integrity} & 0.763 & \textbf{0.749} & 0.764 & 0.763 & 0.744 & 0.764 & 0.731 & 0.718 & 0.734 \\
    \hline
    \textbf{Availability} & \textbf{0.712} & \textbf{0.711} & \textbf{0.711} & 0.700 & 0.696 & 0.702 & 0.660 & 0.657 & 0.660 \\
    \hline
    \textbf{Access Vector} & \textbf{0.914} & \textbf{0.540} & \textbf{0.901} & 0.904 & 0.533 & 0.894 & 0.910 & 0.538 & 0.899 \\
    \hline
    \textbf{Access Complexity} & 0.703 & 0.468 & 0.673 & \textbf{0.718} & \textbf{0.476} & \textbf{0.691} & 0.700 & 0.457 & 0.668 \\
    \hline
    \textbf{Authentication} & \textbf{0.875} & \textbf{0.442} & \textbf{0.844} & 0.864 & 0.425 & 0.832 & 0.866 & 0.441 & 0.840 \\
    \hline
    \textbf{Severity} & 0.668 & \textbf{0.575} & 0.663 & \textbf{0.686} & 0.569 & \textbf{0.675} & 0.661 & 0.549 & 0.652 \\
    \hline
    \end{tabular}
  \label{tab:baseline_comparison_msr19}

\end{table}
\end{landscape}

\noindent It should be noted that WoM is the model in which \emph{concept drift} is not handled. Unfortunately, a direct comparison with the existing WoM~\cite{spanos2018multi} was not possible since they used an older NVD dataset and more importantly, they did not release their source code for reproduction. However, we tried to set up the experiments based on the guidelines and results presented in the previous paper~\cite{spanos2018multi}.

To be more specific, we used BoW predictors and random forest (the best of their three models used) with the following hyperparameters: the number of trees was 100 and the number of features for splitting was 40. For CoM, we used the same optimal classifier of each VC. The comparison results are given in Table~\ref{tab:baseline_comparison_msr19}. CWM performed slightly better than the WoM for four out of seven VCs regarding all evaluation metrics. Also, 4.98\% features of CWM were non-zero, which was nearly five-time denser than 1.03\% of WoM. Also, CoM was the worst model among the three, which had been expected since it contained the least information (smallest number of features). Although CWM did not significantly outperform WoM, its main advantage is to effectively handle the OoV terms (\emph{concept drift}), except new terms without any matching parts. We hope that our solution to \emph{concept drift} will be integrated into practitioner's existing framework/workflow and future research work to perform more robust report-level SV assessment.

\begin{tcolorbox}
\textbf{The summary answer to RQ3}: The WoM does not handle the new cases well, especially those with all zero-value features. Without retraining, the tri-gram character features can still handle the OoV words effectively with no all-zero features for all testing data from 2000 to 2018. Our CWM performs comparably well with the existing WoM and provides nearly five-time richer SV information. Hence, our CWM is better for automated report-level SV assessment with \emph{concept drift}.
\end{tcolorbox}

\subsection{\textbf{RQ4}: To What Extent Can Low-Dimensional Model Retain the Original Performance?}
\label{subsec:rq4_results_msr19}

The n-gram NLP models usually have an issue with the high-dimensional and sparse feature vectors~\cite{kao2007natural}. The large feature sizes of our CWMs in Table~\ref{tab:baseline_comparison_msr19} were 1,649 for Confidentiality, Availability and Access Complexity; 4,154 for Integrity and Access Vector; 3,062 for Authentication; and 5,104 for Severity. To address such challenge in RQ4, we investigated a dimensionality reduction method (i.e., Latent Semantic Analysis (LSA)~\cite{deerwester1990indexing}) and recent sub-word embeddings (e.g., fastText~\cite{bojanowski2017enriching,joulin2016bag}) for SV assessment. fastText is an extension of Word2Vec~\cite{mikolov2013distributed} word embeddings, in which the character-level features are also considered. fastText is different to traditional n-grams in the sense that it determines the meaning of a word/subword based on the surrounding context. Here, we computed the sentence representation as an average fastText embedding of its constituent words and characters. We implemented fastText using the \emph{Gensim}~\cite{rehurek2010software} library in Python.

\begin{table}[t]
  \centering
  \caption[\emph{Weighted F1-Scores} of our original Character-Word Model, 300-dimension Latent Semantic Analysis (LSA-300), fastText trained on SV descriptions (fastText-300) and fastText trained on English Wikipedia pages (fastText-300W).]{\emph{Weighted F1-Scores} of our original CWM (green-colored baseline), 300-dimension Latent Semantic Analysis (LSA-300), fastText trained on SV descriptions (fastText-300) and fastText trained on English Wikipedia pages (fastText-300W).}
    \begin{tabular}{|l|c|c|c|c|}
    \hline
    \makecell[l]{\textbf{SV}\\ \textbf{characteristic}} & \makecell{\textbf{Our}\\ \textbf{CWM}} & \makecell{\textbf{LSA-}\\ \textbf{300}} & \makecell{\textbf{fastText-}\\ \textbf{300}} & \makecell{\textbf{fastText-}\\ \textbf{300W}} \\
    \hline
    \textbf{Confidentiality} & {\color[HTML]{008000}\textbf{0.728}} & 0.656 & \textbf{0.679} & 0.648 \\
    \hline
    \textbf{Integrity} & {\color[HTML]{008000}\textbf{0.764}} & 0.695 & \textbf{0.719} & 0.672 \\
    \hline
    \textbf{Availability} & {\color[HTML]{008000}\textbf{0.711}} & 0.656 & \textbf{0.687} & 0.669 \\
    \hline
    \textbf{Access Vector} & {\color[HTML]{008000}\textbf{0.901}} & 0.892 & \textbf{0.893} & 0.866 \\
    \hline
    \textbf{Access Complexity} & {\color[HTML]{008000}\textbf{0.673}} & 0.611 & \textbf{0.679} & 0.678 \\
    \hline
    \textbf{Authentication} & {\color[HTML]{008000}\textbf{0.844}} & \textbf{0.842} & 0.815 & 0.765 \\
    \hline
    \textbf{Severity} & {\color[HTML]{008000}\textbf{0.663}} & \textbf{0.656} & 0.654 & 0.635 \\
    \hline
    \end{tabular}
  \label{tab:dimension_reduction_msr19}
\end{table}

For LSA, using the elbow method and total explained variances of the principal components, we selected 300 for the dimensions and called it LSA-300. Table~\ref{tab:dimension_reduction_msr19} shows that LSA-300 retained from 90\% to 99\% performance of the original model, but used only 300 dimensions (6-18\% of original model sizes). More remarkably, with the same 300 dimensions, the fastText model trained on SV descriptions was on average better than LSA-300 (97\% vs. 94.5\%). fastText model even slightly outperformed our original CWM for Access Complexity. Moreover, for all seven cases, the fastText model using SV knowledge (fastText-300) had higher \emph{Weighted F1-Scores} than that trained on English Wikipedia pages (fastText-300W)~\cite{fasttest_pretrained}. The result implied that SV descriptions contain specific terms that do not frequently appear in general domains. The domain relevance turns out to be not only applicable to word embeddings~\cite{han2017learning}, but also to character/sub-word embeddings for SV analysis and assessment. Overall, our findings show that LSA and fastText are capable of building efficient report-level SV assessment models without too much performance trade-off.

\begin{tcolorbox}
\textbf{The summary answer to RQ4}: The LSA model with 300 dimensions (6-18\% of the original size) retains from 90\% up to 99\% performance of the original model. With the same feature dimensions, the model with fastText sub-word embeddings provide even more promising results. The fastText model with the SV knowledge outperforms that trained on a general context (e.g., Wikipedia). LSA and fastText can help build efficient models for report-level SV assessment.
\end{tcolorbox}

\section{Threats to Validity}
\label{sec:discussion_msr19}

\noindent \textbf{Internal validity}. We used well-known tools such as the \emph{scikit-learn}~\cite{pedregosa2011scikit} and \emph{nltk}~\cite{loper2002nltk} libraries for ML and NLP. Our optimal models may not guarantee the highest performance for every SV characteristic since there are infinite values of hyperparameters to tune. However, even when the optimal values change, a time-based cross-validation method should still be preferred since we considered the general trend of all SVs. Our models may not provide the state-of-the-art results, but at least they give the baseline performance for handling the \emph{concept drift} of SVs.

\noindent \textbf{External validity}. Our work used NVD -- one of the most comprehensive public repositories of SVs. The size of our dataset is more than 100,000 with the latest SVs in 2018. Our character-word model was demonstrated to consistently handle the OoV words well even with very limited data for all years in the dataset. It is recognized that the model may not work for extreme rare terms in which no existing parts can be found. However, our model is totally re-trainable to deal with such cases or incorporate more sources of SVs' descriptions.

\noindent \textbf{Conclusion validity}. We mitigated the randomness of the results by taking the average value of five-fold cross-validation. The performance comparison of different types of classifiers and NLP representations was also confirmed using statistical hypothesis tests with $p$-values that were much lower than the significance level of 1\%.

\section{Related Work}
\label{sec:related_work_msr19}

\subsection{Report-Level SV Assessment}

It is important to patch critical-first SVs~\cite{hou2017security}. Besides CVSS, there have been many other assessment schemes for SVs~\cite{liu2011vrss,spanos2013wivss,sharma2018improved}. Recently, there has been a detailed Bayesian analysis of various SV scoring systems~\cite{johnson2016can}, which highlights the good overall performance of CVSS. Therefore, we used the well-known CVSS as the ground truth for our approach. We assert that our approach can be generalizable to other SV assessment systems following the same scheme of multi-class classification.

As mentioned in Chapter~\ref{chapter:lit_review}, Bozorgi et al.~\cite{bozorgi2010beyond} pioneered the use of ML models for report-level SV assessment. Their paper used an SVM model and various features (e.g., NVD description, CVSS, published and modified dates) to estimate the likelihood of exploitation and \emph{time-to-exploit} of SVs. Another piece of work analyzed the VCs and trends of SVs by incorporating different SV information from multiple repositories~\cite{murtaza2016mining,almukaynizi2019patch}, security advisories~\cite{edkrantz2015predicting,huang2013novel}, darkweb/deepnet~\cite{almukaynizi2019patch,nunes2016darknet} and social network (Twitter)~\cite{sabottke2015vulnerability}. These efforts assumed that all VCs have been available at the time of analysis. However, our work relaxes this assumption by using only SV descriptions – one of the first pieces of information appearing in new SV reports. As a result, our model can be used for both new and old SVs.

Many studies have built upon the work of Bozorgi et al.~\cite{bozorgi2010beyond} and used NVD descriptions for report-level SV assessment, as reviewed in Chapter~\ref{chapter:lit_review}. Here, we focus on several studies directly related to our current study.
Yamamoto et al.~\cite{yamamoto2015text} used Linear Discriminant Analysis, Na\"ive Bayes and Latent Semantic Indexing combined with an annual effect estimation to determine the CVSS-based VCs of more than 60,000 SVs in NVD. The annual effect focused on recent SVs, but still could not explicitly handle OoV terms in SV descriptions. Spanos et al.~\cite{spanos2018multi} worked on the same task as ours using a multi-target framework with Decision Tree, Random Forest and Gradient Boosting Tree. Note that our approach also contains the word-only model, but we select the optimal models using our time-based cross-validation to better address the \emph{concept drift} issue. SV descriptions were also used to evaluate SV severity~\cite{spanos2017assessment}, associate the frequent terms with each VC~\cite{toloudis2016associating}, determine the type of each SV using topic modeling~\cite{neuhaus2010security} and show SV trends~\cite{murtaza2016mining}. Recently, Han et al.~\cite{han2017learning} have applied deep learning to predict SV severity. The existing literature has demonstrated the usefulness of description for SV analysis and assessment, but has not mentioned how to overcome its \emph{concept drift} challenge. Our work is the first of its kind to provide a robust treatment for SVs' \emph{concept drift}.

\subsection{Temporal Modeling of SVs}

Regarding the temporal relationship of SVs, Roumani et al.~\cite{roumani2015time} proposed a time-series approach using autoregressive integrated moving average and exponential smoothing methods to predict the number of SVs in the future. Another time-series work~\cite{tang2016exploiting} was presented to model the trend in disclosing SVs. A group of researchers published a series of studies~\cite{rajasooriya2017cyber,kaluarachchi2017non,pokhrel2017cybersecurity} on stochastic models such as Hidden Markov Models, Artificial Neural Network, and Support Vector Machine to estimate the occurrence and exploitability of SVs. The focus of the above studies was on the determination of the occurrence of SVs over time. In contrast, our work aims to handle the temporal relationship to build more robust predictive (multi-class classification) models for report-level SV assessment.

\section{Chapter Summary}
\label{sec:conclusions_msr19}

We observe that the existing studies suffer from \emph{concept drift} in SV descriptions that affect both the traditional model selection and prediction steps of report-level SV assessment. We assert that \emph{concept drift} can degrade the robustness of these existing predictive models. We showed that time-based cross-validation should be used for SV analysis to better capture the temporal relationship of SVs. Then, our main contribution is the Character-Word Models (CWMs) to improve the robustness of automated report-level SV assessment with \emph{concept drift}. CWMs were demonstrated to handle \emph{concept drift} of SVs effectively for all the testing data from 2000 to 2018 in NVD even in the case of data scarcity. Our approach also performed comparably well with the existing word-only models. Our CWMs were also much less sparse and thus less prone to overfitting. We also found that Latent Semantic Analysis and sub-word embeddings like fastText help build compact and efficient CWM models (up to 94\% reduction in dimension) with the ability to retain at least 90\% of the predictive performance for all VCs.
Besides the strong performance, we also provide implications on the use of different features and classifiers for building effective report-level SV assessment models.

\section{Appendix - SVs with All Out-of-Vocabulary Words}
\label{sec:Appendix_msr19}
\noindent 64 vulnerabilities along with their CVD-IDs that had all-zero features of word-only model from 2000 to 2018, as mentioned in section~\ref{subsec:rq3_results_msr19}:
CVE-2013-6647, CVE-2015-1000004, CVE-2016-1000113, CVE-2016-1000114, CVE-2016-1000117, CVE-2016-1000118, CVE-2016-1000126, CVE-2016-1000127, CVE-2016-1000128, CVE-2016-1000129, CVE-2016-1000130, CVE-2016-1000131, CVE-2016-1000132, CVE-2016-1000133, CVE-2016-1000134, CVE-2016-1000135, CVE-2016-1000136, CVE-2016-1000137, CVE-2016-1000138, CVE-2016-1000139, CVE-2016-1000140, CVE-2016-1000141, CVE-2016-1000142, CVE-2016-1000143, CVE-2016-1000144, CVE-2016-1000145, CVE-2016-1000146, CVE-2016-1000147, CVE-2016-1000148, CVE-2016-1000149, CVE-2016-1000150, CVE-2016-1000151, CVE-2016-1000152, CVE-2016-1000153, CVE-2016-1000154, CVE-2016-1000155, CVE-2016-1000217, CVE-2017-10798, CVE-2017-10801, CVE-2017-14036, CVE-2017-14536, CVE-2017-15808, CVE-2017-16760, CVE-2017-16785, CVE-2017-17499, CVE-2017-17703, CVE-2017-17774, CVE-2017-6102, CVE-2017-7276, CVE-2017-8783, CVE-2018-10030, CVE-2018-10031, CVE-2018-10382, CVE-2018-11120, CVE-2018-11405, CVE-2018-12501, CVE-2018-13997, CVE-2018-14382, CVE-2018-5285, CVE-2018-5361, CVE-2018-6467, CVE-2018-6834, CVE-2018-8817, CVE-2018-9130

%% file: Chapters/Chapter_4_MSR2022.tex
\chapter{Automated Function-Level Software Vulnerability Assessment}
\label{chap:msr22}

\begin{tcolorbox}
\textbf{Related publication}: This chapter is based on our paper titled ``\textit{On the Use of Fine-grained Vulnerable Code Statements for Software Vulnerability Assessment Models}'' published in the 19\textsuperscript{th} International Conference on Mining Software Repositories (MSR), 2022 (CORE A)~\cite{le2022use}.
\end{tcolorbox}
\bigskip

The proposed approach in Chapter~\ref{chap:msr19} has improved the robustness of report-level Software Vulnerability (SV) assessment models against changing data of SVs in the wild.
Nevertheless, these report-level models still rely on SV descriptions that mostly require significant expertise and manual effort to create, which may cause delays for SV fixing. On the other hand, respective vulnerable code is always available before SVs are fixed.
Many studies have developed Machine Learning (ML) approaches to detect SVs in functions and fine-grained code statements that cause such SVs.
However, as shown in Chapter~\ref{chapter:lit_review}, there is little work on leveraging such detection outputs for data-driven SV assessment to give information about exploitability, impact, and severity of SVs.
The information is important to understand SVs and prioritize their fixing.
Using large-scale data from 1,782 functions of 429 SVs in 200 real-world projects, in Chapter~\ref{chap:msr22}, we investigate ML models for automating function-level SV assessment tasks, i.e., predicting seven Common Vulnerability Scoring System (CVSS) metrics.
We particularly study the value and use of vulnerable statements as inputs for developing the assessment models because SVs in functions are originated in these statements.
We show that vulnerable statements are 5.8 times smaller in size, yet exhibit 7.5-114.5\% stronger assessment performance (Matthews Correlation Coefficient (MCC)) than non-vulnerable statements.
Incorporating context of vulnerable statements further increases the performance by up to 8.9\% (0.64 MCC and 0.75 F1-Score).
Overall, we provide the initial yet promising ML-based baselines for function-level SV assessment, paving the way for further research in this direction.

\newpage

\section{Introduction}

As shown in Chapter~\ref{chapter:lit_review}, previous studies (e.g.,~\mbox{\cite{yamamoto2015text,han2017learning,spanos2018multi,le2019automated,le2021survey}}) have mostly used SV reports to develop data-driven models for assigning the Common Vulnerability Scoring System (CVSS)~\cite{cvss} metrics to Software Vulnerabilities (SVs). Among sources of SV reports, National Vulnerability Database (NVD)~\cite{nvd} has been most commonly used for building SV assessment models~\cite{le2021survey}. The popularity of NVD is mainly because it has SV-specific information (e.g., CVSS metrics) and less noise in SV descriptions than other Issue Tracking Systems (ITSs) like Bugzilla~\cite{bugzilla}. The discrepancy is because NVD reports are vetted by security experts, while ITS reports may be contributed by users/developers with limited security knowledge~\cite{croft2021investigation}.
However, NVD reports are mostly released long after SVs have been fixed.
Our analysis revealed that less than 3\% of the SV reports with the CVSS metrics on NVD had been published before SVs were fixed; on average, these reports appeared 146 days later than the fixes. Note that our findings accord with the previous studies~\cite{li2017large,piantadosi2019fixing}. This delay renders the CVSS metrics required for SV assessment unavailable at fixing time, limiting the adoption of report-level SV assessment for understanding SVs and prioritizing their fixes.

Instead of using SV reports, an alternative and more straight-forward way is to directly take (vulnerable) code as input to enable SV assessment prior to fixing.
Once a code function is confirmed vulnerable, SV assessment models can assign it the CVSS metrics before the vulnerable code gets fixed, even when its report is not (yet) available.
Note that it is non-trivial to use static application security testing tools to automatically create bug/SV reports from vulnerable functions for current SV assessment techniques as these tools often have too many false positives~\mbox{\cite{johnson2013don,aloraini2019empirical}}.
To develop function-level assessment models, it is important to obtain input information about SVs in functions detected by manual debugging or automatic means like data-driven approaches (e.g.,~\mbox{\cite{zhou2019devign,zheng2020impact,lin2020deep}}).
Notably, recent studies (e.g.,~\mbox{\cite{li2021vulnerability,nguyen2021information}}) have shown that an SV in a function usually stems from a very small number of code statements/lines, namely \textit{vulnerable statements}.
Intuitively, these vulnerable statements potentially provide highly relevant information (e.g., causes) for SV assessment models. Nevertheless, a large number of other (non-vulnerable) lines in functions, though do not directly contribute to SVs, can still be useful for SV assessment, e.g., indicating the impacts of an SV on nearby code.
It still remains largely unknown about function-level SV assessment models as well as the extent to which vulnerable and non-vulnerable statements are useful as inputs for these models.

We conduct a large-scale study to fill this research gap. We investigate the usefulness of integrating fine-grained vulnerable statements and different types of code context (relevant/surrounding code) into learning-based SV assessment models. The assessment models employ various feature extraction methods and Machine Learning (ML) classifiers to predict the seven CVSS metrics (Access Vector, Access Complexity, Authentication, Confidentiality, Integrity, Availability, and Severity) for SVs in code functions.

Using 1,782 functions from 429 SVs of 200 real-world projects, we evaluate the use of vulnerable statements and other lines in functions for developing SV assessment models. Despite being up to 5.8 times smaller in size (lines of code), vulnerable statements are more effective for function-level SV assessment, i.e., 7.4-114.5\% higher Matthews Correlation Coefficient (MCC) and 5.5-43.6\% stronger F1-Score, than non-vulnerable lines. Moreover, vulnerable statements with context perform better than vulnerable lines alone.
Particularly, using vulnerable and all the other lines in each function achieves the best performance of 0.64 MCC (8.9\% better) and 0.75 F1-Score (8.5\% better) compared to using only vulnerable statements.
We obtain such improvements when combining vulnerable statements and context as a single input based on their code order, as well as when treating them as two separate inputs.
Having two inputs explicitly provides models with the location of vulnerable statements and context for the assessment tasks, while single input does not.
Surprisingly, we do not obtain any significant improvement of the double-input models over the single-input counterparts.
These results show that function-level SV assessment models can still be effective even without knowing exactly which statements are vulnerable.
Overall, our findings can inform the practice of building function-level SV assessment models.

Our key \textbf{contributions} are summarized as follows:

\begin{contribution}
    \item To the best of our knowledge, we are the first to leverage data-driven models for automating function-level SV assessment tasks that enable SV prioritization/planning prior to fixing.
    \item We study the value of using fine-grained vulnerable statements in functions for building SV assessment models.
	\item We empirically show the necessity and potential techniques of incorporating context of vulnerable statements to improve the assessment performance.
    \item We release our datasets and models for future research at \url{https://github.com/lhmtriet/Function-level-Vulnerability-Assessment}.
\end{contribution}

\noindent \textbf{Chapter organization}. Section~\mbox{\ref{sec:background_msr22}} gives a background on function-level SV assessment. Section~\mbox{\ref{sec:rqs_msr22}} introduces and motivates the three RQs. Section~\mbox{\ref{sec:method_msr22}} describes the methods used for answering these RQs. Section~\mbox{\ref{sec:results_msr22}} presents our empirical results. Section~\mbox{\ref{sec:discussion_msr22}} discusses the findings and threats to validity. Section~\mbox{\ref{sec:related_work_msr22}} mentions the related work. Section~\mbox{\ref{sec:conclusions_msr22}} concludes the study and suggests future directions.

\section{Background and Motivation}
\label{sec:background_msr22}

\input{Chapters/code}

There have been a growing number of studies to detect vulnerable statements in code functions (e.g.,~\cite{li2021vuldeelocator,li2021vulnerability,nguyen2021information}).
Fine-grained detection assumes that not all statements in a function are vulnerable. We confirmed this assumption; i.e., only 14.7\% of the lines in our curated functions were vulnerable (see section~\ref{subsec:data_collection_msr22}). However, it is non-trivial to manually annotate a sufficiently large dataset of vulnerable functions and statements for training SV prediction models. Instead, many studies (e.g.,~\cite{li2021vulnerability,wattanakriengkrai2020predicting,li2020dlfix}) have automatically obtained vulnerable statements from modified lines in Vulnerability Fixing Commits (VFCs) as these lines are presumably removed to fix SVs.
The functions containing such identified statements are considered vulnerable. Note that VFCs are used as they can be relatively easy to retrieve from various sources like National Vulnerability Database (NVD)~\cite{nvd}.
An exemplary function and its vulnerable statement are in \fig~\ref{fig:ex_sv_msr22}. Line 8 ``\code{sb.append(unifyQuotes(dir));}'' is the vulnerable statement; this line was replaced with a non-vulnerable counterpart ``\code{sb.append(quoteOneItem(dir, false));}'' in the VFC. The replacement was made to properly sanitize the input (\code{dir}), preventing OS command injection.

Despite active research in SV detection, there is little work on utilizing the output of such detection for SV assessment.
Previous studies (e.g.,~\mbox{\cite{yamamoto2015text,spanos2017assessment,spanos2018multi,le2019automated,le2021survey}}) have mostly leveraged SV reports, mainly on NVD, to develop SV assessment models that alleviate the need for manually defining complex rules for assessing ever-increasing SVs.
However, these SV reports usually appear long after SV fixing time.
For example, the SV fix in \fig~\ref{fig:ex_sv_msr22} was done 1,533 days before the date it was reported on NVD.
In fact, such a delay, i.e., disclosing SVs after they are fixed, is a recommended practice so that attackers cannot exploit unpatched SVs to compromise systems~\cite{zhou2021finding}. One may argue that internal bug/SV reports in Issue Tracking Systems (ITS) such as JIRA~\cite{jira} or Bugzilla~\cite{bugzilla}
can be released before SV fixing and have severity levels. However, ITS severity levels are often for all bug types, not only SVs. These ITSs also do not readily provide exploitability and impact metrics like CVSS~\cite{cvss} for SVs, limiting assessment information required for fixing prioritization. Moreover, SVs are mostly rooted in source code; thus, it is natural to perform code-based SV assessment.
We propose predicting seven base CVSS metrics (i.e., Access Vector, Access Complexity, Authentication, Confidentiality, Integrity, Availability, and Severity)\footnote{These metrics are from CVSS version 2 and were selected based on the same reasons presented in the study in Chapter~\ref{chap:msr19}. More details can be found in section~\ref{sec:background_msr19}.} after SVs are detected in code functions to enable thorough and prior-fixing SV assessment. We do not perform SV assessment for individual lines as for a given function, like Li et al.~\cite{li2021sysevr}, we observed that there can be more than one vulnerable line and nearly all these lines are strongly related and contribute to the same SV (having the same CVSS metrics).

Vulnerable statements represent the core parts of SVs, but we posit that other (non-vulnerable) parts of a function may also be usable for SV assessment.
Specifically, non-vulnerable statements in a vulnerable function are either \textit{directly} or \textit{indirectly} related to the current SV.
We use program slicing~\cite{weiser1984program} to define directly SV-related statements as the lines affect or are affected by the variables in vulnerable statements.
For example, the blue lines in \fig~\ref{fig:ex_sv_msr22} are directly related to the SV as they define, change, or use the \code{sb} and \code{dir} variables in vulnerable line 8. These SV-related statements can reveal the context/usage of affected variables for analyzing SV exploitability, impact, and severity. For instance, lines 5-6 denote that \code{dir} is a directory and \code{sb} is a string (\code{StringBuilder} object), respectively; line 7 then indicates that a directory change is performed, i.e., the \code{cd} command. This sequence of statements suggests that \code{sb} contains a command changing directory. Line 11 returns the vulnerable command, probably affecting other components.
Besides, indirectly SV-related statements, e.g., the black lines in \fig~\ref{fig:ex_sv_msr22}, are remaining lines in a function excluding vulnerable and directly SV-related statements. These indirectly SV-related lines may still provide information about SVs. For example, lines 3-4 in \fig~\ref{fig:ex_sv_msr22} imply that there is only a \code{null} checking for directory without imposing any privilege requirement to perform the command, potentially reducing the complexity of exploiting the SV. It remains unclear to what extent different types of statements are useful for SV assessment tasks. Therefore, this study aims to unveil the contributions of these statement types to function-level SV assessment models.

\begin{landscape}
\begin{figure*}[t]
    \centering
    \includegraphics[width=\linewidth,keepaspectratio]{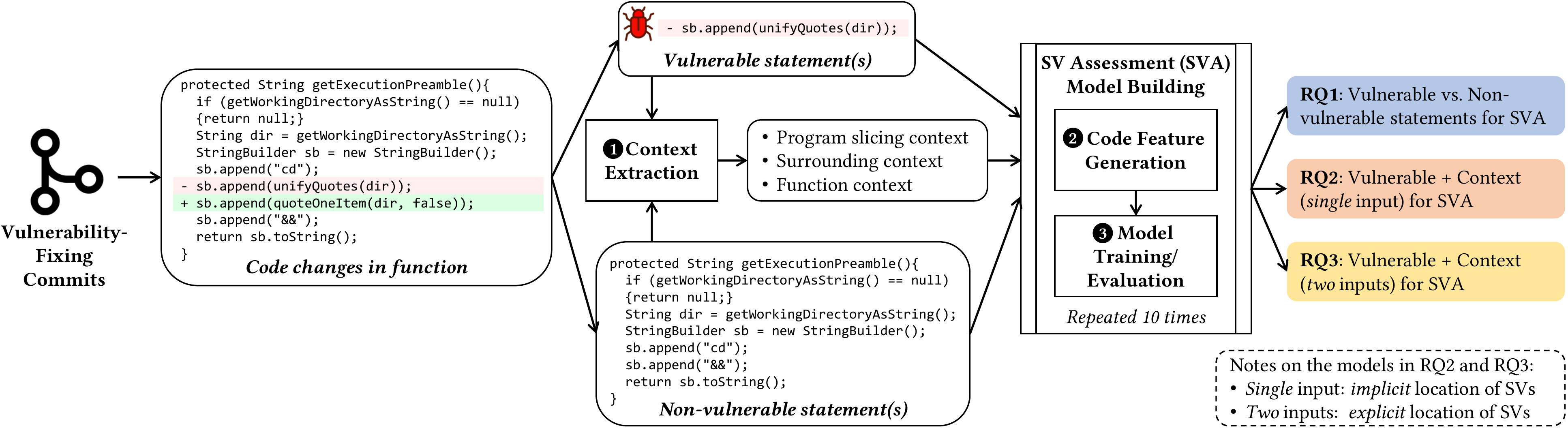}
    \caption[Methodology used to answer the research questions.]{Methodology used to answer the research questions. \textbf{Note}: The vulnerable function is the one described in Fig.~\mbox{\ref{fig:ex_sv_msr22}}.}
    \label{fig:workflow}
\end{figure*}
\end{landscape}

\section{Research Questions}
\label{sec:rqs_msr22}

To demystify the predictive performance of SV assessment models using vulnerable and other statements in code functions, we set out to investigate the following three Research Questions (RQs).

\textbf{RQ1: Are vulnerable code statements more useful than non-vulnerable counterparts for SV assessment models?}
Since vulnerable and non-vulnerable statements are both potentially useful for SV assessment (see section~\mbox{\ref{sec:background_msr22}}), RQ1 compares them for building function-level SV assessment models.
RQ1 tests the hypothesis that vulnerable statements directly causing SVs would provide an advantage in SV assessment performance.
The findings of RQ1 would also inform the practice of leveraging recent advances in fine-grained SV detection for function-level SV assessment.

\textbf{RQ2: To what extent do different types of context of vulnerable statements contribute to SV assessment performance?}
RQ2 studies the impact of using directly and indirectly SV-related statements as context for vulnerable statements, as discussed in section~\ref{sec:background_msr22}, on the performance of SV assessment in functions. We compare the performance of models using different types of context lines (see section~\ref{subsec:code_context_msr22}) that have been commonly used in the literature. RQ2 findings would unveil what types of context (if any) would be beneficial to use alongside vulnerable statements for developing function-level SV assessment models.

\textbf{RQ3: Does separating vulnerable statements and context to provide explicit location of SVs improve assessment performance?}
For SV assessment, RQ2 combines vulnerable statements and their context as a single input following their order in functions, while RQ3 treats these two types of statements as two separate inputs. Separate inputs explicitly specify which statements are vulnerable in each function for assessment models. RQ3 results would give insights into the usefulness of the exact location of vulnerable statements for function-level SV assessment models.

\section{Research Methodology}
\label{sec:method_msr22}

This section presents the experimental setup we used to perform a large-scale study on function-level SV assessment to support prioritization of SVs before fixing. We used a computing cluster with 16 CPU cores and 16GB of RAM to conduct all the experiments.

\noindent \textbf{Workflow overview}. \fig~\ref{fig:workflow} presents the workflow we followed to develop function-level SV assessment models based on various types of code inputs. The workflow has three main steps: (\textit{i}) Collection of vulnerable and non-vulnerable statements from Vulnerability-Fixing Commits (VFCs) (section~\ref{subsec:data_collection_msr22}), (\textit{ii}) Context extraction of vulnerable statements (section~\ref{subsec:code_context_msr22}), and (\textit{iii}) Model building for SV assessment (sections~\ref{subsec:code_features_msr22},~\ref{subsec:assessment_models_msr22}, and~\ref{subsec:model_evaluation_msr22}).
We start with VFCs containing code changes used to fix SVs. As discussed in section~\ref{sec:background_msr22}, we consider the deleted (--) lines in each function of VFCs as vulnerable statements, while the remaining lines are non-vulnerable statements. Details of the extracted VFCs and statements are given in section~\ref{subsec:data_collection_msr22}. Both vulnerable and non-vulnerable statements are used by the \textit{Context Extraction} module (see section~\ref{subsec:code_context_msr22}) to obtain the three types of context with respect to vulnerable statements that potentially provide additional information for SV assessment. The extracted statements along with their context enter the \textit{Model Building} module. The first step in this module is to extract fixed-length feature vectors from code inputs/statements (see section~\ref{subsec:code_features_msr22}). Subsequently, such feature vectors are used to train different data-driven models (see section~\ref{subsec:assessment_models_msr22}) to support automated SV assessment, i.e., predicting the seven CVSS metrics: Access Vector, Access Complexity, Authentication, Confidentiality, Integrity, Availability, and Severity. The model training and evaluation are repeated 10 times to increase the stability of results (see section~\ref{subsec:model_evaluation_msr22}).

\noindent \textbf{RQ-wise method}. The methods to collect data, extract features as well as develop and evaluate models in \fig~\ref{fig:workflow} were utilized for answering all the Research Questions (RQs) in section~\ref{sec:rqs_msr22}.
\textbf{RQ1} developed and compared two types of SV assessment models, namely models using only vulnerable statements and those using only non-vulnerable statements.
In \textbf{RQ2}, for each of the program slicing, surrounding, and function context types, we created a single feature vector by combining the current context and corresponding vulnerable statements, based on their appearance order in the original functions, for model building and performance comparison. In \textbf{RQ3}, for each context type in RQ2, we extracted two separate feature vectors, one from vulnerable statements and another one from the context, and then fed these vectors into SV assessment models. We compared the two-input approach in RQ3 with the single-input counterpart in RQ2.

\subsection{Data Collection}
\label{subsec:data_collection_msr22}

To develop SV assessment models, we need a large dataset of vulnerable functions and statements curated from VFCs, as discussed in section~\ref{sec:background_msr22}. This section describes the collection of such dataset.

\noindent \textbf{VFC identification}. We first scraped VFCs from three popular sources in the literature: NVD~\mbox{\cite{nvd}}, GitHub Advisory Database,\footnote{\url{https://github.com/advisories}} and VulasDB~\mbox{\cite{ponta2019manually}}, a manually curated VFC dataset. The VFCs had dates ranging from July 2000 to September 2021. We only selected VFCs that had the CVSS version 2 metrics as we needed these metrics for SV assessment. Following the recommendation of~\mbox{\cite{mcintosh2017fix}}, we removed any VFCs that had more than 100 files and 10,000 lines of code as these VFCs are likely tangled commits, i.e., not only fixing SVs. We also discarded VFCs that were not written in the Java programming language as Java has been commonly used in both practice\footnote{\url{https://bit.ly/stack-overflow-survey-2021}} and the literature (e.g.,~\mbox{\cite{hoang2019deepjit,alon2019code2vec,mcintosh2017fix,le2021deepcva}}). After the filtering process, we obtained 900 VFCs to extract vulnerable functions/statements for building SV assessment models.

\begin{figure}[t]
    \centering
    \includegraphics[width=\columnwidth,keepaspectratio]{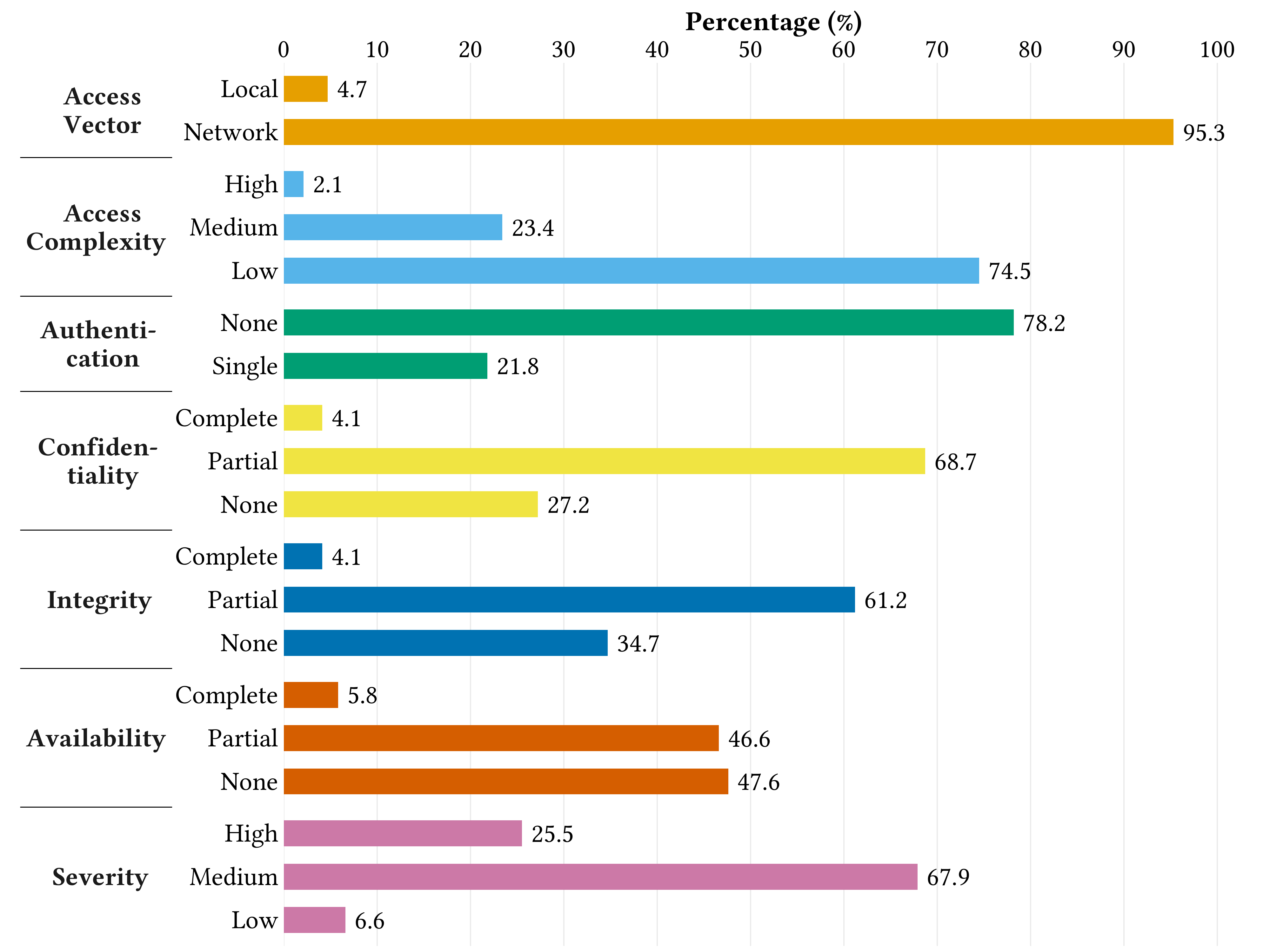}
    \caption{Class distributions of the seven CVSS metrics.}
    \label{fig:cvss_distribution_msr22}
\end{figure}

\noindent \textbf{Extraction of vulnerable functions and statements}. For each VFC, we obtained all the affected files (i.e., containing changed lines), excluding test files because we focused on production code. We followed a common practice~\cite{li2021vulnerability,wattanakriengkrai2020predicting,li2020dlfix} to consider all the functions in each affected file as \textit{vulnerable functions} and the deleted lines in these functions as \textit{vulnerable statements}. We removed functions having only added changes and non-functional/cosmetic changes such as removing/changing inline/multi-line comments, spaces, or empty lines. For the former, added lines only exist in fixed code, making it hard to pinpoint the exact vulnerable statements or root causes leading to such code additions~\cite{rezk2021ghost}. For the latter, cosmetic changes likely do not contribute to SV fixes~\cite{mcintosh2017fix}. We also did not use a function if its entire body was deleted because such a case did not have any non-vulnerable statements for building SV assessment models in our RQs (see section~\ref{sec:rqs_msr22}). After the filtering steps, we retrieved \textbf{1,782 vulnerable functions} and \textbf{5,179 vulnerable statements} of 429 SVs in 200 Java projects. We also obtained the seven CVSS metrics from NVD for each vulnerable function (see \fig~\ref{fig:cvss_distribution_msr22}).

\noindent \textbf{Manual validation of vulnerable functions}.

We randomly selected 317 functions, i.e., with 95\% confidence level and 5\% error~\mbox{\cite{cochran2007sampling}}, from our dataset. The author of this thesis and a PhD student with three-year experience in Software Engineering and Cybersecurity independently validated the functions.
The manual validation was considerably labor-intensive, taking 120 man-hours.
We achieved a substantial agreement with a Cohen's kappa score~\mbox{\cite{viera2005understanding}} of 0.72.
Disagreements were resolved through discussion.
We found that 9\% of the selected functions were not vulnerable, mainly due to tangled fixes/VFCs.
The functions in these VFCs fixed a non-SV related issue, e.g., the \code{nameContainsForbiddenSequence} function in the commit \textit{cefbb94} of the \textit{Eclipse Mojarra} project.
The modifier of this function in the \code{ResourceManager.java} class was changed from \code{private} to \code{package}. This change allowed the reuse of the function in other classes like \code{ClasspathResourceHelper.java} for sanitizing inputs to prevent a path traversal SV (CVE-2020-6950).
We assert that it is challenging to detect all of these cases without manual validation. However, such validation is extremely expensive to scale up to thousands of functions like the dataset we curated in this study.

\subsection{Vulnerable Code Context Extraction}
\label{subsec:code_context_msr22}
This section describes the \textit{Context Extraction} module that takes vulnerable and non-vulnerable statements as inputs and then outputs program slicing, surrounding, or function context.
These context types have been previously used for bug/SV-related tasks~\cite{li2021sysevr,tian2020evaluating,lin2020deep}. However, there is little known about their use and value for function-level SV assessment tasks, which are studied in this work.

\noindent \textbf{Program slicing context}. Program slicing captures relevant statements to a point in a program (line of code) to support software debugging~\cite{weiser1984program}.
This concept has been utilized for SV identification~\cite{li2021sysevr,zheng2021vulspg}. However, using this context for SV assessment is fundamentally different. For SV detection, the location of vulnerable statements is unknown, so program slicing context is usually extracted for all statements including non-vulnerable ones in a function of pre-defined types (e.g., array manipulations, arithmetic operations, and function calls)~\cite{li2021sysevr}. In contrast, for SV assessment, vulnerable statements are known; thus, program slicing context is only obtained for these statements.
With a focus on function-level SV assessment, we considered \textit{intra-procedural program slicing}, i.e., finding relevant lines within the boundary of a function of interest.

Following the common practice in the literature~\cite{salimi2020improving,li2021sysevr}, we used Program Dependence Graph (PDG)~\cite{ferrante1987program} extracted from source code to obtain program slices for vulnerable statements in each function. A PDG contains nodes that represent code statements and directed edges that capture data or control dependencies among nodes. A data dependency exists between two statements when one statement affects/changes the value of a variable used in another statement. For example, ``\code{int b = a + 1;}'' is data-dependent on ``\code{int a = 1;}'' as the variable \code{a} defined in the second statement is used in the first statement. A control dependency occurs when a statement determines whether/how often another statement is executed. For instance, in ``\code{if (b == 2) func(c);}'', ``\code{func(c)}'' only runs if ``\code{b == 2}'', and thus is control-dependent on the former.

Based on data and control dependencies, backward and forward slices are extracted.
\textit{Backward slices} directly change or control the execution of statements affecting the values of variables in vulnerable statements; whereas, \textit{forward slices} are data/control-dependent on vulnerable statements~\cite{dashevskyi2018screening}.
In a PDG, backward slices are nodes that can go to vulnerable nodes through one or more directed edges. In \fig~\ref{fig:ex_sv_msr22}, the \code{dir} variable is defined in line 5 and then used in vulnerable line 8, so line 5 is a backward slice. Forward slices are the nodes that can be reached from vulnerable nodes by following one or more directed edges in a PDG. In \fig~\ref{fig:ex_sv_msr22}, line 11 is data-dependent on vulnerable line 8 as it uses the value of \code{sb}; thus, line 11 is a forward slice. The program slicing context of vulnerable statements in a function is a combination of all backward and forward slices.

\noindent \textbf{Surrounding context}. Another way to define context is to take a fixed number of lines ($n$) before and after a vulnerable statement, which is referred to as \textit{surrounding context}. These surrounding lines may contain relevant information, e.g., values/usage of variables in vulnerable statements. This context is also based on an observation that developers usually first look at nearby code of vulnerable statements to understand how to fix SVs~\cite{tian2020evaluating}.
We discarded surrounding lines that were just code comments or blank lines as these probably do not contribute to the functionality of a function~\cite{mcintosh2017fix}. We also limited surrounding lines to be within-function.

\noindent \textbf{Function context}. Contrary to program slicing and surrounding context that may not use all lines in a function, \textit{function context} uses all function statements, excluding vulnerable ones.
This scope has been commonly used for SV detection models~\cite{zhou2019devign,lin2020deep} because vulnerable statements are unavailable at detection time. This scope contains all the lines of program slicing/surrounding context and other presumably indirectly related lines to vulnerable statements. Accordingly, the performance of using indirectly SV-related lines together with directly SV-relevant lines for SV assessment can be examined.
Note that for a given function, combining function context with vulnerable statements as a single code block (RQ2 in section~\ref{sec:rqs_msr22}) is equivalent to using the whole function, which would result in the same input to SV assessment models regardless of which statements are vulnerable.
This input combination allows us to evaluate the usefulness of the exact location of vulnerable statements for function-level SV assessment models.

\subsection{Code Feature Generation}
\label{subsec:code_features_msr22}

Raw code from vulnerable statements and their context are converted into fixed-length feature vectors to be consumable by learning-based SV assessment models. This step describes five techniques we used to extract features from code inputs.

\noindent \textbf{Bag-of-Tokens}. \textit{Bag-of-Tokens} is based on Bag-of-Words, a popular feature extraction technique in Natural Language Processing (NLP). This technique has been commonly investigated for developing SV assessment models based on textual SV descriptions/reports~\cite{wen2015novel,gawron2017automatic,wang2019intelligent}. We extended this technique to code-based SV assessment by counting the frequency of code tokens. We also applied code-aware tokenization to preserve code syntax and semantics. For instance, \code{var++} was tokenized into \code{var} and \code{++}, explicitly informing a model about incrementing the variable \code{var} by one using the operator (\code{++}).

\noindent \textbf{Bag-of-Subtokens}. \textit{Bag-of-Subtokens} extends Bag-of-Tokens by splitting extracted code tokens into sequences of characters (sub-tokens). These characters help a model learn less frequent tokens better. For instance, an infrequent variable like \code{ThisIsAVeryLongVar} is decomposed into multiple sub-tokens; one of which is \code{Var}, telling a model that this is a potential variable. We extracted sub-tokens of lengths ranging from two to six. Such values have been previously adopted for SV assessment~\cite{nakagawa2019character,le2019automated}. We did not use one-letter characters as they were too noisy, while using more than six characters would significantly increase feature size and computational cost.

\noindent \textbf{Word2vec}. Unlike Bag-of-Tokens and Bag-of-Subtokens that do not consider token context, \textit{Word2vec}~\cite{mikolov2013distributed} extracts features of a token based on its surrounding counterparts. The contextual information from surrounding tokens helps produce similar feature vectors in an embedding space for tokens with (nearly) identical functionality/usage (e.g., \code{average} and \code{mean} variables). Word2vec generates vectors for individual tokens, so we averaged the vectors of all input tokens to represent a code snippet. This averaging method has been demonstrated to be effective for various NLP tasks~\cite{shen2018baseline}.
\tab~\ref{tab:hyperparameter_models_msr22} lists different values for the window and vector sizes of Word2vec used for tuning the performance of learning-based SV assessment models.

\noindent \textbf{fastText}. \textit{fastText}~\cite{bojanowski2017enriching} enhances Word2vec by representing each token with an aggregated feature vector of its constituent sub-tokens. Technically, fastText combines the strengths of semantic representation of Word2vec and subtoken-augmented features of Bag-of-Subtokens. fastText has been shown to build competitive yet compact report-level SV assessment models~\cite{le2019automated}. Like Word2vec, the feature vector of a code snippet was averaged from the vectors of all the input tokens. The length of sub-tokens also ranged from two to six, resembling that of Bag-of-Subtokens.
Other hyperparameters of fastText for optimization are listed in \tab~\ref{tab:hyperparameter_models_msr22}.

\noindent \textbf{CodeBERT}. \textit{CodeBERT}~\cite{feng2020codebert} is an adaptation of BERT~\cite{devlin2018bert}, the current state-of-the-art feature representation technique in NLP, to source code modeling. CodeBERT is a pre-trained model using both natural language and programming language data to produce contextual embedding for code tokens.
The same code token can have different CodeBERT embedding vectors depending on other tokens in an input; whereas, word2vec/fastText produces a single vector for every token regardless of its context. In addition, the source code tokenizer of CodeBERT is built upon Byte-Pair Encoding (BPE)~\cite{sennrich2015neural}. This tokenizer smartly retains sub-tokens that frequently appear in a training corpus rather than keeping all of them as in Bag-of-Subtokens and fastText, balancing between performance and cost. CodeBERT also preserves a special token, \textit{[CLS]}, to represent an entire code input. We leveraged the vector of this \textit{[CLS]} token to extract the features for each code snippet.

We trained all the feature models from scratch, except CodeBERT as it is a pre-trained model. We used CodeBERT's pre-trained vocabulary and embeddings as commonly done in the literature~\cite{zhou2021assessing}. To build the vocabulary for the other feature extraction methods, we considered tokens appearing at least in two samples in a training dataset to avoid vocabulary explosion due to too rare tokens. The exact vocabulary depended on the dataset used in each of the 10 training/evaluation rounds, as described in section~\ref{subsec:model_evaluation_msr22}.
Note that some code snippets, e.g., vulnerable lines extracted from (partial) code changes, were not compilable (i.e., did not contain complete code syntax); thus, we did not use Abstract Syntax Tree (AST) based code representation like Code2vec~\cite{alon2019code2vec} in this study as such representation may not be robust for these cases~\cite{hoang2019deepjit,hoang2020cc2vec}. It is worth noting that Bag-of-Tokens, Bag-of-Subtokens, Word2vec, fastText, and CodeBERT can still work with these cases as these methods operate directly on code tokens.

\subsection{Data-Driven SV Assessment Models}
\label{subsec:assessment_models_msr22}

\begin{table}[t]
  \centering
  \caption[Hyperparameter tuning for software vulnerability assessment models.]{Hyperparameter tuning for SV assessment models.}
    \begin{tabular}{lll}
    \hline
    \multicolumn{1}{l}{\textbf{Step}} & \multicolumn{1}{l}{\textbf{Model}} & \multicolumn{1}{l}{\textbf{Hyperparameters}} \\
    \hline
    \multirow{2}{*}{\makecell[l]{Feature\\ extraction}} & \makecell[l]{Word2vec~\cite{shen2018baseline}} & \makecell[l]{\textit{Vector size}: 150, 300, 500} \\
    & \makecell[l]{fastText~\cite{bojanowski2017enriching}} & \makecell[l]{\textit{Window size}: 3, 4, 5} \\
    \hline
    \multirow{8}{*}{\makecell[l]{CVSS\\ metrics\\ prediction}} & \makecell[l]{Logistic Regression\\ (LR)~\cite{walker1967estimation}} & \multirow{3}{*}{\makecell[l]{\textit{Regularization coefficient}:\\ 0.01, 0.1, 1, 10, 100}} \\
    & \makecell[l]{Support Vector\\ Machine (SVM)~\cite{cortes1995support}} & \\
    \cline{2-3}
     & \multirow{3}{*}{\makecell[l]{K-Nearest\\ Neighbors\\ (KNN)~\cite{altman1992introduction}}} & \makecell[l]{\textit{No. of neighbors}: 5, 11, 31, 51} \\
    & & \makecell[l]{\textit{Weight}: uniform, distance} \\
    & & \textit{Distance norm}: 1, 2 \\
    \cline{2-3}
    & \makecell[l]{Random Forest (RF)~\cite{ho1995random}} & \multirow{3}{*}{\makecell[l]{\textit{No. of estimators}: 100, 200,\\ 300, 400, 500\\ \textit{Max depth}: 3, 5, 7, 9,\\ unlimited\\ \textit{Max. no. of leaf nodes}: 100,\\ 200, 300, unlimited (RF)}} \\
    & \makecell[l]{Extreme Gradient\\ Boosting (XGB)~\cite{chen2016xgboost}} & \\
    & \makecell[l]{Light Gradient\\ Boosting Machine\\ (LGBM)~\cite{ke2017lightgbm}} & \\
    \hline
    \end{tabular}
  \label{tab:hyperparameter_models_msr22}
\end{table}

Features generated from code inputs enter ML models for predicting the CVSS metrics. The predictions of the CVSS metrics are classification problems (see \fig~\ref{fig:cvss_distribution_msr22}). We used six well-known Machine Learning (ML) models for classifying the classes of each CVSS metric: Logistic Regression (LR)~\cite{walker1967estimation}, Support Vector Machine (SVM)~\cite{cortes1995support}, K-Nearest Neighbors (KNN)~\cite{altman1992introduction}, Random Forest (RF)~\cite{ho1995random}, eXtreme Gradient Boosting (XGB)~\mbox{\cite{chen2016xgboost}}, and Light Gradient Boosting Machine (LGBM)~\mbox{\cite{ke2017lightgbm}}. LR, SVM, and KNN are single models, while RF, XGB, and LGBM are ensemble models that leverage multiple single counterparts to reduce overfitting. These classifiers have been used for SV assessment based on SV reports~\mbox{\cite{spanos2018multi,le2019automated}}. We also considered different hyperparameters for tuning the performance of the classifiers, as given in \tab~\ref{tab:hyperparameter_models_msr22}. These hyperparameters have been adapted from the prior studies using similar classifiers~\mbox{\cite{spanos2018multi,le2019automated,le2020puminer}}. Here, we mainly focus on ML techniques, and thus using Deep Learning models~\cite{le2020deep} for the tasks is out of the scope of this study.

\subsection{Model Evaluation}
\label{subsec:model_evaluation_msr22}
\noindent \textbf{Evaluation technique}.
To develop function-level SV assessment models and evaluate their performance, we used 10 rounds of training, validation, and testing. We randomly shuffled the dataset of vulnerable functions in section~\ref{subsec:data_collection_msr22} and then split it into 10 partitions of roughly equal size.\footnote{With 1,782 samples in total, folds 1-9 had 178 samples and fold 10 had 180 samples.} In round $i$, for a model, we used fold $i + 1$ for validation, fold $i + 2$ for testing, and all of the remaining folds for training.
When $i + 1$ or $i + 2$ was larger than 10, its value was wrapped around. For example, if $i = 10$, then $(i + 1)~\text{mod}~10 = 11~\text{mod}~10 = 1$ and $(i + 2)~\text{mod}~10 = 12~\text{mod}~10 = 2$. A grid search of the hyperparameters in \tab~\ref{tab:hyperparameter_models_msr22} was performed using the validation sets to select optimal models. The performance of such optimal models on the test sets was reported.
It is important to note that our evaluation strategy improves upon 10-fold cross-validation and random splitting data into a single training/validation/test set, the two most commonly used evaluation techniques in (fine-grained) SV detection and report-level SV assessment studies~\cite{hanif2021rise,sonnekalb2022deep,li2021vuldeelocator,li2021vulnerability,nguyen2021information,le2021survey}. Our evaluation technique has separate test sets, which cross-validation does not, to objectively measure the performance of tuned/optimal models on unseen data. Using multiple (10) validation/test sets also increases the stability of results compared to a single set~\cite{raschka2018model}.
Moreover, we aim to provide baseline performance for function-level SV assessment in this study, so we did not apply any techniques like class rebalancing or feature selection/reduction to augment the data/features. Such augmentation can be explored in future work.
We also did not compare SV assessment using functions with that using reports as their use cases are different; function-level SV assessment is needed when SV reports are unavailable/unusable. It may be fruitful to compare/combine these two artifacts for SV assessment in the future.

\noindent \textbf{Evaluation measures}.
We used F1-Score\footnote{The macro version of F1-Score was used for multi-class classification.} and Matthews Correlation Coefficient (MCC) measures to quantify how well developed models perform SV assessment tasks. F1-Score values are from 0 to 1, and MCC has values from –1 to 1; 1 is the best value for both measures. These measures have been commonly used for SV assessment (e.g.,~\mbox{\cite{spanos2018multi,kudjo2019improving,le2019automated}}).
We used MCC as the main measure for selecting optimal models because MCC takes into account all classes, i.e., all cells in a confusion matrix, during evaluation~\mbox{\cite{luque2019impact}}.

\noindent \textbf{Statistical analysis}. To confirm the significance of results, we employed the one-sided Wilcoxon signed rank test~\cite{wilcoxon1992individual} and its respective effect size ($r = Z / \sqrt{N}$, where $Z$ is the $Z$-score statistic of the test and $N$ is the total count of samples)~\cite{tomczak2014need}.\footnote{$r \leq 0.1$: negligible, $0.1 < r \leq 0.3$: small, $0.3 < r \leq 0.5$: medium, $r > 0.5$: large~\cite{field2013discovering}} We used the Wilcoxon signed-rank test because it is a non-parametric test that can compare two-paired groups of data, and we considered a test significant if its confidence level was more than 99\% ($p$-value $<$ 0.01).
We did not use the popular Cohen's D~\cite{cohen2013statistical} and Cliff's $\delta$~\cite{macbeth2011cliff} effect sizes as they are not suitable for comparing paired data~\cite{wattanakriengkrai2020predicting}.

\section{Results}
\label{sec:results_msr22}

\subsection{\textbf{RQ1}: Are Vulnerable Code Statements More Useful Than Non-Vulnerable Counterparts for SV Assessment Models?}
\label{subsec:rq1_results_msr22}

\begin{figure}[t]
    \centering
    \includegraphics[width=\columnwidth,keepaspectratio]{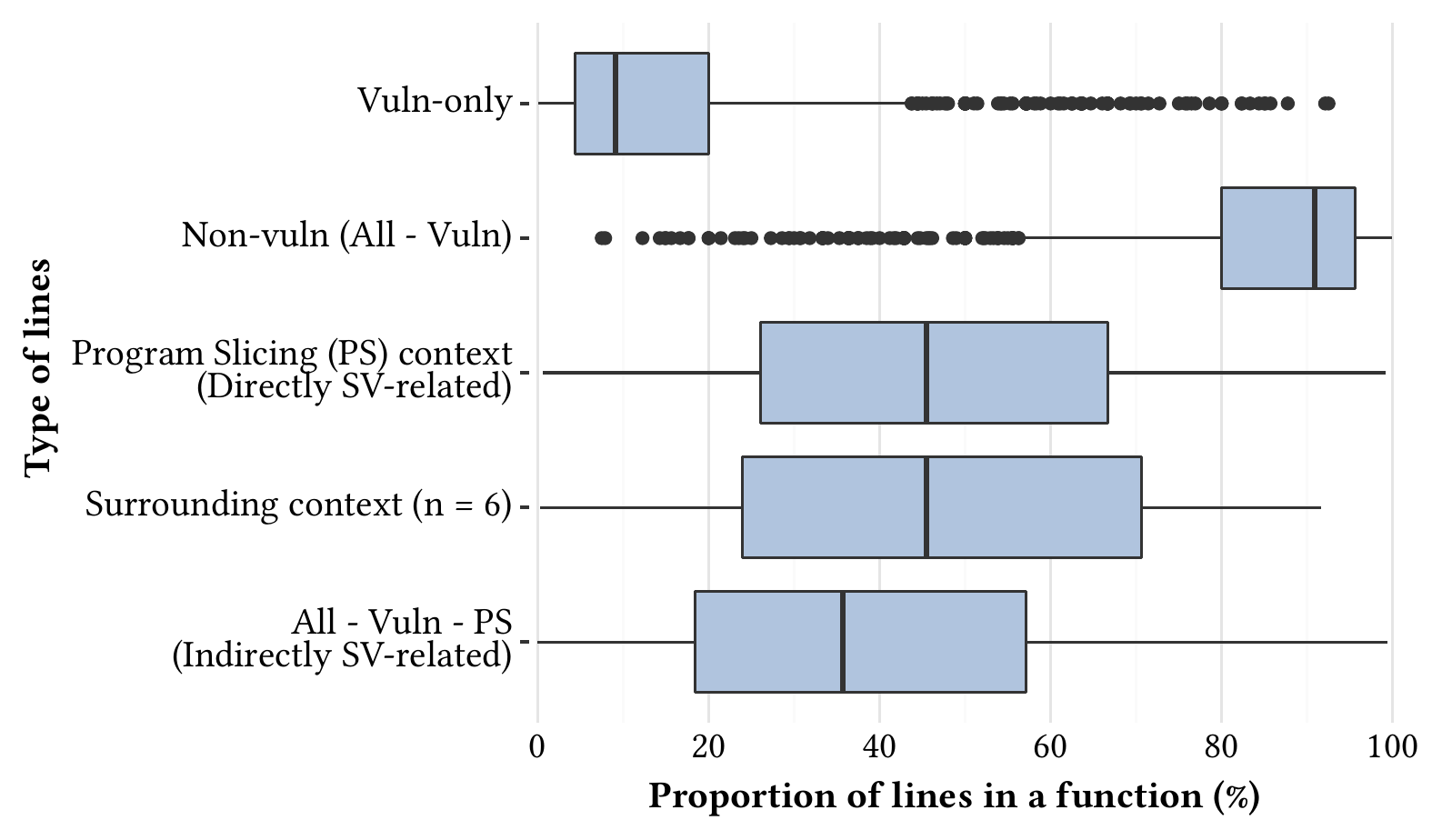}
    \caption[Proportions of different types of lines in a function.]{Proportions of different types of lines in a function. \textbf{Notes}: Min proportion of Vuln-only lines is non-zero (0.03\%); thus, max of Non-vuln line proportion is $<$ 100\%. Cosmetic lines were excluded from the computation of ratios.}
    \label{fig:scope_proportion_msr22}
\end{figure}

Based on the extraction process in section~\ref{subsec:data_collection_msr22}, we collected 1,782 vulnerable functions containing 5,179 vulnerable and 57,633 non-vulnerable statements. The proportions of these two types of statements are given in the first and second boxplots, respectively, in \fig~\ref{fig:scope_proportion_msr22}. On average, 14.7\% of the lines in the selected functions were vulnerable, 5.8 times smaller than that of non-vulnerable lines.
Interestingly, we also observed that 55\% of the functions contained only a single vulnerable statement. These values show that vulnerable statements constitute a very small proportion of functions.

\textbf{\textit{Despite the small size (no. of lines), vulnerable statements contributed more to the predictive performance of the seven assessment tasks than non-vulnerable statements (see \tab~\ref{tab:vuln_vs_nonvuln_msr22})}}.
We considered two variants of non-vulnerable statements for comparison. The first variant, \textit{Non-vuln (random)}, randomly selected the same number of lines as vulnerable statements from non-vulnerable statements in each function.
The second variant, \textit{Non-vuln (all)} aka. Non-vuln (All - Vuln) in \fig~\ref{fig:scope_proportion_msr22}, considered all non-vulnerable statements. Compared to same-sized non-vulnerable statements (Non-vuln (random)), \textit{Vuln-only} (using vulnerable statements solely) produced 116.9\%, 126.6\%, 98.7\%, 90.7\%, 147.9\%, 111.2\%, 116.7\% higher MCC for Access Vector, Access Complexity, Authentication, Confidentiality, Integrity, Availability, and Severity tasks, respectively. On average, Vuln-only was 114.5\% and 43.6\% better than Non-vuln (random) in MCC and F1-Score, respectively. We obtained similar results of Non-vuln (random) when repeating the experiment with differently randomized lines. When using all non-vulnerable statements (Non-vuln (all)), the assessment performance increased significantly, yet was still lower than that of vulnerable statements.
Average MCC and F1-Score of Vuln-only were 7.4\% and 5.5\% higher than Non-vuln (all), respectively.
The improvements of Vuln-only over the two variants of non-vulnerable statements were statistically significant across features/classifiers with $p$-values $<$ 0.01 ($p$-value$_{Non-vuln (random)} = 1.7 \times 10^{-36}$ and $p$-value$_{Non-vuln (all)} = 7.2 \times 10^{-11}$) and non-negligible effect sizes ($r_{Non-vuln (random)} = 0.62$ and $r_{Non-vuln (all)} = 0.32$).
The low performance of Non-vuln (random) implies that SV assessment models likely perform worse if vulnerable statements are incorrectly identified. Moreover, the decent performance of Non-vuln (all) shows that some non-vulnerable statements are potentially helpful for SV assessment, which are studied in detail in RQ2.

\begin{table}[t]
  \centering
  \caption[Testing performance for software vulnerability assessment tasks of vulnerable vs. non-vulnerable statements.]{Testing performance for SV assessment tasks of vulnerable vs. non-vulnerable statements. \textbf{Note}: Bold and grey-shaded values are the best row-wise performance.}
    \begin{tabular}{llccc}
    \hline
    \multirowcell{3}[0ex][l]{\textbf{CVSS metric}} & \multirowcell{3}[0ex][l]{\textbf{Evaluation}\\ \textbf{metric}} & \multicolumn{3}{c}{\textbf{Input type}}\\
    \cline{3-5}
    & & \textbf{Vuln-only} & \makecell{\textbf{Non-vuln}\\ \textbf{(random)}} & \makecell{\textbf{Non-vuln}\\ \textbf{(all)}} \\
    \hline
    \multirowcell{2}[0ex][l]{Access Vector} & F1-Score & \cellcolor[HTML]{C0C0C0} \textbf{0.820} & 0.650 & 0.786 \\
    & MCC & \cellcolor[HTML]{C0C0C0} \textbf{0.681} & \makecell{0.314} & \makecell{0.605}\\
    \hline
    \multirowcell{2}[0ex][l]{Access\\ Complexity} & F1-Score & \cellcolor[HTML]{C0C0C0} \textbf{0.622} & 0.458 & 0.592 \\
    & MCC & \cellcolor[HTML]{C0C0C0} \textbf{0.510} & \makecell{0.225} & \makecell{0.467}\\
    \hline
    \multirowcell{2}[0ex][l]{Authentication} & F1-Score & \cellcolor[HTML]{C0C0C0} \textbf{0.791} & 0.602 & 0.765 \\
    & MCC & \cellcolor[HTML]{C0C0C0} \textbf{0.630} & \makecell{0.317} & \makecell{0.614}\\
    \hline
    \multirowcell{2}[0ex][l]{Confidentiality} & F1-Score & \cellcolor[HTML]{C0C0C0} \textbf{0.645} & 0.411 & 0.625 \\
    & MCC & \cellcolor[HTML]{C0C0C0} \textbf{0.574} & \makecell{0.301} & 0.561\\
    \hline
    \multirowcell{2}[0ex][l]{Integrity} & F1-Score & \cellcolor[HTML]{C0C0C0} \textbf{0.650} & 0.384 & 0.616 \\
    & MCC & \cellcolor[HTML]{C0C0C0} \textbf{0.585} & \makecell{0.236} & \makecell{0.534}\\
    \hline
    \multirowcell{2}[0ex][l]{Availability} & F1-Score & \cellcolor[HTML]{C0C0C0} \textbf{0.647} & 0.417 & 0.624 \\
    & MCC & \cellcolor[HTML]{C0C0C0} \textbf{0.583} & \makecell{0.276} & \makecell{0.551}\\
    \hline
    \multirowcell{2}[0ex][l]{Severity} & F1-Score & \cellcolor[HTML]{C0C0C0} \textbf{0.695} & 0.414 & 0.610 \\
    & MCC & \cellcolor[HTML]{C0C0C0} \textbf{0.583} & \makecell{0.269} & \makecell{0.523}\\
    \hline
    \hline
    \multirowcell{2}[0ex][l]{Average} & F1-Score & \cellcolor[HTML]{C0C0C0} \textbf{0.695} & 0.484 & 0.659\\
    & MCC & \cellcolor[HTML]{C0C0C0} \textbf{0.592} & 0.276 & 0.551 \\
    \hline
    \end{tabular}
    \vspace{-3pt}
  \label{tab:vuln_vs_nonvuln_msr22}
\end{table}

\subsection{\textbf{RQ2}: To What Extent do Different Types of Context of Vulnerable Statements Contribute to SV Assessment Performance?}
\label{subsec:rq2_results_msr22}

In RQ2, we compared the performance of models using \textit{Program Slicing} (PS), \textit{surrounding} and \textit{function} context. Notably, we removed 365 cases for which we could not extract PS context from the dataset in section~\ref{subsec:data_collection_msr22}. Apparently, these cases were the same as using only vulnerable statements, making comparisons biased, especially against Vuln-only itself. The vulnerable statements in these cases mainly did not contain any variable, e.g., using an unsafe function without parameter.\footnote{\url{https://bit.ly/3pg36mp}} In this new dataset, PS and surrounding context approximately constituted 46\%, on average, of lines in vulnerable functions (see \fig~\ref{fig:scope_proportion_msr22}). We used six lines before and after vulnerable statements (n = 6) as surrounding context because this value resulted in the closest average context size to that of PS context. It is impossible to have the exactly same size because PS context is dynamically derived from vulnerable statements, while surrounding context is predefined. The roughly similar size helps test whether directly SV-related lines in PS context would be better than pre-defined surrounding lines for SV assessment.
The training and evaluation processes on the dataset in RQ2 were the same as in RQ1.\footnote{The RQ1 findings still hold when using the new dataset in RQ2, yet with a slight ($\approx$2\%) decrease in absolute model performance.}

\begin{landscape}
\begin{figure*}[t]
     \centering
     \includegraphics[width=0.8\linewidth,keepaspectratio]{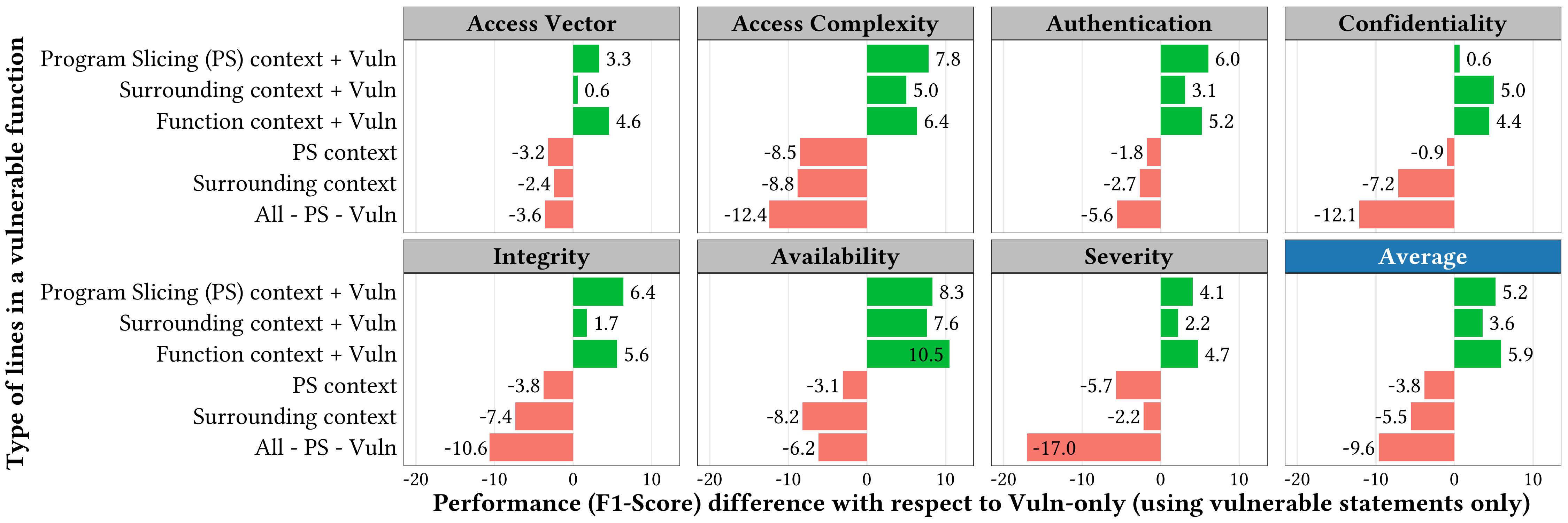}
     \bigskip
     \hfill
     \includegraphics[width=0.8\linewidth,keepaspectratio]{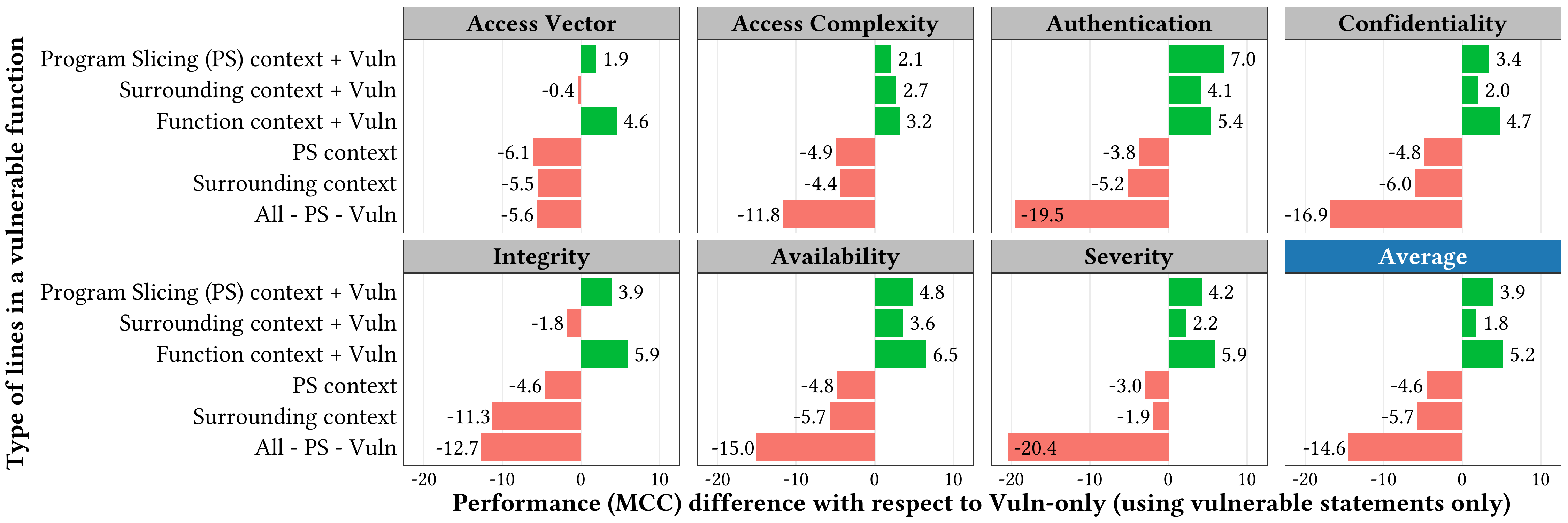}
     \caption[Differences in testing software vulnerability assessment performance (F1-Score and MCC) between models using different types of lines/context and those using only vulnerable statements.]{Differences in testing SV assessment performance (F1-Score and MCC) between models using different types of lines/context and those using only vulnerable statements. \textbf{Note}: The differences were multiplied by 100 to improve readability.}
    \label{fig:context_comparison_msr22}
\end{figure*}
\end{landscape}

\textit{\textbf{Adding context to vulnerable statements led to better SV assessment performance than using vulnerable statements only (see \fig~\ref{fig:context_comparison_msr22}). Among the three, function context was the best, followed by PS and then surrounding context}}. In terms of MCC, function context working together with vulnerable statements beat Vuln-only by 6.4\%, 6.5\%, 9\%, 8.2\%, 11\%, 11.4\%, 9.7\% for Access Vector, Access Complexity, Authentication, Confidentiality, Integrity, Availability, and Severity tasks, respectively. The higher F1-Score values when incorporating function context to vulnerable statements are also evident in \fig~\ref{fig:context_comparison_msr22}.
On average, combining function context and vulnerable statements attained 0.64 MCC and 0.75 F1-Score, surpassing using vulnerable lines solely by 8.9\% in MCC and 8.5\% in F1-Score.
Although PS context + Vuln performed slightly worse than function context + Vuln, MCC and F1-Score of PS context + Vuln were still 6.7\% and 7.5\% ahead of Vuln-only, respectively. The improvements of function and PS context + Vuln over Vuln-only were significant across features/classifiers, i.e., $p$-values of $1.2 \times 10^{-17}$ and $2.1 \times 10^{-13}$ and medium effect sizes of 0.42 and 0.36, respectively. Compared to function/PS context + Vuln, surrounding context + Vuln outperformed Vuln-only by smaller margins, i.e., 3\% for MCC and 5.2\% for F1-Score ($p$-value = $3.7 \times 10^{-8} < 0.01$ with a small effect size ($r = 0.27$)). These findings show the usefulness of directly SV-related (PS) lines for SV assessment models, while six lines surrounding vulnerable statements seemingly contain less related information for the SV assessment tasks.

\noindent Further investigation revealed that only 49\% of lines in PS context overlapped with those in surrounding context (n = 6). Note that the performance of surrounding context tended to approach that of function context as the surrounding context size increased. Using the dataset in RQ1, we also obtained the same patterns, i.e., function context + Vuln > surrounding context + Vuln > Vuln-only.
This result shows that function context is generally better than the other context types, indicating the plausibility of building effective SV assessment models using only the output of function-level SV detection (i.e., requiring no knowledge about which statements are vulnerable in each function).

\textit{\textbf{Although the three context types were useful for SV assessment when combined with vulnerable statements, using these context types alone significantly reduced the performance}}. As shown from RQ1, using only function context (i.e., Non-vuln (all) in \tab~\ref{tab:vuln_vs_nonvuln_msr22}) was 6.9\% inferior in MCC and 5.2\% lower in F1-Score than Vuln-only. Using the new dataset in RQ2, we obtained similar reductions in MCC and F1-Score values. \fig~\ref{fig:context_comparison_msr22} also indicates that using only PS and surrounding context decreased MCC and F1-Score of all the tasks. Particularly, using PS context alone reduced MCC and F1-Score by 7.8\% and 5.5\%, respectively; whereas, such reductions in values for using only surrounding context were 9.8\% and 8\%. These performance drops were confirmed significant with $p$-values < 0.01 and non-negligible effect sizes. Overall, the performance rankings of the context types with and without vulnerable statements were the same, i.e., function > PS > surrounding context. We also observed that all context types were better (increasing 20.1-28.2\% in MCC and 6.9-17.5\% in F1-Score) than non-directly SV-related lines (i.e., All - PS - Vuln in \fig~\ref{fig:context_comparison_msr22}). These findings highlight the need for using context together with vulnerable statements rather than using each of them alone for function-level SV assessment tasks.

\begin{table}[t]
  \centering
  \caption[Differences in testing performance for software vulnerability assessment tasks between double-input models (RQ3) and single-input models (RQ1/2).]{Differences in testing performance for SV assessment tasks between double-input models (RQ3) and single-input models (RQ1/2). \textbf{Notes}: The differences were multiplied by 100 to increase readability. Green and red colors denote value increase and decrease, respectively. Darker color shows a higher magnitude of increase/decrease. Average performance values (multiplied by 100) of double-input models for the three context (ctx) types are in parentheses.
  }
    \begin{tabular}{llP{2.5cm}P{2.5cm}P{2.5cm}}
    \hline
    \multirowcell{3}[0ex][l]{\textbf{CVSS metric}} & \multirowcell{3}[0ex][l]{\textbf{Evaluation}\\ \textbf{metric}} & \multicolumn{3}{c}{\textbf{Input type (double)}}\\
    \cline{3-5}
    & & \makecell{\textbf{PS}\\ \textbf{ctx + Vuln}} & \makecell{\textbf{Surrounding}\\ \textbf{ctx + Vuln}} & \makecell{\textbf{Function}\\ \textbf{ctx + Vuln}} \\
    \hline
    \multirowcell{2}[0ex][l]{Access\\ Vector} & F1-Score & \cellcolor[HTML]{67cf82} 1.3 & \cellcolor[HTML]{6fd187} 1.2 & \cellcolor[HTML]{6fd187} 1.2 \\
    & MCC & \cellcolor[HTML]{07bc3d} 2.6 & \cellcolor[HTML]{00ba38} 2.7 & \cellcolor[HTML]{16bf48} 2.4\\
    \hline
    \multirowcell{2}[0ex][l]{Access\\ Complexity} & F1-Score & \cellcolor[HTML]{ff8a8a} -0.5 & \cellcolor[HTML]{9bdaa7} 0.6 & \cellcolor[HTML]{ff6e6e} -0.7 \\
    & MCC & \cellcolor[HTML]{58cc78} 1.5 & \cellcolor[HTML]{a2dbad} 0.5 & \cellcolor[HTML]{16bf48} 2.4\\
    \hline
    \multirowcell{2}[0ex][l]{Authentication} & F1-Score & \cellcolor[HTML]{a2dbad} 0.5 & \cellcolor[HTML]{93d8a2} 0.7 & \cellcolor[HTML]{b1deb7} 0.3 \\
    & MCC & \cellcolor[HTML]{7dd492} 1.0 & \cellcolor[HTML]{4ac96d} 1.7 & \cellcolor[HTML]{93d8a2} 0.7\\
    \hline
    \multirowcell{2}[0ex][l]{Confidentiality} & F1-Score & \cellcolor[HTML]{ffb3b3} -0.2 & \cellcolor[HTML]{c6e3c6} 0.01 & \cellcolor[HTML]{ff5353} -0.9 \\
    & MCC & \cellcolor[HTML]{ff6e6e} -0.7 & \cellcolor[HTML]{ff8a8a} -0.5 & \cellcolor[HTML]{ff0000} -1.5\\
    \hline
    \multirowcell{2}[0ex][l]{Integrity} & F1-Score & \cellcolor[HTML]{b8e0bc} 0.2 & \cellcolor[HTML]{93d8a2} 0.7 & \cellcolor[HTML]{b8e0bc} 0.2 \\
    & MCC & \cellcolor[HTML]{ffc5c5} -0.07 & \cellcolor[HTML]{76d28d} 1.1 & \cellcolor[HTML]{a2dbad} 0.5\\
    \hline
    \multirowcell{2}[0ex][l]{Availability} & F1-Score & \cellcolor[HTML]{ff8a8a} -0.5 & \cellcolor[HTML]{ffb3b3} -0.2 & \cellcolor[HTML]{ff4545} -1.0\\
    & MCC & \cellcolor[HTML]{aaddb2} 0.4 & \cellcolor[HTML]{aaddb2} 0.4 & \cellcolor[HTML]{c0e1c2} 0.1\\
    \hline
    \multirowcell{2}[0ex][l]{Severity} & F1-Score & \cellcolor[HTML]{93d8a2} 0.7 & \cellcolor[HTML]{85d597} 0.9 & \cellcolor[HTML]{93d8a2} 0.7\\
    & MCC & \cellcolor[HTML]{b8e0bc} 0.2 & \cellcolor[HTML]{c6e3c6} 0.02 & \cellcolor[HTML]{b8e0bc} 0.2\\
    \hline
    \hline
    \multirowcell{2}[0ex][l]{Average} & F1-Score & \cellcolor[HTML]{b8e0bc} 0.2 (74.7) & \cellcolor[HTML]{a2dbad} 0.5 0.5 (73.4) & \cellcolor[HTML]{ffcbcb} -0.03 (75.2)\\
    & MCC & \cellcolor[HTML]{93d8a2} 0.7 (63.1) & \cellcolor[HTML]{8cd79d} 0.8 (61.1) & \cellcolor[HTML]{a2dbad} 0.5 (64.1) \\
    \hline
    \end{tabular}
    \vspace{-3pt}
  \label{tab:single_vs_double_msr22}
\end{table}

\subsection{\textbf{RQ3}: Does Separating Vulnerable Statements and Context to Provide Explicit Location of SVs Improve Assessment Performance?}
\label{subsec:rq3_results_msr22}

RQ3 evaluated the approach of separating vulnerable statements from their context as two inputs for building SV assessment models. Theoretically, this double-input method tells a model the exact vulnerable and context parts in input code, helping the model capture the information from relevant parts for SV assessment tasks more easily. To separate features, feature vectors are generated for each of the two inputs and then concatenated to form a single vector of twice the size of the vector used in RQ2. RQ3 used the same dataset from RQ2 (i.e., excluding cases without PS context) and the respective model evaluation procedure to objectively compare PS context with the other context types.

\textit{\textbf{Overall, double-input models improved the performance for all types of context compared to single-input ones, but the improvements were not substantial ($\approx$1\%)}}. \tab~\ref{tab:single_vs_double_msr22} clearly indicated the improvement trend; i.e., a majority of the cells have green color. We noticed that the rankings of double-input models using different context types still remained the same as in RQ2, i.e., function > PS > surrounding context.
Specifically, double-input models raised the MCC values of single-input models using PS, surrounding, and function context by 1.1\%, 1.4\%, and 0.8\%, respectively. In terms of F1-Score of double-input models, PS and surrounding context had 0.26\% and 0.75\% increase, while function context suffered from a 0.04\% decline. We found the absolute performance differences between double-input and single-input models for the seven tasks were actually small and not statistically significant with negligible effect sizes ($r_{PS~ctx + Vuln} = 0.059$, $r_{Surrounding~ctx + Vuln} = 0.092$, and $r_{Function~ctx + Vuln} = 0.021$). We observed similar changes/patterns of function/surrounding context when using the full dataset in RQ1. The findings suggest that models using function context + Vuln as a single-input in RQ2 still perform competitively. This result strengthens the conclusion in RQ2 that SV assessment models benefit from vulnerable statements along with (in-)directly SV-related lines in functions, yet not necessarily where these lines are located.

\section{Discussion}
\label{sec:discussion_msr22}
\subsection{Function-Level SV Assessment: Baseline Models and Beyond}
\label{subsec:baseline_and_beyond_msr22}

\begin{figure}[t]
    \centering
    \includegraphics[width=\columnwidth,keepaspectratio]{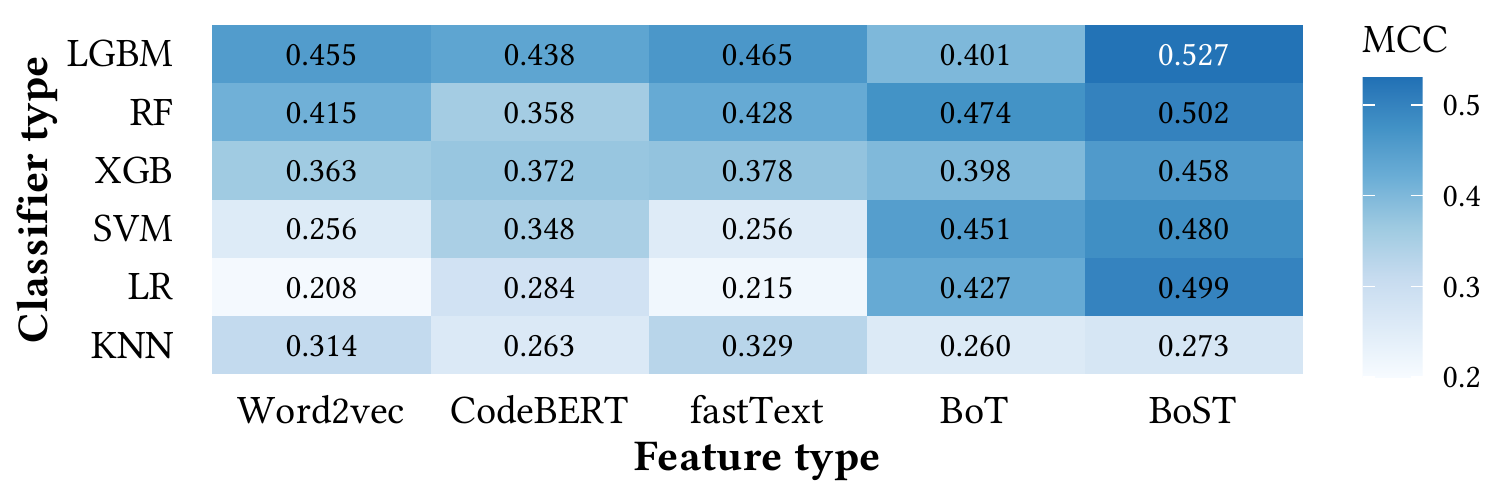}
    \caption[Average performance (MCC) of six classifiers and five features for software vulnerability assessment in functions.]{Average performance (MCC) of six classifiers and five features for SV assessment in functions. Notes:
    BoT and BoST are Bag-of-Tokens and Bag-of-Subtokens, respectively.}
    \label{fig:baselines_msr22}
\end{figure}

From RQ1-RQ3, we have shown that vulnerable statements and their context are useful for SV assessment tasks. In this section, we discuss the performance of various features and classifiers used to develop SV assessment models on the function level. We also explore the patterns of false positives of the models used in this work. Through these discussions, we aim to provide recommendations on building strong baseline models and inspire future data-driven advances in function-level SV assessment.

\noindent \textbf{Practices of building baselines}. \textit{\textbf{Among the investigated features and classifiers, a combination of LGBM classifier and Bag-of-Subtokens features produced the best overall performance for the seven SV assessment tasks (see \fig~\ref{fig:baselines_msr22})}}. In addition, LGBM outperformed the other classifiers, and Bag-of-Subtokens was better than the other features. However, we did not find a single set of hyperparameters that was consistently better than the others, emphasizing the need for hyperparameter tuning for function-level SV assessment tasks, as generally recommended in the literature~\cite{treude2019predicting,tantithamthavorn2018impact}.
Regarding the classifiers, the ensemble ones (LGBM, RF, and XGB) were significantly better than the single counterparts (SVM, LR, and KNN) when averaging across all feature types, aligning with the previous findings for SV assessment using SV reports~\cite{le2019automated,spanos2018multi}. Regarding the features, the ones augmented by sub-tokens (Bag-of-Subtokens, fastText, and CodeBERT) had stronger performance, on average, than the respective feature types using only word-based representation (Bag-of-Tokens and Word2vec). This observation suggests that SV assessment models do benefit from sub-tokens, probably because rare code tokens are more likely to be captured by these features. This result is similar to Le et al.~\cite{le2019automated}'s finding for report-level SV assessment models. All of the aforementioned result comparisons were confirmed statistically significant with $p$-values < 0.01 and non-negligible effect sizes; similar patterns were also obtained for F1-Score.
Surprisingly, the latest feature model, CodeBERT, did not show superior performance in this scenario, likely because the model was originally pre-trained on multiple languages, not only Java (the main language used in this study). Fine-tuning the weights of CodeBERT using SV data in a specific programming language is worthy of exploration for potentially improving the performance of this feature model.
Overall, using the aforementioned baseline features and classifiers, we observed a common pattern in the performance ranking of the seven CVSS metrics across different input types, i.e., Access Vector > Authentication > Severity > Confidentiality -- Integrity -- Availability > Access Complexity. We speculate that the metric-wise class distribution (see \fig~\ref{fig:cvss_distribution_msr22}) can be a potential explanation for this ranking. Specifically, Access Vector and Authentication are binary classifications, which have less uncertainty than the other tasks. In addition, Confidentiality, Integrity, and Availability are all impact metrics with roughly similar distributions, resulting in similar performance as well. Access Complexity suffers the most from imbalanced data among the tasks, and thus this task has the worst performance.

\noindent \textbf{False-positive patterns}.
We manually analyzed the incorrect predictions by the optimal models using the best-performing (function) context from RQ2/RQ3. From these cases, we found two key patterns of false positives.
\textit{\textbf{The first pattern concerned SVs affecting implicit code in function calls}}. For example, a feature requiring authentication, i.e., the synchronous mode, was run before user's password was checked in a function (\code{doFilter}), leading to a potential access-control related SV.\footnote{\url{https://bit.ly/32vyggM} (CVE-2018-1000134)}
The execution of such mode was done by another function, \code{processSync}, but its implementation along with the affected components inside was not visible to the affected function. Such invisibility hinders a model's ability to fully assess SV impacts. A straightforward solution is to employ inter-procedural analysis~\cite{li2018vuldeepecker}, but scalability is a potential issue as SVs can affect multiple functions and even the ones outside of a project (i.e., functions in third-party libraries).
Future work can leverage Question and Answer websites to retrieve and analyze SV-related information of third-party libraries~\cite{le2021large}.
\textit{\textbf{The second type of false positives involved vulnerable variables with obscure context}}. For instance, a function used a potentially vulnerable variable containing malicious inputs from users, but the affected function alone did not contain sufficient information/context about the origin of the variable.\footnote{\url{https://bit.ly/3plU7QP} (CVE-2012-0391)} Without such variable context, a model would struggle to assess the exploitability of an SV; i.e., through which components attackers can penetrate into a system and whether any authentication is required during the penetration. Future work can explore taint analysis~\cite{kim2014survey} to supplement function-level SV assessment models with features about variable origin/flow.

\subsection{Threats to Validity}
\label{subsec:threats_msr22}

The first threat concerns the curation of vulnerable functions and statements for building SV assessment models. We considered the recommendations in the literature to remove noise in the data (e.g., abnormally large and cosmetic changes). We also performed manual validation to double-check the validity of our data.

Another threat is about the robustness of our own implementation of the program slicing extraction. To mitigate the threat, we carefully followed the algorithms and descriptions given in the previous work~\cite{dashevskyi2018screening} to extract the intra-procedural backward and forward slices for a particular code line.

Other threats are related to the selection and optimality of baseline models. We assert that it is nearly impossible to consider all types of available features and models due to limited resources. Hence, we focused on the common techniques and their respective hyperparameters previously used for relevant tasks, e.g., report-level SV assessment. We are also the first to tackle function-level SV assessment using data-driven models, and thus our imperfect baselines can still stimulate the development of more advanced and better-performing techniques in the future.

Regarding the reliability of our study, we confirmed the key findings with $p$-values < 0.01 using non-parametric Wilcoxon signed rank tests and non-negligible effect sizes. Regarding the generalizability of our results, we only performed our study in the Java programming language, yet we mitigated this threat by using 200 real-world projects of diverse domains and scales. The data and models were also released at \url{https://github.com/lhmtriet/Function-level-Vulnerability-Assessment} to support reuse and extension to new languages/applications.

\section{Related Work}
\label{sec:related_work_msr22}
\subsection{Code Granularities of SV Detection}
SV detection has long attracted attention from researchers and there have been many proposed data-driven solutions to automate this task~\cite{ghaffarian2017software}. Neuhaus et al.~\cite{neuhaus2007predicting} were among the first to tackle SV detection in code components/files. This seminal work has inspired many follow-up studies on component/file-level SV detection (e.g.,~\cite{chowdhury2011using,shin2013can,tang2015predicting}).
Over time, function-level SV detection tasks have become more popular~\cite{lin2018cross,zhou2019devign,nguyen2019deep,bilgin2020vulnerability,wang2020combining} as functions are usually much smaller than files, significantly reducing inspection effort for developers. For example, the number of code lines in the methods in our dataset was only 35, on average, nearly 10 times smaller than that (301) of files.
Recently, researchers have begun to predict exact vulnerable statements/lines in functions (e.g.,~\cite{li2021vuldeelocator,li2021vulnerability,nguyen2021information, ding2021velvet}). This emerging research is based on an important observation that only a small number of lines in vulnerable functions contain root causes of SVs. Instead of detecting SVs as in these studies, we focus on SV assessment tasks after SVs are detected. Specifically, utilizing the outputs (vulnerable functions/statements) from these SV detection studies, we perform function-level SV assessment to support SV understanding/prioritization before fixing.

\subsection{Data-Driven SV Assessment}
SV assessment has been an integral step for addressing SVs. CVSS has been shown to provide one of the most reliable metrics for SV assessment~\cite{johnson2016can}. According to Chapter~\ref{chapter:lit_review}, there has been a large and growing body of research work on automating SV assessment tasks, especially predicting the CVSS metrics for ever-increasing SVs. Most of the current studies (e.g.,~\cite{yamamoto2015text,spanos2017assessment,spanos2018multi,le2019automated,elbaz2020fighting,duan2021automated}) have utilized SV descriptions available in bug/SV reports/databases, mostly NVD, to predict the CVSS metrics. However, according to our analysis in section~\ref{sec:background_msr22}, NVD reports of SVs are usually released long (up to 1k days) after SVs have been fixed, rendering report-level SV assessment potentially untimely for SV fixing. Unlike the current studies, we propose shifting the SV assessment tasks to the function level, which can help developers assess functions right after they are found vulnerable and before fixing.
Note that we assess all types of SVs in source code, not only the ones in dependencies~\mbox{\cite{ponta2018beyond,kritikos2019survey}}.
Overall, our study informs the practice of developing strong baselines for function-level SV assessment tasks by combining vulnerable statements and their context.
It is worth noting that function-level SV assessment does not completely replace report-level SV assessment. The latter is still useful for assessing SVs in third-party libraries/software without available source code (e.g., commercial products), to prioritize the application of security patches provided by vendors.

\section{Chapter Summary}
\label{sec:conclusions_msr22}

We motivate the need for function-level SV assessment to provide essential information for developers before fixing SVs. Through large-scale experiments, we studied the use of data-driven models for automatically assigning the seven CVSS assessment metrics to SVs in functions. We demonstrated that strong baselines for these tasks benefited not only from fine-grained vulnerable statements, but also the context of these statements. Specifically, using vulnerable statements with all the other lines in functions produced the best performance of 0.64 MCC and 0.75 F1-Score. These promising results show that function-level SV assessment tasks deserve more attention and contribution from the community, especially techniques that can strongly capture the relations between vulnerable statements and other code lines/components.

%% file: Chapters/code.tex
\captionsetup[lstlisting]{labelsep=period, textfont=footnotesize, labelfont=footnotesize, justification=justified}

\newcommand{\lstbg}[3][0pt]{{\fboxsep#1\colorbox{#2}{\strut #3}}}
\lstdefinelanguage{diff}{
  basicstyle=\ttfamily\small,
  morecomment=[f][\lstbg{red!20}]-,
  morecomment=[f][\lstbg{green!20}]+,
}

\definecolor{difftitle}{HTML}{000099}
\definecolor{diffstart}{HTML}{660099}
\definecolor{diffincl}{HTML}{006600}
\definecolor{diffrem}{HTML}{AA3300}

\definecolor{del_color}{HTML}{FFCCCC}
\definecolor{add_color}{HTML}{CCFFCC}

\lstdefinestyle{lst}{
    numbers=left, 
    numberstyle=\scriptsize, 
    numbersep = 5pt,
    framexleftmargin = 0in,
    framexrightmargin = 0in,
    xleftmargin = 0.18in,
    xrightmargin = 0.1in,
    basicstyle=\ttfamily\scriptsize, 
    frame=lines,
    showtabs=true,
    showspaces=true,
    showstringspaces=false,
    literate={\ }{{\ }}1,
    escapeinside={<@}{@>}
}

\lstset{belowskip=-0.05in}

\begin{figure}
\begin{lstlisting}[language=diff,style=lst]
protected String getExecutionPreamble()
{
    if (getWorkingDirectoryAsString() == null)
    {return null;}
    <@\color{blue}{String dir = getWorkingDirectoryAsString();}@>
    <@\color{blue}{StringBuilder sb = new StringBuilder();}@>
    <@\color{blue}{sb.append("cd");}@>
-   sb.append(unifyQuotes(dir));
+   sb.append(quoteOneItem(dir, false));
    <@\color{blue}{sb.append("\&\&");}@>
    <@\color{blue}{return sb.toString();}@>
}
\end{lstlisting}
     
\caption[A vulnerable function extracted from the fixing commit \textit{b38a1b3} of a software vulnerability (CVE-2017-1000487) in the \textit{Plexus-utils} project.]{A vulnerable function extracted from the fixing commit \textit{b38a1b3} of an SV (CVE-2017-1000487) in the \textit{Plexus-utils} project. \textbf{Notes}: Line 8 is vulnerable. Deleted and added lines are highlighted in red and green, respectively. Blue-colored code lines affect or are affected by line 8 directly.}
\label{fig:ex_sv_msr22}
\end{figure}

%% file: Chapters/Chapter_5_ASE2021.tex
\chapter{Automated Commit-Level Software Vulnerability Assessment}
\label{chap:ase21}

\begin{tcolorbox}
\textbf{Related publication}: This chapter is based on our paper titled ``\textit{DeepCVA: Automated Commit-level Vulnerability Assessment with Deep Multi-task Learning}'' published in the 36\textsuperscript{th} IEEE/ACM International Conference on Automated Software Engineering (ASE), 2021 (CORE A*)~\cite{le2021deepcva}.
\end{tcolorbox}
\bigskip

In Chapter~\ref{chap:msr22}, we have distilled practices of performing Software Vulnerability (SV) assessment on the code function level. While function-level assessment is useful for analyzing SVs before fixing, some SVs may be detected late in codebases and pose significant security risks for a long time. Thus, it is increasingly suggested to identify SV in code commits to give early warnings about potential security risks. However, there is a lack of effort to assess vulnerability-contributing commits right after they are detected to provide timely information about the exploitability, impact and severity of SVs. Such information is important to plan and prioritize the mitigation for the identified SVs. In Chapter~\ref{chap:ase21}, we propose a novel Deep multi-task learning model, DeepCVA, to automate seven Commit-level Vulnerability Assessment tasks simultaneously based on Common Vulnerability Scoring System (CVSS) metrics. We conduct large-scale experiments on 1,229 vulnerability-contributing commits containing 542 different SVs in 246 real-world software projects to evaluate the effectiveness and efficiency of our model. We show that DeepCVA is the best-performing model with 38\% to 59.8\% higher Matthews Correlation Coefficient than many supervised and unsupervised baseline models. DeepCVA also requires 6.3 times less training and validation time than seven cumulative assessment models, leading to significantly less model maintenance cost as well. Overall, DeepCVA presents the first effective and efficient solution to automatically assess SVs early in software systems.

\newpage

\section{Introduction}

As reviewed in Chapter~\ref{chapter:lit_review}, existing techniques (e.g.,~\mbox{\cite{lamkanfi2010predicting,han2017learning,spanos2018multi,le2019automated,le2021survey}}) to automate bug/Software Vulnerability (SV) assessment have mainly operated on bug/SV reports, but these reports may be only available long after SVs appeared in practice.
Our motivating analysis revealed that there were 1,165 days, on average, from when an SV was injected in a codebase until its report was published on National Vulnerability Database (NVD)~\cite{nvd}. Our analysis agreed with the findings of Meneely et al.~\mbox{\cite{meneely2013patch}}.
To tackle late-detected bugs/SVs, recently, Just-in-Time (commit-level) approaches (e.g.,~\mbox{\cite{hoang2019deepjit, kamei2012large,perl2015vccfinder,yang2017vuldigger}}) have been proposed to rely on the changes in code commits to detect bugs/SVs right after bugs/SVs are added to a codebase. Such early commit-level SV detection can also help reduce the delay in SV assessment.

Even when SVs are detected early in commits, we argue that existing automated techniques relying on bug/SV reports still struggle to perform \textit{just-in-time} SV assessment. Firstly, there are significant delays in the availability of SV reports, which render the existing SV assessment techniques unusable. Specifically, SV reports on NVD generally only appear seven days after the SVs were found/disclosed~\mbox{\cite{rodriguez2018analysis}}. Some of the detected SVs may not even be reported on NVD~\mbox{\cite{sawadogo2020learning}}, e.g., because of no disclosure policy. User-submitted bug/SV reports are also only available post-release and more than 82\% of the reports are filed more than 30 days after developers detected the bugs/SVs~\mbox{\cite{thung2012would}}.
Secondly, code review can provide faster SV assessment, but there are still unavoidable delays (from several hours to even days)~\mbox{\cite{bosu2012peer}}.
Delays usually come from code reviewers' late responses and manual analyses depending on the reviewers' workload and code change complexity~\mbox{\cite{thongtanunam2015investigating}}. Thirdly, it is non-trivial to automatically generate bug/SV reports from vulnerable commits as it would require non-code artifacts (e.g., stack traces or program crashes) that are mostly unavailable when commits are submitted~\mbox{\cite{lamkanfi2010predicting,moran2015auto}}.

Performing commit-level SV assessment provides a possibility to inform committers about the exploitability, impact and severity of SVs in code changes and prioritize fixing earlier than current report-level SV assessment approaches. However, to the best of our knowledge, there is no existing work on automating SV assessment in commits. Prior SV assessment techniques that analyze text in SV databases (e.g.,~\mbox{\cite{han2017learning,spanos2018multi,le2019automated}}) also cannot be directly adapted to the commit level. Contrary to text, commits contain deletions and additions of code with specific structure and semantics~\mbox{\cite{hoang2019deepjit,hoang2019patchnet}}. Additionally, we speculate that the expert-based Common Vulnerability Scoring System (CVSS) metrics~\mbox{\cite{cvss}}, which are commonly used to quantify the exploitability, impact and severity level of SVs for SV assessment, can be related. For example, an SQL injection is likely to be highly severe since attackers can exploit it easily via crafted input and compromise data confidentiality and integrity. We posit that these metrics would have common patterns in commits that can be potentially shared between SV assessment models. Predicting related tasks in a shared model has been successfully utilized for various applications~\mbox{\cite{zhang2017survey}}. For instance, an autonomous car is driven with simultaneous detection of vehicles, lanes, signs and pavement~\mbox{\cite{chowdhuri2019multinet}}. These observations motivated us to tackle a new and important research challenge, \textbf{``How can we leverage the common attributes of assessment tasks to perform effective and efficient commit-level SV assessment?''}

We present DeepCVA, a novel \mbox{\underline{\textbf{Deep}}} multi-task learning model, to automate \mbox{\underline{\textbf{C}}}ommit-level \mbox{\underline{\textbf{V}}}ulnerability \mbox{\underline{\textbf{A}}}ssessment. DeepCVA first uses attention-based convolutional gated recurrent units to extract features of code and surrounding context from vulnerability-contributing commits (i.e., commits with vulnerable changes). The model uses these features to predict seven CVSS assessment metrics (i.e., Confidentiality, Integrity, Availability, Access Vector, Access Complexity, Authentication, and Severity) simultaneously using the multi-task learning paradigm. The predicted CVSS metrics can guide SV management and remediation processes.

Our key \textbf{contributions} are summarized as follows:

\begin{contribution}
    \item We are the first to tackle the commit-level SV assessment tasks that enable early security risks estimation and planning for SV remediation.
    \item We propose a unified model, DeepCVA, to automate seven commit-level SV assessment tasks simultaneously.
    \item We extensively evaluate DeepCVA on our curated large-scale dataset of 1,229 vulnerability-contributing commits with 542 SVs from 246 real-world projects.
    \item We demonstrate that DeepCVA has 38\% to 59.8\% higher performance (Matthews Correlation Coefficient (MCC)) than various supervised and unsupervised baseline models using text-based features and software metrics. The proposed context-aware features improve the MCC of DeepCVA by 14.8\%. The feature extractor with attention-based convolutional gated recurrent units, on average, adds 52.9\% MCC for DeepCVA. Multi-task learning also makes DeepCVA 24.4\% more effective and 6.3 times more efficient in training, validation, and testing than separate models for seven assessment tasks.
    \item We release our source code, models and datasets for future research at \url{https://github.com/lhmtriet/DeepCVA}.
\end{contribution}

\noindent \textbf{Chapter organization}. Section~\mbox{\ref{sec:background_ase21}} introduces preliminaries and motivation. Section~\mbox{\ref{sec:deepcva_method_ase21}} proposes the DeepCVA model for commit-level SV assessment. Section~\mbox{\ref{sec:expt_design_ase21}} describes our experimental design and setup. Section~\mbox{\ref{sec:results_ase21}} presents the experimental results. Section~\mbox{\ref{sec:discussion_ase21}} discusses our findings and threats to validity. Section~\mbox{\ref{sec:related_work_ase21}} covers the related work. Section~\mbox{\ref{sec:conclusions_ase21}} concludes the work and proposes future directions.

\section{Background and Motivation}
\label{sec:background_ase21}
\subsection{Vulnerability in Code Commits}
\label{subsec:vuln_commits_ase21}

\begin{figure}[t]
    \centering
    \includegraphics[width=0.83\columnwidth,keepaspectratio]{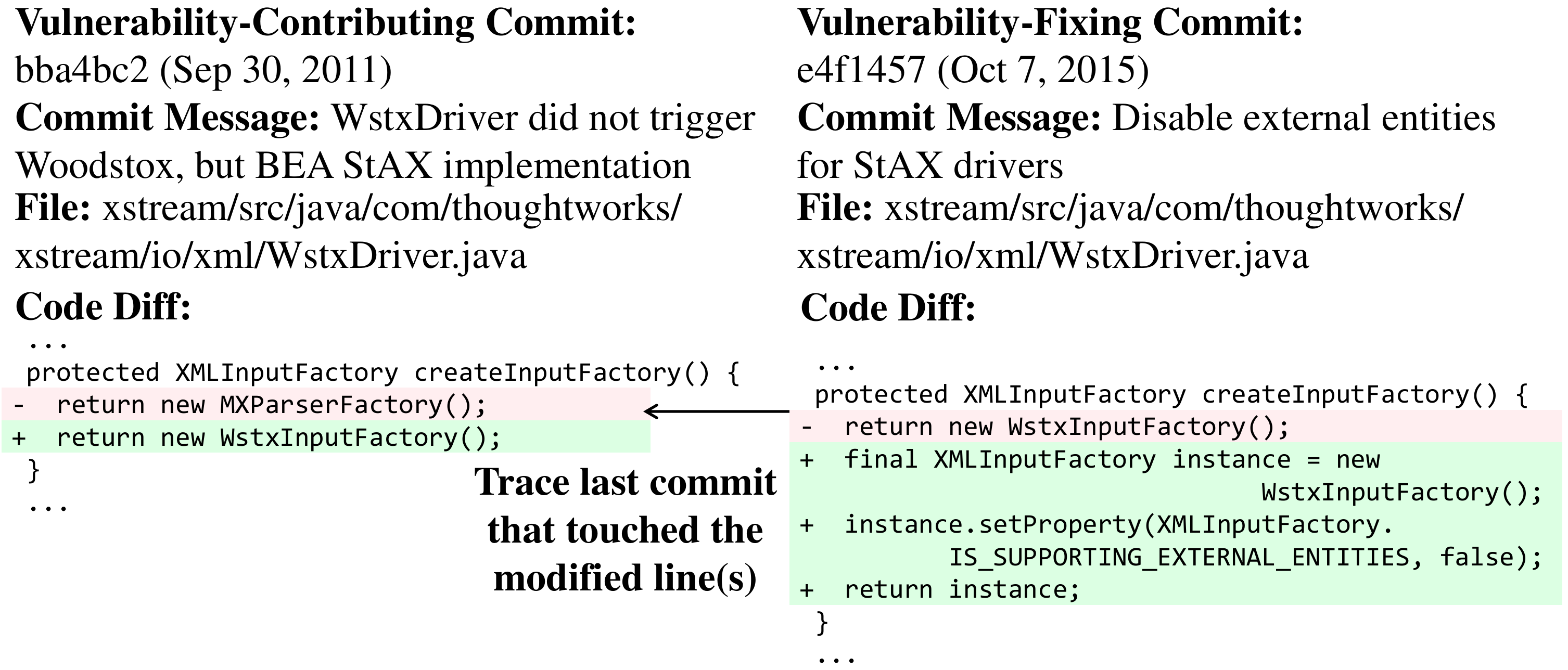}
    \caption[Exemplary software vulnerability fixing commit for the XML external entity injection (XXE) (CVE-2016-3674) and its respective software vulnerability contributing commit in the \textit{xstream} project.]{Exemplary SV fixing commit (right) for the XML external entity injection (XXE) (CVE-2016-3674) and its respective SV contributing commit (left) in the \textit{xstream} project.}
    \label{fig:commit_ex_ase21}
\end{figure}

Commits are an essential unit of any version control system (e.g., Git) and record all the chronological changes made to the codebase of a software project. As illustrated in Fig.~\mbox{\ref{fig:commit_ex_ase21}}, changes in a commit consist of deletion(s) (–) and/or addition(s) (+) in each affected file.

Vulnerability-Contributing Commits (VCCs) are commits whose changes contain SVs~\mbox{\cite{meneely2013patch}}, e.g., using vulnerable libraries or insecure implementation. We focus on VCCs rather than any commits with vulnerable code (in unchanged parts) since addressing VCCs helps mitigate SVs as early as they are added to a project.
VCCs are usually obtained based on Vulnerability-Fixing Commits (VFCs)~\mbox{\cite{perl2015vccfinder, yang2017vuldigger}}. An exemplary VFC and its respective VCC are shown in Fig.~\mbox{\ref{fig:commit_ex_ase21}}. VFCs delete, modify or add code to eliminate an SV (e.g., disabling external entities processing in the XML library in Fig.~\mbox{\ref{fig:commit_ex_ase21}}) and can be found in bug/SV tracking systems. Then, VCCs are commits that last touched the code changes in VFCs.
Our work also leverages VFCs to obtain VCCs for building automated commit-level SV assessment models.

\subsection{Commit-Level SV Assessment with CVSS}
\label{subsec:cva_need_ase21}

Similar to the studies in Chapters~\ref{chap:msr19} and~\ref{chap:msr22}, we use Common Vulnerability Scoring System (CVSS)~\mbox{\cite{cvss}} version 2 of base metrics (i.e., Confidentiality, Integrity, Availability, Access Vector, Access Complexity, Authentication, and Severity) to assess SVs in this study because of their popularity in practice.
Based on CVSS version 2, the VCC (CVE-2016-3674) in Fig.~\mbox{\ref{fig:commit_ex_ase21}} has a considerable impact on Confidentiality. This SV can be exploited with low (Access) complexity with no authentication via a public network (Access Vector), making it an attractive target for attackers.

Despite the criticality of these SVs, there have been delays in reporting, assessing and fixing them. Concretely, the VCC in Fig.~\mbox{\ref{fig:commit_ex_ase21}} required 1,439 and 1,469 days to be reported\footnote{\url{https://github.com/x-stream/xstream/issues/25}} and fixed (in VFC), respectively. Existing SV assessment methods based on bug/SV reports (e.g.,~\mbox{\cite{han2017learning,spanos2018multi,le2019automated}}) would need to wait more than 1,000 days for the report of this SV. However, performing SV assessment right after this commit was submitted can bypass the waiting time for SV reports, enabling developers to realize the exploitability/impacts of this SV and plan to fix it much sooner.
To the best of our knowledge, there has not been any study on automated commit-level SV assessment, i.e., assigning seven CVSS base metrics to a VCC. Our work identifies and aims to bridge this important research gap.

\subsection{Feature Extraction from Commit Code Changes}
\label{subsec:sv_context_motivation_ase21}

The extraction of commit features is important for building commit-level SV assessment models.
Many existing commit-level defect/SV prediction models have only considered commit code changes (e.g.,~\mbox{\cite{hoang2019deepjit,hoang2019patchnet,sabetta2018practical}}). However, we argue that the nearby context of code changes also contributes valuable information to the prediction, as shown in Chapter~\ref{chap:msr22}. For instance, the surrounding code of the changes in Fig.~\mbox{\ref{fig:commit_ex_ase21}} provides extra details; e.g., the method return statement is modified and the return type is \code{XMLInputFactory}. Such a type can help learn patterns of XXE SV that usually occurs with XML processing.

Besides the context, we speculate that SV assessment models can also benefit from the relatedness among the assessment tasks. For example, the XXE SV in Fig.~\mbox{\ref{fig:commit_ex_ase21}} allows attackers to read arbitrary system files, which mainly affects the Confidentiality rather than the Integrity and Availability of a system. This chapter investigates the possibility of incorporating the common features of seven CVSS metrics into a single model using the multi-task learning paradigm~\mbox{\cite{zhang2017survey}} instead of learning seven cumulative individual models. Specifically, multi-task learning leverages the similarities and the interactions of the involved tasks through a shared feature extractor to predict all the tasks simultaneously. Such a unified model can significantly reduce the time and resources to train, optimize and maintain/update the model in the long run.

\begin{landscape}
\begin{figure*}[t]
    \centering
    \includegraphics[width=\linewidth,keepaspectratio]{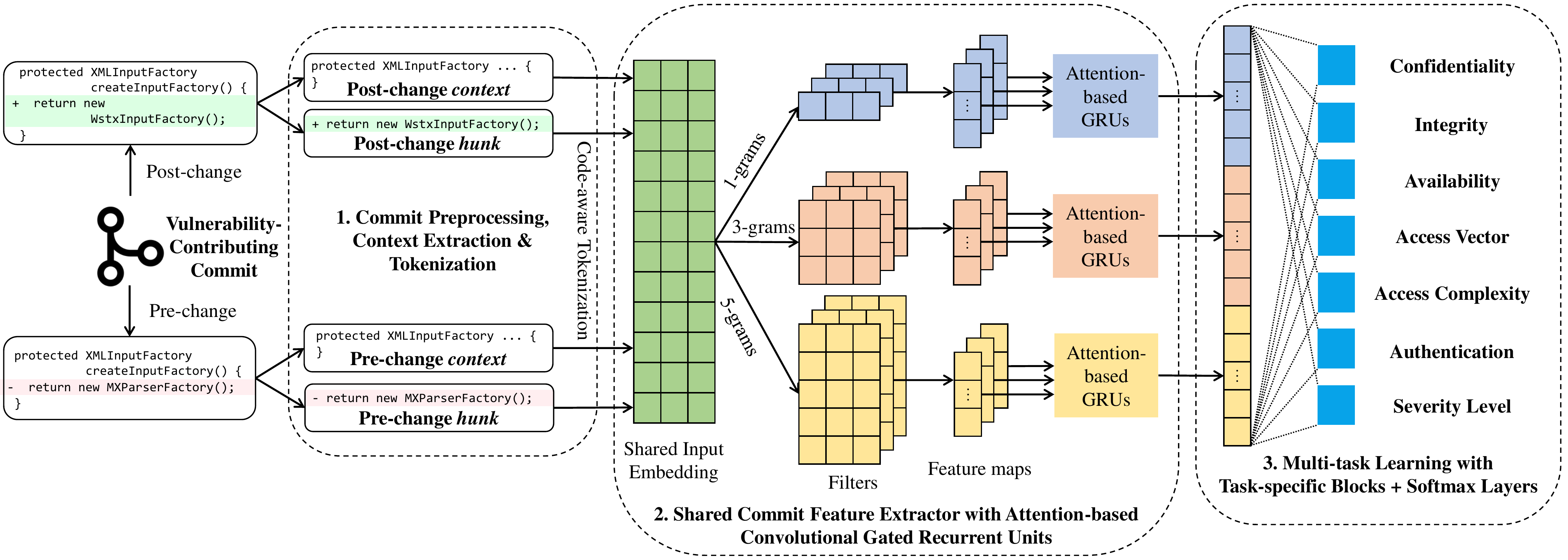}
    \caption[Workflow of DeepCVA for automated commit-level software vulnerability assessment.]{Workflow of DeepCVA for automated commit-level SV assessment. \textbf{Note}: The VCC is the one described in Fig.~\mbox{\ref{fig:commit_ex_ase21}}.}
    \label{fig:workflow_deepcva_ase21}
\end{figure*}
\end{landscape}

\section{The DeepCVA Model}
\label{sec:deepcva_method_ase21}

We propose \textbf{DeepCVA} (see Fig.~\mbox{\ref{fig:workflow_deepcva_ase21}}), a novel \underline{\textbf{Deep}} learning model to automate \underline{\textbf{C}}ommit-level \underline{\textbf{V}}ulnerability \underline{\textbf{A}}ssessment. DeepCVA is a unified and end-to-end trainable model that concurrently predicts seven CVSS metrics (i.e., Confidentiality, Integrity, Availability, Access Vector, Access Complexity, Authentication, and Severity) for a Vulnerability-Contributing Commit (VCC).
DeepCVA contains: (\textit{i}) preprocessing, context extraction and tokenization of code commits (section~\mbox{\ref{subsec:preprocessing_ase21}}), (\textit{ii}) feature extraction from commits shared by seven assessment tasks using attention-based convolutional gated recurrent units (section~\mbox{\ref{subsec:deep_acgru_ase21}}), and (\textit{iii}) simultaneous prediction of seven CVSS metrics using multi-task learning~\mbox{\cite{zhang2017survey}} (section~\mbox{\ref{subsec:multitask_learning_ase21}}).
To assign the CVSS metrics to a new VCC with DeepCVA, we first preprocess the commit, obtain its code changes and respective context and tokenize such code changes/context. Embedding vectors of preprocessed code tokens are then obtained, and the commit feature vector is extracted using the trained feature extractor. This commit feature vector passes through the task-specific blocks and softmax layers to get the seven CVSS outputs with the highest probability values.
Details of each component are given hereafter.

\subsection{Commit Preprocessing, Context Extraction \& Tokenization}
\label{subsec:preprocessing_ase21}

To train DeepCVA, we first obtain and preprocess code changes (hunks) and extract the context of such changes. We then tokenize them to prepare inputs for feature extraction.

\noindent \textbf{Commit preprocessing}. Preprocessing helps remove noise in code changes and reduce computational costs. We remove newlines/spaces and inline/multi-line comments since they do not change code functionality. We do not remove punctuations (e.g., ``\code{;}'', ``\code{(}'', ``\code{)}'') and stop words (e.g., \code{and}/\code{or} operators) to preserve code syntax. We also do not lowercase code tokens since developers can use case-sensitivity for naming conventions of different token types (e.g., variable name: \code{system} vs. class name: \code{System}). Stemming (i.e., reducing a word to its root form such as \code{equals} to \code{equal}) is not applied to code since different names can change code functionality (e.g., the built-in \code{equals} function in Java).

\begin{figure}[t]
    \centering
    \includegraphics[width=\columnwidth,keepaspectratio]{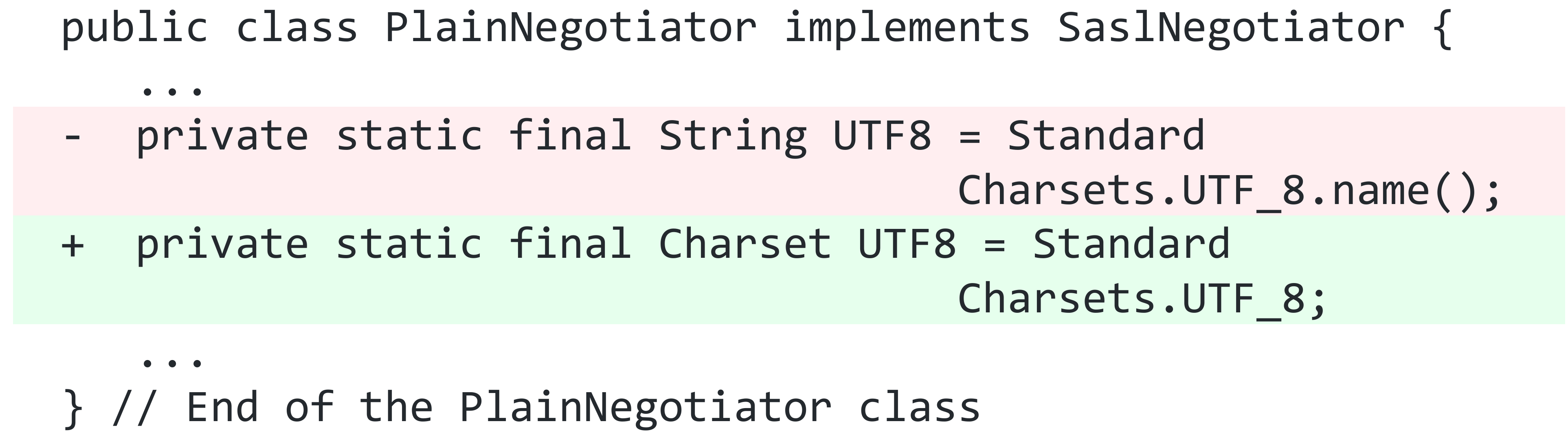}
    \caption{Code changes outside of a method from the commit \textit{4b9fb37} in the \textit{Apache qpid-broker-j} project.}
    \label{fig:method_scope_ase21}
\end{figure}

\noindent \textbf{Context extraction algorithm}. We customize Sahal et al.'s~\mbox{\cite{sahal2018identifying}} \textit{Closest Enclosing Scope} (CES) to identify the context of vulnerable code changes for commit-level SV assessment (see section~\mbox{\ref{subsec:sv_context_motivation_ase21}}). Sahal et al.~\mbox{\cite{sahal2018identifying}} defined an enclosing scope to be the code within a balanced amount of opening and closing curly brackets such as \code{if}/\code{switch}/\code{while}/\code{for} blocks.
Among all enclosing scopes of a hunk, the one with the smallest size (lines of code) is selected as CES to reduce irrelevant code. Sahal et al.~\mbox{\cite{sahal2018identifying}} found CES usually contains hunk-related information (e.g., variable values/types preceding changes). CES also alleviates the need for manually pre-defining the context size as in~\mbox{\cite{perl2015vccfinder,tian2020evaluating}}.
Some existing studies (e.g.,~\mbox{\cite{li2020dlfix,alon2019code2vec}}) only used the method/function scope, but code changes may occur outside of a method. For instance, changes in Fig.~\mbox{\ref{fig:method_scope_ase21}} do not have any enclosing method, but we can still obtain its CES, i.e., the \code{PlainNegotiator} class.

There are still two main limitations with the definition of CES in~\mbox{\cite{sahal2018identifying}}. Firstly, a scope (e.g., \code{for}/\code{while} in Java) with single-line content does not always require curly brackets. Secondly, some programming languages do not use curly brackets to define scopes like Python. To address these two issues, we utilize Abstract Syntax Tree (AST) depth-first traversal (see Algorithm~\mbox{\ref{algo:context_extraction_ase21}}) to obtain CESs of code changes, as AST covers the syntax of all scope types and generalizes to any programming languages.

\begin{algorithm}[t]
    \caption{AST-based extraction of the Closest Enclosing Scopes (CESs) of commit code changes.}
    \label{algo:context_extraction_ase21}
    \DontPrintSemicolon
    \SetAlgoNoLine

    \KwIn{Current Vulnerability-Contributing Commit (VCC): $commit$\\
    Scope type: $scope\_types$}

    \KwOut{CESs of code changes in the current commit: $all\_ces$}
    
    \SetKwFunction{FMain}{extract\_scope}
    \SetKwProg{Fn}{Function}{:}{}
    \Fn{\FMain{$AST, hunk, visited=\emptyset$}}{
    
    \textbf{global} $potential\_scopes$\;
        
    $potential\_scopes \longleftarrow potential\_scopes + AST$\;
    
    $visited \longleftarrow visited + AST$\;
    
    \ForEach{$node \in AST$}{
        \If{$ node \notin visited$ {\bf and} $\text{type}(node) \in scope\_types~$ {\bf and} $start_{node}\leq start_{hunk}$ {\bf and} $end_{node} \geq end_{hunk}$}{
            {$extract\_scope(AST,hunk,visited)$}
        }
    }
    \textbf{return}\; 
    }

    $files \longleftarrow \text{extract\_files}(commit)$\;
    $all\_ces \longleftarrow \emptyset$\;

    \ForEach{${f}_{i}\in files$}{
        $hunks \longleftarrow \text{extract\_hunk}(commit,f_{i})$\;
        
        $AST_{i} \longleftarrow \text{extract\_AST}(f_{i})$\;
        
        \ForEach{${h}_{i}\in hunks$}{
            $potential\_scopes \longleftarrow \emptyset$\;
            
            $\text{extract\_scopes}(AST_{i}, h_{i})$\;
            
            $all\_ces \longleftarrow all\_ces + \underset{\text{size}}{\mathop{\operatorname{argmin}}}(potential\_scopes)$\;    
        }
    }

    \Return $all\_ces$
\end{algorithm}

Algorithm~\mbox{\ref{algo:context_extraction_ase21}} contains: (\textit{i}) the \code{extract\_scope} function for extracting potential scopes of a code hunk (lines 1-8), and (\textit{ii}) the main code to obtain the CES of every hunk in a commit (lines 9-18). The \code{extract\_scope} function leverages depth-first traversal with recursion to go through every node in an AST of a file. Line 3 adds the selected part of an AST to the list of potential scopes (\code{potential\_scopes}) of the current hunk. The first (root) AST is always valid since it encompasses the whole file. Line 6 then checks whether each node (sub-tree) of the current AST has one of the following types: \code{class}, \code{interface}, \code{enum}, \code{method}, \code{if}/\code{else}, \code{switch}, \code{for}/\code{while}/\code{do}, \code{try}/\code{catch}, and is surrounding the current hunk. If the conditions are satisfied, the \code{extract\_scope} function would be called recursively in line 7 until a leaf of the AST is reached. The main code starts to extract the modified files of the current commit in line 9. For each file, we extract code hunks (code deletions/additions) in line 12 and then obtain the AST of the current file using an AST parser in line 13. Line 16 calls the defined \code{extract\_scope} function to generate the potential scopes for each hunk. Among the identified scopes, line 17 adds the one with the smallest size (i.e., the number of code lines excluding empty lines and comments) to the list of CESs (\code{all\_ces}). Finally, line 18 of Algorithm~\mbox{\ref{algo:context_extraction_ase21}} returns all the CESs for the current commit.

We treat deleted (pre-change), added (post-change) code changes and their CESs as four separate inputs to be vectorized by the shared input embedding, as illustrated in Fig.~\mbox{\ref{fig:workflow_deepcva_ase21}}. For each input, we concatenate all the hunks/CESs in all the affected files of a commit to explicitly capture their interactions.

\noindent \textbf{Code-aware tokenization}. The four inputs extracted from a commit are then tokenized with a code-aware tokenizer to preserve code semantics and help prediction models be more generalizable. For example, \code{a++} and \code{b++} are tokenized as \code{a}, \code{b} and \code{++}, explicitly giving a model the information about one-increment operator (\code{++}).
Tokenized code is fed into a shared Deep Learning model, namely Attention-based Convolutional Gated Recurrent Unit (AC-GRU), to extract commit features.

\subsection{Feature Extraction with Deep AC-GRU}
\label{subsec:deep_acgru_ase21}

Deep AC-GRU has a \textit{three-way Convolutional Neural Network} to extract n-gram features and \textit{Attention-based Gated Recurrent Units} to capture dependencies among code changes and their context. This feature extractor is shared by four inputs, i.e., deleted/added code hunks/context. Each input has the size of $N \times L$, where $N$ is the no. of code tokens and $L$ is the vector length of each token. All inputs are truncated or padded to the same length $N$ to support parallelization. The feature vector of each input is obtained from a shared \textit{Input Embedding} layer that maps code tokens into fixed-length arithmetic vectors. The dimensions of this embedding layer are $|V| \times L$, where $|V|$ is the code vocabulary size, and its parameters are learned together with the rest of the model.

\noindent \textbf{Three-way Convolutional Neural Network}. We use a shared three-way Convolutional Neural Network (CNN)~\mbox{\cite{kim2014convolutional}} to extract n-grams (n = 1,3,5) of each input vector. The three-way CNN has filters with three sizes of one, three and five, respectively, to capture common code patterns, e.g., \code{public class Integer}.
The filters are randomly initialized and jointly learned with the other components of DeepCVA. We did not include 2-grams and 4-grams to reduce the required computational resources without compromising the model performance, which has been empirically demonstrated in section~\mbox{\ref{subsec:rq2_results_ase21}}. To generate code features of different window sizes with the three-way CNN, we multiply each filter with the corresponding input rows and apply non-linear ReLU activation function~\mbox{\cite{nair2010rectified}}, i.e., $\operatorname{ReLU}(x) = \operatorname{max}(0, x)$. We repeat the same convolutional process from the start to the end of an input vector by moving the filters down sequentially with a stride of one.
\textcolor{black}{This stride value is the smallest and helps capture the most fine-grained information from input code as compared to larger values.}
Each filter size returns feature maps of the size $(N - K + 1) \times F$, where $K$ is the filter size (one, three or five) and $F$ is the number of filters. Multiple filters are used to capture different semantics of commit data.

\noindent \textbf{Attention-based Gated Recurrent Unit}. The feature maps generated by the three-way CNN sequentially enter a Gated Recurrent Unit (GRU) [31]. GRU, defined in Eq.~\mbox{\eqref{eq:gru_formula_ase21}}, is an efficient version of Recurrent Neural Networks and used to explicitly capture the order and dependencies between code blocks. For example, the \code{return} statement comes after the function declarations of the VCC in Fig.~\mbox{\ref{fig:workflow_deepcva_ase21}}.

\begin{equation}\label{eq:gru_formula_ase21}
\begin{aligned}
  & {{\mathbf{z}}_{t}}=\sigma ({{\mathbf{W}}_{z}}{{\mathbf{x}}_{t}}+{{\mathbf{U}}_{z}}{{\textbf{h}}_{t-1}}+{{b}_{z}}) \\
 & {{\mathbf{r}}_{t}}=\sigma ({{\mathbf{W}}_{r}}{{\mathbf{x}}_{t}}+{{\mathbf{U}}_{r}}{{\mathbf{h}}_{t-1}}+{{b}_{r}}) \\
 & {{{\mathbf{\hat{h}}}}_{t}}=\tanh ({{\mathbf{W}}_{h}}{{\mathbf{x}}_{t}}+{{\mathbf{U}}_{h}}({{\mathbf{r}}_{t}}\odot {{\mathbf{h}}_{t-1}})+{{b}_{h}}) \\
 & {{\mathbf{h}}_{t}}=(1-{\mathbf{z}_{t}})\odot {{\mathbf{h}}_{t-1}}+{\mathbf{z}_{t}}\odot {{{\mathbf{\hat{h}}}}_{t}}
\end{aligned}
\end{equation}
\noindent where $\mathbf{W}_{z}$, $\mathbf{W}_{r}$, $\mathbf{W}_{h}$, $\mathbf{U}_{z}$, $\mathbf{U}_{r}$, $\mathbf{U}_{h}$ are learnable weights, $b_{z}$, $b_{r}$, $b_{h}$ are learnable biases, $\odot$ is element-wise multiplication, $\sigma$ is the sigmoid function and $\tanh()$ is the hyperbolic tangent function.

\noindent To determine the information ($\textbf{h}_{t}$) at each token (time step) $t$, GRU combines the current input ($\textbf{x}_{t}$) and the previous time step ($\textbf{h}_{t-1}$) using the \textit{update} ($\textbf{z}_{t}$) and \textit{reset} ($\textbf{r}_{t}$) gates. $\textbf{h}_{t}$ is then carried on to the next token until the end of the input to maintain the dependencies of the whole code sequence.

The last token output of GRU is often used as the whole sequence representation, yet it suffers from the \textit{information bottleneck} problem~\mbox{\cite{bahdanau2014neural}}, especially for long sequences.
To address this issue, we incorporate the \textit{attention mechanism}~\mbox{\cite{bahdanau2014neural}} into GRU to explicitly capture the contribution of each input token, as formulated in Eq.~\mbox{\eqref{eq:attention_ase21}}.
\begin{equation}\label{eq:attention_ase21}
\begin{aligned}
  & \mathbf{ou}{{\mathbf{t}}_{attention}}=\sum\limits_{i=1}^{m}{{{w}_{i}}{{\mathbf{h}}_{i}}} \\ 
 & {{w}_{i}}=\operatorname{softmax}({{\mathbf{W}}_{s}}\tanh ({{\mathbf{W}}_{a}}{{\mathbf{h}}_{i}}+{{b}_{a}})) \\ 
 & =\frac{\exp ({{\mathbf{W}}_{s}}\tanh ({{\mathbf{W}}_{a}}{{\mathbf{h}}_{i}}+{{b}_{a}}))}{\sum\limits_{j=1}^{m}{\exp ({{\mathbf{W}}_{s}}\tanh ({{\mathbf{W}}_{a}}{{\mathbf{h}}_{j}}+{{b}_{a}}))}}
\end{aligned}
\end{equation}
\noindent where $w_{i}$ is the weight of $\mathbf{h}_{i}$; $\mathbf{W}_{s}$, $\mathbf{W}_{a}$ are learnable weights, $b_{a}$ is learnable bias, and $m$ is the number of code tokens.

The attention-based outputs ($\mathbf{out}_{attention}$) of the three GRUs (see Fig.~\mbox{\ref{fig:workflow_deepcva_ase21}}) are concatenated into a single feature vector to represent each of the four inputs (pre-/post-change hunks/contexts). The commit feature vector is a concatenation of the vectors of all four inputs generated by the shared AC-GRU feature extractor. This feature vector is used for multi-task prediction of seven CVSS metrics.

\subsection{Commit-Level SV Assessment with Multi-task Learning}
\label{subsec:multitask_learning_ase21}

This section describes the multi-task learning layers of DeepCVA for efficient commit-level SV assessment (i.e., learning/predicting seven CVSS tasks simultaneously) using a single model as well as how to train the model end-to-end.

\noindent \textbf{Multi-task learning layers}. The last component of DeepCVA consists of the multi-task learning layers that simultaneously give the predicted CVSS values for seven SV assessment tasks. As illustrated in Fig.~\mbox{\ref{fig:workflow_deepcva_ase21}}, this component contains two main parts: \textit{task-specific blocks} and \textit{softmax layers}. On top of the shared features extracted by AC-GRU, task-specific blocks are necessary to capture the differences among the seven tasks. Each task-specific block is implemented using a fully connected layer with non-linear ReLU activations~\mbox{\cite{nair2010rectified}}. Specifically, the output vector ($\mathbf{task}_{i}$) of the task-specific block for assessment task $i$ is defined in Eq.~\mbox{\eqref{eq:task_specific_block_ase21}}.
\begin{equation}\label{eq:task_specific_block_ase21}
\mathbf{tas}{{\mathbf{k}}_{i}}=\operatorname{ReLU}({{\mathbf{W}}_{t}}{{\mathbf{x}}_{commit}}+{{b}_{t}})
\end{equation}
\noindent where $\mathbf{x}_{commit}$ is the commit feature vector from AC-GRU; $\mathbf{W}_{t}$ is learnable weights and $b_{t}$ is learnable bias.

Each task-specific vector goes through the respective softmax layer to determine the output of each task with the highest predicted probability. The prediction output ($pred_{i}$) of task $i$ is given in Eq.~\mbox{\eqref{eq:softmax_layer_ase21}}.
\begin{equation}\label{eq:softmax_layer_ase21}
\begin{aligned}
  & pre{{d}_{i}}=\operatorname{argmax}(\mathbf{pro}{{\mathbf{b}}_{i}}) \\ 
 & \mathbf{pro}{{\mathbf{b}}_{i}}=\operatorname{softmax}({{\mathbf{W}}_{p}}\mathbf{tas}{{\mathbf{k}}_{i}}+{{b}_{p}}) \\ 
 & \operatorname{softmax}({{z}_{j}})=\frac{\exp ({{z}_{j}})}{\sum\limits_{c=1}^{nlabel{{s}_{i}}}{\exp ({{z}_{c}})}}
\end{aligned}
\end{equation}
\noindent where $\mathbf{prob}_{i}$ contains the predicted probabilities of $nlabels_{i}$ possible outputs of task $i$; $\mathbf{W}_{p}$ is learnable weights and $b_{p}$ is learnable bias.

\noindent \textbf{Training DeepCVA}. To compare DeepCVA's outputs with ground-truth CVSS labels, we define a multi-task loss that averages the cross-entropy losses of seven tasks in Eq.~\mbox{\eqref{eq:multiloss_ase21}}.
\begin{equation}\label{eq:multiloss_ase21}
\begin{aligned}
  & los{{s}_{DeepCVA}}=\sum\limits_{i=1}^{7}{los{{s}_{i}}} \\ 
 & los{{s}_{i}}=-\sum\limits_{c=1}^{nlabel{{s}_{i}}}{y_{i}^{c}\log (prob_{i}^{c})},\,y_{i}^{c}=1\,\text{if}\,c\,\text{is}\,\text{true}\,\text{class}\,\text{else}\,0 \\ 
\end{aligned}
\end{equation}
\noindent where $y_{i}^{c}$, $prob_{i}^{c}$, and $nlabels_{i}$ are the ground-truth value, predicted probability and all labels of CVSS task $i$, respectively.

We minimize this multi-task loss using a stochastic gradient descent method~\mbox{\cite{ruder2016overview}} to optimize the weights of learnable components in DeepCVA. We also use backpropagation~\mbox{\cite{rumelhart1986learning}} to automate partial differentiation with chain-rule and increase the efficiency of gradient computation throughout the model.

\section{Experimental Design and Setup}
\label{sec:expt_design_ase21}

All the experiments ran on a computing cluster that has 16 CPU cores with 16GB of RAM and Tesla V100 GPU.

\subsection{Datasets}
\label{subsec:datasets_ase21}

To develop commit-level SV assessment models, we built a dataset of Vulnerability-Contributing Commits (VCCs) and their CVSS metrics. We used Vulnerability-Fixing Commits (VFCs) to retrieve VCCs, as discussed in section~\mbox{\ref{subsec:vuln_commits_ase21}}.

\noindent \textbf{VFC identification}. We first obtained VFCs from three public sources: NVD~\mbox{\cite{nvd}}, GitHub and its Advisory Database\footnote{\url{https://github.com/advisories}} as well as a manually curated/verified VFC dataset (VulasDB)~\mbox{\cite{ponta2019manually}}. In total, we gathered 13,310 VFCs that had dates ranging from July 2000 to October 2020. We selected VFCs in Java projects as Java has been commonly investigated in the literature (e.g.,~\mbox{\cite{hoang2019deepjit,alon2019code2vec,mcintosh2017fix}}) and also in the top five most popular languages in practice.\footnote{\url{https://insights.stackoverflow.com/survey/2020\#technology-most-loved-dreaded-and-wanted-languages-loved}}
Following the practice of~\mbox{\cite{mcintosh2017fix}}, we discarded VFCs that had more than 100 files and 10,000 lines of code to reduce noise in the data.

\begin{figure}[t]
    \centering
    \includegraphics[width=\columnwidth,keepaspectratio]{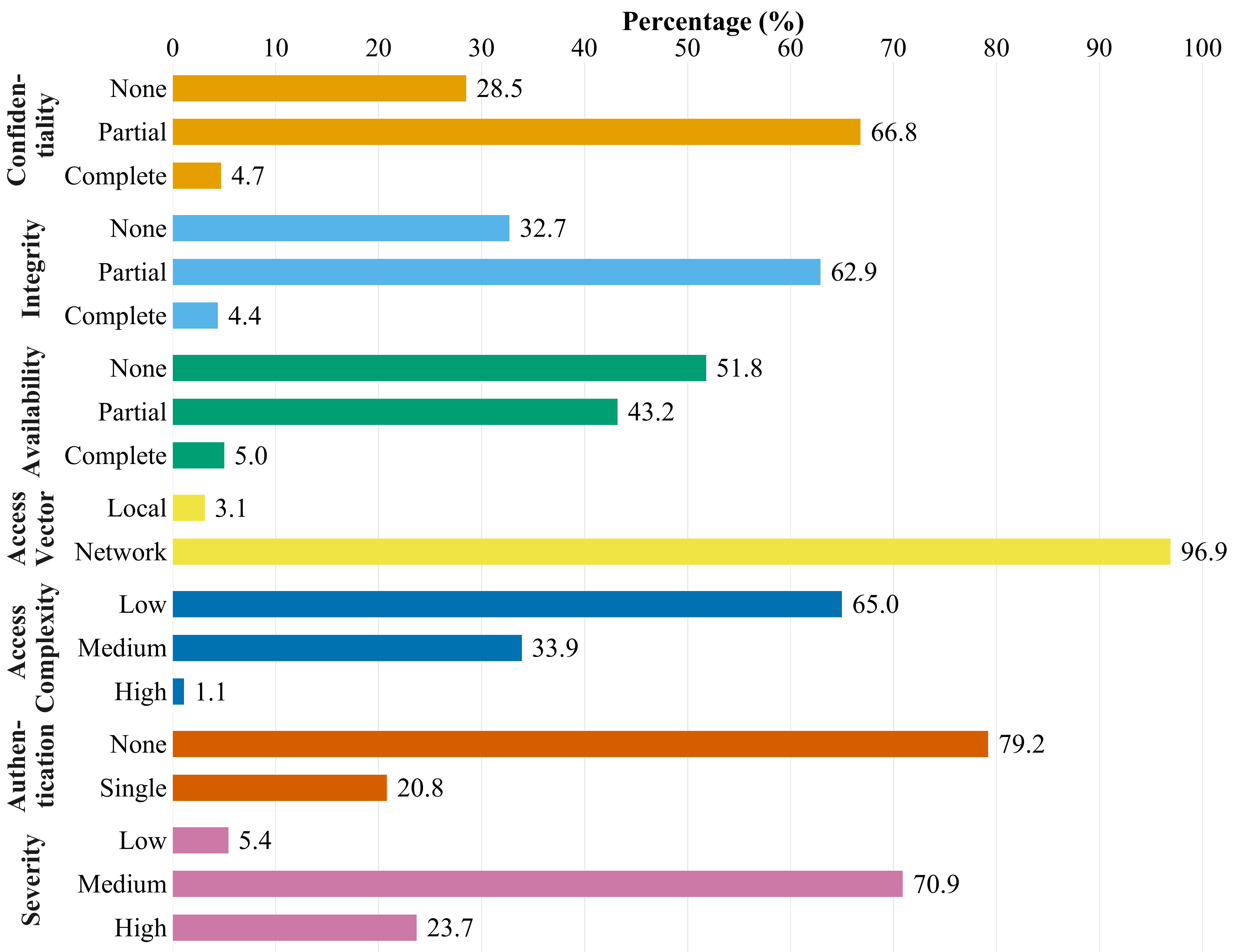}
    \caption[Class distributions of seven software vulnerability assessment tasks.]{Class distributions of seven SV assessment tasks.}
    \label{fig:cvss_distribution_ase21}
\end{figure}

\noindent \textbf{VCC identification with the SZZ algorithm}. After the filtering steps, we had 1,602 remaining unique VFCs to identify VCCs using the SZZ algorithm~\mbox{\cite{sliwerski2005changes}}. This algorithm selects commits that last modified the source code lines deleted or modified to address an SV in a VFC as the respective VCCs of the same SV (see Fig.~\mbox{\ref{fig:commit_ex_ase21}}). As in~\mbox{\cite{sliwerski2005changes}}, we first discarded commits with timestamps after the published dates of the respective SVs on NVD since SVs can only be reported after they were injected in a codebase. We then removed cosmetic changes (e.g., newlines and white spaces) and single-line/multi-line comments in VFCs since these elements do not change code functionality~\mbox{\cite{mcintosh2017fix}}. Like~\mbox{\cite{mcintosh2017fix}}, we also considered copied or renamed files while tracing VCCs.
We obtained \textbf{1,229 unique VCCs}\footnote{The SV reports of all curated VCCs were not available at commit time.} of 542 SVs in 246 real-world Java projects and their corresponding expert-verified CVSS metrics on NVD. Distributions of curated CVSS metrics are illustrated in Fig.~\mbox{\ref{fig:cvss_distribution_ase21}}.
The details of the number of commits and projects retained in each filtering step are also given in Table~\ref{tab:filtering_details_ase21}. Note that some commits and projects were removed during the tracing of VCCs from VFCs due to the issues coined as ghost commits studied by Rezk et al.~\cite{rezk2021ghost}. We did not remove large VCCs (with more than 100 files and 10k lines) as we found several VCCs were large initial/first commits. Our observations agreed with the findings of Meneely et al.~\cite{meneely2013patch}.

\begin{table}[t]

  \centering
  \caption{The number of commits and projects after each filtering step.}
 \begin{tabular}{llll}
    \hline
    \textbf{No.} & \textbf{Filtering step} & \textbf{No. of commits} & \textbf{No. of projects} \\
    \hline
    1 & All unfiltered VFCs & 13,310 & 2,864 \\
    2 & Removing duplicate VFCs & 9,989 & 2,864 \\
    3 & Removing non-Java VFCs  & 1,607 & 361 \\
    4 & \makecell[l]{Removing VFCs with more than\\100 files \& 10k lines} & 1,602 & 358 \\
    5 & \makecell[l]{Tracing VCCs from VFCs using\\the SZZ algorithm} & 3,742 & 342 \\
    6 & \makecell[l]{Removing VCCs with null\\characteristics (CVSS values)} & 2,271 & 246 \\
    7 & Removing duplicate VCCs & 1,229 & 246 \\
    \hline
    \end{tabular}
  \label{tab:filtering_details_ase21}
\end{table}

\noindent \textbf{Manual VCC validation}. To validate our curated VCCs, we randomly selected 293 samples, i.e., 95\% confidence level and 5\% error~\mbox{\cite{cochran2007sampling}}, for two researchers (i.e., the author of this thesis and a PhD student with three-year experience in Software Engineering and Cybersecurity) to independently examine. The manual VCC validation was considerably labor-intensive, which took approximately 120 man-hours.
The Cohen's kappa ($\kappa$) inter-rater reliability score~\mbox{\cite{mchugh2012interrater}} was 0.83, i.e., ``almost perfect'' agreement~\mbox{\cite{hata20199}}.
We also involved another PhD student having two years of experience in Software Engineering and Cybersecurity in the discussion to resolve disagreements. Our validation found that 85\% of the VCCs were valid. In fact, the SZZ algorithm is imperfect~\mbox{\cite{fan2019impact}}, but we assert that it is nearly impossible to obtain near 100\% accuracy without exhaustive manual validation. Specifically, the main source of incorrectly identified VCCs in our dataset was that some files in VFCs were used to update version/documentation or address another issue instead of fixing an SV. One such false positive VCC was the commit \textit{87c89f0} in the \textit{jspwiki} project that last modified the build version in the corresponding VFC.

\begin{figure}[t]
    \centering
    \includegraphics[width=\columnwidth,keepaspectratio]{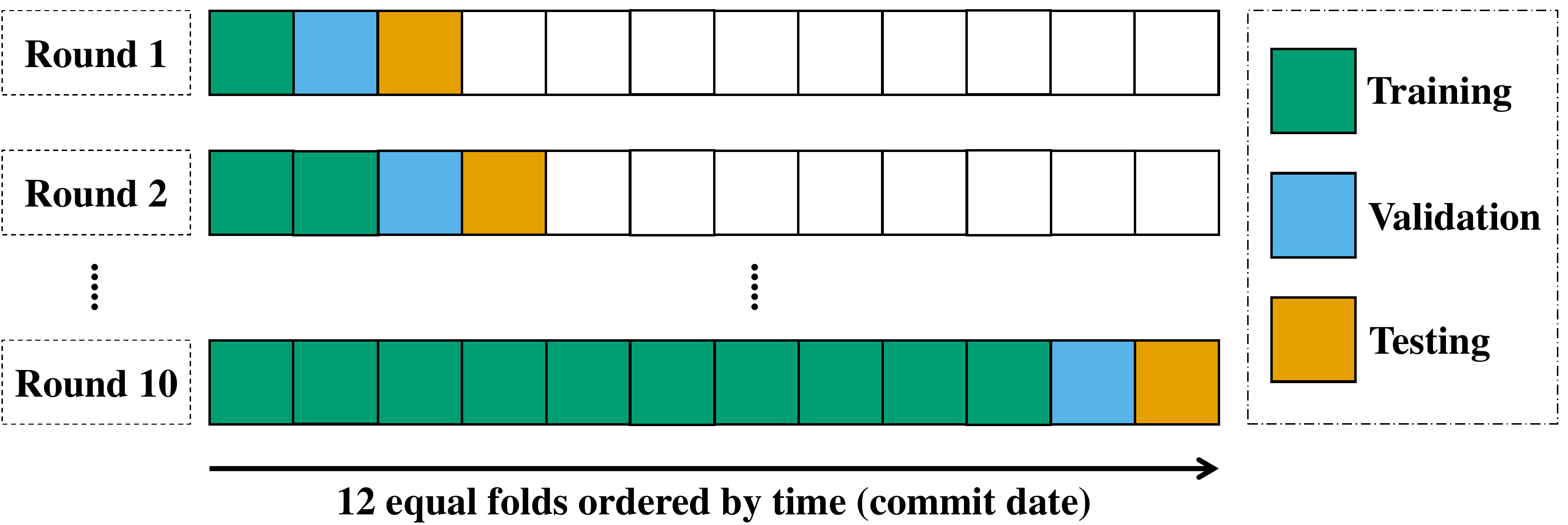}
    \caption{Time-based splits for training, validating \& testing.}
    \label{fig:timebased_splits_ase21}
\end{figure}

\noindent \textbf{Data splitting}. We adopted \textit{time-based} splits~\mbox{\cite{falessi2020need}} for training, validating and testing the models to closely represent real-world scenarios where incoming/future unseen data is \textit{not} present during training~\mbox{\cite{mcintosh2017fix,jimenez2019importance}}. We trained, validated and tested the models in 10 rounds using 12 equal folds split based on commit dates (see Fig.~\mbox{\ref{fig:timebased_splits_ase21}}). Specifically, in round $i$, folds $1 \rightarrow i$, $i+1$ and $i+2$ were used for training, validation and testing, respectively. We chose an optimal model with the highest average \textit{validation} performance and then reported its respective average testing performance over 10 rounds, which helped avoid unstable results of a single testing set~\mbox{\cite{raschka2018model}}.

\subsection{Evaluation Metrics}
\label{subsec:evaluation_metrics_ase21}

To evaluate the performance of automated commit-level SV assessment, we utilized the F1-Score and Matthews Correlation Coefficient (MCC) metrics that have been commonly used in the literature (e.g.,~\mbox{\cite{han2017learning,spanos2018multi,jimenez2019importance}}). These two metrics are suitable for the imbalanced classes~\mbox{\cite{luque2019impact}} in our data (see Fig.~\mbox{\ref{fig:cvss_distribution_ase21}}). F1-Score has a range from 0 to 1, while MCC takes values from –1 to 1, where 1 is the best value for both metrics. MCC was used to select optimal models since MCC explicitly considers all classes~\mbox{\cite{luque2019impact}}.
To evaluate the tasks with more than two classes, we used macro F1-Score~\mbox{\cite{spanos2018multi}} and the multi-class version of MCC~\mbox{\cite{gorodkin2004comparing}}. MCC of the multi-task DeepCVA model was the average MCC of seven constituent tasks. Note that MCC is not directly proportional to F1-score.

\subsection{Hyperparameter and Training Settings of DeepCVA}
\label{subsec:tuning_deepcva_ase21}

\noindent \textbf{Hyperparameter settings}. We used the average \textit{validation} MCC to select optimal hyperparameters for DeepCVA's components. We also ran DeepCVA 10 times each round to reduce the impact of random initialization on model performance. We first chose 1024 for the input length of the pre-/post-change hunks/context (see Fig.~\mbox{\ref{fig:workflow_deepcva_ase21}}), which has been commonly used in the literature (e.g.,~\mbox{\cite{devlin2018bert,radford2019language}}).
Using a shorter input length would likely miss many code tokens, while a longer length would significantly increase the model complexity and training time.
Shorter commits were padded with zeros, and longer ones were truncated to ensure the same input size for parallelization with GPU~\mbox{\cite{hoang2019deepjit,hoang2019patchnet}}. We built a vocabulary of 10k most frequent code tokens in the Input Embedding layer as suggested by~\mbox{\cite{pradel2018deepbugs}}.
Note that using 20k-sized vocabulary only raised the performance by 2\%, yet increased the model complexity by nearly two times.
We selected an input embedding size of 300, i.e., a standard and usually high limit value for many embedding models (e.g.,~\mbox{\cite{mikolov2013distributed,bojanowski2017enriching}}), and we randomly initialized embedding vectors~\mbox{\cite{hoang2019deepjit,kim2014convolutional}}. For the number of filters of the three-way CNN as well as the hidden units of the GRU, Attention and Task-specific blocks, we tried \{32, 64, 128\}, similar to~\mbox{\cite{han2017learning}}. We picked 128 as it had at least 5\% better validation performance than 32 and 64.

\noindent \textbf{Training settings}.
We used the Adam algorithm~\mbox{\cite{kingma2014adam}}, the state-of-the-art stochastic gradient descent method, for training DeepCVA end-to-end with a learning rate of 0.001 and a batch size of 32 as recommended by Hoang et al.~\mbox{\cite{hoang2019deepjit}}. To increase the training stability, we employed Dropout~\mbox{\cite{srivastava2014dropout}} with a dropout rate of 0.2 and Batch Normalization~\mbox{\cite{ioffe2015batch}} between layers. We trained DeepCVA for 50 epochs, and we would stop training if the \textit{validation} MCC did not change in the last five epochs to avoid overfitting~\mbox{\cite{hoang2019deepjit,hoang2019patchnet}}.

\subsection{Baseline Models}
\label{subsec:baselines_ase21}

We considered three types of learning-based baselines for automated commit-level SV assessment, as learning-based models can automatically extract relevant SV patterns/features from input data for prediction without relying on pre-defined rules.
The baselines were (\textit{i}) \textbf{S-CVA}: \textbf{S}upervised single-task model using either software metrics or text-based features including Bag-of-Words (BoW or token count) and Word2vec~\mbox{\cite{mikolov2013distributed}}; (\textit{ii}) \textbf{X-CVA}: supervised e\textbf{X}treme multi-class model that performed a single prediction for all seven tasks using the above feature types; and (\textit{iii}) \textbf{U-CVA}: \textbf{U}nsupervised model using $k$-means clustering~\mbox{\cite{lloyd1982least}} with the same features as S-CVA/X-CVA.
\textcolor{black}{Note that there was no existing technique for automating commit-level SV assessment, so we could only compare DeepCVA with the compatible techniques proposed for related tasks, as described hereafter.}

Software metrics (e.g.,~\mbox{\cite{kamei2012large,perl2015vccfinder,yang2017vuldigger}}) and text-based features (BoW/Word2vec) (e.g., \mbox{\cite{sabetta2018practical,zhou2017automated}}) have been widely used for commit-level prediction.
We used 84 software metrics proposed by~\mbox{\cite{kamei2012large,perl2015vccfinder,yang2017vuldigger}} for defect/SV prediction.
Among these metrics, we converted C/C++ keywords into Java ones to match the language used in our dataset.
The list of software metrics used in this chapter can be found at \url{https://github.com/lhmtriet/DeepCVA}.
As in~\mbox{\cite{kamei2012large}}, in each round in Fig.~\mbox{\ref{fig:timebased_splits_ase21}}, we also removed correlated software metrics that had a Spearman correlation larger than 0.7 based on the training data of that round to avoid performance degradation, e.g., no. of \textit{stars} vs. \textit{forks} of a project.
For BoW and Word2vec, we adopted the same vocabulary size of 10k to extract features from four inputs described in Fig.~\mbox{\ref{fig:workflow_deepcva_ase21}}, as in DeepCVA. Feature vectors of all inputs were concatenated into a single vector.
For Word2vec, we averaged the vectors of all tokens in an input to generate its feature vector, which has been shown to be a strong baseline~\mbox{\cite{shen2018baseline}}.
Like DeepCVA, we also used an embedding size of 300 for each Word2vec token.

Using these feature types, S-CVA trained a separate supervised model for each CVSS task, while X-CVA used a single multi-class model to predict all seven tasks simultaneously. X-CVA worked by concatenating all seven CVSS metrics into a single label.
To extract the results of the individual tasks for X-CVA, we checked whether the ground-truth label of each task was in the concatenated model output. For S-CVA and X-CVA, we applied six popular classifiers: Logistic Regression (LR), Support Vector Machine (SVM), K-Nearest Neighbors (KNN), Random Forest (RF), XGBoost (XGB)~\mbox{\cite{chen2016xgboost}} and Light Gradient Boosting Machine (LGBM)~\mbox{\cite{ke2017lightgbm}}. These classifiers have been used for SV assessment based on SV reports~\mbox{\cite{spanos2018multi,le2019automated}}. The hyperparameters for tuning these classifiers were \textit{regularization}: \{l1, l2\}; \textit{regularization coefficient}: \{0.01, 0.1, 1, 10, 100\} for LR and \{0.01, 0.1, 1, 10, 100, 1,000, 10,000\} for SVM; \textit{no. of neighbors}: \{11, 31, 51\}, \textit{distance norm}: \{1, 2\} and \textit{distance weight}: \{uniform, distance\} for KNN; \textit{no. of estimators}: \{100, 300, 500\}, \textit{max. depth}: \{3, 5, 7, 9, unlimited\}, \textit{max. no. of leaf nodes}: \{100, 200, 300, unlimited\} for RF, XGB and LGBM. These hyperparameters have been adapted from relevant studies~\mbox{\cite{spanos2018multi,le2019automated,le2020puminer}}.

Unlike S-CVA and X-CVA, U-CVA did not require CVSS labels to operate; therefore, U-CVA required less human effort than S-CVA and X-CVA. We tuned U-CVA for each task with the following no. of clusters ($k$): \{2, 3, 4, 5, 6, 7, 8, 9, 10, 15, 20, 25, 30, 35, 40, 45, 50\}. To assess a new commit with U-CVA, we found the cluster with the smallest Euclidean distance to that commit and assigned it the most frequent class of each task in the selected cluster.

\section{Research Questions and Experimental Results}
\label{sec:results_ase21}

\subsection{\textbf{RQ1}: How does DeepCVA Perform Compared to Baseline Models for Commit-level SV Assessment?}
\label{subsec:rq1_results_ase21}

\noindent \textbf{Motivation}. We posit the need for commit-level Software Vulnerability (SV) assessment tasks based on seven CVSS metrics. Such tasks would help developers to understand the SV exploitability and impacts as early as SVs are introduced in a software system and devise remediation plans accordingly. RQ1 evaluates our DeepCVA for this new and important task.

\noindent \textbf{Method}. We compared the effectiveness of our DeepCVA model with the S-CVA, X-CVA and U-CVA baselines (see section~\mbox{\ref{subsec:baselines_ase21}}) on the \textit{testing} sets.
We trained, validated and tested the models using the time-based splits, as described in section~\mbox{\ref{subsec:datasets_ase21}}.
Because of the inherent randomness of GPU-based implementation of DeepCVA,\footnote{\url{https://keras.io/getting\_started/faq/\#how-can-i-obtain-reproducible-results-using-keras-during-development}} we ran DeepCVA 10 times in each round and then averaged its performance. The baselines were not affected by this issue as they did not use GPU. For DeepCVA, we used the hyperparameter/training settings in section~\mbox{\ref{subsec:tuning_deepcva_ase21}}. For each type of baseline, we used grid search on the hyperparameters given in section~\mbox{\ref{subsec:baselines_ase21}} to find the optimal model with the highest \textit{validation} MCC (see section~\mbox{\ref{subsec:evaluation_metrics_ase21}}).

\noindent \textbf{Results}.
\textit{\textbf{DeepCVA outperformed all baselines}}\footnote{MCC values of \textit{random} and \textit{most-frequent-class} baselines were all $<$ 0.01.} \textit{\textbf{(X-CVA, S-CVA and U-CVA) in terms of both MCC and F1-Score}}\footnote{Precision (0.533)/Recall (0.445) of DeepCVA were $>$ than all baselines.}
\textit{\textbf{for all seven tasks (see Table~\mbox{\ref{tab:baseline_comparison_ase21}}).}} DeepCVA got average and best MCC values of 0.247 and 0.286, i.e., 38\% and 59.8\% better than the second-best baseline (X-CVA with Word2vec features), respectively.
Task-wise, DeepCVA had 11.2\%, 42\%, 42.2\%, 20.6\%, 69.2\%, 24.8\% and 39.2\% higher MCC than the best respective baseline models for Confidentiality, Integrity, Availability, Access Vector, Access Complexity, Authentication and Severity tasks, respectively. Notably, the best DeepCVA model achieved stronger performance than all baselines with MCC percentage gaps from 24.1\% (Confidentiality) to 82.5\% (Access Complexity).

\begin{landscape}
\begin{table*}[t]
\fontsize{10}{11}\selectfont
  \centering
  \caption[Testing performance of DeepCVA and baseline models.]{Testing performance of DeepCVA and baseline models. \textbf{Notes}: Optimal classifiers of S-CVA/X-CVA and optimal cluster no. (\textit{k}) of U-CVA are in parentheses. BoW, W2V and SM are Bag-of-Words, Word2vec and Software Metrics, respectively. The best performance of DeepCVA is from the run with the highest MCC in each round. Best row-wise values are in grey.}
    \begin{tabular}{l|l|ccc|ccc|ccc|c}
    \hline
    \multirowcell{3}[0ex][l]{\textbf{CVSS metric}} & \multirowcell{3}[0ex][l]{\textbf{Evaluation}\\ \textbf{metric}} & \multicolumn{10}{c}{\textbf{Model}}\\
    \cline{3-12}
    & & \multicolumn{3}{c|}{\textbf{S-CVA}} & \multicolumn{3}{c|}{\textbf{X-CVA}} & \multicolumn{3}{c|}{\textbf{U-CVA}} & \multirowcell{2}{\textbf{DeepCVA (Best}\\ \textbf{in parentheses)}} \\
    \cline{3-11}
    & & \textbf{BoW} & \textbf{W2V} & \textbf{SM} & \textbf{BoW} & \textbf{W2V} & \textbf{SM} & \textbf{BoW} & \textbf{W2V} & \textbf{SM} &  \\
    \hline
    \multirowcell{2}[0ex][l]{\textbf{Confidentiality}} & \textbf{F1-Score} & 0.416 & 0.406 & 0.423 & 0.420 & 0.434 & 0.429 & 0.292 & 0.332 & 0.313 & \cellcolor[HTML]{C0C0C0} \textbf{0.436 (0.475)} \\
    \hhline{~*{11}{-}}
    & \textbf{MCC} & \makecell{0.174\\ (LR)} & \makecell{0.239\\ (LGBM)} & \makecell{0.232\\ (XGB)} & \makecell{0.188\\ (LR)} & \makecell{0.241\\ (LR)} & \makecell{0.203\\ (XGB)} & \makecell{0.003\\ (50)} & \makecell{0.092\\ (45)} & \makecell{0.017\\ (50)} & \cellcolor[HTML]{C0C0C0} \textbf{0.268 (0.299)}  \\
    \hline
    \multirowcell{2}[0ex][l]{\textbf{Integrity}} & \textbf{F1-Score} & 0.373 & 0.369 & 0.352 & 0.391 & 0.415 & 0.407 & 0.284 & 0.305 & 0.330 & \cellcolor[HTML]{C0C0C0} \textbf{0.430 (0.458)} \\
    \hhline{~*{11}{-}}
    & \textbf{MCC} & \makecell{0.127\\ (LGBM)} & \makecell{0.176\\ (LGBM)} & \makecell{0.146\\ (RF)} & \makecell{0.114\\ (LGBM)} & \makecell{0.160\\ (LR)} & \makecell{0.128\\ (LGBM)} & \makecell{-0.005\\ (25)} & \makecell{0.091\\ (30)} & \makecell{0.084\\ (25)} & \cellcolor[HTML]{C0C0C0} \textbf{0.250 (0.295)} \\
    \hline
    \multirowcell{2}[0ex][l]{\textbf{Availability}} & \textbf{F1-Score} & 0.381 & 0.389 & 0.384 & 0.424 & 0.422 & 0.406 & 0.254 & 0.332 & 0.238 & \cellcolor[HTML]{C0C0C0} \textbf{0.432 (0.475)} \\
    \hhline{~*{11}{-}}
    & \textbf{MCC} & \makecell{0.182\\ (RF)} & \makecell{0.173\\ (LGBM)} & \makecell{0.126\\ (XGB)} & \makecell{0.187\\ (LR)} & \makecell{0.192\\ (LR)} & \makecell{0.123\\ (XGB)} & \makecell{0.064\\ (10)} & \makecell{0.092\\ (45)} & \makecell{0.016\\ (3)} & \cellcolor[HTML]{C0C0C0} \textbf{0.273 (0.303)} \\
    \hline
    \multirowcell{2}[0ex][l]{\textbf{Access Vector}} & \textbf{F1-Score} & 0.511 & 0.487 & 0.440 & 0.499 & 0.532 & 0.487 & 0.477 & 0.477 & 0.477 & \cellcolor[HTML]{C0C0C0} \textbf{0.554 (0.578)} \\
    \hhline{~*{11}{-}}
    & \textbf{MCC} & \makecell{0.07\\ (XGB)} & \makecell{0.051\\ (LR)} & \makecell{0.018\\ (LR)} & \makecell{0.044\\ (LGBM)} &  \makecell{0.107\\ (LR)} & \makecell{0.012\\ (LGBM)} & \makecell{0.000\\ (9)} & \makecell{0.000\\ (40)} & \makecell{0.000\\ (6)} & \cellcolor[HTML]{C0C0C0} \textbf{0.129 (0.178)} \\
    \hline
    \multirowcell{2}[0ex][l]{\textbf{Access Complexity}} & \textbf{F1-Score} & 0.437 & 0.448 & 0.417 & 0.412 & 0.445 & 0.361 & 0.315 & 0.365 & 0.385 & \cellcolor[HTML]{C0C0C0} \textbf{0.464 (0.475)} \\
    \hhline{~*{11}{-}}
    & \textbf{MCC} & \makecell{0.119\\ (LR)} & \makecell{0.143\\ (XGB)} & \makecell{0.111\\ (LGBM)} & \makecell{0.131\\ (LR)} & \makecell{0.121\\ (XGB)} & \makecell{0.088\\ (SVM)} & \makecell{0.000\\ (4)} & \makecell{0.022\\ (30)} & \makecell{0.119\\ (15)} & \cellcolor[HTML]{C0C0C0} \textbf{0.242 (0.261)} \\
    \hline
    \multirowcell{2}[0ex][l]{\textbf{Authentication}} & \textbf{F1-Score} & 0.601 & 0.584 & 0.593 & 0.541 & 0.618 & 0.586 & 0.458 & 0.526 & 0.492 & \cellcolor[HTML]{C0C0C0} \textbf{0.657 (0.677)} \\
    \hhline{~*{11}{-}}
    & \textbf{MCC} & \makecell{0.258\\ (SVM)} & \makecell{0.264\\ (XGB)} & \makecell{0.268\\ (LGBM)} & \makecell{0.212\\ (RF)} & \makecell{0.282\\ (SVM)} & \makecell{0.208\\ (XGB)} & \makecell{0.062\\ (50)} & \makecell{0.162\\ (30)} & \makecell{0.089\\ (50)} & \cellcolor[HTML]{C0C0C0} \textbf{0.352 (0.388)} \\
    \hline
    \multirowcell{2}[0ex][l]{\textbf{Severity}} & \textbf{F1-Score} & 0.407 & 0.357 & 0.345 & 0.382 & 0.381 & 0.358 & 0.283 & 0.288 & 0.287 & \cellcolor[HTML]{C0C0C0} \textbf{0.424 (0.460)} \\
    \hhline{~*{11}{-}}
    & \textbf{MCC} & \makecell{0.144\\ (LR)} & \makecell{0.153\\ (XGB)} & \makecell{0.057\\ (XGB)} & \makecell{0.130\\ (LR)} & \makecell{0.149\\ (LGBM)} & \makecell{0.058\\ (XGB)} & \makecell{-0.018\\ (4)} & \makecell{0.010\\ (15)} & \makecell{0.026\\ (4)} & \cellcolor[HTML]{C0C0C0} \textbf{0.213 (0.277)} \\
    \hline
    \hline
    \multirowcell{2}[0ex][l]{\textbf{Average}} & \textbf{F1-Score} & 0.447 & 0.434 & 0.422 & 0.438 & 0.464 & 0.433 & 0.338 & 0.375 & 0.360 & \cellcolor[HTML]{C0C0C0} \textbf{0.485 (0.514)} \\
    & \textbf{MCC} & 0.153 & 0.171 & 0.137 & 0.144 & 0.179 & 0.117 & 0.015 & 0.067 & 0.050 & \cellcolor[HTML]{C0C0C0} \textbf{0.247 (0.286)} \\
    \hline
    \end{tabular}
  \label{tab:baseline_comparison_ase21}
\end{table*}
\end{landscape}

\noindent The average and task-wise F1-Score values of DeepCVA also beat those of the best baseline (X-CVA with Word2vec features) by substantial margins.
We found that DeepCVA significantly outperformed the best baseline models in terms of both MCC and F1-score averaging across all seven tasks, confirmed with $p$-values $<$ 0.01 using the non-parametric Wilcoxon signed-rank tests~\mbox{\cite{wilcoxon1992individual}}.
These results show the effectiveness of the novel design of DeepCVA.

An example to qualitatively demonstrate the effectiveness of DeepCVA is the VCC \textit{ff655ba} in the \textit{Apache xerces2-j} project, in which a hashing algorithm was added. This algorithm was later found vulnerable to hashing collision that could be exploited with timing attacks in the fixing commit \textit{992b5d9}. This SV was caused by the order of items being added to the hash table in the \code{put(String key, int value)} function. Such an order could not be easily captured by baseline models whose features did not consider the sequential nature of code (i.e., BoW, Word2vec and software metrics)~\cite{le2020deep}. More details about the contributions of different components to the overall performance of DeepCVA are covered in section~\mbox{\ref{subsec:rq2_results_ase21}}.

Regarding the baselines, the average MCC value (0.147) of X-CVA was on par with that (0.154) of S-CVA. This result reinforces the benefits of leveraging the common attributes among seven CVSS metrics to develop effective commit-level SV assessment models. However, X-CVA was still not as strong as DeepCVA mainly because of its much lower training data utilization per output. For X-CVA, there was an average of 39 output combinations of CVSS metrics in the training folds, i.e., 31 commits per output. In contrast, DeepCVA had 13.2 times more data per output as there were at most three classes for each task (see Fig.~\mbox{\ref{fig:cvss_distribution_ase21}}). Finally, we found supervised learning (S-CVA, X-CVA and DeepCVA) to be at least 74.6\% more effective than the unsupervised approach (U-CVA). This result shows the usefulness of using CVSS metrics to guide the extraction of commit features.

\subsection{\textbf{RQ2}: What are the Contributions of the Main Components in DeepCVA to Model Performance?}
\label{subsec:rq2_results_ase21}

\noindent \textbf{Motivation}. We have shown in RQ1 that DeepCVA significantly outperformed all the baselines for seven commit-level SV assessment tasks. RQ2 aims to give insights into the contributions of the key components to such a strong performance of DeepCVA. Such insights can help researchers and practitioners to build effective SV assessment models.

\noindent \textbf{Method}. We evaluated the performance contributions of the main components of DeepCVA: (\textit{i}) Closest Enclosing Scope (CES) of code changes, (\textit{ii}) CNN filter size, (\textit{iii}) Three-way CNN, (\textit{iv}) Attention-based GRU, (\textit{v}) Attention mechanism, (\textit{vii}) Task-specific blocks and (\textit{vi}) Multi-task learning. For each component, we first removed it from DeepCVA, retrained the model variant and reported its \textit{testing} result. When we removed Attention-based GRU, we used max-pooling~\mbox{\cite{hoang2019deepjit,kim2014convolutional}} after the three-way CNN to generate the commit vector. When we removed Multi-task learning, we trained a separate model for each of the seven CVSS metrics. We also investigated an Abstract Syntax Tree (AST) variant of DeepCVA, in which we complemented input code tokens with their syntax (e.g., \code{int a = 1} is a \code{VariableDeclarationStatement}, where \code{a} is an \code{Identifier} and \code{1} is a \code{NumberLiteral}). This AST-based variant explored the usefulness of syntactical information for commit-level SV assessment. We extracted the nodes in an AST that contained code changes and their CES. If more than two nodes contained the code of interest, we chose the one at a lower depth in the AST. We then flattened the nodes with depth-first traversal for feature extraction~\mbox{\cite{lin2018cross}}.

\begin{landscape}
\begin{figure*}[t]
    \centering
    \includegraphics[width=\linewidth,keepaspectratio]{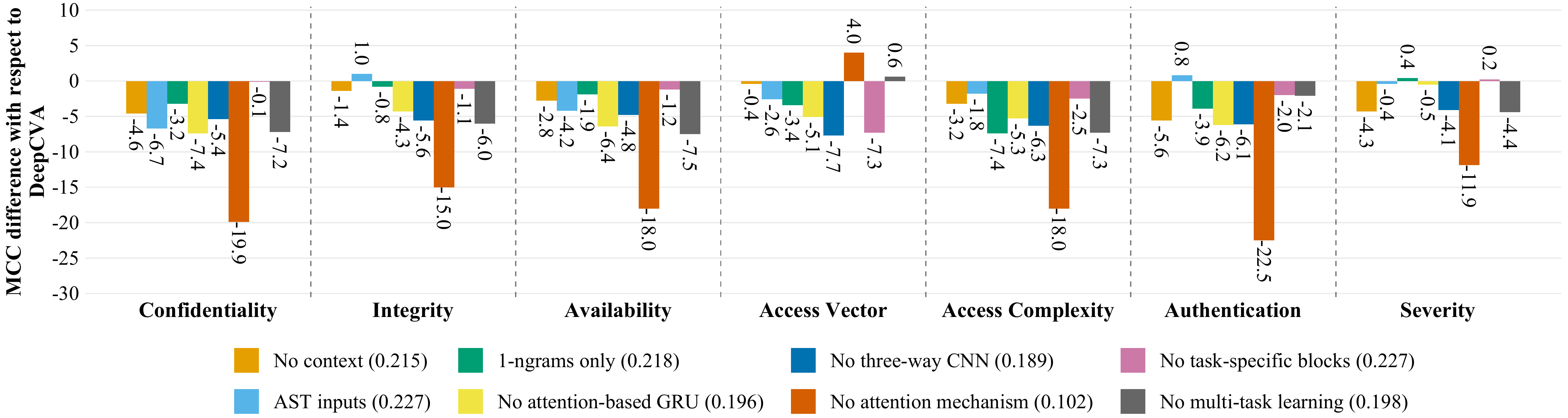}
    \caption[Differences of testing MCC of the model variants compared to the proposed DeepCVA.]{Differences of testing MCC (multiplied by 100 for readability) of the model variants compared to the proposed DeepCVA in section~\mbox{\ref{sec:deepcva_method_ase21}}. \textbf{Note}: The average MCC values (without multiplying by 100) of the model variants are in parentheses.}
    \label{fig:deepcva_components_ase21}
\end{figure*}
\end{landscape}

\noindent \textbf{Results}. \textit{\textbf{As depicted in Fig.~\mbox{\ref{fig:deepcva_components_ase21}}, the main components}}\footnote{\label{fn:att_mechanism_results}We excluded the DeepCVA variant with no attention mechanism as its performance was abnormally low, affecting the overall trend of other variants.} \textit{\textbf{uplifted the average MCC of DeepCVA by 25.9\% for seven tasks}}. Note that 7/8 model variants (except the model with no attention mechanism) outperformed the best baseline model from RQ1. These results were confirmed with $p$-values $<$ 0.01 using Wilcoxon signed-rank tests~\mbox{\cite{wilcoxon1992individual}}. Specifically, the components\textsuperscript{\ref{fn:att_mechanism_results}} of DeepCVA increased the MCC values by 25.3\%, 20.8\%, 21.5\%, 35.8\%, 35.5\%, 18.9\% and 23.6\% for Confidentiality, Integrity, Availability, Access Vector, Access Complexity, Authentication and Severity, respectively.

For the inputs, using the Smallest Enclosing Scope (CES) of code changes resulted in a 14.8\% increase in MCC compared to using hunks only, while using AST inputs had 8.8\% lower performance.
This finding suggests that code context is important for assessing SVs in commits. In contrast, syntactical information is not as necessary since code structure can be implicitly captured by code tokens and their sequential order using our AC-GRU.

The key components of the AC-GRU feature extractor boosted the performance by 13.2\% (3-grams vs. 1-grams), 25.6\% (Attention-based GRU), 30.2\% (Three-way CNN) and 142\% (Attention). Note that DeepCVA surpassed the state-of-the-art 3-gram~\mbox{\cite{han2017learning}} and 1-gram~\mbox{\cite{hoang2019deepjit}} CNN-only architectures for (commit-level) SV/defect prediction. These results show the importance of combining the (1,3,5)-gram three-way CNN with attention-based GRUs rather than using them individually. We also found that 1-5 grams did not significantly increase the performance ($p$-value = 0.186), confirming our decision in section~\mbox{\ref{subsec:deep_acgru_ase21}} to only use 1,3,5-sized filters.

For the prediction layers, we raised 8.8\% and 24.4\% MCC of DeepCVA with Task-specific blocks and Multi-task learning, respectively. Multi-task DeepCVA took 8,988 s (2.5 hours) and 25.7 s to train/validate and test in 10 rounds $\times$ 10 runs, which were 6.3 and 6.2 times faster compared to those of seven single-task DeepCVA models, respectively. DeepCVA was only 11.3\% and 12.7\% slower in training/validating and testing than one single-task model on average, respectively. These values highlight the efficiency of training and maintaining the multi-task DeepCVA model. Finally, obtaining Severity using the CVSS formula~\mbox{\cite{spanos2018multi}} from the predicted values of the other six metrics dropped MCC by 17.4\% for this task. This result supports predicting Severity directly from commit data.

\subsection{\textbf{RQ3}: What are the Effects of Class Rebalancing Techniques on Model Performance?}
\label{subsec:rq3_results}

\noindent \textbf{Motivation}. Recent studies (e.g.,~\mbox{\cite{tantithamthavorn2018impact,li2019comparative}}) have shown that class rebalancing techniques (i.e., equalizing the class distributions in the training set) can improve model effectiveness for defect/SV prediction. However, these rebalancing techniques can only be applied to single-task models, not multi-task ones. The reason is that each task has a unique class distribution (see Fig.~\mbox{\ref{fig:cvss_distribution_ase21}}), and thus balancing class distribution of one task will not balance classes of the others. RQ3 is important to test whether multi-task DeepCVA still outperforms single-task baselines in RQ1/RQ2 using rebalancing techniques.

\noindent \textbf{Method}. We compared the \textit{testing} performance of multi-task DeepCVA with baselines in RQ1/RQ2 using two popular oversampling techniques~\mbox{\cite{tantithamthavorn2018impact}}: \textit{Random OverSampling} (\textit{ROS}) and \textit{SMOTE}~\mbox{\cite{chawla2002smote}}. ROS randomly duplicates the existing samples of minority classes, while SMOTE randomly generates synthetic samples between the existing minority-class samples and their nearest neighbor(s) based on Euclidean distance. We did not consider undersampling, as such models performed poorly because of some very small minority classes (e.g., \textit{Low} Access Complexity had only 14 samples). We applied ROS and SMOTE to \textit{only the training set} and then optimized all baseline models again. Like~\mbox{\cite{tantithamthavorn2018impact}}, we also tuned SMOTE using grid search with different values of nearest neighbors: \{1, 5, 10, 15, 20\}. We could not apply SMOTE to single-task DeepCVA as features were trained end-to-end and unavailable prior training for finding nearest neighbors. We also did not apply SMOTE to X-CVA as there was always a single-sample class in each round, producing no nearest neighbor.

\noindent \textbf{Results}. \textit{\textbf{ROS and SMOTE increased the average performance (MCC) of 3/4 baselines except X-CVA (see Table~\mbox{\ref{tab:rebalancing_results_ase21}}). However, the average MCC of our multi-task DeepCVA was still 14.4\% higher than that of the best oversampling-augmented baseline (single-task DeepCVA with ROS)}}. Overall, MCC increased by 8\%, 6.9\% and 9.1\% for S-CVA (ROS), S-CVA (SMOTE) and single-task DeepCVA (ROS), respectively. These improvements were confirmed significant with $p$-values $<$ 0.01 using Wilcoxon signed-rank tests~\mbox{\cite{wilcoxon1992individual}}. We did not report oversampling results of U-CVA as they were still much worse compared to others. We found single-task DeepCVA benefited the most from oversampling, probably since Deep Learning usually performs better with more data~\mbox{\cite{zheng2020impact}}. In contrast, oversampling did not improve X-CVA as oversampling did not generate as many samples for X-CVA per class as for S-CVA (i.e., X-CVA had 13 times, on average, more classes than S-CVA). These results further strengthen the effectiveness and efficiency of multi-task learning of DeepCVA for commit-level SV assessment even without the overheads of rebalancing/oversampling data.

\begin{table}[t]
  \centering
  \caption[Testing performance (MCC) of optimal baselines using oversampling techniques and multi-task DeepCVA.]{Testing performance (MCC) of optimal baselines using oversampling techniques and multi-task DeepCVA. \textbf{Note}: \textsuperscript{\textdagger}denotes that the oversampled models outperformed the non-oversampled one reported in RQ1/RQ2.}
    \begin{tabular}{l|ccccc}
    \hline
    \textbf{CVSS Task} & \makecell[c]{\textbf{S-CVA}\\ \textbf{(ROS)}} & \makecell[c]{\textbf{S-CVA}\\ \textbf{(SMOTE)}} & \makecell[c]{\textbf{X-CVA}\\ \textbf{(ROS)}} & \makecell[c]{\textbf{Single-task}\\ \textbf{DeepCVA}\\ \textbf{(ROS)}} & \makecell[c]{\textbf{Multi-task}\\ \textbf{DeepCVA}}\\
    \hline
    \textbf{Confidentiality} & 0.220 & 0.203 & 0.185 & 0.250\textsuperscript{\textdagger} & \cellcolor[HTML]{C0C0C0} \textbf{0.268} \\
    \textbf{Integrity} & 0.174 & 0.168 & 0.179\textsuperscript{\textdagger} & 0.206\textsuperscript{\textdagger} & \cellcolor[HTML]{C0C0C0} \textbf{0.250} \\
    \textbf{Availability} & 0.195\textsuperscript{\textdagger} & 0.187\textsuperscript{\textdagger} & 0.182 & 0.209\textsuperscript{\textdagger} & \cellcolor[HTML]{C0C0C0} \textbf{0.273} \\
    \textbf{Access Vector} & 0.115\textsuperscript{\textdagger} & 0.110\textsuperscript{\textdagger} & 0.092 & \cellcolor[HTML]{C0C0C0} \textbf{0.156}\textsuperscript{\textdagger} & 0.129 \\
   \textbf{Access Comp.} & 0.172\textsuperscript{\textdagger} & 0.186\textsuperscript{\textdagger} & 0.144\textsuperscript{\textdagger} & 0.190\textsuperscript{\textdagger} & \cellcolor[HTML]{C0C0C0} \textbf{0.242} \\
    \textbf{Authentication} & 0.325\textsuperscript{\textdagger} & 0.340\textsuperscript{\textdagger} & 0.299\textsuperscript{\textdagger} & 0.318 & \cellcolor[HTML]{C0C0C0} \textbf{0.352} \\
    \textbf{Severity} & 0.132 & 0.124 & 0.141 & 0.186\textsuperscript{\textdagger} & \cellcolor[HTML]{C0C0C0} \textbf{0.213} \\
    \hline
    \hline
    \textbf{Average} & 0.190\textsuperscript{\textdagger} & 0.188\textsuperscript{\textdagger} & 0.175 & 0.216\textsuperscript{\textdagger} & \cellcolor[HTML]{C0C0C0} \textbf{0.247} \\
    \hline
    \end{tabular}
  \label{tab:rebalancing_results_ase21}
\end{table}

\section{Discussion}
\label{sec:discussion_ase21}

\subsection{DeepCVA and Beyond}

DeepCVA has been shown to be effective for commit-level SV assessment in the three RQs, but our model still has false positives. We analyze several representative patterns of such false positives to help further advance this task and solutions for researchers and practitioners in the future.

Some commits were too complex and large (tangled) to be assessed correctly. For example, the VCC \textit{015f7ef} in the \textit{Apache Spark} project contained 1,820 additions and 146 deletions across 29 files; whereas, the untrusted deserialization SV occurred in just one line 56 in \code{LauncherConnection.java}. Recent techniques (e.g.,~\mbox{\cite{wattanakriengkrai2020predicting,pascarella2019fine}}) pinpoint more precise locations (e.g., individual files or lines in commits) of defects, especially in tangled changes.
Such techniques can be adapted to remove irrelevant code in VCCs (i.e., changes that do not introduce or contain SVs). More relevant code potentially gives more fine-grained information for the SV assessment tasks. Note that DeepCVA provides a strong baseline for comparing against fine-grained approaches.

DeepCVA also struggled to accurately predict assessment metrics for SVs related to external libraries. For instance, the SV in the commit \textit{015f7ef} above occurs with the \code{ObjectInputStream} class from the \code{java.io} package, which sometimes prevented DeepCVA from correctly assessing an SV. If an SV happens frequently with a package in the training set, (e.g., the XML library of the VCC \textit{bba4bc2} in Fig.~\mbox{\ref{fig:commit_ex_ase21}}), DeepCVA still can infer correct CVSS metrics. Pre-trained code models on large corpora~\mbox{\cite{alon2019code2vec,devlin2018bert,hoang2020cc2vec}} along with methods to search/generate code~\mbox{\cite{gu2018deep}} and documentation~\mbox{\cite{hu2018deep}} as well as (SV-related) information from developer Q\&A forums~\cite{le2021large} can be investigated to provide enriched context of external libraries, which would in turn support more reliable commit-level SV assessment with DeepCVA.

We also observed that DeepCVA, alongside the considered baseline models, performed significantly worse, in terms of MCC, for Access Vector compared to the remaining tasks (see Table~\ref{tab:baseline_comparison_ase21}). We speculate that such low performance is mainly because Access Vector contains the most significant class imbalance among the tasks, as shown in Fig.~\ref{fig:cvss_distribution_ase21}. For single-task models, we found that using class rebalancing techniques such as ROS or SMOTE can help improve the performance, as demonstrated in RQ3 (see section~\ref{subsec:rq3_results}). However, it is still unclear how to apply the current class rebalancing techniques for multi-task learning models such as DeepCVA. Thus, we suggest that more future work should investigate specific class rebalancing and/or data augmentation to address such imbalanced data in the context of multi-task learning.

\subsection{Threats to Validity}

The first threat is the collection of VCCs. We followed the practices in the literature to reduce the false positives of the SZZ algorithm. We further mitigated this threat by performing independent manual validation with three researchers with at least two years of experience in Software Engineering and Cybersecurity.

Another concern is the potential suboptimal tuning of baselines and DeepCVA. However, it is impossible to try the entire hyperparameter space within a reasonable amount of time. For the baseline models, we lessened this threat by using a wide range of hyperparameters from the previous studies to reoptimize these models from scratch on our data. For DeepCVA, we adapted the best practices recommended in the relevant literature to our tasks.

The reliability and generalizability of our findings are also potential threats. We ran DeepCVA 10 times to mitigate the experimental randomness. We confirmed our results using non-parametric statistical tests with a confidence level $>$ 99\%. Our results may not generalize to all software projects. However, we reduced this threat by conducting extensive experiments on 200+ real-world projects of different scales and domains.

\section{Related Work}
\label{sec:related_work_ase21}

\subsection{Data-Driven SV Prediction and Assessment}

As reviewed in Chapter~\ref{chapter:lit_review}, many studies have developed data-driven approaches that can harness large-scale SV data from public security databases like NVD to determine different characteristics of SVs. Specifically, SV information on NVD has been utilized to infer the types~\mbox{\cite{neuhaus2010security}}, exploitation~\mbox{\cite{bozorgi2010beyond,bullough2017predicting}}, time-to-exploit~\cite{bozorgi2010beyond} and various CVSS assessment metrics~\mbox{\cite{le2019automated,spanos2018multi,elbaz2020fighting,gong2019joint}} of SVs.
Other studies~\mbox{\cite{ponta2018beyond, ponta2020detection}} have leveraged code patterns in fixing commits of third-party libraries to assess SVs in such libraries. Our work is fundamentally different from these previous studies since we are the first to investigate the potential of performing assessment of all SV types (not only vulnerable libraries) using commit changes rather than bug/SV reports/fixes. Our approach allows practitioners to realize the exploitability/impacts of SVs in their systems much earlier, e.g., up to 1,000 days before (see section~\mbox{\ref{subsec:cva_need_ase21}}), as compared to using bug/SV reports/fixes. Less delay in SV assessment helps practitioners to plan/prioritize SV fixing with fresh design and implementation in their minds. Moreover, we have shown that multi-task learning, i.e., predicting all CVSS metrics simultaneously, can significantly increase the effectiveness and reduce the model development and maintenance efforts in commit-level SV assessment.
It should be noted that report-level prediction is still necessary for assessing SVs in third-party libraries/software, especially the ones without available code (commits), to prioritize vendor-provided patch application, as well as SVs missed by commit-level detection.

\subsection{SV Analytics in Code Changes}

Commit-level prediction (e.g.,~\mbox{\cite{kamei2012large,hoang2019patchnet,yang2015deep}}) has been explored to provide \textit{just-in-time} information for developers about code issues, but such studies mainly focused on generic software defects. However, SV is a special type of defects~\mbox{\cite{camilo2015bugs}} that can threaten the security properties of a software project. Thus, SV requires special treatment~\mbox{\cite{peters2017text}} and domain knowledge~\mbox{\cite{gegick2010identifying}}. Meneely et al.~\mbox{\cite{meneely2013patch}} and Bosu et al.~\mbox{\cite{bosu2014identifying}} conducted in-depth studies on how code and developer metrics affected the introduction and review of VCCs. Besides analyzing the characteristics of VCCs, other studies~\mbox{\cite{perl2015vccfinder,yang2017vuldigger,chen2019large}} also developed commit-level SV detection models that leveraged software and text-based metrics. Different from the previous studies that have detected VCCs, we focus on the assessment of such VCCs. SV assessment is as important as the detection step since assessment metrics help early plan and prioritize remediation for the identified SVs. It is worth noting that the existing SV detection techniques can be used to flag VCCs that would then be assessed by our DeepCVA model.

\section{Chapter Summary}
\label{sec:conclusions_ase21}

We introduce DeepCVA, a novel deep multi-task learning model, to tackle a new task of commit-level SV assessment. DeepCVA promptly informs practitioners about the CVSS severity level, exploitability, and impact of SVs in code changes after they are committed, enabling more timely and informed remediation. DeepCVA substantially outperformed many baselines (even the ones enhanced with rebalanced data) for the seven commit-level SV assessment tasks. Notably, multi-task learning utilizing the relationship of assessment tasks helped our model be 24.4\% more effective and 6.3 times more efficient than single-task models.
With the reported performance, DeepCVA realizes the first promising step towards a holistic solution to assessing SVs as early as they appear.

%% file: Chapters/Chapter_6_EASE2021.tex
\chapter{Collection and Analysis of Developers' Software Vulnerability Concerns on Question and Answer Websites}
\label{chap:ease21}

\vspace{-9pt}

\begin{tcolorbox}
\textbf{Related publications}: This chapter is based on two of our papers: (1) ``\textit{PUMiner: Mining Security Posts from Developer Question and Answer Websites with PU Learning}'' published in the 17\textsuperscript{th} International Conference on Mining Software Repositories (MSR), 2020 (CORE A)~\cite{le2020puminer}, and (2) ``\textit{A Large-scale Study of Security Vulnerability Support on Developer Q\&A Websites}'' published in the 25\textsuperscript{th} International Conference on Evaluation and Assessment in Software Engineering (EASE), 2021 (CORE A)~\cite{le2021large}.
\end{tcolorbox}
\bigskip

In the previous Chapters~\ref{chap:msr19},~\ref{chap:msr22}, and~\ref{chap:ase21}, we have proposed different automated solutions to perform Software Vulnerability (SV) assessment using the knowledge gathered from software artifacts, i.e., SV reports and source code. In practice, besides relying on these software artifacts for SV assessment, developers also seek information about SVs, e.g., experience/solutions of fixing similar SVs, on developer Question and Answer (Q\&A) websites. The SV information provided on Q\&A sites also sheds light on developers' real-world concerns/challenges with addressing SVs, which can complement other assessment metrics like CVSS to support better SV understanding and fixing. However, there is still little known about these SV-specific discussions on different Q\&A sites.
Chapter~\ref{chap:ease21} presents a large-scale empirical study to understand developers' SV discussions and how these discussions are being supported by Q\&A sites.
We curate 71,329 SV posts from two large Q\&A sites, namely Stack Overflow (SO) and Security StackExchange (SSE), and then use topic modeling to uncover the key developers' SV topics of concern.
We then analyze the popularity, difficulty, and level of expertise for each topic. We also perform a qualitative analysis to identify the types of solutions to SV-related questions.
We identify 13 main SV discussion topics. Many topics do not follow the distributions and trends in expert-based security sources, e.g., Common Weakness Enumeration (CWE) and Open Web Application Security Project (OWASP). We also discover that SV discussions attract more experts to answer than many other domains. Nevertheless, some difficult SV topics/types still receive quite limited support from experts, which may suggest it is challenging to fix them in practice. Moreover, we identify seven key types of answers given to SV questions, in which SO often provides code and instructions, while SSE usually gives experience-based advice and explanations. These solutions on Q\&A sites may be reused to fix similar SVs, reducing the SV fixing effort.
Overall, the findings of this chapter enable researchers and practitioners to effectively leverage SV knowledge on Q\&A sites for (data-driven) SV assessment.

\newpage

\section{Introduction}
It is important to constantly track and resolve Software Vulnerabilities (SVs) to ensure the availability, confidentiality and integrity of software systems~\mbox{\cite{ghaffarian2017software}}. Developers can seek assessment information for resolving SVs from sources verified by security experts such as Common Weakness Enumeration (CWE)~\cite{cwe}, National Vulnerability Database (NVD)~\cite{nvd} and Open Web Application Security Project (OWASP)~\cite{owasp_website}.
However, these expert-based SV sources do not provide any mechanisms for developers to promptly ask and answer questions about issues in implementing/understanding the reported SV solutions/concepts.
On the other hand, developer Questions and Answer (Q\&A) websites contain a plethora of such SV-related discussions.
Stack Overflow (SO)\footnote{\url{https://stackoverflow.com/}} and Security StackExchange (SSE)\footnote{\url{https://security.stackexchange.com/}} contain some of the largest numbers of SV-related discussions among developer Q\&A sites with contributions from millions of users~\mbox{\cite{le2020puminer}}.

The literature has analyzed different aspects of discussions on Q\&A sites, but there is still no investigation of how SO and SSE are supporting SV-related discussions. Specifically, the main concepts~\mbox{\cite{yang2016security}}, the top languages/technologies and user demographics~\mbox{\cite{bayati2016information}}, as well as user perceptions and interactions~\mbox{\cite{lopez2019anatomy}} of general security discussions on SO have been studied. However, from our analysis (see section~\mbox{\ref{subsec:post_collection_ease21}}), only about 20\% of the available SV posts on SO were investigated in the previous studies, limiting a thorough understanding of SV topics (developers' concerns when tackling SVs in practice) on Q\&A sites. 
Moreover, the prior studies only focused on SO, and little insight has been given into the support of SV discussions on different Q\&A sites. Such insight would potentially affect the use of a suitable site (e.g., SO vs. SSE) to obtain necessary SV assessment information for SV prioritization and fixing.

To fill these gaps, we conduct a large-scale empirical study using 71,329 SV posts curated from SO and SSE. Specifically, we use Latent Dirichlet Allocation (LDA)~\mbox{\cite{blei2003latent}} topic modeling and qualitative analysis to answer the following four Research Questions (RQs) that measure the support of Q\&A sites for different SV discussion topics:

\noindent \textbf{RQ1}: What are SV discussion topics on Q\&A sites?

\noindent\textbf{RQ2}: What are the popular and difficult SV topics?

\noindent \textbf{RQ3}: What is the level of expertise for supporting SV questions?

\noindent \textbf{RQ4}: What types of answers are given to SV questions?

\noindent Our findings to these RQs can help raise developers' awareness of common SVs and enable them to seek solutions to such SVs more effectively on Q\&A sites. We also identify the areas to which experts can contribute to assist the secure software engineering community. Moreover, these common developers' SV concerns and their characteristics can be leveraged for making data-driven SV assessment models more practical (closer to developers' real-world needs) and enabling more effective understanding and fixing prioritization of commonly encountered SVs. Furthermore, we release one of the largest datasets of SV discussions that we have carefully curated from Q\&A sites for replication and future work at \url{https://github.com/lhmtriet/SV_Empirical_Study}.

\noindent \textbf{Chapter organization}. Section~\mbox{\ref{sec:related_work_ease21}} covers the related work. Section~\mbox{\ref{sec:case_study_ease21}} describes the four research questions along with the methods and data used in this chapter. Sections~\mbox{\ref{sec:results_ease21}} presents the results of each research question. Section~\mbox{\ref{sec:discussion_ease21}} discusses the findings including how they can be used for SV assessment, and then mentions the threats to validity. Section~\mbox{\ref{sec:conclusions_ease21}} concludes and suggests several future directions.

\section{Related Work}
\label{sec:related_work_ease21}
\subsection{Topic Modeling on Q\&A Websites}

Q\&A websites such as SO and SSE contain a large number of discussion posts. LDA~\cite{blei2003latent} has been frequently used to extract the taxonomy/topics of various software-related domains from such posts. In 2014, a seminal work of Barua et al.~\cite{barua2014developers} discovered the topics of all SO posts. They also found that LDA could find more consistent topics than the tags on SO. Many subsequent studies have leveraged LDA to investigate discussions of specific domains, such as general security~\cite{yang2016security}, concurrent computing~\cite{ahmed2018concurrency}, mobile computing~\cite{rosen2016mobile}, big data~\cite{bagherzadeh2019going}, machine learning~\cite{bangash2019developers} and deep learning~\cite{han2020programmers}. Among the aforementioned studies, Yang et al.~\mbox{\cite{yang2016security}} is the closest to our work. However, our work is still fundamentally different from this previous study. Despite sharing a similar security context to Yang et al.~\mbox{\cite{yang2016security}}, we focus specifically on the flaws of security implementation/features since exploitation of such flaws can disclose user's data and interrupt system operations.
Moreover, we consider the content of both questions and answers of SV posts on two Q\&A sites (SO and SSE) rather than just questions on SO as in~\mbox{\cite{yang2016security}}. This gives more in-depth insights into how different Q\&A sites are supporting on-going SV discussions. Detailed discussion on these differences is given in section~\mbox{\ref{subsec:vs_literature}}.

\vspace{-6pt}

\subsection{SV Assessment Using Open Sources}

SV assessment has long been of interest to researchers.
Shahzad et al.~\cite{shahzad2012large} conducted a large-scale study on the characteristics (e.g., risk metrics, exploitation, affected vendors and products) of reported SVs on NVD. Besides empirical study, there is another active research trend to build data-driven models to analyze SVs, as mentioned in Chapter~\ref{chapter:lit_review}. Bozorgi et al.~\cite{bozorgi2010beyond} used Support Vector Machine to predict the probability and time-to-exploit of SVs. There have been many follow-up studies since then on developing learning-based models (e.g.,~\cite{le2019automated,han2017learning,sahin2019conceptual}) to determine various properties of SVs using expert-based SV sources (e.g., CWE and NVD). A recent study~\cite{horawalavithana2019mentions} leveraged security mentions on social media (i.e., Twitter and Reddit) to forecast the SV-related activities on GitHub. Unlike the above studies, we focus on SV analytics on developer Q\&A sites. Several studies (e.g.,~\mbox{\cite{meng2018secure,rahman2019snakes}}) analyzed SVs of different programming languages using code snippets on SO. Contrary to these studies, we do not limit our investigation to any specific programming language, and we consider every type of SV-related posts, not just the ones with code snippets.

\vspace{-6pt}

\section{Research Method}
\label{sec:case_study_ease21}

\subsection{Research Questions}
\label{subsec:rq_method_ease21}

We investigated four RQs to study the support of Q\&A websites for SV-related discussions.
To answer these RQs, we retrieved 71,329 SV posts from a general Q\&A website (SO) and a security-centric one (SSE) using both the tags and content of posts (see section~\ref{subsec:post_collection_ease21}).

\noindent \textbf{RQ1: What are SV discussion topics on Q\&A sites?}\label{rq1_method}

\noindent \underline{\textit{Motivation}}: To provide fine-grained information about the support of SO and SSE for different types of SV discussions, we first needed to identify the taxonomy of commonly discussed SV topics in RQ1. Our taxonomy does not aim to replace the existing ones provided by experts (e.g., CWE or OWASP), but rather helps to highlight the important aspects of SVs from developers' perspectives.
Such taxonomy would also provide developers' real-world needs to help make SV assessment effort more practical.

\noindent \underline{\textit{Method}}: Following the standard practice in~\mbox{\cite{barua2014developers,yang2016security,rosen2016mobile,ahmed2018concurrency,bagherzadeh2019going,bangash2019developers,han2020programmers}}, RQ1 used the Latent Dirichlet Allocation (LDA) \cite{blei2003latent} topic modeling technique (see section~\ref{subsec:lda_ease21}) to select SV discussion topics based on the titles, questions and answers of SV posts on both SO and SSE. LDA is commonly used since it can produce topic distribution (assigning multiple topics with varying relevance) for a post, providing more flexibility/scalability than manual coding. We also used the topic share metric~\cite{barua2014developers} in Eq.~\eqref{eq:share_ease21} to compute the proportion ($\text{share}_{i}$) of each SV topic and their trends over time.
\begin{myequation}\label{eq:share_ease21}
\text{share}_{i}=\frac{1}{N}\sum\limits_{p\,\in \,D}{\text{LDA}(p,\,\,{{\text{T}}_{i}})}
\end{myequation}

\noindent where $p$, $D$ and $N$ are a single SV post, the list of all SV posts and the number of such posts, respectively; $\text{T}_{i}$ is the $i^{\text{th}}$ topic and LDA is the trained LDA model.

\noindent \textbf{RQ2: What are the popular and difficult SV topics?}\label{rq2_method}

\noindent \underline{\textit{Motivation}}: After the SV topics were identified, RQ2 identified the popular and difficult topics on Q\&A websites. The results of RQ2 can aid the selection of a suitable (i.e., more popular and less difficult) Q\&A site for respective SV topics.

\noindent \underline{\textit{Method}}: To quantify the topic popularity, we used four metrics from~\cite{rosen2016mobile,yang2016security, bagherzadeh2019going,ahmed2018concurrency}, namely the average values of (\textit{i}) views, (\textit{ii}) scores (upvotes minus downvotes), (\textit{iii}) favorites and (\textit{iv}) comments. Intuitively, a more popular topic would attract more attention (views), interest (scores/favorites) and activities (comments) per post from users. We also obtained the geometric mean of the popularity metrics to produce a more consistent result across different topics. Geometric mean was used instead of arithmetic mean here since the metrics could have different units/scales. To measure the topic difficulty, we used the three metrics from~\cite{rosen2016mobile,yang2016security, bagherzadeh2019going,ahmed2018concurrency}: (\textit{i}) percentage of getting accepted answers, (\textit{ii}) median time (hours) to receive an accepted answer since posted, and (\textit{iii}) average ratio of answers to views. A more difficult topic would, on average, have a lower number of accepted answers and ratio of answers to views, but a higher amount of time to obtain accepted answers. To achieve this, we took reciprocals of the difficulty metrics (\textit{i}) and (\textit{iii}) so that a more difficult topic had a higher geometric mean of the metrics.

\noindent \textbf{RQ3: What is the level of expertise to answer SV questions?}\label{rq3_method}

\noindent \underline{\textit{Motivation}}: RQ3 checked the expertise level available on Q\&A websites to answer SV questions, especially the ones of difficult topics.
The findings of RQ3 can shed light on the amount of support each topic receives from experienced users/experts on Q\&A sites and which topic may require more attention from experts.
Note that experts here are users who frequently contribute helpful (accepted) answers/knowledge.

\noindent \underline{\textit{Method}}: We measured both users' general and specific expertise for SV topics on Q\&A sites. For the general expertise, we leveraged the commonly used metric, the reputation points~\mbox{\cite{hanrahan2012modeling,meng2018secure,rahman2019snakes}}, of users who got accepted answers since reputation is gained through one's active participation and appreciation from the Q\&A community in different topics. A higher reputation received for a topic usually implies that the questions of that topic are of more interest to experts. Similar to~\mbox{\cite{meng2018secure}}, we did not normalize the reputation by user's participation time since reputation may not increase linearly, e.g., due to users leaving the sites.
However, reputation is not specific to any topic; thus, it does not reflect whether a user is experienced with a topic. Hence, we represented developers' specific expertise with the SV content in their answers on Q\&A sites. This was inspired by Dey et al.'s findings that developers' expertise/knowledge could be expressed through their generated content~\mbox{\cite{dey2020representation}}. We determined a user's expertise in SV topics using the topic distribution generated by LDA applied to the concatenation of all answers to SV questions given by that user. The specific expertise of an SV topic (see Eq.~\mbox{\eqref{eq:spec_exp_ease21}}) was then the total correlation between LDA outputs of the current topic in SV questions and the specific expertise of users who got the respective accepted answers. The correlation of LDA values could reveal the knowledge (SV topics) commonly used to answer questions of a certain (SV) topic~\mbox{\cite{barua2014developers}}.

\begin{myequation}\label{eq:spec_exp_ease21}
\begin{aligned}
Specific\_Expertis{{e}_{i}}=\sum\limits_{p\in D}{\text{LDA}(Q(p),\,\,{{\text{T}}_{i}})\odot \text{LDA}(K({{U}_{\text{Accept}\text{.}}}))} \\ 
K({{U}_{\text{Accept}\text{.}}})=A_{{{U}_{\text{Accept}\text{.}}}}^{1}+A_{{{U}_{\text{Accept}\text{.}}}}^{2}+...+A_{{{U}_{\text{Accept}\text{.}}}}^{k}(k=\left| {{A}_{{{U}_{\text{Accept}\text{.}}}}} \right|)
\end{aligned}
\end{myequation}

\noindent where $D$ is the list SV posts and $\text{T}_{i}$ is the $i^{\text{th}}$ topic, while $Q(p)$ and $K({{U}_{\text{Accept}\text{.}}})$ are the question content and SV knowledge of the user ${U}_{\text{Accept}\text{.}}$ who gave the accepted answer of the post $p$, respectively. $\odot$ is the topic-wise multiplication. $\left| {{A}_{{{U}_{\text{Accept}\text{.}}}}} \right|$ is all SV-related answers given by user ${U}_{\text{Accept}\text{.}}$. Note that we only considered posts with accepted answers to make it consistent with the general expertise.

\noindent Specifically, for each question, we first extracted the user that gave the accepted answer (${U}_{\text{Accept}\text{.}}$). We then gathered all answers, not necessarily accepted, of that user in SV posts ($\left| {{A}_{{{U}_{\text{Accept}\text{.}}}}} \right|$). Such answer list was the SV knowledge of ${U}_{\text{Accept}\text{.}}$ ($K({{U}_{\text{Accept}\text{.}}})$).
Finally, we computed the LDA topic-wise correlation between the topic $\text{T}_{i}$ in the current SV question ($\text{LDA}(Q(p),\,\,{{\text{T}}_{i}})$) and the user knowledge ($\text{LDA}(K({{U}_{\text{Accept}\text{.}}}))$) to determine the specific expertise for post $p$.

\noindent \textbf{RQ4: What types of answers are given to SV questions?}\label{rq4_method}

\noindent \underline{\textit{Motivation}}: RQ4 extended RQ2 in terms of the solution types given if an SV question is satisfactorily answered. We do not aim to provide solutions for every single SV. Rather, we analyze and compare the types of support for different SV topics on SO and SSE, which can guide developers to a suitable site depending on their needs (e.g., looking for certain artefacts). To the best of our knowledge, we are the first to study answer types of SVs on Q\&A sites.

\noindent \underline{\textit{Method}}: We employed an open coding procedure~\cite{seaman1999qualitative} to inductively identify answer types. LDA is \textit{not suitable} for this purpose since it relies on word co-occurrences to determine categories. In contrast, the same type of solutions may not share any similar words. In RQ4, we only considered the posts with accepted answer to ensure the high quality and relevance of the answers. We then used stratified sampling to randomly select 385 posts (95\% confidence level with 5\% margin error~\cite{cochran2007sampling}) each from SO and SSE to categorize the answer types. Stratification ensured the proportion of each topic was maintained.
Following~\mbox{\cite{chen2020comprehensive}}, the author of this thesis and a PhD student with three years of experience in Software Engineering and Cybersecurity first conducted a pilot study to assign initial codes to 30\% of the selected posts and grouped similar codes into answer types. For example, the accepted answers of SO posts 32603582 (PostgreSQL code), 20763476 (MySQL code) and 12437165 (Android/Java code) were grouped into the \textit{Code Sample} category. Similarly to~\mbox{\cite{treude2011programmers}}, we also allowed one post to have more than one answer type.
The two same people then independently assigned the identified categories to the remaining 70\% of the posts. The Kappa inter-rater score ($\kappa$)~\cite{mchugh2012interrater} was 0.801 (strong agreement), showing the reliability of our coding. Another PhD student with two-year experience in Software Engineering and Cybersecurity was involved to discuss and resolve the disagreements. We also correlated the answer types with the question types on Q\&A sites~\cite{treude2011programmers}.

\vspace{-6pt}

\subsection{Software Vulnerability Post Collection}
\label{subsec:post_collection_ease21}

\begin{landscape}
\begin{figure*}[t]
    \centering
    \includegraphics[width=\linewidth,keepaspectratio]{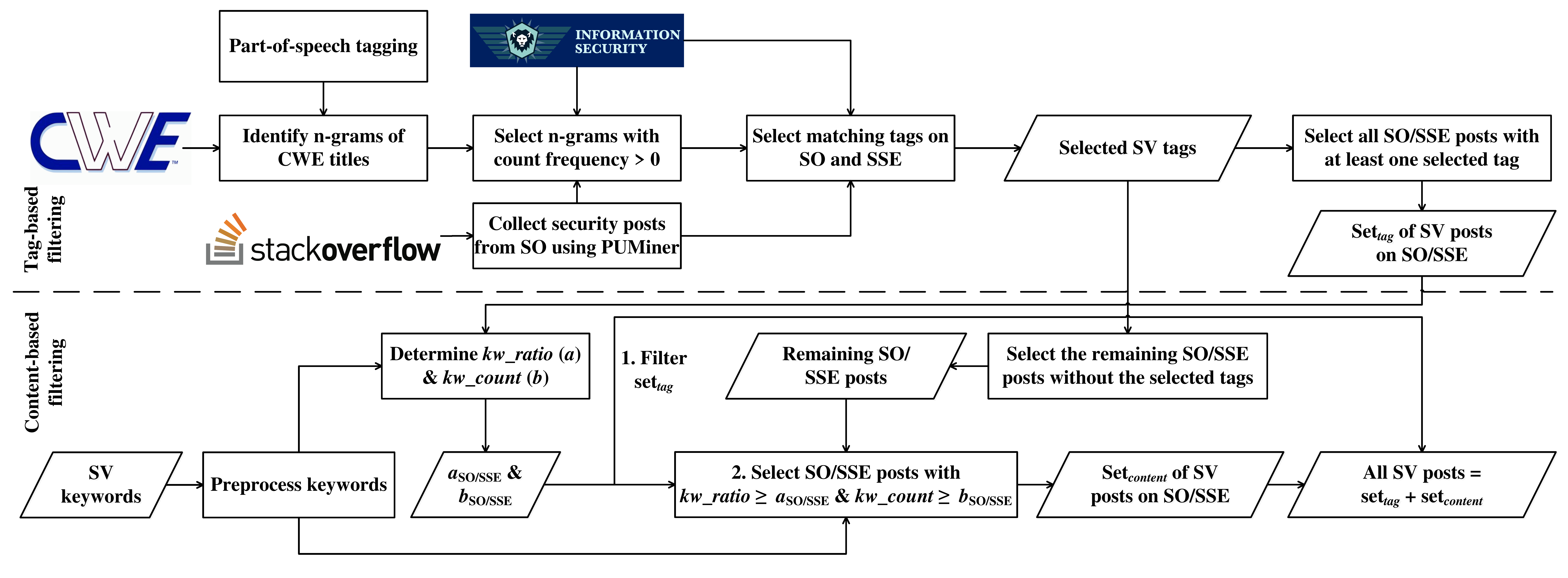}
    \caption[Workflow of retrieving posts related to software vulnerability on Q\&A websites using tag-based and content-based filtering heuristics.]{Workflow of retrieving posts related to SV on Q\&A websites using tag-based and content-based filtering heuristics.}
    \label{fig:workflow_ease21}
\end{figure*}
\end{landscape}

To study the support of Q\&A sites for SV discussions, we proposed a workflow (see Fig.~\ref{fig:workflow_ease21}) to obtain, to the best of our knowledge, the largest and most contemporary set of SV posts on both SO and SSE.
To identify SV-related posts, we started from security posts on Q\&A sites as every SV is a security issue by definition. This decision helped to increase the relevance of our retrieved SV posts. Specifically, all the posts on SSE were presumably relevant to security as SSE is a security-centric site; whereas, on SO, we used security posts automatically collected using our novel tool, PUMiner~\cite{le2020puminer}. PUMiner alleviates the need for the non-security (negative) class to predict security posts on Q\&A sites based on a two-stage PU learning framework~\cite{bekker2018learning}. Retrieving non-security posts in practice is challenging since these posts should not contain any security context, which requires significant human effort to define and verify. It is also worth noting that manual selection of security/non-security posts also does not scale to millions of posts on Q\&A sites. PUMiner has been demonstrated to be more effective in retrieving security posts on SO than many learning-based baselines such as one-class SVM~\cite{scholkopf2000support,manevitz2001one} and positive-similarity filtering~\cite{mendsaikhan2019identification} on unseen posts. PUMiner can also successfully predict the cases where keyword matching totally missed with an MCC of 0.745. Notably, with only 1\% labeled positive posts, PUMiner is still 160\% better than fully-supervised learning. More details of PUMiner can be found in Appendix~\ref{sec:appendix_ease21}.
Using the curated security posts on SO and SSE, we then employed \textit{tag-based} and \textit{content-based filtering} to retrieve SV posts based on their tags and content of other parts (i.e., title, body and answers), respectively. We considered a post to be related to SV when it mainly discussed a security flaw and/or exploitation/testing/fixing of such flaw to compromise a software system (e.g., SO post 29098142\footnote{\url{stackoverflow.com/questions/29098142} (postid: 29098142). SSE format is security.stackexchange.com/questions/postid. Posts in our paper follow these formats.}). A post was not SV-related if it just asked how to implement/use a security feature (e.g., SO post 685855) without any explicit mention of a flaw. All the tags, keywords and posts collected were released at \url{https://github.com/lhmtriet/SV_Empirical_Study}.

\noindent \textbf{Tag-based filtering}.
We had a \textit{vulnerability} tag on SSE but not on SO to obtain SV-related posts, and the \textit{security} tag on SO used by~\mbox{\cite{yang2016security}} was too coarse-grained for the SV domain. Many posts with the \textit{security} tag did not explicitly mention SV (e.g., SO post 65983245 about privacy or SO post 66066267 about how to obtain security-relevant commits). Therefore, we used Common Weakness Enumeration (CWE), which contains various SV-related terms, to define relevant SV tags. However, the full CWE titles were usually long and uncommonly used in Q\&A discussions. For example, the fully-qualified CWE name of SQL-injection (CWE-89) is ``\emph{Improper Neutralization of Special Elements used in an SQL Command (`SQL Injection')}'', which appeared only nine times on SO and SSE. Therefore, we needed to extract shorter and more common terms from the full CWE titles. We adopted Part-of-Speech (POS) tagging for this purpose, in which we only considered consecutive (n-grams of) verbs, nouns and adjectives since most of them conveyed the main meaning of a title. For instance, we obtained the following 2-grams for CWE-89: \textit{improper neutralization}, \textit{special elements}, \textit{elements used}, \textit{sql command}, \textit{sql injection}. We obtained 2,591 n-gram (1 $\leq$ n $\leq$ 3) terms that appeared at least once on either SO or SSE. To ensure the relevance of these terms, we manually removed the irrelevant terms without any specific SV context (e.g., \textit{special elements}, \textit{elements used} and \textit{sql command} in the above example). We found 60 and 63 SV-related tags on SO and SSE that matched the above n-grams, respectively. We then obtained the initial $\text{set}_{tag}$ of SV posts that had at least one of these selected tags.

\noindent \textbf{Content-based filtering}. As recommended by some recent studies~\cite{haque2020challenges,le2020puminer}, tag-based filtering was not sufficient for selecting posts due to wrong tags (e.g., non SV-post 38539393 on SO with \textit{stack-overflow} tag) or general tags (e.g., SV post 15029849 on SO with only \textit{php} tag). Therefore, as depicted in Fig.~\ref{fig:workflow_ease21}, we customized content-based filtering, which was based on keyword matching, to refine the $\text{set}_{tag}$ obtained from the tag-based filtering step and select missing SV posts that were not associated with SV tags. First, we presented the up-to-date list of 643 SV keywords for matching at \url{https://github.com/lhmtriet/SV_Empirical_Study}. These keywords were preprocessed with stemming and augmented with American/British spellings, space/hyphen to better handle various types of (mis-)spellings/plurality.

\noindent For instance, we considered the following variants: \emph{input(-)sanitization/sanit/sanitisation/ sanitis} for ``\emph{input sanitization}''. Similar to~\cite{haque2020challenges,le2020puminer}, we also performed exact matching for three-character keywords and subword matching for longer ones to reduce false positives. Subsequently, for each $\text{set}_{tag}$ (SO and SSE) obtained in the tag-based filtering step, we computed two content-based metrics (see Eq.~\eqref{eq:content_heuristics_ease21})~\cite{haque2020challenges,le2020puminer}: $kw\_count$ and $kw\_ratio$, denoting the count and appearance proportion of SV keywords in a post, respectively. $Kw\_count$ ensured diverse SV-related content in a post, while $kw\_ratio$ increased the confidence that these relevant words did not appear by chance.

\begin{myequation}\label{eq:content_heuristics_ease21}
kw\_coun{{t}_{p}}=\left| SV\_KW{{s}_{p}} \right|\,,\,\,\,kw\_rati{{o}_{p}}=\frac{\left| SV\_KW{{s}_{p}} \right|}{\left| Word{{s}_{p}} \right|}
\end{myequation}

\noindent where $\left| SV\_KW{{s}_{p}} \right|$ and $\left| Word{{s}_{p}} \right|$ are the numbers of SV keywords and total number of words in post $p$, respectively.

\noindent Based on the post content and human inspection, the thresholds $a_{\text{SO/SSE}}$ and $b_{\text{SO/SSE}}$ for filtering $\text{set}_{tag}$ (step 1) as well as selecting extra posts based on their content (step 2) were found, as given in Table~\ref{tab:content_thresholds_ease21}. Using these thresholds, we obtained $\text{set}_{tag}$ and $\text{set}_{content}$ of SV posts on SO and SSE, respectively, as shown in Fig.~\ref{fig:workflow_ease21}.

\begin{table}[t]
  \centering
  \caption[Content-based thresholds ($a_{\text{SO/SSE}}$ \& $b_{\text{SO/SSE}}$) for the two steps of the content-based filtering.]{Content-based thresholds ($a_{\text{SO/SSE}}$ \& $b_{\text{SO/SSE}}$) for the two steps of the content-based filtering as shown in Fig.~\ref{fig:workflow_ease21}.}
    \begin{tabular}{cP{2.3cm}P{2.3cm}P{2.3cm}P{2.3cm}}
    \hline
    \multirowcell{2}{\textbf{Thres-}\\ \textbf{hold}} & \multicolumn{2}{c}{\textbf{Stack Overflow (SO)}} & \multicolumn{2}{c}{\makecell{\textbf{Security}\\ \textbf{StackExchange (SSE)}}} \\
    \cline{2-5}
    & \textbf{Step 1} & \textbf{Step 2} & \textbf{Step 1} & \textbf{Step 2} \\
    \hline
    \textbf{a} & 1 & 3 & 2 & 3 \\
    \hline
    \textbf{b} & 0.011 & 0.017 & 0.017 & 0.025 \\
    \hline
    \end{tabular}
  \label{tab:content_thresholds_ease21}
\end{table}

\noindent \textbf{SV datasets and validation}.
As of June 2020, we retrieved 285,720 and 58,912 security posts from SO using PUMiner~\cite{le2020puminer} and SSE using Stack Exchange Data Explorer, respectively.
We then applied the tag-based and content-based filtering steps in Fig.~\ref{fig:workflow_ease21} and obtained \textbf{71,329 SV posts} (see Table~\ref{tab:sv_post_stats_ease21}) in total including 55,883 and 15,436 ones for $\text{set}_{tag}$ and $\text{set}_{content}$, respectively. We manually validated four different sets of SV posts, i.e., $\text{set}_{tag}$ and $\text{set}_{content}$ for SO and SSE, respectively. Specifically, we randomly sampled 385 posts (significant size~\cite{cochran2007sampling}) in each set for two researchers (i.e., the author of this thesis and a PhD student with three years of experience in Software Engineering and Cybersecurity) to examine independently.

\begin{table}[t]
  \centering
  \caption[The obtained software vulnerability posts using our tag-based and content-based filtering heuristics.]{The obtained SV posts using our tag-based and content-based filtering heuristics.}
    \begin{tabular}{lccc}
    \hline
     & \makecell{\textbf{Stack Over-}\\ \textbf{flow (SO)}} & \makecell{\textbf{Security Stack-}\\ \textbf{Exchange (SSE)}} & \textbf{SO + SSE}\\
    \hline
    \textbf{Set\textsubscript{\emph{tag}}} & 46,212 & 9,677 & 55,889 \\
    \hline
    \textbf{Set\textsubscript{\emph{content}}} & 12,660 & 2,780 & 15,440 \\
    \hline
    \textbf{Total} & 58,872 & 12,457 & 71,329 \\
    \hline
    \end{tabular}
  \label{tab:sv_post_stats_ease21}
\end{table}

For $\text{set}_{tag}$, we disagreed on 7/770 cases and only two posts were not related to SV. The main issue was still the incorrect tag assignment (e.g., SSE post 175264 was about dll injection but tagged with \textit{malware}\footnote{This post was short yet contained many SV keywords (e.g., ``\textit{injection}'' and ``\textit{hijack}''), resulting in high $kw\_count$ and $kw\_ratio$ of the content-based filtering.}), though this issue had been significantly reduced by the content-based filtering. For $\text{set}_{content}$, the relevance of the posts was very high as there was no discrepant case.

Our SV dataset was only 20\% overlapping with the existing security dataset~\cite{yang2016security}, implying that there were significant differences in the nature of the two studies. Note that we followed the settings in~\cite{yang2016security} to retrieve the updated security posts from the same SO data we used in our study. We also reported the top tags of SV posts (see Table~\ref{tab:top_tags_ease21}) and compared them with the ones of security posts~\cite{yang2016security} and a subset of all the posts containing an equal number of posts to the SV posts on SO and SSE. SV posts were associated with many SV-related tags (e.g., \emph{memory-leaks}, \emph{malware}, \emph{segmentation-fault}, \emph{xss}, \emph{exploit} and \emph{penetration-test}). Conversely, security posts were tagged with general terms that may not explicitly discuss security flaws such as \emph{encryption}, \emph{authentication} and \emph{passwords}. The tags of general posts were mostly programming languages on SO and general security terms on SSE. These findings highlight the importance of obtaining SV-specific posts instead of reusing the security posts to study the support of Q\&A sites for SV-related discussions.

\begin{table}[t]
  \centering
  \caption[Top-5 tags of software vulnerability, security and general posts on Stack Overflow and Security StackExchange.]{Top-5 tags of SV, security and general posts on SO and SSE (in parentheses).}
    \begin{tabular}{cccc}
    \hline
    \textbf{No.} & \multicolumn{1}{c}{\textbf{SV posts}} & \multicolumn{1}{c}{\textbf{Security posts}} & \multicolumn{1}{c}{\textbf{General posts}} \\
    \hline
    1     & \makecell{memory-leaks\\ (malware)} & \makecell{security\\ (encryption)} & \makecell{javascript\\ (encryption)} \\
    \hline
    2     & \makecell{segmentation-\\fault (web-appli-\\cation)} & encryption (tls) & java (tls) \\
    \hline
    3     & php (xss) & \makecell{php (authentication)} & \makecell{python\\(authentication)} \\
    \hline
    4     & c (exploit) & java (passwords) & \makecell{c\#\\(passwords)} \\
    \hline
    5     & \makecell{security\\(penetration-test)} & \makecell{cryptography\\(web-application)} & \makecell{php\\ (certificates)} \\
    \hline
    \end{tabular}
  \label{tab:top_tags_ease21}
\end{table}

\subsection{Topic Modeling with LDA}
\label{subsec:lda_ease21}

Following the common practice of the existing work (e.g.,~\cite{yang2016security,barua2014developers,bagherzadeh2019going}), we extracted the topics of the identified SV-related posts on both SO and SSE using Latent Dirichlet Allocation (LDA)~\cite{blei2003latent}.

\noindent \textbf{Preprocessing of SV posts}.
Following the previous practices of~\cite{yang2016security,ahmed2018concurrency}, we first removed the HTML tags and code snippets in each post as these elements were not informative for topic modeling.
We also converted the text to lowercase, removed punctuations, and then eliminated stop words and performed stemming (reducing a word to its root form) to avoid irrelevant and multi-form words.

\noindent \textbf{Topic modeling with LDA}. We applied LDA to the title, question body and all answers of each Q\&A post.
Regarding the number of topics ($k$) of LDA, we examined an inclusive range from 2 to 80, with an increment of one topic at a time. As suggested in~\mbox{\cite{barua2014developers,rosen2016mobile,ahmed2018concurrency,bagherzadeh2019going}}, alongside $k$, we also tried different values of $\alpha$ (1/$k$ or 50/$k$) and $\beta$ (0.01 or same as $\alpha$) hyperparameters to optimize the performance of LDA. $\alpha$ controls the sparsity of the topic-distribution per post and $\beta$ determines the sparsity of the word-distribution per topic. For each tuple of ($k$, $\alpha$ and $\beta$), we ran LDA with 1,000 iterations, then evaluated the coherence metric~\mbox{\cite{roder2015exploring}} of the identified topics. Coherence metric has been recommended by many previous studies (e.g.,~\cite{abdellatif2020challenges,zahedi2020mining}) to select the optimal number of LDA topics since it usually highly correlates with human understandability. Topic coherence is the average correlation between pairs of words that appear in the same topic. The higher value of the coherence metric means the more coherent content of the posts within the same topic.
To avoid insignificant topics like~\cite{barua2014developers}, we only considered topics with a probability of at least 0.1 in a post.
We manually read the top-20 most frequent words and 15 random posts of each topic per site (SO/SSE) obtained by the trained LDA models to label the name of that topic as done in~\cite{bagherzadeh2019going,ahmed2018concurrency}.
The LDA model with the most relevant set of topics would be used for answering the four RQs.

\section{Results}
\label{sec:results_ease21}

\subsection{\textbf{RQ1}: What are SV Discussion Topics on Q\&A Sites?}
\label{subsec:rq1_results_ease21}

Following the procedure in section~\ref{subsec:lda_ease21}, we identified \textit{13} SV topics (see Table~\ref{tab:topic_list_ease21}) on SO and SSE using the optimal LDA model with $\alpha=\beta=0.08$. We found LDA models having from 11 to 17 topics produced similar coherence metrics. Three of the authors manually examined these cases, as in~\cite{abdellatif2020challenges}. Duplicate and/or platform-specific topics (e.g., web and mobile) appeared from 14 topics, making the taxonomy less generalizable. 11 and 12 topics also had high-level topics (e.g., combining XSS and CSRF). Thus, 13 was chosen as the optimal number of SV topics. All the terms/posts of each SV topic can be found at \url{https://github.com/lhmtriet/SV_Empirical_Study}. We describe each topic hereafter with example SO/SSE posts. We examined 15 random posts per topic per site. If we identified some common patterns of discussions (e.g., attack vectors or assets) on a site, we would extract another 15 random posts of the respective site to confirm our observations. If a pattern was no longer evident in the latter 15 posts, we would not report it.

\begin{table}[t]
  \centering
  \caption[Software vulnerability topics on Stack Overflow and Security StackExchange identified by Latent Dirichlet Allocation along with their proportions and trends over time.]{SV topics on SO and SSE identified by LDA along with their proportions and trends over time. \textbf{Notes}: The topic proportions on SSE are in parenthesis. The trends of SO are the top solid sparklines, while the trends of SSE are the bottom dashed sparklines. Unit of proportion: \%.}
    \begin{tabular}{lcc}
    \hline
    \textbf{Topic Name} & \textbf{Proportion} & \textbf{Trend} \\
    \hline
    Malwares (T1) & 1.39 (8.18) & \parbox[c]{0.7cm}{\includegraphics[width=0.8cm,keepaspectratio]{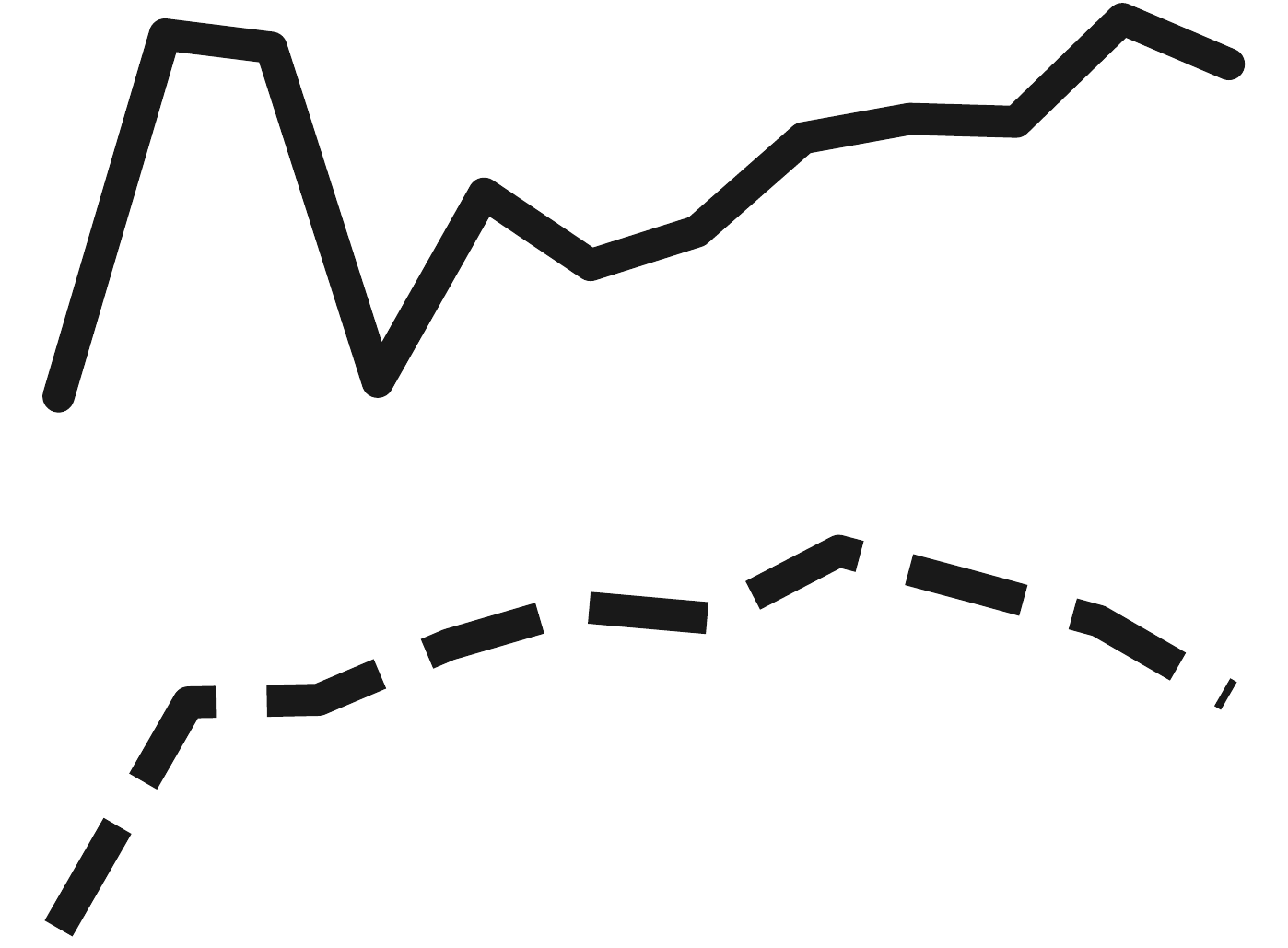}} \\
    \hline
    SQL Injection (T2) & 11.0 (4.17) & \parbox[c]{0.7cm}{\includegraphics[width=0.8cm,keepaspectratio]{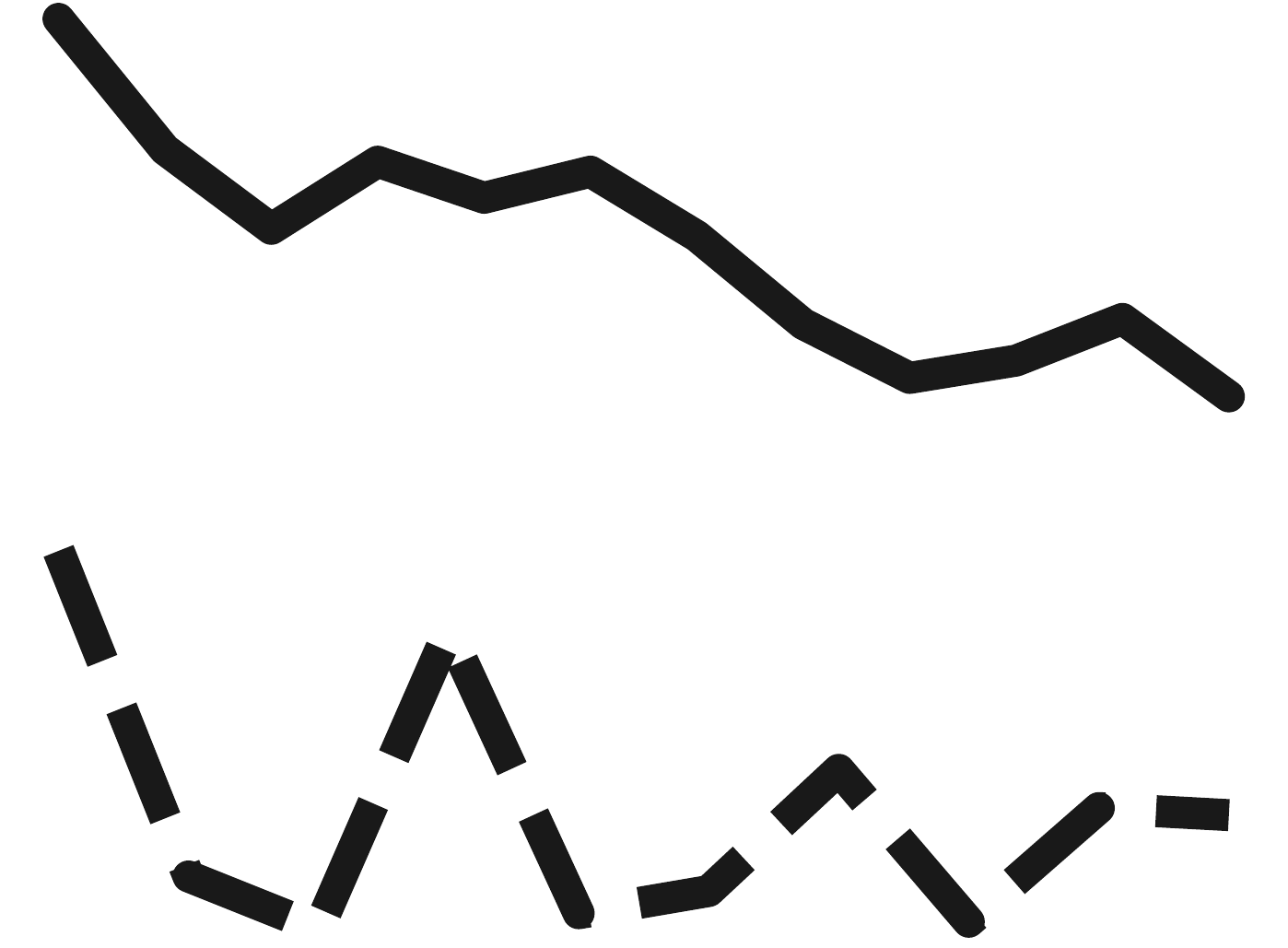}} \\
    \hline
    Vulnerability Scanning Tools (T3) & 5.42 (3.15) & \parbox[c]{0.7cm}{\includegraphics[width=0.8cm,keepaspectratio]{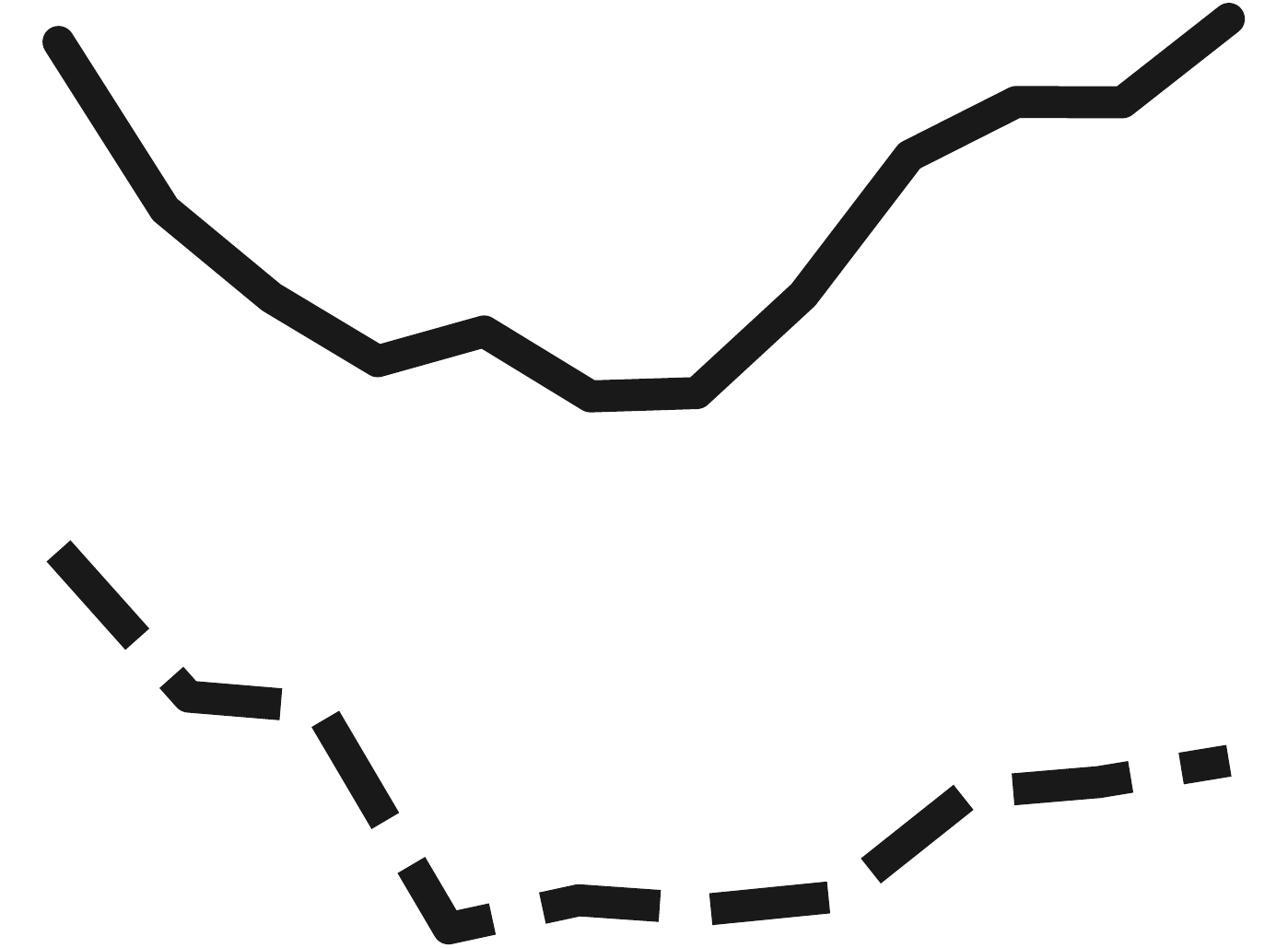}}\\
    \hline
    Cross-site Request Forgery (CSRF) (T4) & 9.49 (5.09) & \parbox[c]{0.7cm}{\includegraphics[width=0.8cm,keepaspectratio]{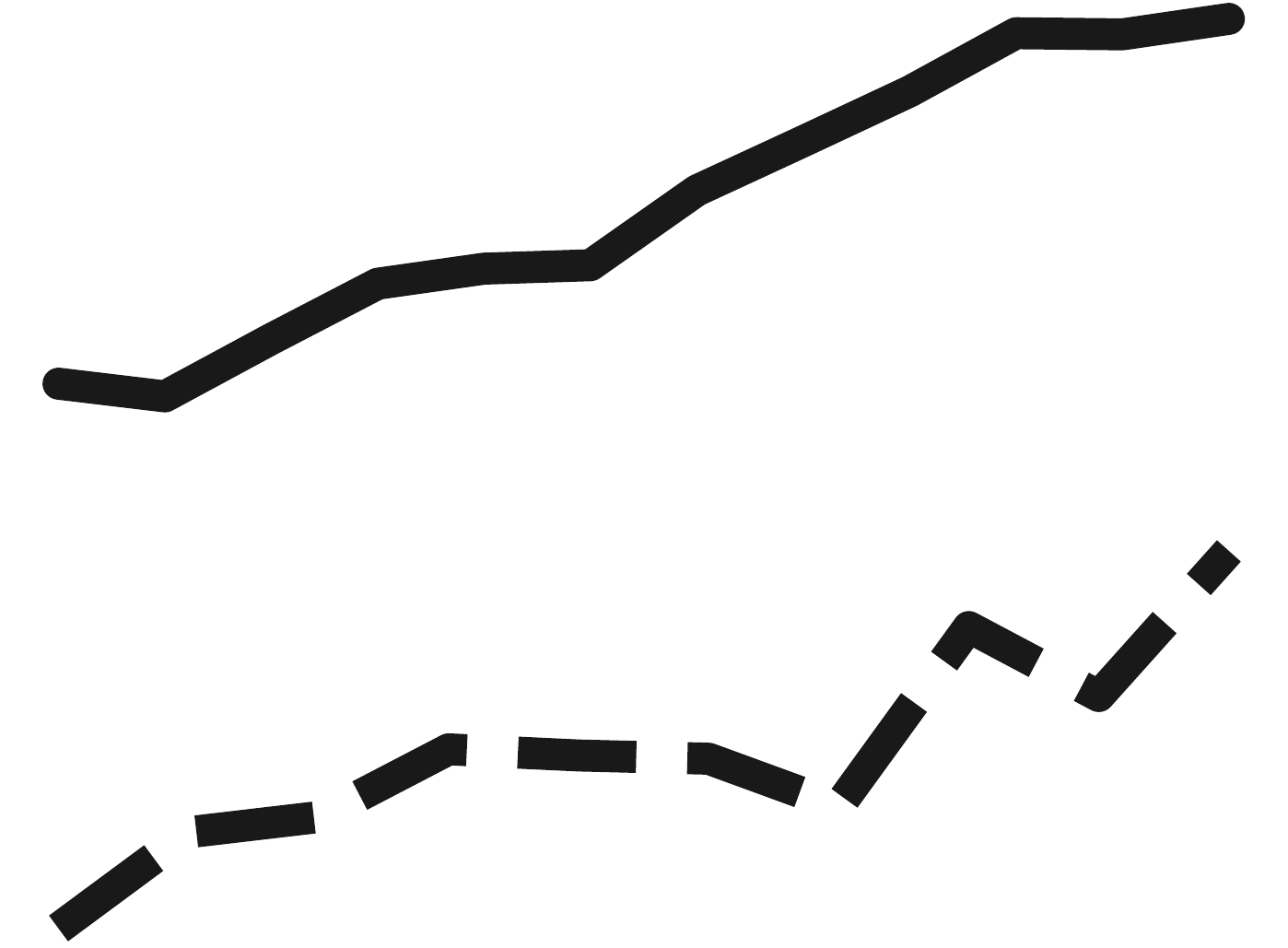}} \\
    \hline
    File-related Vulnerabilities (T5) & 2.88 (3.24) & \parbox[c]{0.7cm}{\includegraphics[width=0.8cm,keepaspectratio]{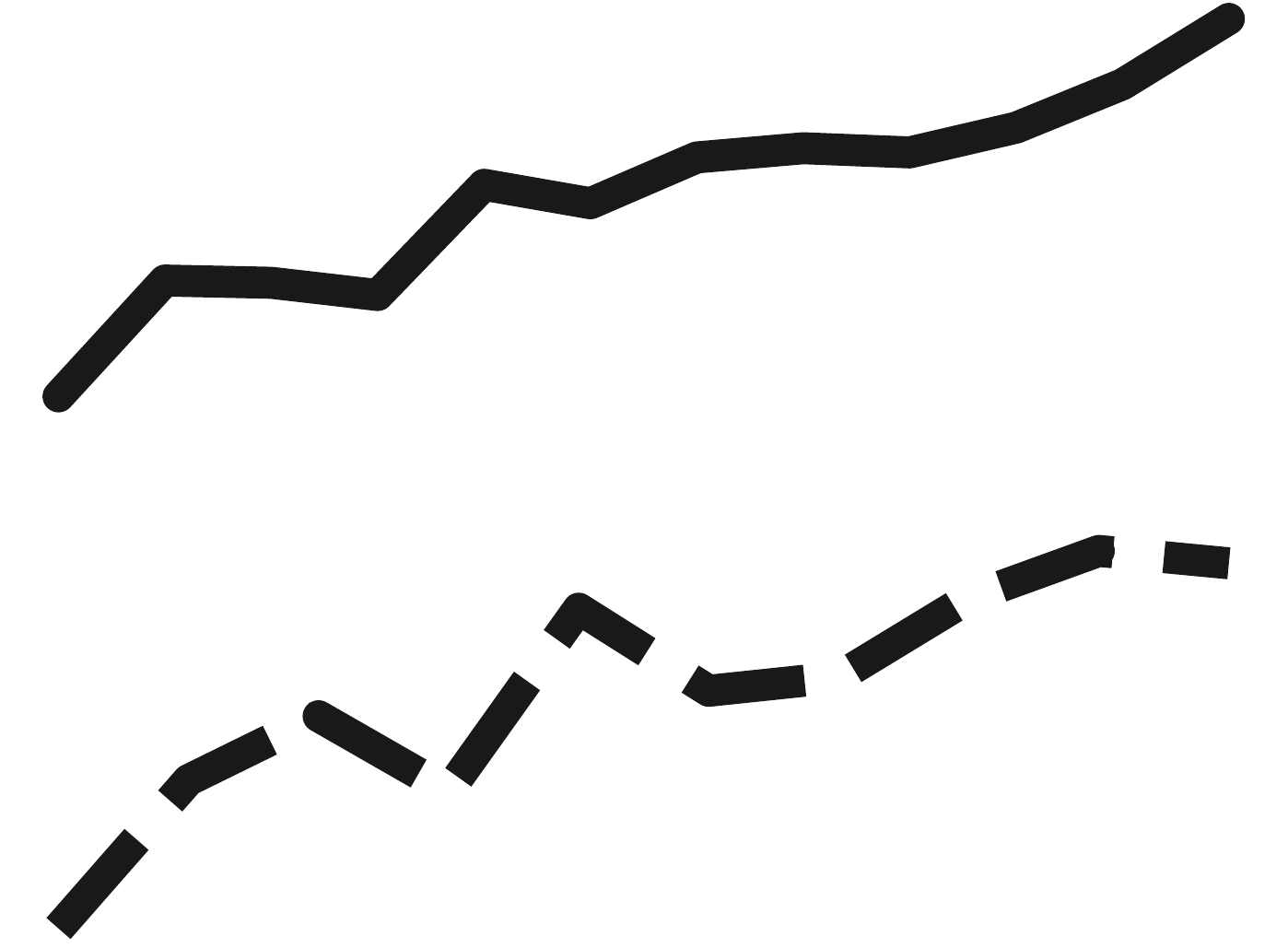}} \\
    \hline
    Synchronization Errors (T6) & 3.79 (0.47) & \parbox[c]{0.7cm}{\includegraphics[width=0.8cm,keepaspectratio]{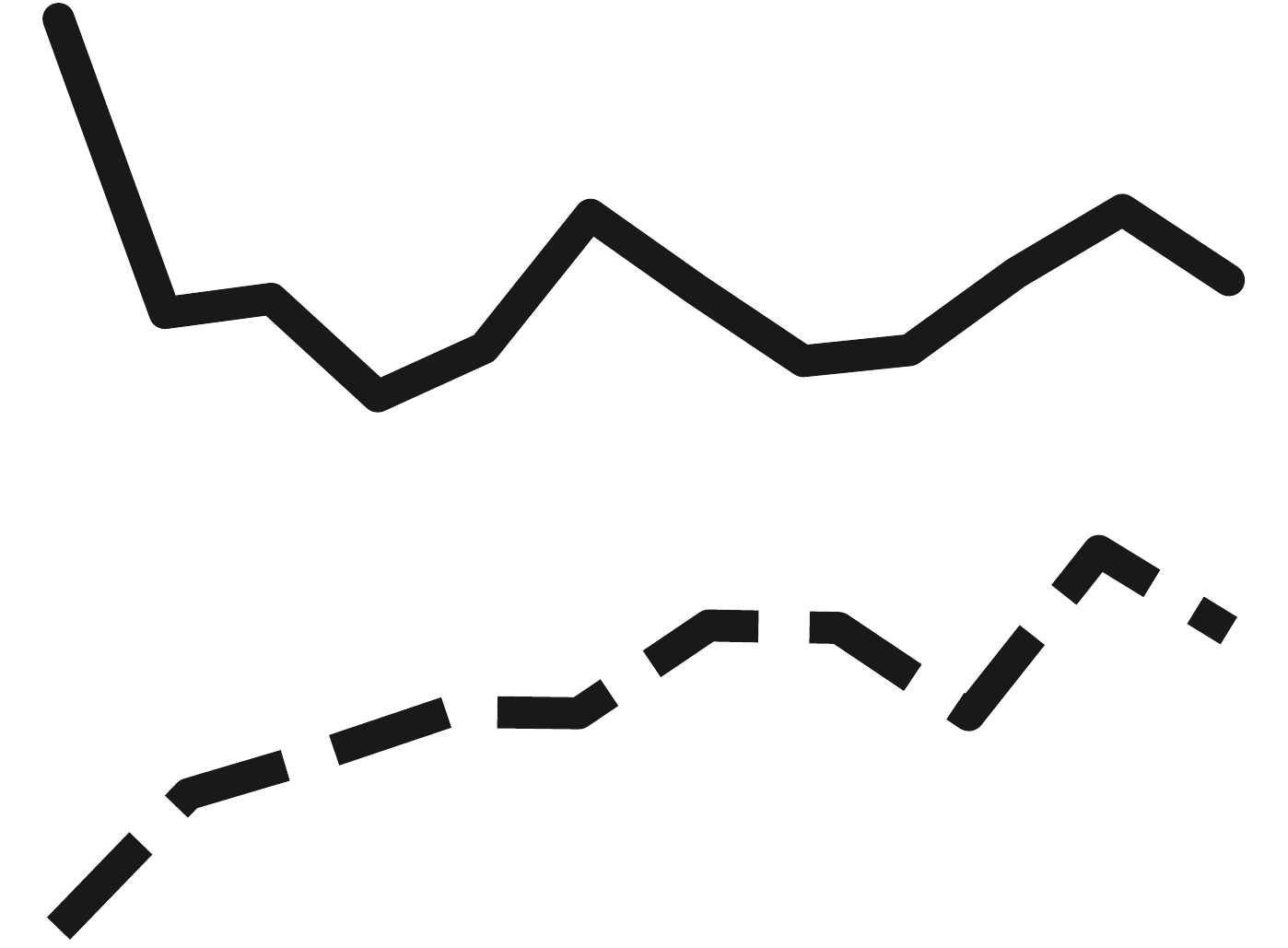}} \\
    \hline
    Encryption Errors (T7) & 1.82 (7.81) & \parbox[c]{0.7cm}{\includegraphics[width=0.8cm,keepaspectratio]{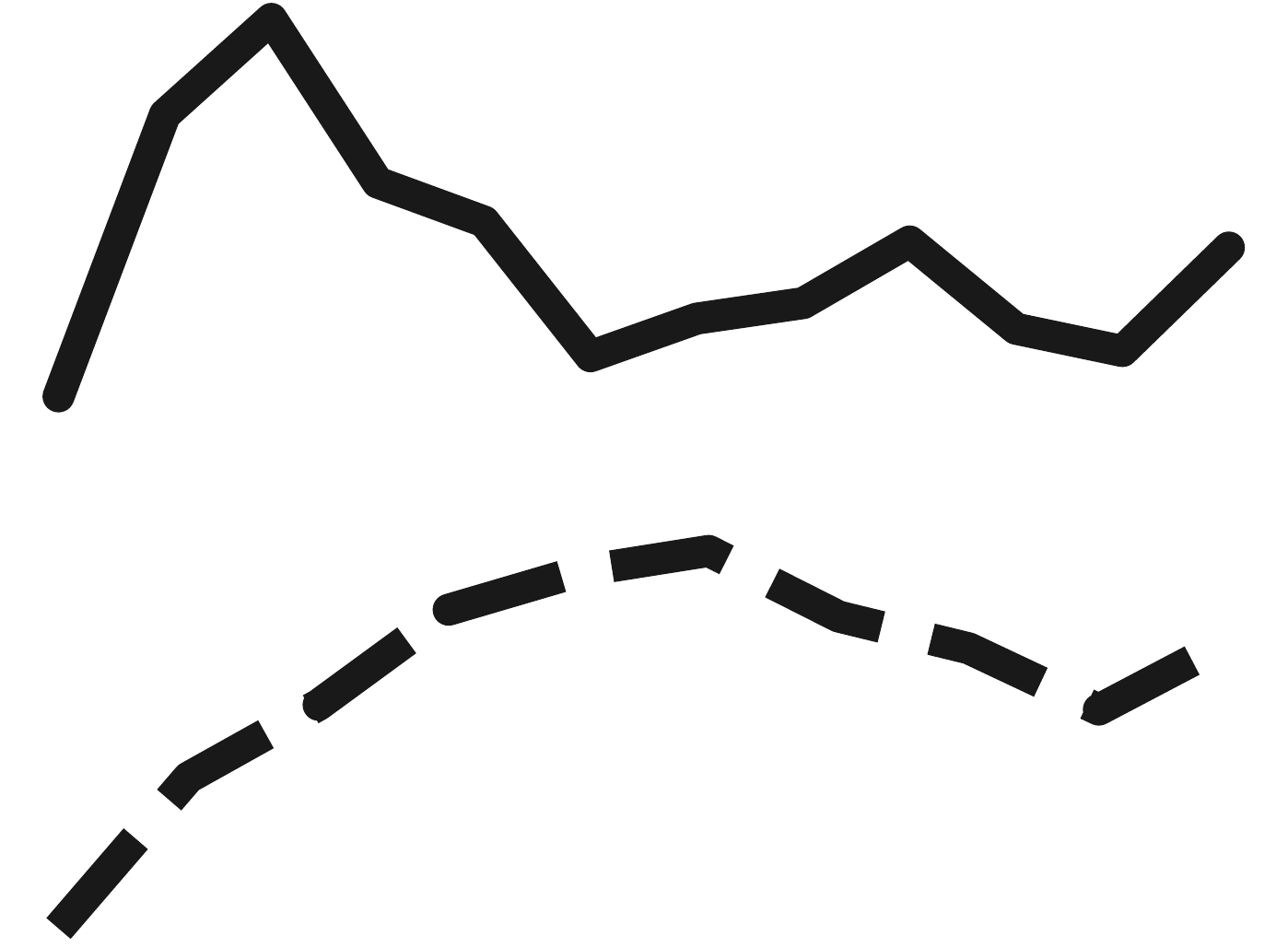}} \\
    \hline
    Resource Leaks (T8) & 10.6 (0.42) & \parbox[c]{0.7cm}{\includegraphics[width=0.8cm,keepaspectratio]{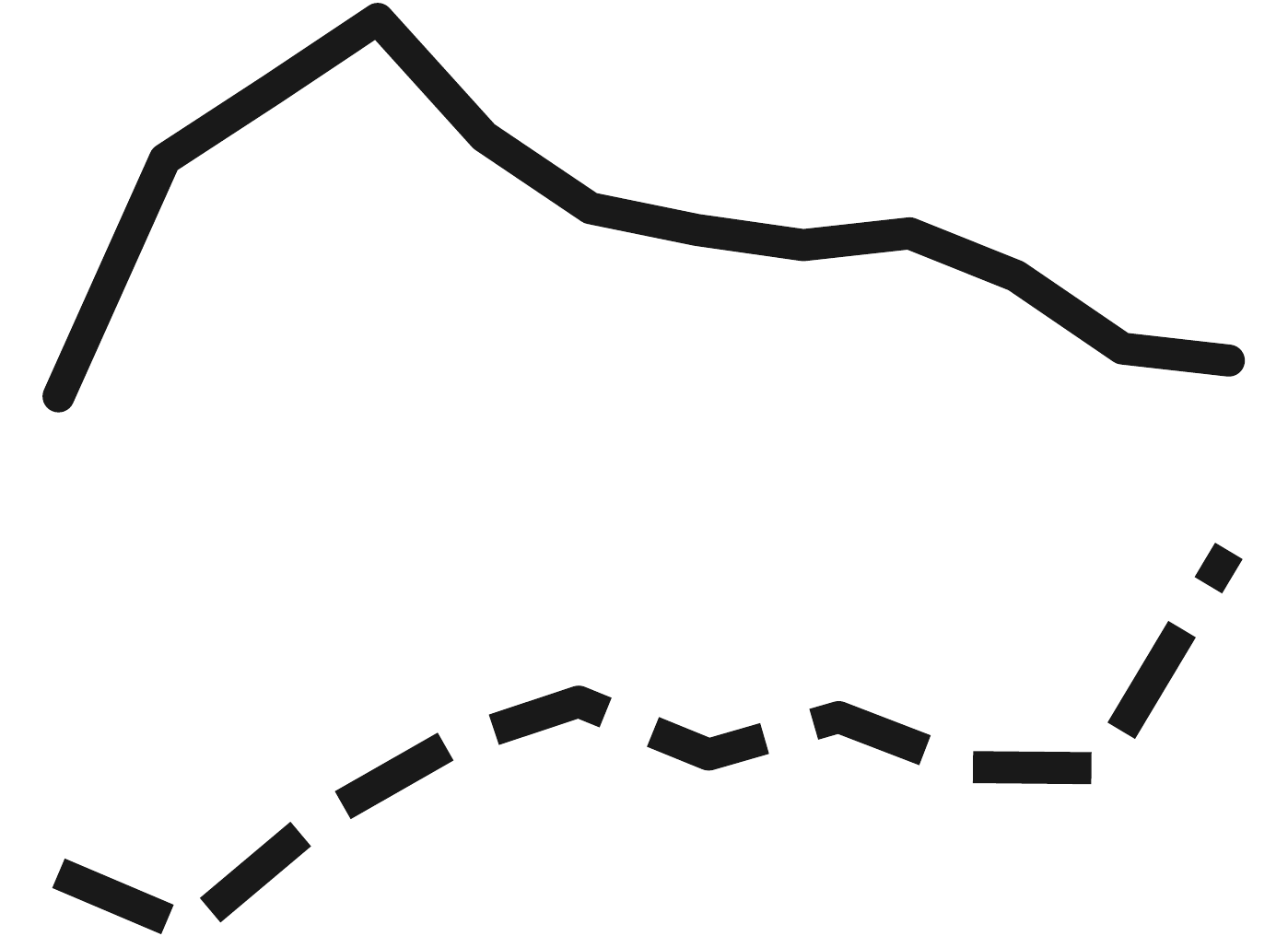}} \\
    \hline
    Network Attacks (T9) & 1.37 (8.79) & \parbox[c]{0.7cm}{\includegraphics[width=0.8cm,keepaspectratio]{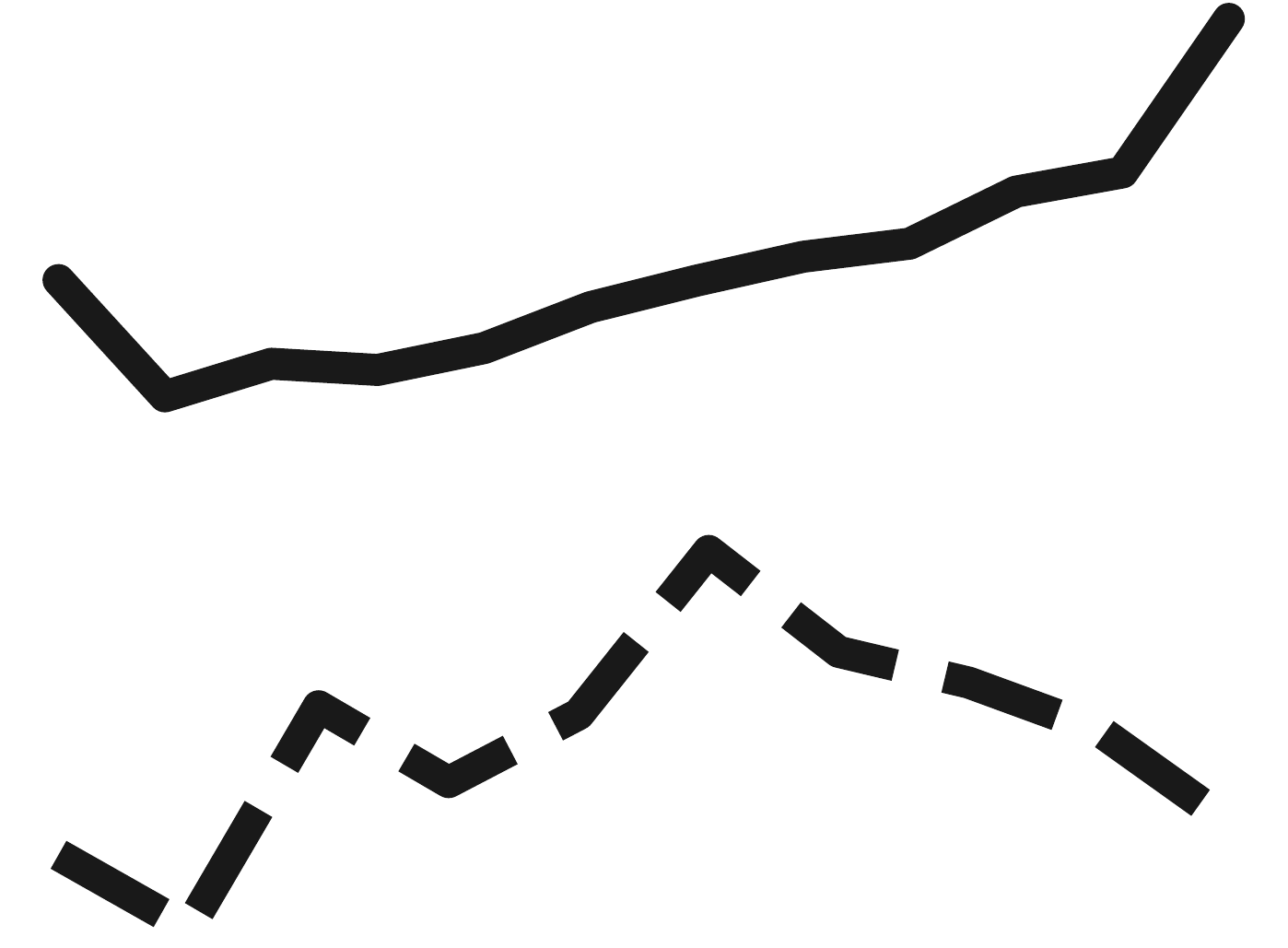}} \\
    \hline
    Memory Allocation Errors (T10) & \textbf{21.6} (2.82) & \parbox[c]{0.7cm}{\includegraphics[width=0.8cm,keepaspectratio]{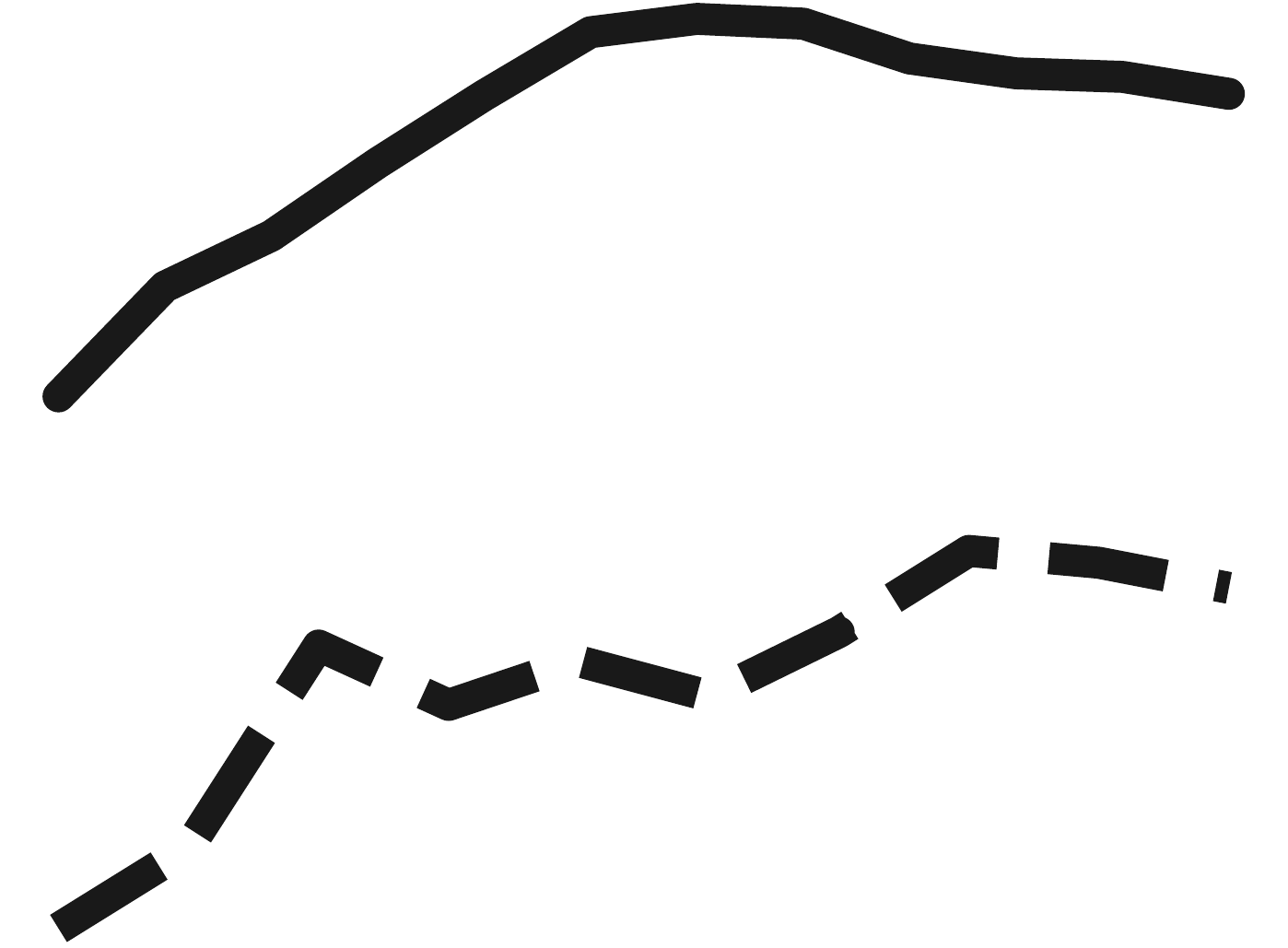}} \\
    \hline
    Cross-site Scripting (XSS) (T11) & 7.73 (8.09) & \parbox[c]{0.7cm}{\includegraphics[width=0.8cm,keepaspectratio]{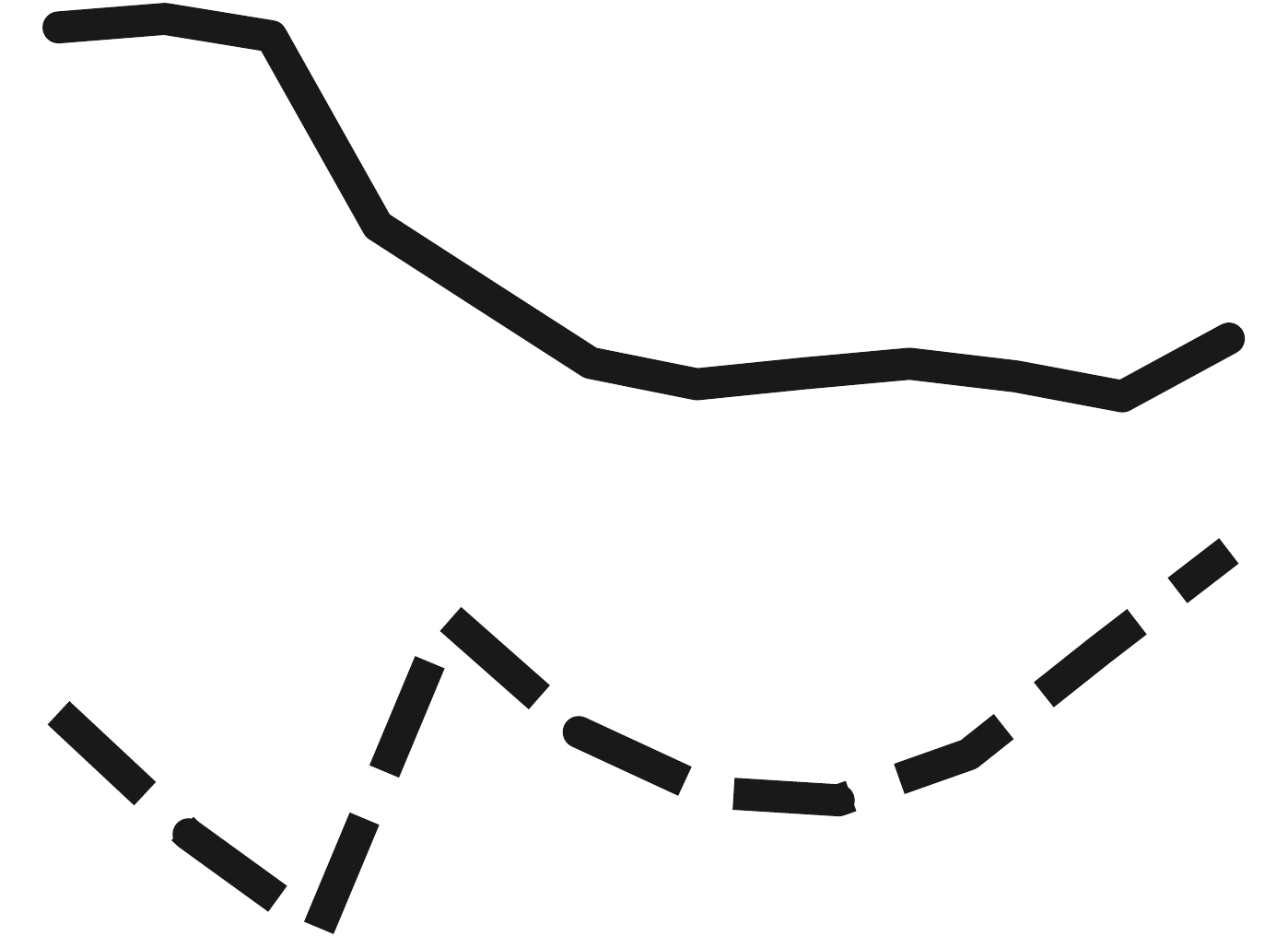}}  \\
    \hline
    Vulnerability Theory (T12) & 10.7 (\textbf{33.7}) & \parbox[c]{0.7cm}{\includegraphics[width=0.8cm,keepaspectratio]{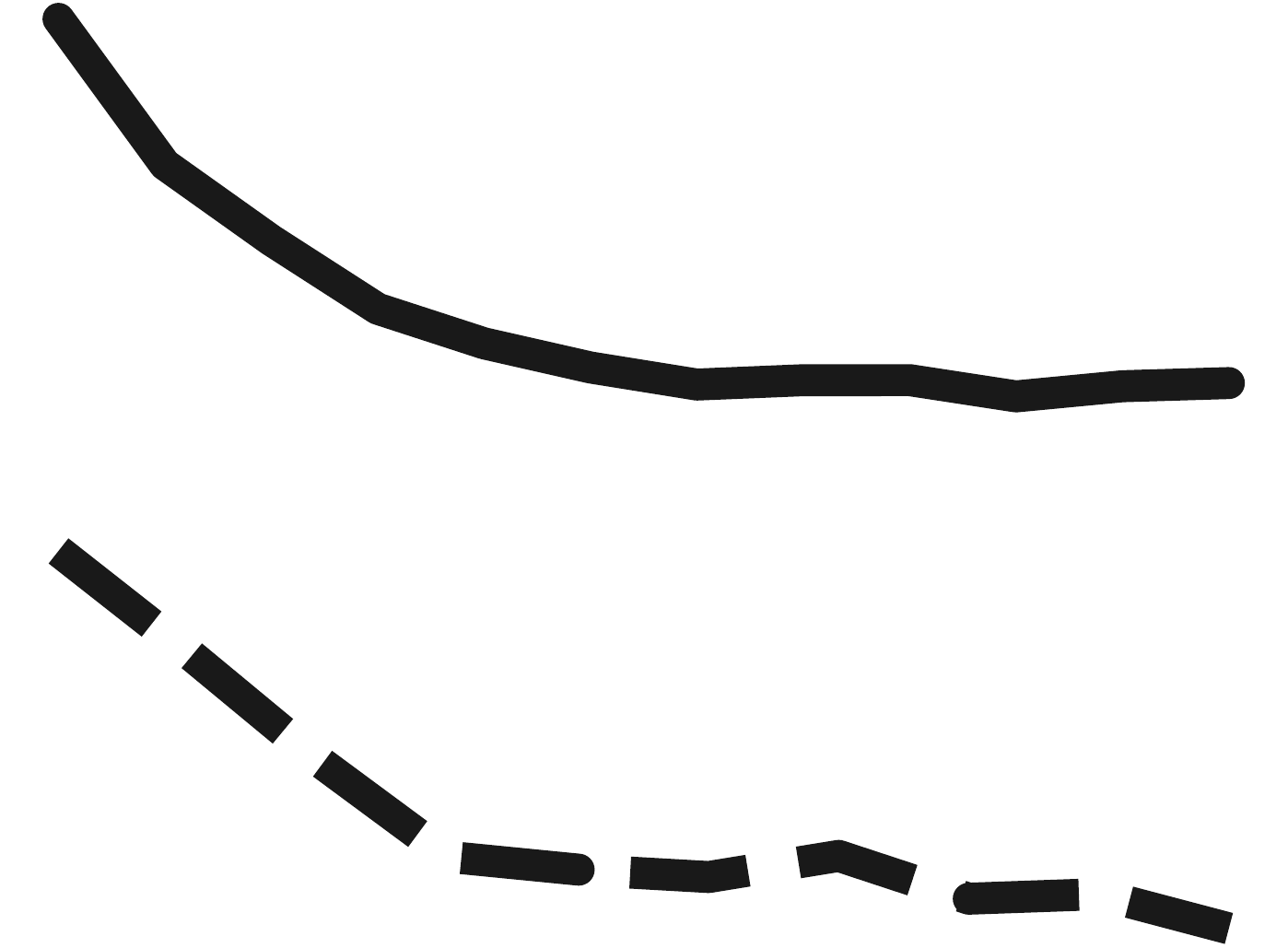}} \\
    \hline
    Brute-force/Timing Attacks (T13) & 1.08 (1.28) & \parbox[c]{0.7cm}{\includegraphics[width=0.8cm,keepaspectratio]{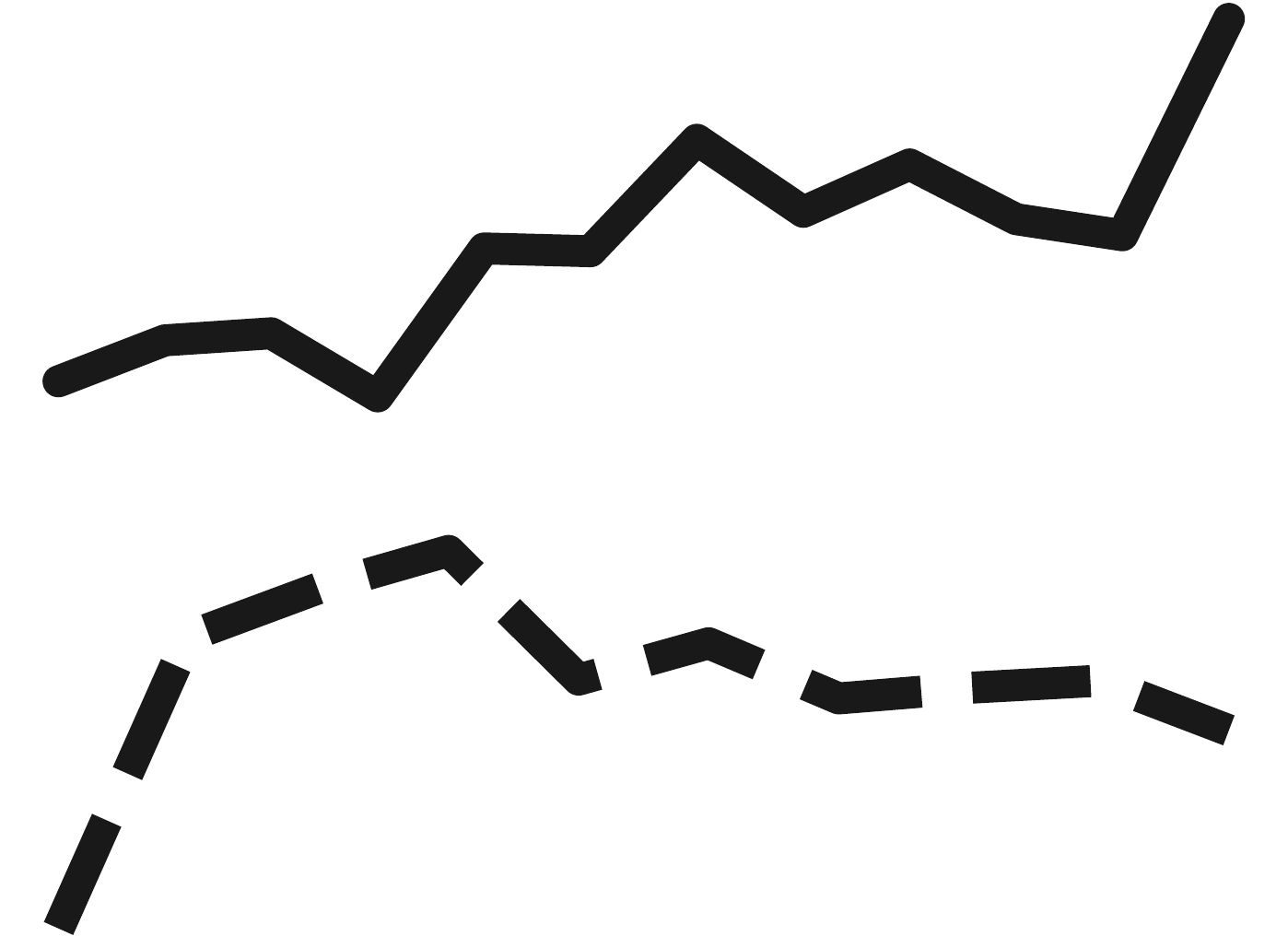}} \\
    \hline
    \end{tabular}
  \label{tab:topic_list_ease21}
\end{table}

\textbf{Malwares (T1)}. This topic referred to the detection and removal of malicious software.
T1 posts on SO were usually about malwares in content management systems such as Wordpress or Joomla (e.g., post 16397854: ``\textit{How to remove wp-stats malware in wordpress}'' or post 11464297: ``\textit{How to remove .htaccess virus}'). In contrast, SSE often discussed malwares/viruses coming from storage devices such as SSD (e.g., post 227115: ``\textit{Can viruses of one ssd transfer to another ssd?}'') or USB (e.g., post 173804: ``\textit{Can Windows 10 bootable USB drive get infected while trying to reinstall Windows?}'').

\textbf{SQL Injection (T2)}. This topic concerned tactics to properly sanitize malicious inputs that could modify SQL commands and pose threats (e.g., stealing or changing data) to databases in various programming languages (e.g., PHP, Java, C\#). A commonly discussed tactic was to use prepared statements, which also helped increase the efficiency of query processing. For example, developers asked questions like ``\textit{How to parameterize complex oledb queries?}'' (SO post 9650292) or ``\textit{How to make this code safe from SQL injection and use bind parameters}'' (SSE post 138385).

\textbf{Vulnerability Scanning Tools (T3)}. This topic was about issues related to tools for automated detection/assessment of potential SVs in an application. Discussions of T3 mentioned different tools, and OWASP ZAP was a commonly discussed one. For example, post 62570277 on SO discussed ``\textit{Jenkins-zap installation failed}'', while post 126851 on SSE asked ``\textit{How do I turn off automated testing in OWASP ZAP?}'' One possible explanation is that OWASP ZAP is a free and easy-to-use tool for detecting and assessing SVs that appear in the well-known top-10 OWASP list for web applications.

\textbf{Cross-site Request Forgery (CSRF) (T4)}. This topic contained discussions on proper setup and configuration of web application frameworks to prevent CSRF SVs. These SVs could be exploited to send requests to perform unauthorized actions from an end-user that a web application trusts. Discussions covered various issues in implementing different CSRF prevention techniques recommended by OWASP.\footnote{\mbox{\label{fn:csrf_owasp}}\url{https://cheatsheetseries.owasp.org/cheatsheets/Cross-Site\_Request\_Forgery\_Prevention\_Cheat\_Sheet.html}}
Some commonly discussed techniques were anti-CSRF token (e.g., SO post 59664094: ``\textit{Why Laravel 4 CSRF token is not working?}''), double submit cookie (e.g., SSE post 203996: ``\textit{What is double submit cookie? And how it is used in the prevention of CSRF attack?}''), and SameSite cookie attribute (e.g., SO post 41841880: ``\textit{What is the benefit of blocking cookie for clicked link? (SameSite=strict)}'').

\textbf{File-related Vulnerabilities (T5)}. Discussions of this topic were about SVs in files that could be exploited to gain unauthorized access. The common SV types were Path/Directory Traversal via Symlink (e.g., SSE post 165860: ``\textit{Symlink file name - possible exploit?}''), XML External Entity (XXE) Injection (e.g., SO post 51860873: ``\textit{Is SAXParserFactory susceptible to XXE attacks?}''), and Unrestricted File Upload (e.g., SSE post 111935: ``\textit{Exploiting a PHP server with a .jpg file upload}''). These SVs usually occurred for Linux-based systems, suggesting that Linux is more popular for servers.

\textbf{Synchronization Errors (T6)}. This topic involved SVs produced through errors in synchronization logic (usually related to threads), which could slow down system performance. Some common SV types being discussed were deadlocks (e.g., SO post 38960765: ``\textit{How to avoid dead lock due to multiple oledb command for same table in ssis}'') and race conditions (e.g., SSE post 163209: ``\textit{What's the meaning of `the some sort of race condition' here?}'').

\textbf{Encryption Errors (T7)}. This topic included cryptographic issues leading to falsified authentication or retrieval of sensitive data, e.g., Man-in-the-middle (MITM) attack. Many posts discussed public/private keys for encryption/decryption, especially using SSL/TLS certificates to defend against MITM attacks (attempts to steal information sent between browsers and servers). Some example discussions are post 23406005 on SO (``\textit{Man In Middle Attack for HTTPS}'') or post 105773 on SSE (``\textit{How is it that SSL/TLS is so secure against password stealing?}''). This may imply that many developers are still not familiar with these certificates in practice.

\textbf{Resource Leaks (T8)}. This topic considered SVs arising from improper releases of unused memory which could deplete resources and decrease system performance. Many discussions of T8 were about memory leaks in mobile app development. Issues were usually related to Android (e.g., SO post 58180755: ``\textit{Deal with Activity Destroying and Memory leaks in Android}'') or IOS (e.g., SO post 47564784: ``\textit{iOS dismissing a view controller doesn't release memory}'').

\textbf{Network Attacks (T9)}. This topic discussed attacks carried out over an online computer network, e.g., Denial of Service (DoS) and IP/ARP Spoofing, and potential mitigations. These network attacks directly affected the availability of a system. For instance, SSE post 86440 discussed ``\textit{VPN protection against DDoS}'' or SO post 31659468 asked ``\textit{How to prevent ARP spoofing attack in college?}''.

\textbf{Memory Allocation Errors (T10)}. T10 and T8 were both related to memory issues, but T10 did not consider memory release. Rather, this topic more focused on SVs caused by accessing or using memory outside of what allocated that could be exploited to access restricted memory location or crash an application. In this topic, segmentation faults (e.g., SO post 31260018: ``\textit{Segmentation fault removal duplicate elements in unsorted linked list}'') and buffer overflows (e.g., SSE post 190714: ``\textit{buffer overflow 64 bit issue}'') were commonly discussed by developers.

\textbf{Cross-site Scripting (XSS) (T11)}. This topic mentioned tactics to properly neutralize user inputs to a web page to prevent XSS attacks. These attacks could exploit users' trust in web servers/pages to trick them to execute malicious scripts and perform unwanted actions. XSS (T11) and CSRF (T4) are both client-side SVs, but XSS is more dangerous since it can bypass all countermeasures of T4.\mbox{\textsuperscript{\mbox{\ref{fn:csrf_owasp}}}} On SO and SSE, discussions covered all three types of XSS: (\textit{i}) reflected XSS (e.g., SSE post 57268: ``\textit{How does the anchor tag (<a>) let you do an Reflected XSS?}''), (\textit{ii}) stored/persistent XSS (e.g., SO post 54771897: ``\textit{How to defend against stored XSS inside a JSP attribute value in a form}''), and (\textit{iii}) DOM-based XSS (e.g., SO post 44673283: ``\textit{DOM XSS detection using javascript(source and sink detection)}'').

\textbf{Vulnerability Theory (T12)}. This topic focused on theoretical/social aspects and best practices in the SV life cycle. Many posts compared different SV-related terminologies, e.g., SSE post 103018 asked about ``\textit{In CIA triad of information security, what's the difference between confidentiality and availability?}'' or SO post 402936 discussed ``\textit{Bugs versus vulnerabilities?}''. Several other posts asked about internal SV reporting process (e.g., SO post 3018198: ``\textit{How best to present a security vulnerability to a web development team in your own company?}'') or public SV disclosure policy (e.g., SSE post: ``\textit{How to properly disclose a security vulnerability anonymously?}'').

\textbf{Brute-force/Timing Attacks (T13)}. T13 and T7 both exploited cryptographic flaws, but these two topics used different attack vectors/methods. T7 focused on MITM attacks, while T13 was about attacks making excessive attempts or capturing the timing of a process to gain unauthorized access. Some example posts of T13 are SO post 3009988 (``\textit{What's the big deal with brute force on hashes like MD5}'') or SSE post 9192 (``\textit{Timing attacks on password hashes}'').

\noindent \textbf{Proportion and Evolution of SV Topics}. We analyzed the proportion (share metric in Eq.~\eqref{eq:share_ease21}) and the evolution trend of SV topics from their inception on SO (2008) and SSE (2010) to 2020 (see Table~\ref{tab:topic_list_ease21}). The topic patterns and dynamics of SO were different from those of SSE. Specifically, Memory Allocation Errors (T10) had the greatest number of posts on SO, while Vulnerability Theory (T12) had the largest proportion on SSE. Apart from XSS (T11) and Brute-force/Timing Attacks (T13), topics with many posts in one source were not common in the other source. Moreover, we discovered three consistent topic trends on both SO and SSE: Malwares (T1) ($\nearrow$), CSRF (T4) ($\nearrow$), File-related SVs (T5) ($\nearrow$) and Vulnerability Theory (T12) ($\searrow$). Among them, CSRF had the fastest changing pace.
These trends were confirmed significant with $p$-values $<$ 0.01 using Mann-Kendall non-parametric trend test~\cite{mann1945nonparametric}.

\begin{figure}[t]
    \centering
    \includegraphics[width=\columnwidth,keepaspectratio]{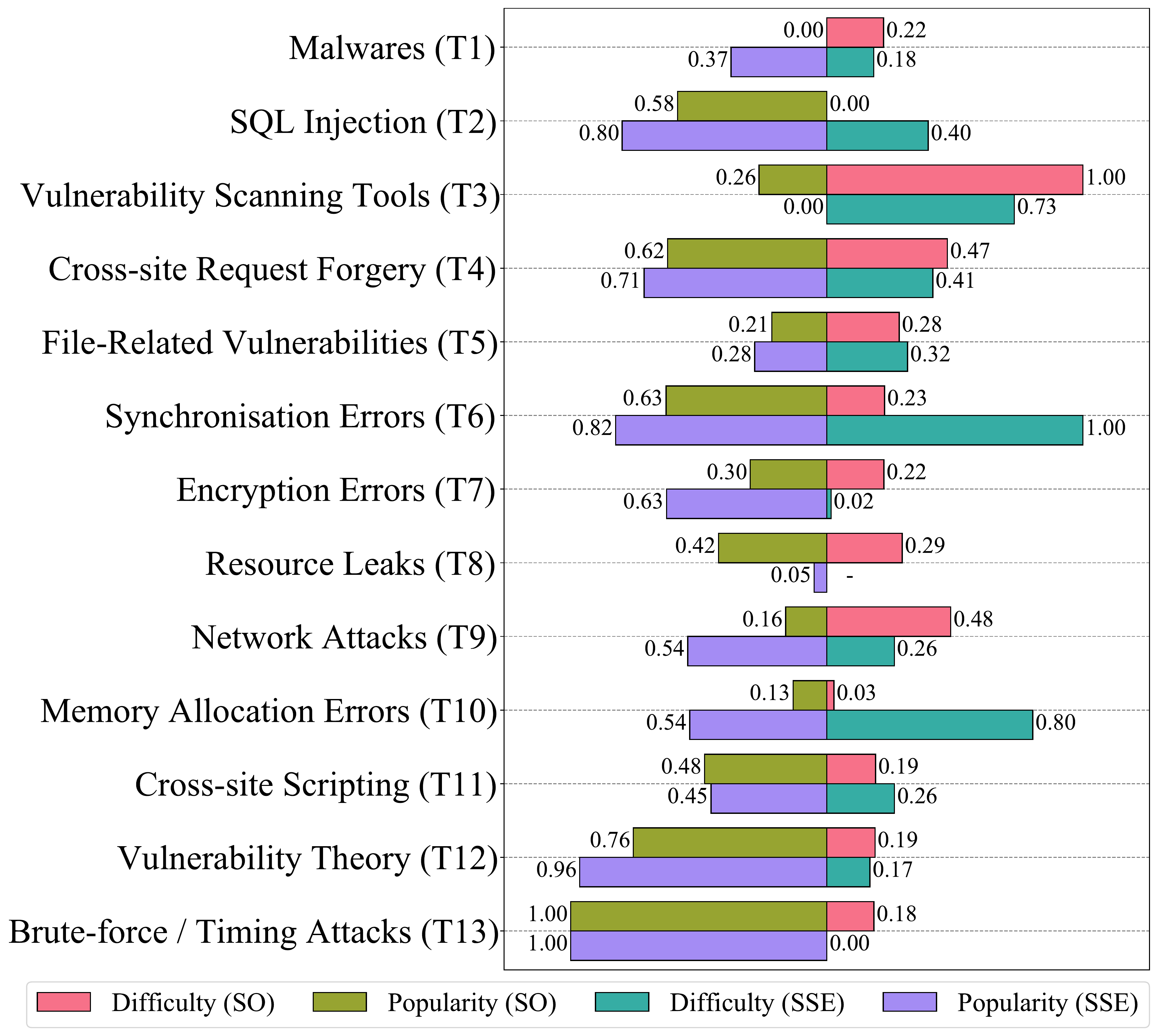}
    \caption[Popularity and difficulty of 13 software vulnerability topics on Stack Overflow and Security StackExchange.]{Popularity and difficulty of 13 SV topics on SO and SSE. \textbf{Notes}: The values were normalized by the max and min values of each category. Difficulty of T8 on SSE was excluded since it did not have any accepted answer.}
    \label{fig:pop_diff_ease21}
\end{figure}

\subsection{\textbf{RQ2}: What are the Popular and Difficult SV Topics on Q\&A Sites?}
\label{subsec:rq2_results_ease21}

As shown in Fig.~\ref{fig:pop_diff_ease21}, the popularity and difficulty of 13 identified SV topics were different between SO and SSE. For conciseness, we only report the geometric means of the popularity and difficulty metrics in this section. The complete values of individual metrics (see section~\mbox{\ref{rq2_method}}) can be found at \url{https://github.com/lhmtriet/SV_Empirical_Study}.

\textbf{Topic Popularity}. Brute-force/Timing attacks (T13) and Vulnerability Theory (T12) were the top-2 most popular topics. Despite being the most popular topic, T13 only had 1.1\% and 1.3\% posts on SO and SSE, respectively. Conversely, Memory Allocation Errors (T10) had the most posts on SO (RQ2), but T10 was only the second least popular topic. We found no significant correlation between the topic popularity and share metric with Kendall's Tau correlation test~\cite{knight1966computer} at 95\% confidence level. These findings suggest that the share metric does not necessarily reflect the topic popularity since it does not consider users' activities on Q\&A sites.

\textbf{Topic Difficulty}. The most difficult topics were not popular or associated with many posts, i.e., Vulnerability Scanning Tools (T3) and Network Attacks (T9) on SO as well as T3, Synchronization Errors (T6) and Memory Allocation Errors (T10) on SSE. The high difficulty of T3 on both sites was potentially caused by the low familiarity with a wide array of vendors and tools available for SV detection and assessment~\cite{kritikos2019survey}. Some topics with many posts (high share metric) like Memory Allocation Errors (T10) and SQL Injection (T2) were the two easiest ones on SO despite being significantly more difficult on SSE. On the contrary, Malwares (T1) and Network Attacks (T9) were more popular yet easier on SSE. These numbers suggest that it may be better to ask the topics T2, T8 (only a few posts on SSE) and T10 on SO to obtain answers faster, while asking T1 and T9 on SSE would be more optimal. However, the topic difficulty did not correlate with either the topic popularity or share metric on both SO and SSE, confirmed using Kendall's Tau test~\cite{knight1966computer} with a confidence level of 95\%. With the same confidence level, no significant differences in terms of average topic-wise popularity and difficulty between SO and SSE were recorded using non-parametric Mann-Whitney U-test~\cite{mann1947test}.

\subsection{\textbf{RQ3}: What is the Level of Expertise to Answer SV Questions on Q\&A Sites?}
\label{subsec:rq3_results_ease21}

\textbf{General Expertise}. The average general expertise (reputation) of the accepted answerers in SV posts was 1.3 to 5.8 times higher than those of generic posts~\cite{barua2014developers}, general security~\cite{yang2016security}, mobile development~\cite{rosen2016mobile}, concurrency~\cite{ahmed2018concurrency}, machine learning~\cite{bangash2019developers} and deep learning~\cite{han2020programmers}. The higher reputation values were confirmed with $p$-values $<$ 0.01 (significance level) using non-parametric Mann-Whitney U-test~\mbox{\cite{mann1947test}}. However, the average percentage of the same users who got accepted answers on both SO and SSE was quite small across topics, i.e., 1\% to 18\%, implying not much SV knowledge sharing between the two sites. The average topic-wise reputation on SO was higher than that of SSE with a $p$-value of 0.001 (Mann-Whitney U-test). This might be because SO users could engage in many more posts of different topics (not only security). Table~\ref{tab:gen_exp_ease21} reports the general expertise of 13 SV topics. On SO, Brute-force/Timing Attacks (T13), SQL Injection (T2), Synchronization Errors (T6) and XSS (T11) were the topics that experts focused on the most. On SSE, T13 again and Encryption Errors (T7) were the topics of interest for experts. In contrast, Malwares (T1), Vulnerability Scanning Tools (T3) and Network Attacks (T9) on SO did not attract as much attention from experts.
On SSE, T3 was also of the least interest to experts. Overall, experts on Q\&A sites tended to favor the SV topics with high popularity and low difficulty, confirmed with $p$-values $<$ 0.01 using Kendall's Tau correlation test~\cite{knight1966computer}.

\begin{figure}[t]
    \centering
    \includegraphics[width=\columnwidth,keepaspectratio]{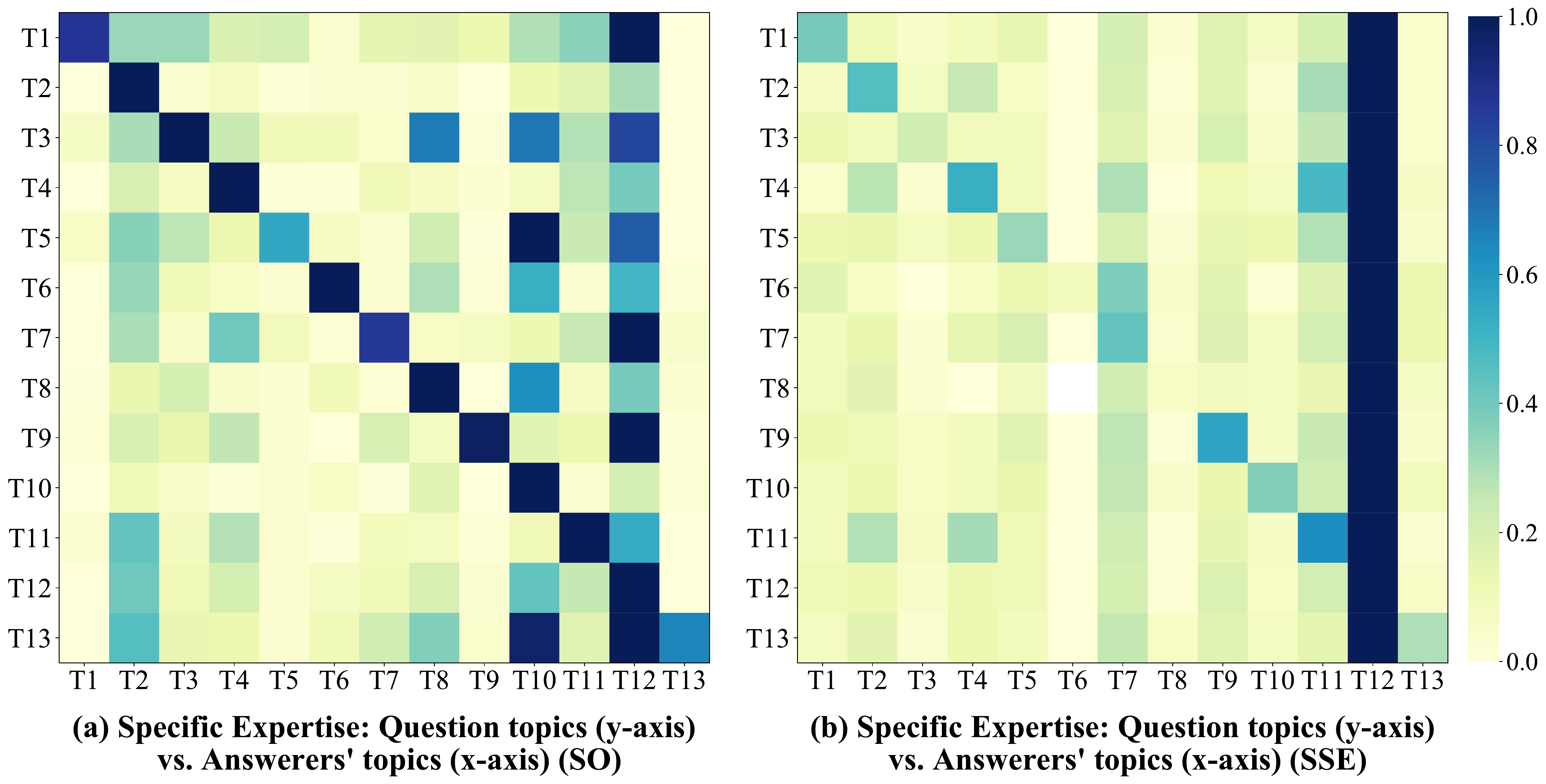}
    \caption[Topic correlations between software vulnerability questions \& answerers' software vulnerability specific knowledge on Stack Overflow \& Security StackExchange.]{Topic correlations between SV questions \& answerers' SV specific knowledge on SO (a) \& SSE (b). \textbf{Notes}: Light to dark color shows weak to strong correlation. Each cell was normalized by the max and min values of each question topic.}
    \label{fig:qa_topic_correlation_ease21}
\end{figure}

\textbf{Specific Expertise}.
Fig.~\mbox{\ref{fig:qa_topic_correlation_ease21}} shows the correlation between the pairs of question SV topics and answerers' SV topics (see Eq.~\mbox{\eqref{eq:spec_exp_ease21}}). The most frequent answerers' SV topic was Vulnerability Theory (T12). On SSE, frequent answerers for T12 could answer every question topic. On SO, besides T12, users specialized in Memory-related Errors (T8 and T10) also answered the questions of other SV topics. These patterns might be because of the prevalence (RQ2) of topics T8 and T10 on SO as well as T12 on SSE. Conversely, Malwares (T1), Network Attacks (T9) and Brute-force/Timing Attacks (T13) on SO as well as Synchronization Errors (T6), Resource Leaks (T8) and T13 on SSE had unique answerers (i.e., users who usually answered questions of only one topic in the SV domain).
Furthermore, on SO, most answerers were relevant for each SV topic (dark color on the diagonal in Fig.~\mbox{\ref{fig:qa_topic_correlation_ease21}}a), but it was not always the case on SSE (see Fig.~\mbox{\ref{fig:qa_topic_correlation_ease21}}b). Such results suggest that it may be easier to find relevant answerers for different SV topics on SO.

\begin{landscape}
\begin{table*}[t]
  \centering
  \caption[General expertise in terms of average reputation of each topic on Stack Overflow and Security StackExchange.]{General expertise in terms of average reputation of each topic on SO and SSE (in parentheses). \textbf{Notes}: The values were normalized by the max and min values of each category. T8 on SSE was excluded since it did not have any accepted answer.}
    \begin{tabular}{lccccccccccccc}
    \hline
    \makecell{\textbf{General}\\ \textbf{expertise}} & \textbf{T1} & \textbf{T2} & \textbf{T3} & \textbf{T4} & \textbf{T5} & \textbf{T6} & \textbf{T7} & \textbf{T8} & \textbf{T9} & \textbf{T10} & \textbf{T11} & \textbf{T12} & \textbf{T13} \\
    \hline
    \textbf{Reputation} & \makecell{\textbf{0.00}\\ (0.16)} & \makecell{0.93\\ (0.05)} & \makecell{\textbf{0.01}\\ (\textbf{0.00})} & \makecell{0.28\\ (0.18)} & \makecell{0.43\\ (0.14)} & \makecell{0.88\\ (0.04)} & \makecell{0.51\\ (0.72)} & \makecell{0.49\\ (--)} & \makecell{0.07\\ (0.24)} & \makecell{0.62\\ (0.34)} & \makecell{0.84\\ (0.22)} & \makecell{0.59\\ (0.29)} & \makecell{\textbf{1.00}\\ (\textbf{1.00})} \\
    \hline
    \end{tabular}
  \label{tab:gen_exp_ease21}
\end{table*}
\end{landscape}

\subsection{\textbf{RQ4}: What Types of Answers are Given to SV Questions on Q\&A Sites?}
\label{subsec:rq4_results_ease21}

\begin{table*}[t]
\fontsize{8}{9}\selectfont
  \centering
  \caption[Answer types of software vulnerability discussions identified on Q\&A websites.]{Answer types of SV discussions identified on Q\&A websites. \textbf{Note}: An answer can have more than one solution type. Proportions of SSE (the last column) are in parenthesis.}
    \begin{tabular}{p{0.17\textwidth}p{0.3\textwidth}p{0.2\textwidth}p{0.16\textwidth}}
    \hline
    \makecell[l]{\textbf{Answer type of} \\ \textbf{SV discussions}} & \multicolumn{1}{c}{\textbf{Description \& Example Posts}} & \makecell[c]{\textbf{Top-3 related}\\ \textbf{question types~\cite{treude2011programmers}}} & \makecell[c]{\textbf{Proportion (\%)}\\ \textbf{on SO \& SSE}} \\
    \hline
    (Dis-)Confirmation (DC/Co) & Confirm/agree or refute/disagree with a major point or concept made by the asker (e.g., SO post 16155188 or SSE post 31306) & Decision Help, How-to, Conceptual & \makecell[c]{11.5 (23.7)} \\
    \hline
    Explanation (Ex) & Explain concepts, definitions and ``why'' to take certain actions (e.g., SO post 53446941 or SSE post 157240) & Decision Help, Conceptual, Discrepancy & \makecell[c]{14.6 (27.1)} \\
    \hline
    Error (Er) & Point out an error in the source code or another attachment of the initial question (e.g., SO post 29750534 or SSE post 159907) & Discrepancy, Error, How-to & \makecell[c]{13.0 (2.3)} \\
    \hline
    Action to Take (AT) & Describe step/action(s) (``how-to'') to solve a problem (e.g., SO post 22860382 or SSE post 180053) & How-to, Discrepancy, Decision Help & \makecell[c]{\textbf{22.5} (15.4)} \\
    \hline
    External Source (ES) & Provide reference/link to external source(s) (e.g., SO post 445177 or SSE post 107498) & Decision Help, How-to, Discrepancy & \makecell[c]{18.1 (\textbf{28.7})} \\
    \hline
    Code Sample (CS) & Provide an explicit example of code snippet (e.g., SO post 20763476 or SSE post 36804) & Discrepancy, How-to, Error & \makecell[c]{16.6 (2.0)} \\
    \hline
    Self-Answer (SA) & Answer given by the same user who submitted the question (e.g., SO post 55784402 or SSE post 100761) & Discrepancy, Error, How-to & \makecell[c]{3.7 (1.0)} \\
    \hline
    \end{tabular}
  \label{tab:answer_types_ease21}
\end{table*}

Our open coding process in RQ4 identified \textit{seven} answer categories of SV discussions on Q\&A sites, as shown in Table~\ref{tab:answer_types_ease21}. Some answer types provided experience (DC/Co and Er) or language/platform-specific support (AT, ES and CS), which is hardly found on expert-based security sites (e.g., CWE or OWASP). We correlated such answer types with the question categories of Treude et al.~\cite{treude2011programmers}.
We found reasonable matches between answer and question types, e.g., (dis)agreeing (DC/Co) with a decision (Decision Help), explaining (Ex) a concept (Conceptual), and providing different solutions to resolve unexpected situations (Discrepancy). Discrepancy and Error were also among the most frequent question types, supporting that our posts were about issues/errors in addressing SVs.

\textbf{Site-wise answer types}. According to Table~\ref{tab:answer_types_ease21}, Action to Take (AT) and External Sources (ES) were the most common answer types on SO and SSE, respectively; whereas, Self-Answer (SA) was the least frequent one on both sites. We also noticed that both sites usually referred to external sources (ES). The most common sources included Wikipedia, other posts (e.g., related answers), GitHub issues/commits, product documentation (e.g., PHP, MySQL and Android) and SV sources (e.g., CVE (Common Vulnerabilities and Exposures)~\cite{cve}, NVD~\cite{nvd}, CWE~\cite{cwe}, OWASP~\cite{owasp_website}, CVE Details~\cite{cve_details} and Exploit-DB~\cite{exploit_db}). Note that some provided links were unavailable or no longer maintained (e.g., CVE Details). Overall, the answers to SV questions on SO frequently provided detailed instructions (AT) and/or code samples (CS), while SSE tended to share more experience (DC/Co and Ex) to help.

\textbf{Topic-wise answer types}. We extend the site-wise findings to individual topics to enable developers to select the respective site (SO vs. SSE) based on their preferable SV solution types, as shown in  Table~\mbox{\ref{tab:answers_per_topic_ease21}}. Specifically, SO highlighted steps (AT) to fix Malwares (T1), Memory Allocation Errors (T10) and XSS (T11) as well as provided code snippets for SQL Injection (T2). On the other hand, SSE gave more relevant sources and explanations for such topics. One may argue that these different answer types were because of the different question types between SO and SSE, but we did not find any such significant differences for these topics. Instead, the fact that SO and SSE had quite different accepted answerers, as shown in RQ3, probably led to such different solution types.
There were four topics, namely SV Scanning Tools (T3), Synchronization \& Encryption Errors (T6 \& T7), Brute-force/Timing Attacks (T13), sharing similar top solutions on both SO and SSE. The remaining topics (T4, T5, T8 and T9) were mostly answered with explanation (Ex) or external sources (ES) on both SO and SSE.

\begin{table}[t]
  \centering
  \caption[Top-1 answer types of 13 software vulnerability topics on Stack Overflow \& Security StackExchange.]{Top-1 answer types of 13 SV topics on SO \& SSE (in parenthesis). \textbf{Note}: T8 on SSE was excluded since it did not have any accepted answer.}
    \begin{tabular}{cc|cc}
    \hline
    \textbf{Topic} & \textbf{Top-1 Answer Type} & \textbf{Topic} & \textbf{Top-1 Answer Type}\\
    \hline
    T1 & AT/ES (Ex) & T8 & ES (--) \\
    \hline
    T2 & CS (ES) & T9 & DC/Co/ES/SA (Ex) \\
    \hline
    T3 & ES (ES) & T10 & AT (Ex)  \\
    \hline
    T4 & ES (DC/Co) & T11 & AT (DC/Co) \\
    \hline
    T5 & Ex (ES) & T12 & Ex (ES)  \\
    \hline
    T6 & Ex/ES (DC/Co/ES) & T13 & Ex/ES (Ex) \\
    \hline
    T7 & Ex/ES (Ex) & -- & -- \\
    \hline
    \end{tabular}
  \label{tab:answers_per_topic_ease21}
\end{table}

\section{Discussion}
\label{sec:discussion_ease21}

\subsection{SV Discussion Topics on Q\&A Sites vs. Existing Security Taxonomies} \label{subsec:vs_literature}

\textbf{SV-specific topics and their support on Q\&A sites}. Compared to Yang et al.'s taxonomy~\mbox{\cite{yang2016security}}, we found related topics: T2, T3, T5, T10, T11 and T13, but we still had the following important differences. Firstly, our topics were emphasized more on security flaws, e.g., issues with encryption/decryption algorithms (T7) than how to implement/use them as in~\mbox{\cite{yang2016security}}. Secondly, we identified SV-specific topics previously unreported in~\mbox{\cite{yang2016security}}: Malwares (T1), CSRF (T4), Synchronization Errors (T6), Resource Leaks (T8), Network Attacks (T9) and Vulnerability Theory (T12). These SV topics show the necessity of focusing on SV-specific posts instead of general security ones. Thirdly, unlike~\mbox{\cite{yang2016security}}, we did not consider language-dependent topics (i.e., PHP, Flash, Javascript, Java and ASP.NET), helping our topics be more generalizable (e.g., XSS can occur in both PHP and ASP.NET). Zahedi et al.~\mbox{\cite{zahedi2018empirical}} also devised a security taxonomy for GitHub issues; however, they focused on security features and implementation instead of any specific SV types. Note that we studied SV posts on both SO and SSE, while the existing studies only used one source of data (SO), enhancing the generalizability of our study. Specifically, we shed light on the differences between SV discussion topics on SO and those on SSE in terms of their proportions (RQ1), popularity/difficulty (RQ2), level of expertise (RQ3) and types of answers (RQ4). Our findings can be leveraged to select a suitable site (i.e., more popular/experts, less 
difficult or having certain answer types) for asking different SV questions.

\sethlcolor{white}\hl{\textbf{Disconnection between SV discussions and expert-based SV sources}. Two researchers (i.e., the author of this thesis and a PhD student with three years of experience in Software Engineering and Cybersecurity) manually mapped 13 SV topics with CWEs, as shown in Table~\mbox{\ref{tab:topic_cwe}}. The agreement between the two people was strong (Kappa score~\mbox{\cite{mchugh2012interrater}} was 0.892), and disagreements were resolved during a discussion section with another researcher (i.e., a PhD student with two-year experience in Software Engineering and Cybersecurity).
We found that only seven of them were overlapping with the two well-known expert-based SV taxonomies: top-25 CWE\footnote{\url{https://cwe.mitre.org/top25/}} and top-10 OWASP\footnote{\url{https://owasp.org/www-project-top-ten/}}. The overlapping topics were T2 (SQL Injection), T4 (CSRF), T5 (i.e., Path-traversal and Unrestricted File Upload), T7 (i.e., Improper Certificate Validation), T8 (Resource Leaks), T10 (Memory Allocation Errors) and T11 (XSS). There was no CWE for T3 and T12 since they mainly discussed SV scanning tools and/or socio-technical issues, respectively. In fact, using keyword matching, we found that only 159 and 71 out of a total of 839 CWEs were mentioned on SO and SSE, respectively; and only 20 and two CWEs appeared more than 100 times on SO and SSE, respectively. Other popular SV sources (i.e., CVE, NVD and OWASP) were also mentioned less than 10\% on these two sites. Moreover, the fast increase of CSRF (T2), as reported in RQ1, is noteworthy given that this SV type has been removed from the top-10 OWASP since 2013. Our investigation suggested that many developers were aware of CSRF standard prevention techniques, but it was not always easy to apply these techniques and/or use built-in CSRF protection of a web framework (e.g., Spring Security) in practice. These results imply a strong disconnection in the SV patterns between expert-based sources and discussions on Q\&A sites, supporting our motivation to study developers' real-life concerns in addressing SVs.}

\begin{table}[t]
  \centering
  \caption[The mapping between 13 software vulnerability topics and their Common Weakness Enumeration (CWE) values.]{The mapping between 13 SV topics and their Common Weakness Enumeration (CWE) values.}
    \begin{tabular}{lc}
    \hline
    \textbf{Topic Name} & \textbf{CWE values} \\
    \hline
    Malwares (T1) & 506-512, 904 \\
    \hline
    SQL Injection (T2) & 20, 74, 89, 707, 943 \\
    \hline
    Vulnerability Scanning Tools (T3) & -- \\
    \hline
    Cross-site Request Forgery (CSRF) (T4) & 352, 1173 \\
    \hline
    File-related Vulnerabilities (T5) & 23, 34-36, 61, 434 \\
    \hline
    Synchronization Errors (T6) & 362, 662, 667, 820, 821, 833 \\
    \hline
    Encryption Errors (T7) & 295, 300, 310 \\
    \hline
    Resource Leaks (T8) & 400, 401, 404, 772 \\
    \hline
    Network Attacks (T9) & 290, 291, 400 \\
    \hline
    Memory Allocation Errors (T10) & 119-127, 787, 822-825, 835 \\
    \hline
    Cross-site Scripting (XSS) (T11) & 20, 79-87, 707  \\
    \hline
    Vulnerability Theory (T12) & -- \\
    \hline
    Brute-force/Timing Attacks (T13) & 208, 261, 307, 385, 799, 916  \\
    \hline
    \end{tabular}
  \label{tab:topic_cwe}
\end{table}

\subsection{Implications for (Data-Driven)  SV Assessment}
From RQ1, we found that Q\&A sites (i.e., SO and SSE) do not report zero-day SVs, but instead they contain the key SV-related issues that developers have been facing in practice. We have shown that these issues constitute only a small percentage of SVs and SV types reported in expert-based SV sources such as NVD and CWE. This finding suggests that more effort should be directed towards performing assessment, e.g., predicting CVSS metrics, for these commonly encountered types of SVs. It is also recommended to develop assessment models for each of these types rather than generic models for all SV types to improve assessment performance.

Moreover, from RQ1 and RQ2, we found the prevalent, popular and increasing SV types like Brute-force/Timing Attacks, Memory/File-related SVs, Malwares and CSRF. In the context of SV assessment, these common SV types can be prioritized as \textit{custom SV types} for prediction using data-driven models (see section~\ref{subsubsec:custom_types} in Chapter~\ref{chapter:lit_review}) because SVs of these types are commonly encountered by developers and have a high possibility of being exploited in the wild.

From RQ2 and RQ3, the SV types (topics) with low difficulty and high expertise may also indicate that it is relatively easy to address these SVs in practice, and vice versa. The difficulty and expertise metrics can be added to the list of currently used technical metrics (e.g., CVSS metrics and time-to-fix SVs in section~\ref{subsubsec:fixing_effort}) to approximate the amount of effort required to fix the identified SVs of such types. Such approximation using such external data from Q\&A sites is particularly useful for projects that do not have sufficient project-specific SV data for training reliable models to predict SV fixing time.

RQ4 revealed different types of solutions that have been provided to address the SVs. We found that Action to Take and Code Sample are among the most actionable solution types as they can be directly applied or customized to fix SVs found in a codebase. If no solution of these two types can be found for a detected SV, it may suggest that the SV is unique and requires more effort to fix (i.e., developing a solution from scratch).

Overall, the findings demonstrate the potential of incorporating SV data from Q\&A sites into (data-driven) SV assessment. However, it is also worth noting that most of the tools (topic T3 in Table~\ref{tab:topic_list_ease21}) currently used by developers for detecting and assessing SVs are based on static analysis techniques rather than data-driven models. One reason can be that data-driven approaches are usually black-box compared to static analysis counterparts. For example, many questions related to these tools on Q\&A sites were about configuring these tools in specific ways to suite developers' needs, but this is mostly not possible with the current data-driven models. This implies a potential disconnection between state-of-the-art and the state-of-the-practice techniques for SV assessment. Future work is required to make data-driven SV assessment models more interpretable (i.e., why a model makes certain decisions) and configurable (i.e., controling/changing a rule learned by a model) so that these models can be more widely adopted in practice.

\subsection{Threats to Validity}

Our data collection is the first threat. We might have missed some SV posts, but we followed standard techniques in the literature. It is hard to guarantee 100\% relevance of the retrieved posts without exhaustive manual validation, which is nearly impossible with more than 70k posts. However, this threat was greatly reduced since the selected posts were carefully checked by three researchers with at least two years of experience in Software Engineering and Cybersecurity.

The identified taxonomies can be another concern. Topic modeling with LDA has been shown effective for processing a large amount of textual posts, but there is still subjectivity in labeling the topics. We mitigated this threat by manually examining at least 30 posts per topic and cross-checking with three of the authors. We also performed a similar manual checking for the answer types in RQ4.

The generalizability of our study may be a threat as well. The patterns we found may not be the same for other Q\&A sites and domains. However, the reported patterns for SV discussions on SO and SSE were at least confirmed significant using statistical tests with $p$-values $<$ 0.01. We also released our code and data at \url{https://github.com/lhmtriet/SV_Empirical_Study} for replication and extension to other domains.

\section{Chapter Summary}
\label{sec:conclusions_ease21}

Through a large-scale study of 71,329 posts on SO and SSE, we have revealed the support of SV-focused discussions on Q\&A sites.
Using LDA, we devised 13 commonly discussed SV topics on Q\&A sites.
We also characterized these topics in terms of their popularity, difficulty, level expertise and solutions received on these sites.
Overall, Q\&A sites do support SV discussions on SO and SSE, and these discussions shed light on the key SV concerns/types that are of particular interest to developers.
Given their importance/prevalence in practice, more priority should be given to assess these SV types.
Moreover, data about the popularity, difficulty, expertise level and solution availability on these Q\&A sites can be considered for (automatically) assessing the effort required to fix SVs (e.g., more relevant resources/support on Q\&A sites may make an SV fix easier). Overall, rich data on crowdsourcing Q\&A sites open up various opportunities for the next generation of data-driven SV assessment methods that are better tailored to developers' real-world needs.

\section{Appendix - PUMiner Overview}
\label{sec:appendix_ease21}

In this Appendix, we briefly introduce the PUMiner approach that we used in section~\ref{subsec:post_collection_ease21} to retrieve security-related posts on SO that can be further refined to obtain the SV-related posts for the presented empirical study in this chapter. We hereby describe the context-aware two-stage PU learning model that forms the core of PUMiner.
The complete details and results of PUMiner can be found in the original publication~\cite{le2020puminer}.

\subsection{PUMiner - A Context-aware Two-stage PU Learning Model for Retrieving Security Q\&A Posts}
\label{subsec:two_stage_pu_model}

PUMiner is a learning model to distinguish security from non-security posts on Q\&A websites. The novelty of PUMiner is that this model does not require negative (non-security) posts to perform the prediction, saving significant effort for practitioners as it is non-trivial to define and obtain such non-security posts in practice. Thus, PUMiner operates on security-related and unlabeled posts. There are two main components in PUMiner: (\textit{i}) feature generation using doc2vec~\cite{le2014distributed} and (\textit{ii}) PU model building using generated features for retrieving security posts. These two components are described hereafter.

PUMiner uses doc2vec~\cite{le2014distributed} to represent input features of posts on Q\&A sites for training the PU model.
Doc2vec~\cite{le2014distributed} jointly learns the representation/embedding of a document (paragraph vector) with its constituent words.
We choose doc2vec to generate embeddings as it can capture the semantic relationship of a word, unlike traditional feature extraction methods such as BoW, n-grams, or tf-idf~\cite{han2017learning, white2015toward}.
Specifically, our work adopts the distributed memory architecture to train doc2vec models as suggested in~\cite{le2014distributed}. Regarding the label of a document/post, a post may contain multiple tags to describe complex concepts, e.g., \emph{php} and \emph{security} tags represent security issues (sql-injection) in PHP. Therefore, besides single tags, we also combine all tags to handle the topic mixture. For example, a post with \emph{php} and \emph{security} tags would have \emph{php\_security} alongside \emph{php}, \emph{security} and \emph{post id} as labels. We also sort labels alphabetically to avoid duplicates (e.g., \emph{php\_security} and \emph{security\_php} are the same).

\begin{algorithm}[t]
    \caption{Context-aware two-stage PU learning model building.}
    \label{algo:sec_pu_learning}
    \DontPrintSemicolon
    \SetAlgoNoLine

    \KwIn{List of posts: $P_{in}$\\
    Labels of posts: $labels\in \{positive, unlabeled\}$\\
    Size of embeddings and Window size: $sz$, $ws$\\
    Classifier and its model configurations: $C$, $config$}

    \KwOut{The trained feature and PU models: feature\_model,
    $\text{model}_{PU}$}

    $word\_list, tag\_list \longleftarrow \emptyset, \emptyset$\;

    \ForEach{${p}_{i}\in {P}_{in}$}{
        $words, tags \longleftarrow \text{tokenize}({p}_{i}), \text{extract\_tags}({p}_{i})$\;
        $word\_list, tag\_list \longleftarrow word\_list + \{words\}, tag\_list + \{tags\}$\;
    }

    $\text{feature\_model} \longleftarrow \text{train\_doc2vec}(word\_list, tag\_list, sz, ws)$\;
    ${{\mathbf{X}}_{in}} \longleftarrow \text{obtain\_feature}(\text{feature\_model}, word\_list)$\;

    $P \longleftarrow \{{\mathbf{x}}_{p}|{\mathbf{x}}_{p}\in {\mathbf{X}}_{in}\wedge label(p)=positive\}$\;
    $U \longleftarrow \{{\mathbf{x}}_{p}|{\mathbf{x}}_{p}\in {\mathbf{X}}_{in}\wedge label(p)=unlabeled\}$\;

    ${\textbf{centroid}}_{P}, {\textbf{centroid}}_{U} \longleftarrow \frac{\sum\limits_{p\in P}{{{\mathbf{x}}_{p}}}}{\left| P \right|}, \frac{\sum\limits_{p\in U}{{{\mathbf{x}}_{p}}}}{\left| U \right|}$

    $RN \longleftarrow \emptyset$\;

    \ForEach(\tcp*[h]{Stage-1 PU: Identify reliable negatives}){${\mathbf{x}}_{i}\in {\mathbf{X}}_{in}$}{
        \If{$d({\mathbf{x}}_{i},{\mathbf{centroid}}_{U})<\alpha *d({\mathbf{x}}_{i},{\mathbf{centroid}}_{P})$}{
            $RN \longleftarrow RN + \{{\mathbf{x}}_{i}\}$\;
        }
    }

    ${\text{model}}_{PU} \longleftarrow \text{train\_classifier}(C, P, RN, config)$ \tcp{Stage-2 PU}

    \Return feature\_model, $\text{model}_{PU}$
\end{algorithm}

Next, we present the context-aware two-stage PU learning model (see Algorithm~\ref{algo:sec_pu_learning}) to retrieve security-related posts -- the core of our PUMiner framework. This algorithm requires a list of discussion posts with their respective labels (positive or unlabeled), along with the configurations of doc2vec (size of embeddings and window size) and classification models (model hyperparameters). Details of Algorithm~\ref{algo:sec_pu_learning} are given hereafter.

\noindent \textbf{Lines 1-8: Learning doc2vec context-aware feature vectors}. Lines 1-4 tokenize text and extract tags of each post to prepare the data for training doc2vec models. Line 5 trains a doc2vec embedding model to learn the context of words and posts. Line 6 then obtains the embedding vector of each post using the trained doc2vec model. Lines 7 and 8 extract the features of both positive and unlabeled posts, respectively.

\noindent \textbf{Lines 9-13: Stage one of PU learning model}. Inspired by PU learning in other domains~\cite{elkan2008learning,mordelet2014bagging,li2003learning,fusilier2015detecting}, we assume that the context-aware doc2vec embeddings can make posts of the same class stay in close proximity in the embedding space.
Stage-one PU learning first identifies (reliable/pure) negative (non-security) posts in the unlabeled set that are as different as possible from the positive set. A traditional binary classifier would work poorly in this case since the negative class is not pure~\cite{li2003learning}.
In line 9 of Algorithm~\ref{algo:sec_pu_learning}, we propose to approximately locate unknown negative posts using the centroid (average) vectors of the known positive (${\textbf{centroid}}_{P}$) and unlabeled (${\textbf{centroid}}_{U}$) sets, respectively. Since the number of non-security posts is dominant (i.e., up to 96\% on Stack Overflow as per our analysis), ${\textbf{centroid}}_{U}$ would represent the negative class more than the positive one. Lines 11-13 compute and compare the cosine distances~\cite{mikolov2013distributed,le2014distributed,li2003learning} (see Eq.~\eqref{eq:cosine_dist}) from each post to the two centroids. If the current post is closer to ${\textbf{centroid}}_{U}$ (i.e., more towards the negative class), it would be selected as a reliable negative.
\begin{equation}\label{eq:cosine_dist}
  \text{cosine}\_\text{distance}=d(i,\,j)=1-\frac{{{\mathbf{p}}_{i}}\centerdot {{\mathbf{p}}_{j}}}{\left\| {{\mathbf{p}}_{i}} \right\|\times \left\| {{\mathbf{p}}_{j}} \right\|}
\end{equation}
where ${\mathbf{p}}_{i}$ and ${\mathbf{p}}_{j}$ are the embedding vectors of the posts $i^{\text{th}}$ and $j^{\text{th}}$, respectively. The range of $\text{cosine}\_\text{distance}$ is [0, 2].

\noindent We also propose a scaling factor ($\alpha$) to increase the flexibility of our centroid-based approach, which would be jointly optimized with other hyperparameters of binary classifiers in the second stage. Besides having only one hyperparameter for tuning, this centroid-based approach can incrementally learn new post(s) very fast with a complexity time of $O(1)$, as given in Eq.~\eqref{eq:new_centroid}.

\begin{equation}\label{eq:new_centroid}
  {\mathbf{centroid}}_{new}=
  \frac{{\mathbf{centroid}}_{old} \times  N+{\mathbf{x}}_{new}}{N+\text{size}({{\mathbf{x}}_{new}})}
\end{equation}
where ${\mathbf{centroid}}_{old}$ , ${\mathbf{centroid}}_{new}$ are the centroid vectors before and after learning new post(s) (${\mathbf{x}}_{new}$), while $N$ is the original number of posts in the positive or unlabeled set.

\noindent \textbf{Lines 14-15: Stage two of PU learning model}. Using positive and (reliable) negative posts from the first stage, the second stage of PU learning (line 14) trains a binary classifier with its hyperparameters. In the model building process, a PU model is trained with the optimal classifier and its configurations obtained from a validation process (e.g., k-fold cross-validation). Finally, line 15 saves the trained feature and PU models to disk for future inference of whether a post is related to security.

%% file: Chapters/Chapter_7_Conclusions.tex
\chapter{Conclusions and Future Work}
\label{chap:conclusions_future_work}

Software Vulnerability (SV) assessment is an integral step of software development. There has been an increasing number of studies on data-driven SV assessment. Data-driven approaches have the benefits of extracting patterns and rules from large-scale historical SV data in the wild without manual definition from experts. Such capability of these approaches enables researchers to automate various SV assessment tasks that were previously not possible with static analysis and rule-based counterparts.
Despite the rising interest in data-driven SV assessment, the application of these approaches in practice has been hindered by three key issues: (\textit{i}) concept drift in SV data used for report-level SV assessment, (\textit{ii}) lack of code-based SV assessment, and (\textit{iii}) SV assessment without considering developers' real-world SV needs.
This thesis has tackled these three challenges to improve the real-world applicability of data-driven SV assessment.
We have first systematized the knowledge and have identified the key practices of the field.
Then, we have developed various solutions to enhance the performance and applicability of SV assessment models in real-world settings.
These solutions have been based on the customization of recent advances in Machine Learning, Deep Learning (DL) and Natural Language Processing (NLP), as well as the utilization of relevant software artifacts (e.g., source code) and open sources (e.g., Q\&A sites) that have been little explored in the literature.
We have demonstrated the potential effectiveness and value of our solutions using real-world SV data. Chapter~\ref{chap:conclusions_future_work} first summarizes the key contributions and findings of this thesis and then suggests several avenues for future research\footnote{Some of the future research directions presented in this chapter are based on our paper ``\textit{A Survey on Data-Driven Software Vulnerability Assessment and Prioritization}'', ACM Computing Surveys journal (CORE A*)~\cite{le2021survey}.} in the emerging area of data-driven SV assessment.

\newpage

\section{Summary of Contributions and Findings}
\label{sec:findings_contributions}

The significant contributions and key findings of the thesis are summarized as follows.

\subsection{A Systematization of Knowledge of Data-Driven SV Assessment}

In Chapter~\ref{chapter:lit_review}, we have reviewed 84 representative primary studies on data-driven SV assessment to systematize the knowledge of the field. Our review has identified a taxonomy of five key SV assessment tasks/themes that have been automated using data-driven approaches. The themes are predictions of SV (\textit{i}) exploitation (probability, time, and characteristics), (\textit{ii}) impact (confidentiality, integrity, availability, scope, and custom impacts), (\textit{iii}) severity (probability, levels, and score), (\textit{iv}) type (CWE and custom types), as well as (\textit{v}) other tasks (e.g., retrieval of affected product names/vendors, SV knowledge graph construction, and SV fixing effort prediction).

The review has also shed light on the common data-driven practices adopted by these studies to automate SV assessment tasks. Regarding the data sources, NVD~\cite{nvd}/CVE~\cite{cve} has been most frequently used to provide inputs (mostly SV descriptions) and outputs (e.g., CVSS metrics) for SV assessment models. Moreover, Bag-of-words, Word2vec~\cite{mikolov2013distributed} and more recently BERT~\cite{devlin2018bert} have been common methods for extracting features from text-based descriptions of SVs. In addition, linear Support Vector Machine has been the most commonly used for building prediction models, while ensemble Machine Learning (ML) and Deep Learning (DL) models are also on the rise. These models have been mainly evaluated on random-based splits like k-fold cross-validation using common classification metrics (e.g., F1-Score and/or Accuracy) or regression metrics (e.g., Mean absolute error and/or Correlation coefficient).

While data-driven approaches have shown great potential for various SV assessment tasks, we have identified three key issues affecting their practical applicability.
Firstly, current studies have mostly used SV descriptions/reports, but they have not investigated the impacts of changing data (concept drift) of these descriptions/reports due to ever-increasing SVs on their models. Such impacts can affect the robustness and even lower the performance of these models over time when deployed in practice.
Secondly, existing studies have hardly leveraged (vulnerable) source code in which SVs are usually rooted as inputs for developing data-driven SV assessment models. In practice, using source code directly for SV assessment can alleviate the need for SV reports, which in turn supports more timely SV assessment.
Thirdly, the current literature on SV assessment has mainly focused on characteristics of SVs (e.g., exploitability and impacts) rather than real-world concerns encountered by developers while addressing SVs (e.g., difficulty of implementing solutions). Such real-world concerns can facilitate more thorough assessment of SVs by incorporating developers' needs.
These three challenges have motivated us to devise respective solutions presented in Chapters~\ref{chap:msr19},~\ref{chap:msr22},~\ref{chap:ase21}, and~\ref{chap:ease21}.

\subsection{Automated Report-Level Assessment for Ever-Increasing SVs}

In Chapter~\ref{chap:msr19}, we have conducted a large-scale analysis of the concept drift issue on report-level SV assessment models (i.e., predicting seven CVSS version 2 base metrics~\cite{cvss_v2}). Using data of 105,124 SVs on NVD, we have first shown that concept drift is indeed an issue in SV descriptions over the years due to releases or discoveries of new software products and cyber-attacks. Concept drift can make k-fold cross-validation, the most commonly used evaluation technique, inflate the validation performance up to 4.7 times compared to that of time-based validation (year-based splits). The main reason is that k-fold cross-validation mixes future SV data during training (i.e., using data of SVs that have not yet been reported at training time), while time-based validation does not. Thus, we strongly recommend time-based validation instead of k-fold cross-validation for future work on report-level SV assessment.

We have also proposed using subwords extracted from SV descriptions as features for report-level SV assessment models to help these models be robust against Out-of-Vocabulary words caused by concept drift. We have shown that the proposed subword-based models are much more resilient to all the changes in SV descriptions over time, while exhibiting competitive or better performance compared to existing word-only models. We have also demonstrated the possibility of building compact concept-drift-aware SV assessment models (up to 94\% reduction in model size while retaining at least 90\% of the performance) by using fastText~\cite{bojanowski2017enriching,joulin2016bag}. Overall, we have raised the awareness of the concept drift issue in SV data/reports and proposed an effective data-driven solution to addressing such issue for report-level SV assessment.

\subsection{Automated Early SV Assessment using Code Functions}

In Chapter~\ref{chap:msr22}, we have shown the benefits of performing function-level SV assessment (automatically assigning the seven CVSS version 2 metrics to SVs in functions) over report-level SV assessment. We have found that SV reports collected from NVD required for SV assessment usually appear on average 146 days after the fixing time of respective SVs. Such a delay makes report-level SV assessment likely untimely. On the other hand, vulnerable code is available at fixing time by default, thus could be leveraged to support SV assessment even when SV reports are not (yet) available.

Using 1,782 functions of 429 SVs in 200 real-world projects, we have explored the use of vulnerable statements containing root causes of SVs together with other (non-vulnerable) lines in functions as inputs for developing ML models to automate SV assessment. The optimal function-level SV assessment models, on average, have achieved a promising performance of 0.64 Matthews Correlation Coefficient (MCC) and 0.75 F1-Score using all statements in each function (i.e., vulnerable statements alongside all the other context lines). We have highlighted the possibility of performing function-level SV assessment without knowing exactly where vulnerable statements are located in functions. We have also recommended high-performing features and classifiers for this task.
Overall, we have distilled the first practices of using data-driven approaches to automate function-level SV assessment to enable earlier fixing prioritization without waiting for SV reports.

\subsection{Automated Just-in-Time SV Assessment using Code Commits}

In Chapter~\ref{chap:ase21}, we have motivated the need for SV assessment in code commits to avoid assessment and remediation latencies caused by hidden SVs. We have pointed out that in practice, SVs can stay hidden in codebases for a long time, up to 1,469 days, before being reported. As a result, performing SV assessment in commits, where vulnerable changes are first added to a project, provides just-in-time information about SVs for remediation planning and prioritization. This new assessment task requires a suitable approach that can directly operate with code changes in commits.

We have proposed DeepCVA, a novel deep multi-task learning model, to tackle commit-level SV assessment. Specifically, DeepCVA simultaneously predicts the seven base metrics of CVSS version 2 in a single model using shared context-aware commit features extracted by attention-based convolutional gated recurrent units.
Through large-scale experiments on 1,229 vulnerability-contributing commits of 542 different SVs in 246 real-world software projects, we have demonstrated the substantially better performance (38\% to 59.8\% higher MCC) of DeepCVA than that of many supervised and unsupervised baseline models. Multi-task learning has also enabled DeepCVA to require 6.3 times less time for training and maintenance than seven cumulative assessment models.
With the reported effectiveness and efficiency of DeepCVA, we have made the first promising step towards a holistic solution to assessing SVs as early as they appear, which also contributes to the industrial movement of shift-left security in DevSecOps (securing software adopting the DevOps paradigm)~\cite{rajapakse2022challenges}.

\subsection{Insights of Developers' Real-World Concerns on Question and Answer Websites for Data-Driven SV Assessment}

As shown in Chapter~\ref{chapter:lit_review}, there has been a deficiency in considering developers' real-world SV-related concerns for SV assessment. Most of the current studies have used expert-defined taxonomies, e.g., CVSS or CWE, as SV assessment outputs. However, these outputs usually do not put an emphasis on the challenges that developers commonly encounter when addressing SVs in practice, e.g., difficulty in implementing a solution proposed by experts. Such lacking considerations can lead to incomplete assessment of SVs, affecting the decision making in prioritizing SV remediation.

Given that developers usually seek solutions to these concerns on Question and Answer (Q\&A) sites, in Chapter~\ref{chap:ease21}, we have collected 71,329 SV-related posts from two large Q\&A sites, namely Stack Overflow (SO) and Security StackExchange (SSE), to identify developers' real-world SV concerns and analyze the support these concerns receive on these sites for SV assessment. We have used Latent Dirichlet Allocation (LDA)~\cite{blei2003latent} to semi-automatically identify 13 commonly discussed SV topics/concerns on SO and SSE. We have discovered that these concerns are only a subset of all SVs/SV types reported by experts and do not follow the patterns of standard rankings like top-10 OWASP or top-25 CWE.
Such differences ask for higher priorities and more specific techniques in assessing these commonly encountered SV types.
We have also characterized these concerns in terms of their popularity, difficulty, expertise level and solutions received on these sites, which can be used to estimate the complexity/effort required to fix SVs. For instance, an identified SV whose similar SVs have received little support on Q\&A sites would imply high remediation difficulty/effort.
Overall, crowdsourcing Q\&A sites like SO and SSE have much potential for providing supplementary SV assessment metrics from developers' perspectives.

\section{Opportunities for Future Research}
\label{sec:future_directions}

As shown in section~\ref{sec:findings_contributions}, this thesis has made significant contributions to improve the practicality of data-driven SV assessment. However, there are still many opportunities for future research to further advance the field. Note that this section does not cover straightforward extensions of our work in Chapters~\ref{chap:msr19},~\ref{chap:msr22},~\ref{chap:ase21}, and~\ref{chap:ease21} such as using more projects, features, models, and/or programming languages. Rather, we focus on the future research opportunities that have received little/no attention so far in this area.

\subsection{Integration of SV Data on Issue Tracking Systems}\label{subsubsec:bug_vs_sv}

Existing studies, including ours in Chapters~\ref{chap:msr19}, have mainly utilized NVD/CVE for collecting SV reports; whereas, \textit{bug/issue tracking systems} like JIRA,\footnote{\url{https://www.atlassian.com/software/jira}} Bugzilla\footnote{\url{https://www.bugzilla.org/}} or GitHub issues\footnote{\url{https://docs.github.com/en/issues}} also contain an abundance of SV reports, yet have been underexplored for data-driven SV assessment.
Besides providing SV descriptions like CVE/NVD, these issue tracking systems also contain other artifacts such as steps to reproduce, stack traces and test cases that give extra information about SVs~\cite{zimmermann2010makes}.
However, it is not trivial to obtain and integrate these SV-related bug reports with the ones on SV databases.

One way to retrieve SVs on issue tracking systems is to use security bug reports~\cite{bhuiyan2021security}. Much research work has been put into developing effective models to automatically retrieve security bug reports (e.g.,~\cite{gegick2010identifying,peters2017text,wu2021data}). Among these studies, Wu et al.~\cite{wu2021data} manually verified and cleaned the security bug reports to provide a clean dataset for automated security bug report identification. However, more of such manual effort is still required to obtain up-to-date data because the original security bug reports in~\cite{wu2021data} were actually a part of the dataset collected back in 2014~\cite{ohira2015dataset}.

It is worth noting that \textit{not} all security bug reports are related to SVs such as issues/improvements in implementing security features.\footnote{The security bug report AMBARI-1373 on JIRA (\url{https://issues.apache.org/jira/browse/AMBARI-1373}) was about improving the front-end of AMBARI Web by displaying the current logged in user.} Thus, future studies need to filter out these cases before using security bug reports for SV assessment.
We also emphasize that some SV-related bug reports are overlapping with the ones on NVD (e.g., the SV report AMBARI-14780\footnote{\url{https://issues.apache.org/jira/browse/AMBARI-14780}} on JIRA refers to CVE-2016-0731 on CVE/NVD). Such overlaps would require data cleaning during the integration of reports on issue tracking systems and SV databases to avoid data duplication (e.g., similar SV descriptions) when developing SV assessment models.

\subsection{Improving Data Efficiency for Data-Driven SV Assessment}\label{subsec:data_efficiency}
As shown in Chapter~\ref{chapter:lit_review}, many of the SV assessment tasks being automated by data-driven approaches, including the prediction of CVSS base metrics in Chapters~\ref{chap:msr19},~\ref{chap:msr22}, and~\ref{chap:ase21}, suffer from the data imbalance and data scarcity issues. Moreover, existing work as well as our studies in Chapters~\ref{chap:msr19},~\ref{chap:msr22}, and~\ref{chap:ase21} have mainly used fully-supervised learning models for automating these tasks, but these models require sufficiently large and fully labeled data to perform well. To address the data-hungriness of these fully-supervised learning models, future studies can approach the SV assessment tasks with \textit{low-shot learning} and/or \textit{semi-supervised learning}.

\textit{Low-shot learning} a.k.a. \textit{few-shot learning} is designed to perform supervised learning using only a few examples per class, significantly reducing the labeling effort and increasing model robustness against imbalanced data~\cite{wang2020generalizing}.
According to Chapter~\ref{chapter:lit_review}, so far, only one study in this area utilized low-shot learning with a deep Siamese network~\cite{das2021v2w} (i.e., a shared feature model with similarity learning) to effectively predict SV types (CWE) and even generalize to unseen classes (i.e., zero-shot learning). There are still many opportunities for investigating different few-shot learning techniques for other SV assessment tasks, e.g., predicting CVSS metrics. Note that the shared features in few-shot learning can also be enhanced with pre-trained models (e.g., BERT~\cite{devlin2018bert}) on another domain/task/project with more labeled data than the current task/project in the SV domain.

\textit{Semi-supervised learning} enables training models with limited labeled data yet a large amount of unlabeled data~\cite{van2020survey}, potentially leveraging hidden/unlabeled SVs in the wild. Recently, we have seen an increasing interest in using different techniques of this learning paradigm in the SV domain such as collecting SV patches using \textit{multi-view co-training}~\cite{sawadogo2020learning}, retrieving SV discussions on developer Q\&A sites using \textit{positive-unlabeled learning}~\cite{le2020puminer}, curating SVs from multiple sources in the wild using \textit{self-training}~\cite{chen2020machine}. However, it is still little known about the effectiveness of semi-supervised learning for SV assessment tasks.

\subsection{Customized Data-Driven SV Assessment}
Similar to the existing studies reviewed in Chapter~\ref{chapter:lit_review}, in Chapters~\ref{chap:msr19},~\ref{chap:msr22}, and~\ref{chap:ase21}, we have used the standard CVSS metrics~\cite{cvss} for assessing the exploitability, impact and severity levels/score of SVs, but there are increasing concerns that these CVSS outputs are still generic.
Specifically, Spring et al.~\cite{spring2021time} argued that CVSS tends to provide one-size-fits-all assessment metrics regardless of the context of SVs; i.e., the same SVs in different domains/environments are assigned the same metric values. For instance, banking systems may consider the confidentiality and integrity of databases more important than the availability of web/app interfaces.

In the future, alongside CVSS, prediction models should also incorporate the domain/business knowledge to customize the assessment of SVs to a system of interest (e.g., the impact of SVs on critical component(s) and/or the readiness of developers/solutions for mitigating such SVs in the current system).
Development environments (e.g., programming language, tools or frameworks) being used are among the key factors that affect the fixing of a specific SV in an organization. Thus, SV-related issues that developers have encountered with these environments discussed on Q\&A sites can be considered to enrich SV assessment information.
Particularly, future work can leverage the taxonomy we identified in Chapter~\ref{chap:ease21} to automatically match detected SVs with similar issues on Q\&A sites, and then follow our framework to indicate the level of support (popularity, difficulty, expertise level, and solution type) these issues have received. The higher the support is, the more likely developers will find relevant information to fix the current SV, which in turn reduces the fixing complexity/effort.
In the future, case studies with practitioners will also be fruitful to correlate the quantitative performance of models and their usability/usefulness in real-world systems (e.g., reducing more critical SVs yet using fewer resources).

\subsection{Enhancing Interpretability of SV Assessment Models}
As discussed in Chapter~\ref{chap:ease21}, the lack of interpretability is one of the key reasons impeding the adoption of data-driven approaches compared to static analysis tools for SV assessment in practice. Model interpretability is important to increase the transparency of the predictions made by a model, allowing practitioners to adjust the model/data to meet certain requirements~\cite{zhang2020survey}. According to Chapter~\ref{chapter:lit_review}, very few reviewed papers in this area (e.g.,~\cite{toloudis2016associating,han2017learning}) have explicitly discussed important features and/or explained why/when their models worked/failed for a task.

SV assessment can draw inspiration from the related SV detection area where the interpretability of (DL-based) prediction models has been actively explored mainly by using (\textit{i}) specific model architectures/parameters or (\textit{ii}) external interpretation models/techniques~\cite{zhang2020survey}. In the first approach, prior studies successfully used the feature activation maps in a CNN model~\cite{russell2018automated} or leveraged attention-based neural network~\cite{duan2019vulsniper} to highlight and visualize the important code tokens that contribute to SVs. The second approach uses separate interpretation models on top of trained SV detectors. The interpretation models are either domain/model-agnostic~\cite{warnecke2020evaluating}, domain-agnostic yet specific to a model type (graph neural network~\cite{li2021vulnerability}) or SV-specific~\cite{zou2021interpreting}. The aforementioned approaches produce local/sample-wise interpretation, which can be aggregated to obtain global/task-wise interpretation. The global interpretation is similar to the feature importance of traditional ML models~\cite{chandrashekar2014survey} such as the weights of linear models (e.g., Logistic regression) or the (im)purity of nodes split by each feature in tree-based models (e.g., Random forest). However, it is still unclear about the applicability/effectiveness of these approaches for interpreting ML/DL-based SV assessment models, requiring further investigations.

\subsection{Data-Driven SV Assessment in Data-Driven Systems}\label{subsec:sv_ap_data_driven_systems}

Like the current literature (see Chapter~\ref{chapter:lit_review}), this thesis has mainly focused on SV assessment for traditional software, but we envision there is an impending need for SV assessment in data-driven/Artificial Intelligence (AI)-based systems.
Data-driven/AI-based systems (e.g., smart recommender systems, chatbots, robots, and autonomous cars) are an emerging breed of systems whose cores are powered by AI technologies, e.g., ML and DL models built on data, rather than human-defined instructions as in traditional systems~\cite{kumar2020legal}. However, it is challenging to adapt the current practices of SV assessment to data-driven/AI-based systems. Some of these challenges are presented hereafter.

CVSS~\cite{cvss} is currently the most popular SV assessment framework for traditional systems, but its compatibility with data-driven systems still requires more investigation. The current CVSS documentation lacks instructions on how to assign metrics/score for SVs in data-driven systems. For example, it is unclear how to assign static CVSS metrics to systems with automatically updated data-driven models~\cite{chen2020machine} because adversarial examples for exploitation would likely change after the models are updated.
Such ambiguities should be clarified/resolved in future CVSS versions as data-driven systems become more prevalent. The types of SVs in ML/DL models in data-driven systems are also mostly different from the ones provided by CWE~\cite{cwe}. The difference is mainly because these new SVs do not only emerge from configurations/code as in traditional systems, but also from training data and/or trained models~\cite{rosenberg2021adversarial}. Thus, we recommend that a new category of these SVs should be studied and potentially incorporated into CWE, similar to the newly added category for architectural SVs.\footnote{\url{https://cwe.mitre.org/data/definitions/1008.html}}

Existing SV assessment models for traditional systems have not considered unique data/model-related characteristics/features of data-driven systems~\cite{wan2019does}. Specifically, data-driven systems also encompass information about data (e.g., format, type, size and distribution) and ML/DL model(s) (e.g., configurations, parameters and performance). It is worth noting that SVs of ML/DL models in data-driven systems can also come from the frameworks used to develop such models (e.g., Tensorflow\footnote{\url{https://github.com/tensorflow/tensorflow}} or Keras\footnote{\url{https://github.com/keras-team/keras}}). However, developers of data-driven systems may not be aware of the (security) issues in the used ML/DL frameworks~\cite{liu2020using}. Thus, besides currently used features, future work should also consider the information about underlying data/models and ML/DL development frameworks to improve the SV representation for building models to assess SVs in data-driven systems.

%% file: main.bbl
\begin{thebibliography}{100}
\providecommand{\url}[1]{#1}
\csname url@samestyle\endcsname
\providecommand{\newblock}{\relax}
\providecommand{\bibinfo}[2]{#2}
\providecommand{\BIBentrySTDinterwordspacing}{\spaceskip=0pt\relax}
\providecommand{\BIBentryALTinterwordstretchfactor}{4}
\providecommand{\BIBentryALTinterwordspacing}{\spaceskip=\fontdimen2\font plus
\BIBentryALTinterwordstretchfactor\fontdimen3\font minus
  \fontdimen4\font\relax}
\providecommand{\BIBforeignlanguage}[2]{{%
\expandafter\ifx\csname l@#1\endcsname\relax
\typeout{** WARNING: IEEEtran.bst: No hyphenation pattern has been}%
\typeout{** loaded for the language `#1'. Using the pattern for}%
\typeout{** the default language instead.}%
\else
\language=\csname l@#1\endcsname
\fi
#2}}
\providecommand{\BIBdecl}{\relax}
\BIBdecl

\bibitem{andreessen2011software}
M.~Andreessen, ``Why software is eating the world,'' \emph{Wall Street
  Journal}, vol.~20, no. 2011, p.~C2, 2011.

\bibitem{google_codebase}
\BIBentryALTinterwordspacing
Wired, ``Google is 2 billion lines of code—and it's all in one place.''
  [Online]. Available:
  \url{https://www.wired.com/2015/09/google-2-billion-lines-codeand-one-place/}
\BIBentrySTDinterwordspacing

\bibitem{ghaffarian2017software}
S.~M. Ghaffarian and H.~R. Shahriari, ``Software vulnerability analysis and
  discovery using machine-learning and data-mining techniques: A survey,''
  \emph{ACM Computing Surveys (CSUR)}, vol.~50, no.~4, pp. 1--36, 2017.

\bibitem{nvd_vuln}
\BIBentryALTinterwordspacing
NIST, ``Vulnerability definition on nvd.'' [Online]. Available:
  \url{https://nvd.nist.gov/vuln}
\BIBentrySTDinterwordspacing

\bibitem{heartbleed}
\BIBentryALTinterwordspacing
I.~Synopsys, ``Heartbleed bug.'' [Online]. Available:
  \url{https://heartbleed.com/}
\BIBentrySTDinterwordspacing

\bibitem{log4j}
\BIBentryALTinterwordspacing
T.~Conversation, ``What is log4j?'' [Online]. Available:
  \url{https://bit.ly/log4j_the_conversation}
\BIBentrySTDinterwordspacing

\bibitem{cyber_loss_2021}
\BIBentryALTinterwordspacing
A.~C.~S. Centre, ``Acsc annual cyber threat report 2020-21.'' [Online].
  Available:
  \url{https://www.cyber.gov.au/acsc/view-all-content/reports-and-statistics/acsc-annual-cyber-threat-report-2020-21}
\BIBentrySTDinterwordspacing

\bibitem{nayak2014some}
K.~Nayak, D.~Marino, P.~Efstathopoulos, and T.~Dumitra{\c{s}}, ``Some
  vulnerabilities are different than others,'' in \emph{International Workshop
  on Recent Advances in Intrusion Detection}.\hskip 1em plus 0.5em minus
  0.4em\relax Springer, 2014, pp. 426--446.

\bibitem{khan2018review}
S.~Khan and S.~Parkinson, ``Review into state of the art of vulnerability
  assessment using artificial intelligence,'' in \emph{Guide to Vulnerability
  Analysis for Computer Networks and Systems}.\hskip 1em plus 0.5em minus
  0.4em\relax Springer, 2018, pp. 3--32.

\bibitem{smyth2017software}
V.~Smyth, ``Software vulnerability management: how intelligence helps reduce
  the risk,'' \emph{Network Security}, vol. 2017, no.~3, pp. 10--12, 2017.

\bibitem{le2021survey}
T.~H.~M. Le, H.~Chen, and M.~A. Babar, ``A survey on data-driven software
  vulnerability assessment and prioritization,'' \emph{ACM Computing Surveys
  (CSUR)}, 2021.

\bibitem{foreman2019vulnerability}
P.~Foreman, \emph{Vulnerability management}.\hskip 1em plus 0.5em minus
  0.4em\relax CRC Press, 2019.

\bibitem{kritikos2019survey}
K.~Kritikos, K.~Magoutis, M.~Papoutsakis, and S.~Ioannidis, ``A survey on
  vulnerability assessment tools and databases for cloud-based web
  applications,'' \emph{Array}, vol.~3, p. 100011, 2019.

\bibitem{bell1985expert}
M.~Z. Bell, ``Why expert systems fail,'' \emph{Journal of the Operational
  Research Society}, vol.~36, no.~7, pp. 613--619, 1985.

\bibitem{han2011data}
J.~Han, M.~Kamber, and J.~Pei, ``Data mining concepts and techniques third
  edition,'' \emph{The Morgan Kaufmann Series in Data Management Systems},
  vol.~5, no.~4, pp. 83--124, 2011.

\bibitem{nvd}
\BIBentryALTinterwordspacing
NIST, ``National vulnerability database.'' [Online]. Available:
  \url{https://nvd.nist.gov}
\BIBentrySTDinterwordspacing

\bibitem{vuln_stat}
\BIBentryALTinterwordspacing
------, ``Number of vulnerabilities reported on nvd in 2021.'' [Online].
  Available:
  \url{http://nvd.nist.gov/vuln/search/statistics?form\_type=Basic\&results\_type=statistics\&search\_type=all&isCpeNameSearch=false}
\BIBentrySTDinterwordspacing

\bibitem{cruzes2011research}
D.~S. Cruzes and T.~Dyb{\aa}, ``Research synthesis in software engineering: A
  tertiary study,'' \emph{Information and Software Technology}, vol.~53, no.~5,
  pp. 440--455, 2011.

\bibitem{gama2014survey}
J.~Gama, I.~{\v{Z}}liobait{\.e}, A.~Bifet, M.~Pechenizkiy, and A.~Bouchachia,
  ``A survey on concept drift adaptation,'' \emph{ACM computing surveys
  (CSUR)}, vol.~46, no.~4, pp. 1--37, 2014.

\bibitem{spanos2017assessment}
G.~Spanos, L.~Angelis, and D.~Toloudis, ``Assessment of vulnerability severity
  using text mining,'' in \emph{the 21st Pan-Hellenic Conference on
  Informatics}, 2017, pp. 1--6.

\bibitem{han2017learning}
Z.~Han, X.~Li, Z.~Xing, H.~Liu, and Z.~Feng, ``Learning to predict severity of
  software vulnerability using only vulnerability description,'' in \emph{2017
  IEEE International Conference on Software Maintenance and Evolution
  (ICSME)}.\hskip 1em plus 0.5em minus 0.4em\relax IEEE, 2017, pp. 125--136.

\bibitem{spanos2018multi}
G.~Spanos and L.~Angelis, ``A multi-target approach to estimate software
  vulnerability characteristics and severity scores,'' \emph{Journal of Systems
  and Software}, vol. 146, pp. 152--166, 2018.

\bibitem{le2019automated}
T.~H.~M. Le, B.~Sabir, and M.~A. Babar, ``Automated software vulnerability
  assessment with concept drift,'' in \emph{the 16th International Conference
  on Mining Software Repositories (MSR)}.\hskip 1em plus 0.5em minus
  0.4em\relax IEEE, 2019, pp. 371--382.

\bibitem{rajapakse2022challenges}
R.~N. Rajapakse, M.~Zahedi, M.~A. Babar, and H.~Shen, ``Challenges and
  solutions when adopting devsecops: A systematic review,'' \emph{Information
  and Software Technology}, vol. 141, p. 106700, 2022.

\bibitem{meneely2013patch}
A.~Meneely, H.~Srinivasan, A.~Musa, A.~R. Tejeda, M.~Mokary, and B.~Spates,
  ``When a patch goes bad: Exploring the properties of
  vulnerability-contributing commits,'' in \emph{2013 ACM/IEEE International
  Symposium on Empirical Software Engineering and Measurement}.\hskip 1em plus
  0.5em minus 0.4em\relax IEEE, 2013, pp. 65--74.

\bibitem{kamei2012large}
Y.~Kamei, E.~Shihab, B.~Adams, A.~E. Hassan, A.~Mockus, A.~Sinha, and
  N.~Ubayashi, ``A large-scale empirical study of just-in-time quality
  assurance,'' \emph{IEEE Transactions on Software Engineering}, vol.~39,
  no.~6, pp. 757--773, 2012.

\bibitem{zhang2017survey}
Y.~Zhang and Q.~Yang, ``A survey on multi-task learning,'' \emph{arXiv preprint
  arXiv:1707.08114}, 2017.

\bibitem{cwe}
\BIBentryALTinterwordspacing
MITRE, ``Common weakness enumeration.'' [Online]. Available:
  \url{https://cwe.mitre.org}
\BIBentrySTDinterwordspacing

\bibitem{cvss}
\BIBentryALTinterwordspacing
FIRST, ``Common vulnerability scoring system.'' [Online]. Available:
  \url{https://www.first.org/cvss}
\BIBentrySTDinterwordspacing

\bibitem{blei2003latent}
D.~M. Blei, A.~Y. Ng, and M.~I. Jordan, ``Latent dirichlet allocation,''
  \emph{Journal of machine learning research}, vol.~3, no. Jan, pp. 993--1022,
  2003.

\bibitem{kamongi2013vulcan}
P.~Kamongi, S.~Kotikela, K.~Kavi, M.~Gomathisankaran, and A.~Singhal, ``Vulcan:
  Vulnerability assessment framework for cloud computing,'' in \emph{2013 IEEE
  7th International Conference on Software Security and Reliability}.\hskip 1em
  plus 0.5em minus 0.4em\relax IEEE, 2013, pp. 218--226.

\bibitem{bozorgi2010beyond}
M.~Bozorgi, L.~K. Saul, S.~Savage, and G.~M. Voelker, ``Beyond heuristics:
  learning to classify vulnerabilities and predict exploits,'' in \emph{the
  16th ACM SIGKDD international conference on Knowledge discovery and data
  mining}, 2010, pp. 105--114.

\bibitem{sabottke2015vulnerability}
C.~Sabottke, O.~Suciu, and T.~Dumitraș, ``Vulnerability disclosure in the age
  of social media: Exploiting twitter for predicting real-world exploits,'' in
  \emph{24th $\{$USENIX$\}$ Security Symposium}, 2015, pp. 1041--1056.

\bibitem{bullough2017predicting}
B.~L. Bullough, A.~K. Yanchenko, C.~L. Smith, and J.~R. Zipkin, ``Predicting
  exploitation of disclosed software vulnerabilities using open-source data,''
  in \emph{the 3rd ACM on International Workshop on Security And Privacy
  Analytics}, 2017, pp. 45--53.

\bibitem{zeng2020software}
P.~Zeng, G.~Lin, L.~Pan, Y.~Tai, and J.~Zhang, ``Software vulnerability
  analysis and discovery using deep learning techniques: A survey,'' \emph{IEEE
  Access}, 2020.

\bibitem{singh2020applying}
S.~K. Singh and A.~Chaturvedi, ``Applying deep learning for discovery and
  analysis of software vulnerabilities: A brief survey,'' \emph{Soft Computing:
  Theories and Applications}, pp. 649--658, 2020.

\bibitem{semasaba2020literature}
A.~O.~A. Semasaba, W.~Zheng, X.~Wu, and S.~A. Agyemang, ``Literature survey of
  deep learning-based vulnerability analysis on source code,'' \emph{IET
  Software}, 2020.

\bibitem{lin2020software}
G.~Lin, S.~Wen, Q.-L. Han, J.~Zhang, and Y.~Xiang, ``Software vulnerability
  detection using deep neural networks: a survey,'' \emph{the IEEE}, vol. 108,
  no.~10, pp. 1825--1848, 2020.

\bibitem{pastor2020not}
J.~Pastor-Galindo, P.~Nespoli, F.~G. M{\'a}rmol, and G.~M. P{\'e}rez, ``The not
  yet exploited goldmine of osint: Opportunities, open challenges and future
  trends,'' \emph{IEEE Access}, vol.~8, pp. 10\,282--10\,304, 2020.

\bibitem{evangelista2020systematic}
J.~R.~G. Evangelista, R.~J. Sassi, M.~Romero, and D.~Napolitano, ``Systematic
  literature review to investigate the application of open source intelligence
  (osint) with artificial intelligence,'' \emph{Journal of Applied Security
  Research}, pp. 1--25, 2020.

\bibitem{sun2018data}
N.~Sun, J.~Zhang, P.~Rimba, S.~Gao, L.~Y. Zhang, and Y.~Xiang, ``Data-driven
  cybersecurity incident prediction: A survey,'' \emph{IEEE Communications
  Surveys \& Tutorials}, vol.~21, no.~2, pp. 1744--1772, 2018.

\bibitem{goodfellow2016deep}
I.~Goodfellow, Y.~Bengio, and A.~Courville, \emph{Deep learning}.\hskip 1em
  plus 0.5em minus 0.4em\relax MIT press, 2016.

\bibitem{dissanayake2020software}
N.~Dissanayake, A.~Jayatilaka, M.~Zahedi, and M.~A. Babar, ``Software security
  patch management--a systematic literature review of challenges, approaches,
  tools and practices,'' \emph{arXiv preprint arXiv:2012.00544}, 2020.

\bibitem{keele2007guidelines}
S.~Keele, ``Guidelines for performing systematic literature reviews in software
  engineering,'' Technical report, Ver. 2.3 EBSE Technical Report. EBSE, Tech.
  Rep., 2007.

\bibitem{wohlin2014guidelines}
C.~Wohlin, ``Guidelines for snowballing in systematic literature studies and a
  replication in software engineering,'' in \emph{the 18th international
  conference on evaluation and assessment in software engineering}, 2014, pp.
  1--10.

\bibitem{edkrantz2015predictingthesis}
M.~Edkrantz, ``Predicting exploit likelihood for cyber vulnerabilities with
  machine learning,'' Master's thesis, 2015.

\bibitem{edkrantz2015predicting}
M.~Edkrantz, S.~Truv{\'e}, and A.~Said, ``Predicting vulnerability exploits in
  the wild,'' in \emph{2015 IEEE 2nd International Conference on Cyber Security
  and Cloud Computing}.\hskip 1em plus 0.5em minus 0.4em\relax IEEE, 2015, pp.
  513--514.

\bibitem{almukaynizi2017proactive}
M.~Almukaynizi, E.~Nunes, K.~Dharaiya, M.~Senguttuvan, J.~Shakarian, and
  P.~Shakarian, ``Proactive identification of exploits in the wild through
  vulnerability mentions online,'' in \emph{2017 International Conference on
  Cyber Conflict (CyCon US)}.\hskip 1em plus 0.5em minus 0.4em\relax IEEE,
  2017, pp. 82--88.

\bibitem{almukaynizi2019patch}
------, ``Patch before exploited: An approach to identify targeted software
  vulnerabilities,'' in \emph{AI in Cybersecurity}.\hskip 1em plus 0.5em minus
  0.4em\relax Springer, 2019, pp. 81--113.

\bibitem{xiao2018patching}
C.~Xiao, A.~Sarabi, Y.~Liu, B.~Li, M.~Liu, and T.~Dumitras, ``From patching
  delays to infection symptoms: Using risk profiles for an early discovery of
  vulnerabilities exploited in the wild,'' in \emph{27th $\{$USENIX$\}$
  Security Symposium ($\{$USENIX$\}$ Security 18)}, 2018, pp. 903--918.

\bibitem{tavabi2018darkembed}
N.~Tavabi, P.~Goyal, M.~Almukaynizi, P.~Shakarian, and K.~Lerman, ``Darkembed:
  Exploit prediction with neural language models,'' in \emph{the AAAI
  Conference on Artificial Intelligence}, vol.~32, no.~1, 2018.

\bibitem{de2020evaluating}
D.~A. de~Sousa, E.~R. de~Faria, and R.~S. Miani, ``Evaluating the performance
  of twitter-based exploit detectors,'' \emph{arXiv preprint arXiv:2011.03113},
  2020.

\bibitem{fang2020fastembed}
Y.~Fang, Y.~Liu, C.~Huang, and L.~Liu, ``Fastembed: Predicting vulnerability
  exploitation possibility based on ensemble machine learning algorithm,''
  \emph{Plos one}, vol.~15, no.~2, p. e0228439, 2020.

\bibitem{huang2020poster}
S.-Y. Huang and Y.~Wu, ``Dynamic software vulnerabilities threat prediction
  through social media contextual analysis,'' in \emph{the 15th Asia Conference
  on Computer and Communications Security}, 2020, pp. 892--894.

\bibitem{jacobs2020improving}
J.~Jacobs, S.~Romanosky, I.~Adjerid, and W.~Baker, ``Improving vulnerability
  remediation through better exploit prediction,'' \emph{Journal of
  Cybersecurity}, vol.~6, no.~1, p. tyaa015, 2020.

\bibitem{yin2020apply}
J.~Yin, M.~Tang, J.~Cao, and H.~Wang, ``Apply transfer learning to
  cybersecurity: Predicting exploitability of vulnerabilities by description,''
  \emph{Knowledge-Based Systems}, vol. 210, p. 106529, 2020.

\bibitem{bhatt2021exploitability}
N.~Bhatt, A.~Anand, and V.~Yadavalli, ``Exploitability prediction of software
  vulnerabilities,'' \emph{Quality and Reliability Engineering International},
  vol.~37, no.~2, pp. 648--663, 2021.

\bibitem{suciu2021expected}
O.~Suciu, C.~Nelson, Z.~Lyu, T.~Bao, and T.~Dumitras, ``Expected
  exploitability: Predicting the development of functional vulnerability
  exploits,'' \emph{arXiv preprint arXiv:2102.07869}, 2021.

\bibitem{younis2014using}
A.~A. Younis and Y.~K. Malaiya, ``Using software structure to predict
  vulnerability exploitation potential,'' in \emph{2014 IEEE Eighth
  International Conference on Software Security and
  Reliability-Companion}.\hskip 1em plus 0.5em minus 0.4em\relax IEEE, 2014,
  pp. 13--18.

\bibitem{yan2017exploitmeter}
G.~Yan, J.~Lu, Z.~Shu, and Y.~Kucuk, ``Exploitmeter: Combining fuzzing with
  machine learning for automated evaluation of software exploitability,'' in
  \emph{2017 IEEE Symposium on Privacy-Aware Computing (PAC)}.\hskip 1em plus
  0.5em minus 0.4em\relax IEEE, 2017, pp. 164--175.

\bibitem{tripathi2017exniffer}
S.~Tripathi, G.~Grieco, and S.~Rawat, ``Exniffer: Learning to prioritize
  crashes by assessing the exploitability from memory dump,'' in \emph{2017
  24th Asia-Pacific Software Engineering Conference (APSEC)}.\hskip 1em plus
  0.5em minus 0.4em\relax IEEE, 2017, pp. 239--248.

\bibitem{cha2012unleashing}
S.~K. Cha, T.~Avgerinos, A.~Rebert, and D.~Brumley, ``Unleashing mayhem on
  binary code,'' in \emph{2012 IEEE Symposium on Security and Privacy}.\hskip
  1em plus 0.5em minus 0.4em\relax IEEE, 2012, pp. 380--394.

\bibitem{grieco2016toward}
G.~Grieco, G.~L. Grinblat, L.~Uzal, S.~Rawat, J.~Feist, and L.~Mounier,
  ``Toward large-scale vulnerability discovery using machine learning,'' in
  \emph{the Sixth ACM Conference on Data and Application Security and Privacy},
  2016, pp. 85--96.

\bibitem{dolan2016lava}
B.~Dolan-Gavitt, P.~Hulin, E.~Kirda, T.~Leek, A.~Mambretti, W.~Robertson,
  F.~Ulrich, and R.~Whelan, ``Lava: Large-scale automated vulnerability
  addition,'' in \emph{2016 IEEE Symposium on Security and Privacy (SP)}.\hskip
  1em plus 0.5em minus 0.4em\relax IEEE, 2016, pp. 110--121.

\bibitem{zhang2018assisting}
L.~Zhang and V.~L. Thing, ``Assisting vulnerability detection by prioritizing
  crashes with incremental learning,'' in \emph{TENCON 2018-2018 IEEE Region 10
  Conference}.\hskip 1em plus 0.5em minus 0.4em\relax IEEE, 2018, pp.
  2080--2085.

\bibitem{cve}
\BIBentryALTinterwordspacing
MITRE, ``Common vulnerabilities and exposures.'' [Online]. Available:
  \url{https://cve.mitre.org/}
\BIBentrySTDinterwordspacing

\bibitem{exploitdb}
\BIBentryALTinterwordspacing
O.~Security, ``Exploit database.'' [Online]. Available:
  \url{https://www.exploit-db.com}
\BIBentrySTDinterwordspacing

\bibitem{symantec}
\BIBentryALTinterwordspacing
Broadcom, ``Symantec attack signatures.'' [Online]. Available:
  \url{https://bit.ly/symantec_att_sign}
\BIBentrySTDinterwordspacing

\bibitem{ms_security}
\BIBentryALTinterwordspacing
Microsoft, ``Microsoft security advisories.'' [Online]. Available:
  \url{https://bit.ly/ms_sec_advisories}
\BIBentrySTDinterwordspacing

\bibitem{zeroday_initiative}
\BIBentryALTinterwordspacing
T.~Micro, ``Zeroday initiative security advisories.'' [Online]. Available:
  \url{https://bit.ly/zeroday_sec}
\BIBentrySTDinterwordspacing

\bibitem{metasploit}
\BIBentryALTinterwordspacing
Rapid7, ``Metasploit security advisories.'' [Online]. Available:
  \url{https://www.rapid7.com/db/modules}
\BIBentrySTDinterwordspacing

\bibitem{securityfocus}
\BIBentryALTinterwordspacing
S.~Inc, ``Bugtraq vulnerability database.'' [Online]. Available:
  \url{http://www.securityfocus.com}
\BIBentrySTDinterwordspacing

\bibitem{recorded_future}
\BIBentryALTinterwordspacing
R.~Future, ``Recorded future security advisories.'' [Online]. Available:
  \url{https://bit.ly/rf_sec}
\BIBentrySTDinterwordspacing

\bibitem{kenna_security}
\BIBentryALTinterwordspacing
I.~Kenna~Security, ``Kenna security.'' [Online]. Available:
  \url{http://www.kennasecurity.com}
\BIBentrySTDinterwordspacing

\bibitem{eset}
\BIBentryALTinterwordspacing
ESET, ``Eset security advisories.'' [Online]. Available:
  \url{https://bit.ly/eset_virus}
\BIBentrySTDinterwordspacing

\bibitem{trend_micro}
\BIBentryALTinterwordspacing
T.~Micro, ``Trend micro security advisories.'' [Online]. Available:
  \url{https://bit.ly/trend_micro_sec}
\BIBentrySTDinterwordspacing

\bibitem{nunes2016darknet}
E.~Nunes, A.~Diab, A.~Gunn, E.~Marin, V.~Mishra, V.~Paliath, J.~Robertson,
  J.~Shakarian, A.~Thart, and P.~Shakarian, ``Darknet and deepnet mining for
  proactive cybersecurity threat intelligence,'' in \emph{2016 IEEE Conference
  on Intelligence and Security Informatics (ISI)}.\hskip 1em plus 0.5em minus
  0.4em\relax IEEE, 2016, pp. 7--12.

\bibitem{chen2016xgboost}
T.~Chen and C.~Guestrin, ``Xgboost: A scalable tree boosting system,'' in
  \emph{the 22nd International Conference on Knowledge Discovery and Data
  Mining}, 2016, pp. 785--794.

\bibitem{ke2017lightgbm}
G.~Ke, Q.~Meng, T.~Finley, T.~Wang, W.~Chen, W.~Ma, Q.~Ye, and T.-Y. Liu,
  ``Lightgbm: A highly efficient gradient boosting decision tree,''
  \emph{Advances in neural information processing systems}, vol.~30, pp.
  3146--3154, 2017.

\bibitem{devlin2018bert}
J.~Devlin, M.-W. Chang, K.~Lee, and K.~Toutanova, ``Bert: Pre-training of deep
  bidirectional transformers for language understanding,'' \emph{arXiv preprint
  arXiv:1810.04805}, 2018.

\bibitem{zhu2015aligning}
Y.~Zhu, R.~Kiros, R.~Zemel, R.~Salakhutdinov, R.~Urtasun, A.~Torralba, and
  S.~Fidler, ``Aligning books and movies: Towards story-like visual
  explanations by watching movies and reading books,'' in \emph{the IEEE
  international conference on computer vision}, 2015, pp. 19--27.

\bibitem{wikipedia}
\BIBentryALTinterwordspacing
W.~Foundation, ``Wikipedia pages.'' [Online]. Available:
  \url{https://www.wikipedia.org}
\BIBentrySTDinterwordspacing

\bibitem{cvss_v31}
\BIBentryALTinterwordspacing
FIRST, ``Cvss version 3.1.'' [Online]. Available:
  \url{https://www.first.org/cvss/v3.1/specification-document}
\BIBentrySTDinterwordspacing

\bibitem{bernaschi2002remus}
M.~Bernaschi, E.~Gabrielli, and L.~V. Mancini, ``Remus: A security-enhanced
  operating system,'' \emph{ACM Transactions on Information and System Security
  (TISSEC)}, vol.~5, no.~1, pp. 36--61, 2002.

\bibitem{manadhata2010attack}
P.~K. Manadhata and J.~M. Wing, ``An attack surface metric,'' \emph{IEEE
  Transactions on Software Engineering}, vol.~37, no.~3, pp. 371--386, 2010.

\bibitem{horwitz1990interprocedural}
S.~Horwitz, T.~Reps, and D.~Binkley, ``Interprocedural slicing using dependence
  graphs,'' \emph{ACM Transactions on Programming Languages and Systems
  (TOPLAS)}, vol.~12, no.~1, pp. 26--60, 1990.

\bibitem{basic_fuzzing_framework}
\BIBentryALTinterwordspacing
CERT, ``Basic fuzzing framework.'' [Online]. Available:
  \url{https://bit.ly/basic_fuzzing_framework}
\BIBentrySTDinterwordspacing

\bibitem{ofuzz}
\BIBentryALTinterwordspacing
S.~K. Cha, ``Ofuzz.'' [Online]. Available:
  \url{https://github.com/sangkilc/ofuzz}
\BIBentrySTDinterwordspacing

\bibitem{crammer2006online}
K.~Crammer, O.~Dekel, J.~Keshet, S.~Shalev-Shwartz, and Y.~Singer, ``Online
  passive aggressive algorithms,'' 2006.

\bibitem{jacobs2019exploit}
J.~Jacobs, S.~Romanosky, B.~Edwards, M.~Roytman, and I.~Adjerid, ``Exploit
  prediction scoring system (epss),'' \emph{arXiv preprint arXiv:1908.04856},
  2019.

\bibitem{jacobs2021exploit}
J.~Jacobs, S.~Romanosky, B.~Edwards, I.~Adjerid, and M.~Roytman, ``Exploit
  prediction scoring system (epss),'' \emph{Digital Threats: Research and
  Practice}, vol.~2, no.~3, pp. 1--17, 2021.

\bibitem{chen2019using}
H.~Chen, R.~Liu, N.~Park, and V.~Subrahmanian, ``Using twitter to predict when
  vulnerabilities will be exploited,'' in \emph{the 25th International
  Conference on Knowledge Discovery \& Data Mining}, 2019, pp. 3143--3152.

\bibitem{chen2019vest}
H.~Chen, J.~Liu, R.~Liu, N.~Park, and V.~Subrahmanian, ``Vest: A system for
  vulnerability exploit scoring \& timing,'' in \emph{IJCAI}, 2019, pp.
  6503--6505.

\bibitem{kivela2014multilayer}
M.~Kivel{\"a}, A.~Arenas, M.~Barthelemy, J.~P. Gleeson, Y.~Moreno, and M.~A.
  Porter, ``Multilayer networks,'' \emph{Journal of complex networks}, vol.~2,
  no.~3, pp. 203--271, 2014.

\bibitem{yamamoto2015text}
Y.~Yamamoto, D.~Miyamoto, and M.~Nakayama, ``Text-mining approach for
  estimating vulnerability score,'' in \emph{2015 4th International Workshop on
  Building Analysis Datasets and Gathering Experience Returns for Security
  (BADGERS)}.\hskip 1em plus 0.5em minus 0.4em\relax IEEE, 2015, pp. 67--73.

\bibitem{wen2015novel}
T.~Wen, Y.~Zhang, Y.~Dong, and G.~Yang, ``A novel automatic severity
  vulnerability assessment framework,'' \emph{Journal of Communications},
  vol.~10, no.~5, pp. 320--329, 2015.

\bibitem{toloudis2016associating}
D.~Toloudis, G.~Spanos, and L.~Angelis, ``Associating the severity of
  vulnerabilities with their description,'' in \emph{International Conference
  on Advanced Information Systems Engineering}.\hskip 1em plus 0.5em minus
  0.4em\relax Springer, 2016, pp. 231--242.

\bibitem{ognawala2018automatically}
S.~Ognawala, R.~N. Amato, A.~Pretschner, and P.~Kulkarni, ``Automatically
  assessing vulnerabilities discovered by compositional analysis,'' in
  \emph{the 1st International Workshop on Machine Learning and Software
  Engineering in Symbiosis}, 2018, pp. 16--25.

\bibitem{ognawala2016macke}
S.~Ognawala, M.~Ochoa, A.~Pretschner, and T.~Limmer, ``Macke: Compositional
  analysis of low-level vulnerabilities with symbolic execution,'' in \emph{the
  31st IEEE/ACM International Conference on Automated Software Engineering},
  2016, pp. 780--785.

\bibitem{elbaz2020fighting}
C.~Elbaz, L.~Rilling, and C.~Morin, ``Fighting n-day vulnerabilities with
  automated cvss vector prediction at disclosure,'' in \emph{the 15th
  International Conference on Availability, Reliability and Security}, 2020,
  pp. 1--10.

\bibitem{jiang2020approach}
Y.~Jiang and Y.~Atif, ``An approach to discover and assess vulnerability
  severity automatically in cyber-physical systems,'' in \emph{13th
  International Conference on Security of Information and Networks}, 2020, pp.
  1--8.

\bibitem{gawron2017automatic}
M.~Gawron, F.~Cheng, and C.~Meinel, ``Automatic vulnerability classification
  using machine learning,'' in \emph{International Conference on Risks and
  Security of Internet and Systems}.\hskip 1em plus 0.5em minus 0.4em\relax
  Springer, 2017, pp. 3--17.

\bibitem{gong2019joint}
X.~Gong, Z.~Xing, X.~Li, Z.~Feng, and Z.~Han, ``Joint prediction of multiple
  vulnerability characteristics through multi-task learning,'' in \emph{2019
  24th International Conference on Engineering of Complex Computer Systems
  (ICECCS)}.\hskip 1em plus 0.5em minus 0.4em\relax IEEE, 2019, pp. 31--40.

\bibitem{chen2010categorization}
Z.~Chen, Y.~Zhang, and Z.~Chen, ``A categorization framework for common
  computer vulnerabilities and exposures,'' \emph{The Computer Journal},
  vol.~53, no.~5, pp. 551--580, 2010.

\bibitem{ruohonen2017classifying}
J.~Ruohonen, ``Classifying web exploits with topic modeling,'' in \emph{2017
  28th International Workshop on Database and Expert Systems Applications
  (DEXA)}.\hskip 1em plus 0.5em minus 0.4em\relax IEEE, 2017, pp. 93--97.

\bibitem{aksu2018automated}
M.~U. Aksu, K.~Bicakci, M.~H. Dilek, A.~M. Ozbayoglu, and E.~{\i}. Tatli,
  ``Automated generation of attack graphs using nvd,'' in \emph{the 8th
  Conference on Data and Application Security and Privacy}, 2018, pp. 135--142.

\bibitem{stanley2002evolving}
K.~O. Stanley and R.~Miikkulainen, ``Evolving neural networks through
  augmenting topologies,'' \emph{Evolutionary computation}, vol.~10, no.~2, pp.
  99--127, 2002.

\bibitem{liu2019automated}
H.~Liu and B.~Li, ``Automated classification of attacker privileges based on
  deep neural network,'' in \emph{International Conference on Smart Computing
  and Communication}.\hskip 1em plus 0.5em minus 0.4em\relax Springer, 2019,
  pp. 180--189.

\bibitem{kanakogi2021tracing}
K.~Kanakogi, H.~Washizaki, Y.~Fukazawa, S.~Ogata, T.~Okubo, T.~Kato, H.~Kanuka,
  A.~Hazeyama, and N.~Yoshioka, ``Tracing capec attack patterns from cve
  vulnerability information using natural language processing technique,'' in
  \emph{the 54th Hawaii International Conference on System Sciences}, 2021, p.
  6996.

\bibitem{cvss_v2}
\BIBentryALTinterwordspacing
FIRST, ``Cvss version 2.'' [Online]. Available:
  \url{https://www.first.org/cvss/v2/guide}
\BIBentrySTDinterwordspacing

\bibitem{cvss_v3}
\BIBentryALTinterwordspacing
------, ``Cvss version 3.'' [Online]. Available:
  \url{https://www.first.org/cvss/v3.0/specification-document}
\BIBentrySTDinterwordspacing

\bibitem{blei2007supervised}
D.~M. Blei and J.~D. McAuliffe, ``Supervised topic models,'' in \emph{the 20th
  International Conference on Neural Information Processing Systems}, 2007, p.
  121–128.

\bibitem{xforce}
\BIBentryALTinterwordspacing
I.~S. Services, ``Online database x-force.'' [Online]. Available:
  \url{http://www.iss.net/xforce}
\BIBentrySTDinterwordspacing

\bibitem{wold1987principal}
S.~Wold, K.~Esbensen, and P.~Geladi, ``Principal component analysis,''
  \emph{Chemometrics and intelligent laboratory systems}, vol.~2, no. 1-3, pp.
  37--52, 1987.

\bibitem{kipf2016semi}
T.~N. Kipf and M.~Welling, ``Semi-supervised classification with graph
  convolutional networks,'' \emph{arXiv preprint arXiv:1609.02907}, 2016.

\bibitem{zhang2021survey}
Y.~Zhang and Q.~Yang, ``A survey on multi-task learning,'' \emph{IEEE
  Transactions on Knowledge and Data Engineering}, 2021.

\bibitem{bahdanau2014neural}
D.~Bahdanau, K.~Cho, and Y.~Bengio, ``Neural machine translation by jointly
  learning to align and translate,'' \emph{arXiv preprint arXiv:1409.0473},
  2014.

\bibitem{secunia}
\BIBentryALTinterwordspacing
S.~Inc., ``Secunia vulnerability advisories.'' [Online]. Available:
  \url{http://secunia.com}
\BIBentrySTDinterwordspacing

\bibitem{hastie2009elements}
T.~Hastie, R.~Tibshirani, and J.~Friedman, \emph{The elements of statistical
  learning: data mining, inference, and prediction}.\hskip 1em plus 0.5em minus
  0.4em\relax Springer Science \& Business Media, 2009.

\bibitem{kim2014convolutional}
Y.~Kim, ``Convolutional neural networks for sentence classification,'' in
  \emph{the 2014 Conference on Empirical Methods in Natural Language Processing
  ({EMNLP})}.\hskip 1em plus 0.5em minus 0.4em\relax Association for
  Computational Linguistics, 2014, pp. 1746--1751.

\bibitem{le2014distributed}
Q.~Le and T.~Mikolov, ``Distributed representations of sentences and
  documents,'' in \emph{International conference on machine learning}.\hskip
  1em plus 0.5em minus 0.4em\relax PMLR, 2014, pp. 1188--1196.

\bibitem{capec}
\BIBentryALTinterwordspacing
MITRE, ``Common attack pattern enumeration and classification.'' [Online].
  Available: \url{https://capec.mitre.org}
\BIBentrySTDinterwordspacing

\bibitem{pan2009survey}
S.~J. Pan and Q.~Yang, ``A survey on transfer learning,'' \emph{IEEE
  Transactions on knowledge and data engineering}, vol.~22, no.~10, pp.
  1345--1359, 2009.

\bibitem{kudjo2019improving}
P.~K. Kudjo, J.~Chen, M.~Zhou, S.~Mensah, and R.~Huang, ``Improving the
  accuracy of vulnerability report classification using term frequency-inverse
  gravity moment,'' in \emph{2019 IEEE 19th International Conference on
  Software Quality, Reliability and Security (QRS)}.\hskip 1em plus 0.5em minus
  0.4em\relax IEEE, 2019, pp. 248--259.

\bibitem{chen2020automatic}
J.~Chen, P.~K. Kudjo, S.~Mensah, S.~A. Brown, and G.~Akorfu, ``An automatic
  software vulnerability classification framework using term frequency-inverse
  gravity moment and feature selection,'' \emph{Journal of Systems and
  Software}, vol. 167, p. 110616, 2020.

\bibitem{kudjo2020effect}
P.~K. Kudjo, J.~Chen, S.~Mensah, R.~Amankwah, and C.~Kudjo, ``The effect of
  bellwether analysis on software vulnerability severity prediction models,''
  \emph{Software Quality Journal}, pp. 1--34, 2020.

\bibitem{malhotra2021severity}
R.~Malhotra, ``Severity prediction of software vulnerabilities using textual
  data,'' in \emph{International Conference on Recent Trends in Machine
  Learning, IoT, Smart Cities and Applications}.\hskip 1em plus 0.5em minus
  0.4em\relax Springer, 2021, pp. 453--464.

\bibitem{chen2016turning}
K.~Chen, Z.~Zhang, J.~Long, and H.~Zhang, ``Turning from tf-idf to tf-igm for
  term weighting in text classification,'' \emph{Expert Systems with
  Applications}, vol.~66, pp. 245--260, 2016.

\bibitem{wang2019intelligent}
P.~Wang, Y.~Zhou, B.~Sun, and W.~Zhang, ``Intelligent prediction of
  vulnerability severity level based on text mining and xgbboost,'' in
  \emph{2019 Eleventh International Conference on Advanced Computational
  Intelligence (ICACI)}.\hskip 1em plus 0.5em minus 0.4em\relax IEEE, 2019, pp.
  72--77.

\bibitem{liu2019vulnerability}
K.~Liu, Y.~Zhou, Q.~Wang, and X.~Zhu, ``Vulnerability severity prediction with
  deep neural network,'' in \emph{2019 5th International Conference on Big Data
  and Information Analytics (BigDIA)}.\hskip 1em plus 0.5em minus 0.4em\relax
  IEEE, 2019, pp. 114--119.

\bibitem{sharma2021software}
R.~Sharma, R.~Sibal, and S.~Sabharwal, ``Software vulnerability prioritization
  using vulnerability description,'' \emph{International Journal of System
  Assurance Engineering and Management}, vol.~12, no.~1, pp. 58--64, 2021.

\bibitem{sahin2019conceptual}
S.~E. Sahin and A.~Tosun, ``A conceptual replication on predicting the severity
  of software vulnerabilities,'' in \emph{the Evaluation and Assessment on
  Software Engineering}, 2019, pp. 244--250.

\bibitem{nakagawa2019character}
S.~Nakagawa, T.~Nagai, H.~Kanehara, K.~Furumoto, M.~Takita, Y.~Shiraishi,
  T.~Takahashi, M.~Mohri, Y.~Takano, and M.~Morii, ``Character-level
  convolutional neural network for predicting severity of software
  vulnerability from vulnerability description,'' \emph{IEICE Transactions on
  Information and Systems}, vol. 102, no.~9, pp. 1679--1682, 2019.

\bibitem{zhang2020general}
X.~Zhang, H.~Xie, H.~Yang, H.~Shao, and M.~Zhu, ``A general framework to
  understand vulnerabilities in information systems,'' \emph{IEEE Access},
  vol.~8, pp. 121\,858--121\,873, 2020.

\bibitem{khazaei2016automatic}
A.~Khazaei, M.~Ghasemzadeh, and V.~Derhami, ``An automatic method for cvss
  score prediction using vulnerabilities description,'' \emph{Journal of
  Intelligent \& Fuzzy Systems}, vol.~30, no.~1, pp. 89--96, 2016.

\bibitem{spanos2013wivss}
G.~Spanos, A.~Sioziou, and L.~Angelis, ``Wivss: a new methodology for scoring
  information systems vulnerabilities,'' in \emph{the 17th panhellenic
  conference on informatics}, 2013, pp. 83--90.

\bibitem{lai2015recurrent}
S.~Lai, L.~Xu, K.~Liu, and J.~Zhao, ``Recurrent convolutional neural networks
  for text classification,'' in \emph{the AAAI Conference on Artificial
  Intelligence}, vol.~29, no.~1, 2015.

\bibitem{hochreiter1997long}
S.~Hochreiter and J.~Schmidhuber, ``Long short-term memory,'' \emph{Neural
  computation}, vol.~9, no.~8, pp. 1735--1780, 1997.

\bibitem{zhang2015character}
X.~Zhang, J.~Zhao, and Y.~LeCun, ``Character-level convolutional networks for
  text classification,'' \emph{arXiv preprint arXiv:1509.01626}, 2015.

\bibitem{chen2019vase}
H.~Chen, J.~Liu, R.~Liu, N.~Park, and V.~Subrahmanian, ``Vase: A twitter-based
  vulnerability analysis and score engine,'' in \emph{2019 IEEE International
  Conference on Data Mining (ICDM)}.\hskip 1em plus 0.5em minus 0.4em\relax
  IEEE, 2019, pp. 976--981.

\bibitem{anwar2020cleaning}
A.~Anwar, A.~Abusnaina, S.~Chen, F.~Li, and D.~Mohaisen, ``Cleaning the nvd:
  Comprehensive quality assessment, improvements, and analyses,'' \emph{arXiv
  preprint arXiv:2006.15074}, 2020.

\bibitem{anwar2021cleaning}
------, ``Cleaning the nvd: Comprehensive quality assessment, improvements, and
  analyses,'' \emph{IEEE Transactions on Dependable and Secure Computing},
  2021.

\bibitem{wang2010vulnerability}
J.~A. Wang and M.~Guo, ``Vulnerability categorization using bayesian
  networks,'' in \emph{the sixth annual workshop on cyber security and
  information intelligence research}, 2010, pp. 1--4.

\bibitem{shuai2013automatic}
B.~Shuai, H.~Li, M.~Li, Q.~Zhang, and C.~Tang, ``Automatic classification for
  vulnerability based on machine learning,'' in \emph{2013 IEEE International
  Conference on Information and Automation (ICIA)}.\hskip 1em plus 0.5em minus
  0.4em\relax IEEE, 2013, pp. 312--318.

\bibitem{na2016study}
S.~Na, T.~Kim, and H.~Kim, ``A study on the classification of common
  vulnerabilities and exposures using na{\"\i}ve bayes,'' in
  \emph{International Conference on Broadband and Wireless Computing,
  Communication and Applications}.\hskip 1em plus 0.5em minus 0.4em\relax
  Springer, 2016, pp. 657--662.

\bibitem{ruohonen2018toward}
J.~Ruohonen and V.~Lepp{\"a}nen, ``Toward validation of textual information
  retrieval techniques for software weaknesses,'' in \emph{International
  Conference on Database and Expert Systems Applications}.\hskip 1em plus 0.5em
  minus 0.4em\relax Springer, 2018, pp. 265--277.

\bibitem{huang2019automatic}
G.~Huang, Y.~Li, Q.~Wang, J.~Ren, Y.~Cheng, and X.~Zhao, ``Automatic
  classification method for software vulnerability based on deep neural
  network,'' \emph{IEEE Access}, vol.~7, pp. 28\,291--28\,298, 2019.

\bibitem{aota2020automation}
M.~Aota, H.~Kanehara, M.~Kubo, N.~Murata, B.~Sun, and T.~Takahashi,
  ``Automation of vulnerability classification from its description using
  machine learning,'' in \emph{2020 IEEE Symposium on Computers and
  Communications (ISCC)}.\hskip 1em plus 0.5em minus 0.4em\relax IEEE, 2020,
  pp. 1--7.

\bibitem{aghaei2020threatzoom}
E.~Aghaei, W.~Shadid, and E.~Al-Shaer, ``Threatzoom: Cve2cwe using hierarchical
  neural network,'' \emph{arXiv preprint arXiv:2009.11501}, 2020.

\bibitem{das2021v2w}
S.~S. Das, E.~Serra, M.~Halappanavar, A.~Pothen, and E.~Al-Shaer, ``V2w-bert: A
  framework for effective hierarchical multiclass classification of software
  vulnerabilities,'' in \emph{2021 IEEE 8th International Conference on Data
  Science and Advanced Analytics (DSAA)}.\hskip 1em plus 0.5em minus
  0.4em\relax IEEE, 2021, pp. 1--12.

\bibitem{zou2019muvuldeepecker}
D.~Zou, S.~Wang, S.~Xu, Z.~Li, and H.~Jin, ``$\mu$vuldeepecker: A deep
  learning-based system for multiclass vulnerability detection,'' \emph{IEEE
  Transactions on Dependable and Secure Computing}, 2019.

\bibitem{murtaza2016mining}
S.~S. Murtaza, W.~Khreich, A.~Hamou-Lhadj, and A.~B. Bener, ``Mining trends and
  patterns of software vulnerabilities,'' \emph{Journal of Systems and
  Software}, vol. 117, pp. 218--228, 2016.

\bibitem{lin2017machine}
Z.~Lin, X.~Li, and X.~Kuang, ``Machine learning in vulnerability databases,''
  in \emph{2017 10th International Symposium on Computational Intelligence and
  Design (ISCID)}, vol.~1.\hskip 1em plus 0.5em minus 0.4em\relax IEEE, 2017,
  pp. 108--113.

\bibitem{han2018deepweak}
Z.~Han, X.~Li, H.~Liu, Z.~Xing, and Z.~Feng, ``Deepweak: Reasoning common
  software weaknesses via knowledge graph embedding,'' in \emph{2018 IEEE 25th
  International Conference on Software Analysis, Evolution and Reengineering
  (SANER)}.\hskip 1em plus 0.5em minus 0.4em\relax IEEE, 2018, pp. 456--466.

\bibitem{venter2008standardising}
H.~S. Venter, J.~H. Eloff, and Y.~Li, ``Standardising vulnerability
  categories,'' \emph{Computers \& Security}, vol.~27, no. 3-4, pp. 71--83,
  2008.

\bibitem{neuhaus2010security}
S.~Neuhaus and T.~Zimmermann, ``Security trend analysis with cve topic
  models,'' in \emph{2010 IEEE 21st International Symposium on Software
  Reliability Engineering}.\hskip 1em plus 0.5em minus 0.4em\relax IEEE, 2010,
  pp. 111--120.

\bibitem{mounika2019analyzing}
V.~Mounika, X.~Yuan, and K.~Bandaru, ``Analyzing cve database using
  unsupervised topic modelling,'' in \emph{2019 International Conference on
  Computational Science and Computational Intelligence}, 2019, pp. 72--77.

\bibitem{vanamala2020topic}
M.~Vanamala, X.~Yuan, and K.~Roy, ``Topic modeling and classification of common
  vulnerabilities and exposures database,'' in \emph{2020 International
  Conference on Artificial Intelligence, Big Data, Computing and Data
  Communication Systems (icABCD)}.\hskip 1em plus 0.5em minus 0.4em\relax IEEE,
  2020, pp. 1--5.

\bibitem{aljedaani2020lda}
W.~Aljedaani, Y.~Javed, and M.~Alenezi, ``Lda categorization of security bug
  reports in chromium projects,'' in \emph{the 2020 European Symposium on
  Software Engineering}, 2020, pp. 154--161.

\bibitem{williams2018analyzing}
M.~A. Williams, S.~Dey, R.~C. Barranco, S.~M. Naim, M.~S. Hossain, and
  M.~Akbar, ``Analyzing evolving trends of vulnerabilities in national
  vulnerability database,'' in \emph{2018 IEEE International Conference on Big
  Data (Big Data)}.\hskip 1em plus 0.5em minus 0.4em\relax IEEE, 2018, pp.
  3011--3020.

\bibitem{williams2020vulnerability}
M.~A. Williams, R.~C. Barranco, S.~M. Naim, S.~Dey, M.~S. Hossain, and
  M.~Akbar, ``A vulnerability analysis and prediction framework,''
  \emph{Computers \& Security}, vol.~92, p. 101751, 2020.

\bibitem{russo2019summarizing}
E.~R. Russo, A.~D~Sorbo, C.~A. Visaggio, and G.~Canfora, ``Summarizing
  vulnerabilities’ descriptions to support experts during vulnerability
  assessment activities,'' \emph{Journal of Systems and Software}, vol. 156,
  pp. 84--99, 2019.

\bibitem{kursa2010boruta}
M.~B. Kursa, A.~Jankowski, and W.~R. Rudnicki, ``Boruta--a system for feature
  selection,'' \emph{Fundamenta Informaticae}, vol. 101, no.~4, pp. 271--285,
  2010.

\bibitem{sard}
\BIBentryALTinterwordspacing
NIST, ``Software assurance reference dataset (sard).'' [Online]. Available:
  \url{https://samate.nist.gov/SRD}
\BIBentrySTDinterwordspacing

\bibitem{han2000mining}
J.~Han, J.~Pei, and Y.~Yin, ``Mining frequent patterns without candidate
  generation,'' \emph{ACM sigmod record}, vol.~29, no.~2, pp. 1--12, 2000.

\bibitem{kohonen1990self}
T.~Kohonen, ``The self-organizing map,'' \emph{the IEEE}, vol.~78, no.~9, pp.
  1464--1480, 1990.

\bibitem{owasp_website}
\BIBentryALTinterwordspacing
OWASP, ``Open web application security project.'' [Online]. Available:
  \url{https://bit.ly/owasp_main}
\BIBentrySTDinterwordspacing

\bibitem{naim2017scalable}
S.~M. Naim, A.~P. Boedihardjo, and M.~S. Hossain, ``A scalable model for
  tracking topical evolution in large document collections,'' in \emph{2017
  IEEE International Conference on Big Data (Big Data)}.\hskip 1em plus 0.5em
  minus 0.4em\relax IEEE, 2017, pp. 726--735.

\bibitem{barranco2019analyzing}
R.~C. Barranco, A.~P. Boedihardjo, and M.~S. Hossain, ``Analyzing evolving
  stories in news articles,'' \emph{International Journal of Data Science and
  Analytics}, vol.~8, no.~3, pp. 241--256, 2019.

\bibitem{weerawardhana2014automated}
S.~Weerawardhana, S.~Mukherjee, I.~Ray, and A.~Howe, ``Automated extraction of
  vulnerability information for home computer security,'' in
  \emph{International Symposium on Foundations and Practice of Security}.\hskip
  1em plus 0.5em minus 0.4em\relax Springer, 2014, pp. 356--366.

\bibitem{dong2019towards}
Y.~Dong, W.~Guo, Y.~Chen, X.~Xing, Y.~Zhang, and G.~Wang, ``Towards the
  detection of inconsistencies in public security vulnerability reports,'' in
  \emph{28th $\{$USENIX$\}$ Security Symposium}, 2019, pp. 869--885.

\bibitem{gonzalez2019automated}
D.~Gonzalez, H.~Hastings, and M.~Mirakhorli, ``Automated characterization of
  software vulnerabilities,'' in \emph{2019 IEEE International Conference on
  Software Maintenance and Evolution (ICSME)}.\hskip 1em plus 0.5em minus
  0.4em\relax IEEE, 2019, pp. 135--139.

\bibitem{vdo}
\BIBentryALTinterwordspacing
NIST, ``Vulnerability description ontology.'' [Online]. Available:
  \url{https://bit.ly/nist_vdo}
\BIBentrySTDinterwordspacing

\bibitem{binyamini2020automated}
H.~Binyamini, R.~Bitton, M.~Inokuchi, T.~Yagyu, Y.~Elovici, and A.~Shabtai,
  ``An automated, end-to-end framework for modeling attacks from vulnerability
  descriptions,'' \emph{arXiv preprint arXiv:2008.04377}, 2020.

\bibitem{binyamini2021framework}
------, ``A framework for modeling cyber attack techniques from security
  vulnerability descriptions,'' in \emph{the 27th ACM SIGKDD Conference on
  Knowledge Discovery \& Data Mining}, 2021, pp. 2574--2583.

\bibitem{ou2005mulval}
X.~Ou, S.~Govindavajhala, and A.~W. Appel, ``Mulval: A logic-based network
  security analyzer,'' in \emph{USENIX security symposium}, vol.~8.\hskip 1em
  plus 0.5em minus 0.4em\relax Baltimore, MD, 2005, pp. 113--128.

\bibitem{guo2020predicting}
H.~Guo, Z.~Xing, and X.~Li, ``Predicting missing information of key aspects in
  vulnerability reports,'' \emph{arXiv preprint arXiv:2008.02456}, 2020.

\bibitem{guo2021detecting}
H.~Guo, S.~Chen, Z.~Xing, X.~Li, Y.~Bai, and J.~Sun, ``Detecting and augmenting
  missing key aspects in vulnerability descriptions,'' \emph{ACM Transactions
  on Software Engineering and Methodology}, 2021.

\bibitem{waareus2020automated}
E.~W{\aa}reus and M.~Hell, ``Automated cpe labeling of cve summaries with
  machine learning,'' in \emph{International Conference on Detection of
  Intrusions and Malware, and Vulnerability Assessment}.\hskip 1em plus 0.5em
  minus 0.4em\relax Springer, 2020, pp. 3--22.

\bibitem{yitagesu2021vulnpos}
S.~Yitagesu, X.~Zhang, Z.~Feng, X.~Li, and Z.~Xing, ``Automatic part-of-speech
  tagging for security vulnerability descriptions,'' in \emph{2021 IEEE/ACM
  18th International Conference on Mining Software Repositories (MSR)}.\hskip
  1em plus 0.5em minus 0.4em\relax IEEE, 2021, pp. 29--40.

\bibitem{marcus1993building}
M.~Marcus, B.~Santorini, and M.~A. Marcinkiewicz, ``Building a large annotated
  corpus of english: The penn treebank,'' 1993.

\bibitem{sun2021generating}
J.~Sun, Z.~Xing, H.~Guo, D.~Ye, X.~Li, X.~Xu, and L.~Zhu, ``Generating
  informative cve description from exploitdb posts by extractive
  summarization,'' \emph{arXiv preprint arXiv:2101.01431}, 2021.

\bibitem{horawalavithana2019mentions}
S.~Horawalavithana, A.~Bhattacharjee, R.~Liu, N.~Choudhury, L.~O.~Hall, and
  A.~Iamnitchi, ``Mentions of security vulnerabilities on reddit, twitter and
  github,'' in \emph{IEEE/WIC/ACM International Conference on Web
  Intelligence}, 2019, pp. 200--207.

\bibitem{xiao2019embedding}
H.~Xiao, Z.~Xing, X.~Li, and H.~Guo, ``Embedding and predicting software
  security entity relationships: A knowledge graph based approach,'' in
  \emph{International Conference on Neural Information Processing}.\hskip 1em
  plus 0.5em minus 0.4em\relax Springer, 2019, pp. 50--63.

\bibitem{othmane2017time}
L.~B. Othmane, G.~Chehrazi, E.~Bodden, P.~Tsalovski, and A.~D. Brucker, ``Time
  for addressing software security issues: Prediction models and impacting
  factors,'' \emph{Data Science and Engineering}, vol.~2, no.~2, pp. 107--124,
  2017.

\bibitem{finkel2005incorporating}
J.~R. Finkel, T.~Grenager, and C.~D. Manning, ``Incorporating non-local
  information into information extraction systems by gibbs sampling,'' in
  \emph{the 43rd Annual Meeting of the Association for Computational
  Linguistics (ACL’05)}, 2005, pp. 363--370.

\bibitem{cve_details}
\BIBentryALTinterwordspacing
S.~Özkan, ``Cve details.'' [Online]. Available:
  \url{https://www.cvedetails.com}
\BIBentrySTDinterwordspacing

\bibitem{securitytracker}
\BIBentryALTinterwordspacing
SecurityTracker, ``Securitytracker vulnerability database.'' [Online].
  Available: \url{https://securitytracker.com}
\BIBentrySTDinterwordspacing

\bibitem{openwall}
\BIBentryALTinterwordspacing
O.~Project, ``Openwall security advisories.'' [Online]. Available:
  \url{https://bit.ly/sec_openwall}
\BIBentrySTDinterwordspacing

\bibitem{peters2018deep}
M.~E. Peters, M.~Neumann, M.~Iyyer, M.~Gardner, C.~Clark, K.~Lee, and
  L.~Zettlemoyer, ``Deep contextualized word representations,'' \emph{arXiv
  preprint arXiv:1802.05365}, 2018.

\bibitem{cpe}
\BIBentryALTinterwordspacing
MITRE, ``Common platform enumeration.'' [Online]. Available:
  \url{https://cpe.mitre.org}
\BIBentrySTDinterwordspacing

\bibitem{ben2015factors}
L.~ben Othmane, G.~Chehrazi, E.~Bodden, P.~Tsalovski, A.~D. Brucker, and
  P.~Miseldine, ``Factors impacting the effort required to fix security
  vulnerabilities,'' in \emph{International Conference on Information
  Security}.\hskip 1em plus 0.5em minus 0.4em\relax Springer, 2015, pp.
  102--119.

\bibitem{kumar2020legal}
R.~S.~S. Kumar, J.~Penney, B.~Schneier, and K.~Albert, ``Legal risks of
  adversarial machine learning research,'' \emph{arXiv preprint
  arXiv:2006.16179}, 2020.

\bibitem{zhang2013predicting}
H.~Zhang, L.~Gong, and S.~Versteeg, ``Predicting bug-fixing time: an empirical
  study of commercial software projects,'' in \emph{2013 35th International
  Conference on Software Engineering (ICSE)}.\hskip 1em plus 0.5em minus
  0.4em\relax IEEE, 2013, pp. 1042--1051.

\bibitem{akbarinasaji2018predicting}
S.~Akbarinasaji, B.~Caglayan, and A.~Bener, ``Predicting bug-fixing time: A
  replication study using an open source software project,'' \emph{journal of
  Systems and Software}, vol. 136, pp. 173--186, 2018.

\bibitem{bojanowski2017enriching}
P.~Bojanowski, E.~Grave, A.~Joulin, and T.~Mikolov, ``Enriching word vectors
  with subword information,'' \emph{Transactions of the Association for
  Computational Linguistics}, vol.~5, pp. 135--146, 2017.

\bibitem{sabir2021machine}
B.~Sabir, F.~Ullah, M.~A. Babar, and R.~Gaire, ``Machine learning for detecting
  data exfiltration: A review,'' \emph{ACM Computing Surveys (CSUR)}, vol.~54,
  no.~3, pp. 1--47, 2021.

\bibitem{hommersom2021automated}
D.~Hommersom, A.~Sabetta, B.~Coppola, and D.~A. Tamburri, ``Automated mapping
  of vulnerability advisories onto their fix commits in open source
  repositories,'' \emph{arXiv preprint arXiv:2103.13375}, 2021.

\bibitem{symantec_threat}
\BIBentryALTinterwordspacing
Broadcom, ``Symantec threat explorer.'' [Online]. Available:
  \url{https://bit.ly/symantec_threats}
\BIBentrySTDinterwordspacing

\bibitem{van2020survey}
J.~E. Van~Engelen and H.~H. Hoos, ``A survey on semi-supervised learning,''
  \emph{Machine Learning}, vol. 109, no.~2, pp. 373--440, 2020.

\bibitem{wang2020generalizing}
Y.~Wang, Q.~Yao, J.~T. Kwok, and L.~M. Ni, ``Generalizing from a few examples:
  A survey on few-shot learning,'' \emph{ACM Computing Surveys (CSUR)},
  vol.~53, no.~3, pp. 1--34, 2020.

\bibitem{mikolov2013distributed}
T.~Mikolov, I.~Sutskever, K.~Chen, G.~Corrado, and J.~Dean, ``Distributed
  representations of words and phrases and their compositionality,''
  \emph{arXiv preprint arXiv:1310.4546}, 2013.

\bibitem{le2020deep}
T.~H.~M. Le, H.~Chen, and M.~A. Babar, ``Deep learning for source code modeling
  and generation: Models, applications, and challenges,'' \emph{ACM Computing
  Surveys (CSUR)}, vol.~53, no.~3, pp. 1--38, 2020.

\bibitem{zhang2020survey}
Y.~Zhang, P.~Ti{\v{n}}o, A.~Leonardis, and K.~Tang, ``A survey on neural
  network interpretability,'' \emph{arXiv preprint arXiv:2012.14261}, 2020.

\bibitem{dzeroski2002combining}
S.~Dzeroski and B.~Zenko, ``Is combining classifiers better than selecting the
  best one?'' in \emph{ICML}, vol. 2002.\hskip 1em plus 0.5em minus 0.4em\relax
  Citeseer, 2002, p. 123e30.

\bibitem{cortes1995support}
C.~Cortes and V.~Vapnik, ``Support-vector networks,'' \emph{Machine learning},
  vol.~20, no.~3, pp. 273--297, 1995.

\bibitem{reimers2019sentence}
N.~Reimers and I.~Gurevych, ``Sentence-bert: Sentence embeddings using siamese
  bert-networks,'' \emph{arXiv preprint arXiv:1908.10084}, 2019.

\bibitem{mazuera2021shallow}
A.~Mazuera-Rozo, A.~Mojica-Hanke, M.~Linares-Vasquez, and G.~Bavota, ``Shallow
  or deep? an empirical study on detecting vulnerabilities using deep
  learning,'' in \emph{the 2021 IEEE/ACM 29th International Conference on
  Program Comprehension (ICPC)}, 2021, pp. 276--287.

\bibitem{spring2021time}
J.~Spring, E.~Hatleback, A.~Householder, A.~Manion, and D.~Shick, ``Time to
  change the cvss?'' \emph{IEEE Security \& Privacy}, vol.~19, no.~2, pp.
  74--78, 2021.

\bibitem{menzies2018500+}
T.~Menzies, S.~Majumder, N.~Balaji, K.~Brey, and W.~Fu, ``500+ times faster
  than deep learning:(a case study exploring faster methods for text mining
  stackoverflow),'' in \emph{2018 IEEE/ACM 15th International Conference on
  Mining Software Repositories (MSR)}.\hskip 1em plus 0.5em minus 0.4em\relax
  IEEE, 2018, pp. 554--563.

\bibitem{palacio2019evaluation}
J.-O. Palacio-Ni{\~n}o and F.~Berzal, ``Evaluation metrics for unsupervised
  learning algorithms,'' \emph{arXiv preprint arXiv:1905.05667}, 2019.

\bibitem{raschka2018model}
S.~Raschka, ``Model evaluation, model selection, and algorithm selection in
  machine learning,'' \emph{arXiv preprint arXiv:1811.12808}, 2018.

\bibitem{de2019evolution}
F.~G. de~Oliveira~Neto, R.~Torkar, R.~Feldt, L.~Gren, C.~A. Furia, and
  Z.~Huang, ``Evolution of statistical analysis in empirical software
  engineering research: Current state and steps forward,'' \emph{Journal of
  Systems and Software}, vol. 156, pp. 246--267, 2019.

\bibitem{jiao2016performance}
Y.~Jiao and P.~Du, ``Performance measures in evaluating machine learning based
  bioinformatics predictors for classifications,'' \emph{Quantitative Biology},
  vol.~4, no.~4, pp. 320--330, 2016.

\bibitem{li2017large}
F.~Li and V.~Paxson, ``A large-scale empirical study of security patches,'' in
  \emph{2017 ACM SIGSAC Conference on Computer and Communications Security},
  2017, pp. 2201--2215.

\bibitem{piantadosi2019fixing}
V.~Piantadosi, S.~Scalabrino, and R.~Oliveto, ``Fixing of security
  vulnerabilities in open source projects: A case study of apache http server
  and apache tomcat,'' in \emph{2019 12th IEEE Conference on Software Testing,
  Validation and Verification (ICST)}.\hskip 1em plus 0.5em minus 0.4em\relax
  IEEE, 2019, pp. 68--78.

\bibitem{ampel2021linking}
B.~Ampel, S.~Samtani, S.~Ullman, and H.~Chen, ``Linking common vulnerabilities
  and exposures to the mitre att\&ck framework: A self-distillation approach,''
  \emph{arXiv preprint arXiv:2108.01696}, 2021.

\bibitem{kuehn2021ovana}
P.~Kuehn, M.~Bayer, M.~Wendelborn, and C.~Reuter, ``Ovana: An approach to
  analyze and improve the information quality of vulnerability databases,'' in
  \emph{The 16th International Conference on Availability, Reliability and
  Security}, 2021, pp. 1--11.

\bibitem{kekul2021multiclass}
H.~Kek{\"u}l, B.~Ergen, and H.~Arslan, ``A multiclass hybrid approach to
  estimating software vulnerability vectors and severity score,'' \emph{Journal
  of Information Security and Applications}, vol.~63, p. 103028, 2021.

\bibitem{shahid2021cvss}
M.~R. Shahid and H.~Debar, ``Cvss-bert: Explainable natural language processing
  to determine the severity of a computer security vulnerability from its
  description,'' in \emph{2021 20th IEEE International Conference on Machine
  Learning and Applications (ICMLA)}.\hskip 1em plus 0.5em minus 0.4em\relax
  IEEE, 2021, pp. 1600--1607.

\bibitem{lyu2021character}
J.~Lyu, Y.~Bai, Z.~Xing, X.~Li, and W.~Ge, ``A character-level convolutional
  neural network for predicting exploitability of vulnerability,'' in
  \emph{2021 International Symposium on Theoretical Aspects of Software
  Engineering (TASE)}.\hskip 1em plus 0.5em minus 0.4em\relax IEEE, 2021, pp.
  119--126.

\bibitem{babalau2021severity}
I.~Babalau, D.~Corlatescu, O.~Grigorescu, C.~Sandescu, and M.~Dascalu,
  ``Severity prediction of software vulnerabilities based on their text
  description,'' in \emph{2021 23rd International Symposium on Symbolic and
  Numeric Algorithms for Scientific Computing (SYNASC)}.\hskip 1em plus 0.5em
  minus 0.4em\relax IEEE, 2021, pp. 171--177.

\bibitem{charmanas2021predicting}
K.~Charmanas, N.~Mittas, and L.~Angelis, ``Predicting the existence of
  exploitation concepts linked to software vulnerabilities using text mining,''
  in \emph{25th Pan-Hellenic Conference on Informatics}, 2021, pp. 352--356.

\bibitem{yin2022vulnerability}
J.~Yin, M.~Tang, J.~Cao, H.~Wang, M.~You, and Y.~Lin, ``Vulnerability
  exploitation time prediction: an integrated framework for dynamic imbalanced
  learning,'' \emph{World Wide Web}, vol.~25, no.~1, pp. 401--423, 2022.

\bibitem{bulut2021nl2vul}
M.~F. Bulut and J.~Hwang, ``Nl2vul: Natural language to standard vulnerability
  score for cloud security posture management,'' in \emph{2021 IEEE 14th
  International Conference on Cloud Computing (CLOUD)}.\hskip 1em plus 0.5em
  minus 0.4em\relax IEEE, 2021, pp. 566--571.

\bibitem{ramesh2021automatic}
V.~Ramesh, S.~Abraham, P.~Vinod, I.~Mohamed, C.~A. Visaggio, and S.~Laudanna,
  ``Automatic classification of vulnerabilities using deep learning and machine
  learning algorithms,'' in \emph{2021 International Joint Conference on Neural
  Networks (IJCNN)}.\hskip 1em plus 0.5em minus 0.4em\relax IEEE, 2021, pp.
  1--8.

\bibitem{aivatoglou2021tree}
G.~Aivatoglou, M.~Anastasiadis, G.~Spanos, A.~Voulgaridis, K.~Votis, and
  D.~Tzovaras, ``A tree-based machine learning methodology to automatically
  classify software vulnerabilities,'' in \emph{2021 IEEE International
  Conference on Cyber Security and Resilience (CSR)}.\hskip 1em plus 0.5em
  minus 0.4em\relax IEEE, 2021, pp. 312--317.

\bibitem{vishnu2022deep}
P.~Vishnu, P.~Vinod, and S.~Y. Yerima, ``A deep learning approach for
  classifying vulnerability descriptions using self attention based neural
  network,'' \emph{Journal of Network and Systems Management}, vol.~30, no.~1,
  pp. 1--27, 2022.

\bibitem{wang2022automatic}
Q.~Wang, Y.~Li, Y.~Wang, and J.~Ren, ``An automatic algorithm for software
  vulnerability classification based on cnn and gru,'' \emph{Multimedia Tools
  and Applications}, pp. 1--22, 2022.

\bibitem{yosifova2021predicting}
V.~Yosifova, A.~Tasheva, and R.~Trifonov, ``Predicting vulnerability type in
  common vulnerabilities and exposures (cve) database with machine learning
  classifiers,'' in \emph{2021 12th National Conference with International
  Participation (ELECTRONICA)}.\hskip 1em plus 0.5em minus 0.4em\relax IEEE,
  2021, pp. 1--6.

\bibitem{yang2021few}
G.~Yang, S.~Dineen, Z.~Lin, and X.~Liu, ``Few-sample named entity recognition
  for security vulnerability reports by fine-tuning pre-trained language
  models,'' in \emph{International Workshop on Deployable Machine Learning for
  Security Defense}.\hskip 1em plus 0.5em minus 0.4em\relax Springer, 2021, pp.
  55--78.

\bibitem{yitagesu2021unsupervised}
S.~Yitagesu, Z.~Xing, X.~Zhang, Z.~Feng, X.~Li, and L.~Han, ``Unsupervised
  labeling and extraction of phrase-based concepts in vulnerability
  descriptions,'' in \emph{2021 36th IEEE/ACM International Conference on
  Automated Software Engineering (ASE)}.\hskip 1em plus 0.5em minus 0.4em\relax
  IEEE, 2021, pp. 943--954.

\bibitem{yuan2021predicting}
L.~Yuan, Y.~Bai, Z.~Xing, S.~Chen, X.~Li, and Z.~Deng, ``Predicting entity
  relations across different security databases by using graph attention
  network,'' in \emph{2021 IEEE 45th Annual Computers, Software, and
  Applications Conference (COMPSAC)}.\hskip 1em plus 0.5em minus 0.4em\relax
  IEEE, 2021, pp. 834--843.

\bibitem{guo2021key}
H.~Guo, Z.~Xing, S.~Chen, X.~Li, Y.~Bai, and H.~Zhang, ``Key aspects
  augmentation of vulnerability description based on multiple security
  databases,'' in \emph{2021 IEEE 45th Annual Computers, Software, and
  Applications Conference (COMPSAC)}.\hskip 1em plus 0.5em minus 0.4em\relax
  IEEE, 2021, pp. 1020--1025.

\bibitem{luong2014addressing}
M.-T. Luong, I.~Sutskever, Q.~V. Le, O.~Vinyals, and W.~Zaremba, ``Addressing
  the rare word problem in neural machine translation,'' \emph{arXiv preprint
  arXiv:1410.8206}, 2014.

\bibitem{huang2011using}
C.-C. Huang, H.-C. Yen, P.-C. Yang, S.-T. Huang, and J.~S. Chang, ``Using
  sublexical translations to handle the oov problem in machine translation,''
  \emph{ACM Transactions on Asian Language Information Processing (TALIP)},
  vol.~10, no.~3, pp. 1--20, 2011.

\bibitem{liu2018context}
A.~Liu and K.~Kirchhoff, ``Context models for oov word translation in
  low-resource languages,'' \emph{arXiv preprint arXiv:1801.08660}, 2018.

\bibitem{razmara2013graph}
M.~Razmara, M.~Siahbani, R.~Haffari, and A.~Sarkar, ``Graph propagation for
  paraphrasing out-of-vocabulary words in statistical machine translation,'' in
  \emph{the 51st Annual Meeting of the Association for Computational
  Linguistics (Volume 1: Long Papers)}, 2013, pp. 1105--1115.

\bibitem{old_sv_exploit}
\BIBentryALTinterwordspacing
R.~Future, ``Exploiting old vulnerabilities.'' [Online]. Available:
  \url{https://www.recordedfuture.com/exploiting-old-vulnerabilities/}
\BIBentrySTDinterwordspacing

\bibitem{kao2007natural}
A.~Kao and S.~R. Poteet, \emph{Natural language processing and text
  mining}.\hskip 1em plus 0.5em minus 0.4em\relax Springer Science \& Business
  Media, 2007.

\bibitem{pedregosa2011scikit}
F.~Pedregosa, G.~Varoquaux, A.~Gramfort, V.~Michel, B.~Thirion, O.~Grisel,
  M.~Blondel, P.~Prettenhofer, R.~Weiss, V.~Dubourg \emph{et~al.},
  ``Scikit-learn: Machine learning in python,'' \emph{the Journal of machine
  Learning research}, vol.~12, pp. 2825--2830, 2011.

\bibitem{loper2002nltk}
E.~Loper and S.~Bird, ``Nltk: the natural language toolkit,'' \emph{arXiv
  preprint cs/0205028}, 2002.

\bibitem{porter1980algorithm}
M.~F. Porter, ``An algorithm for suffix stripping.'' \emph{Program}, vol.~14,
  no.~3, pp. 130--137, 1980.

\bibitem{bergmeir2012use}
C.~Bergmeir and J.~M. Ben{\'\i}tez, ``On the use of cross-validation for time
  series predictor evaluation,'' \emph{Information Sciences}, vol. 191, pp.
  192--213, 2012.

\bibitem{deerwester1990indexing}
S.~Deerwester, S.~T. Dumais, G.~W. Furnas, T.~K. Landauer, and R.~Harshman,
  ``Indexing by latent semantic analysis,'' \emph{Journal of the American
  society for information science}, vol.~41, no.~6, pp. 391--407, 1990.

\bibitem{joulin2016bag}
A.~Joulin, E.~Grave, P.~Bojanowski, and T.~Mikolov, ``Bag of tricks for
  efficient text classification,'' \emph{arXiv preprint arXiv:1607.01759},
  2016.

\bibitem{kaggle_website}
\BIBentryALTinterwordspacing
Kaggle, ``Kaggle.'' [Online]. Available: \url{https://www.kaggle.com/}
\BIBentrySTDinterwordspacing

\bibitem{russell2002artificial}
S.~Russell and P.~Norvig, ``Artificial intelligence: a modern approach,'' 2002.

\bibitem{walker1967estimation}
S.~H. Walker and D.~B. Duncan, ``Estimation of the probability of an event as a
  function of several independent variables,'' \emph{Biometrika}, vol.~54, no.
  1-2, pp. 167--179, 1967.

\bibitem{basu2003support}
A.~Basu, C.~Walters, and M.~Shepherd, ``Support vector machines for text
  categorization,'' in \emph{36th Annual Hawaii International Conference on
  System Sciences, 2003. the}.\hskip 1em plus 0.5em minus 0.4em\relax IEEE,
  2003, pp. 7--pp.

\bibitem{ho1995random}
T.~K. Ho, ``Random decision forests,'' in \emph{3rd international conference on
  document analysis and recognition}, vol.~1.\hskip 1em plus 0.5em minus
  0.4em\relax IEEE, 1995, pp. 278--282.

\bibitem{blumer1987occam}
A.~Blumer, A.~Ehrenfeucht, D.~Haussler, and M.~K. Warmuth, ``Occam's razor,''
  \emph{Information processing letters}, vol.~24, no.~6, pp. 377--380, 1987.

\bibitem{le2018identification}
T.~H.~M. Le, T.~T. Tran, and L.~K. Huynh, ``Identification of hindered internal
  rotational mode for complex chemical species: A data mining approach with
  multivariate logistic regression model,'' \emph{Chemometrics and Intelligent
  Laboratory Systems}, vol. 172, pp. 10--16, 2018.

\bibitem{wilcoxon1992individual}
F.~Wilcoxon, ``Individual comparisons by ranking methods,'' in
  \emph{Breakthroughs in Statistics}.\hskip 1em plus 0.5em minus 0.4em\relax
  Springer, 1992, pp. 196--202.

\bibitem{marutho2018determination}
D.~Marutho, S.~H. Handaka, E.~Wijaya \emph{et~al.}, ``The determination of
  cluster number at k-mean using elbow method and purity evaluation on headline
  news,'' in \emph{2018 international seminar on application for technology of
  information and communication}.\hskip 1em plus 0.5em minus 0.4em\relax IEEE,
  2018, pp. 533--538.

\bibitem{rehurek2010software}
R.~Rehurek and P.~Sojka, ``Software framework for topic modelling with large
  corpora,'' in \emph{In the LREC 2010 Workshop on New Challenges for NLP
  Frameworks}.\hskip 1em plus 0.5em minus 0.4em\relax Citeseer, 2010.

\bibitem{fasttest_pretrained}
\BIBentryALTinterwordspacing
Facebook, ``Pre-trained word vectors of fasttext.'' [Online]. Available:
  \url{https://github.com/facebookresearch/fastText/blob/master/pretrained-vectors.md}
\BIBentrySTDinterwordspacing

\bibitem{hou2017security}
Y.~Hou, X.~Ren, Y.~Hao, T.~Mo, and W.~Li, ``A security vulnerability threat
  classification method,'' in \emph{International Conference on Broadband and
  Wireless Computing, Communication and Applications}.\hskip 1em plus 0.5em
  minus 0.4em\relax Springer, 2017, pp. 414--426.

\bibitem{liu2011vrss}
Q.~Liu and Y.~Zhang, ``Vrss: A new system for rating and scoring
  vulnerabilities,'' \emph{Computer Communications}, vol.~34, no.~3, pp.
  264--273, 2011.

\bibitem{sharma2018improved}
R.~Sharma and R.~Singh, ``An improved scoring system for software vulnerability
  prioritization,'' in \emph{Quality, IT and Business Operations}.\hskip 1em
  plus 0.5em minus 0.4em\relax Springer, 2018, pp. 33--43.

\bibitem{johnson2016can}
P.~Johnson, R.~Lagerstr{\"o}m, M.~Ekstedt, and U.~Franke, ``Can the common
  vulnerability scoring system be trusted? a bayesian analysis,'' \emph{IEEE
  Transactions on Dependable and Secure Computing}, vol.~15, no.~6, pp.
  1002--1015, 2016.

\bibitem{huang2013novel}
C.-C. Huang, F.-Y. Lin, F.~Y.-S. Lin, and Y.~S. Sun, ``A novel approach to
  evaluate software vulnerability prioritization,'' \emph{Journal of Systems
  and Software}, vol.~86, no.~11, pp. 2822--2840, 2013.

\bibitem{roumani2015time}
Y.~Roumani, J.~K. Nwankpa, and Y.~F. Roumani, ``Time series modeling of
  vulnerabilities,'' \emph{Computers \& Security}, vol.~51, pp. 32--40, 2015.

\bibitem{tang2016exploiting}
M.~Tang, M.~Alazab, and Y.~Luo, ``Exploiting vulnerability disclosures:
  statistical framework and case study,'' in \emph{2016 Cybersecurity and
  Cyberforensics Conference (CCC)}.\hskip 1em plus 0.5em minus 0.4em\relax
  IEEE, 2016, pp. 117--122.

\bibitem{rajasooriya2017cyber}
S.~M. Rajasooriya, C.~P. Tsokos, and P.~K. Kaluarachchi, ``Cyber security:
  Nonlinear stochastic models for predicting the exploitability,''
  \emph{Journal of information Security}, vol.~8, no.~2, pp. 125--140, 2017.

\bibitem{kaluarachchi2017non}
P.~K. Kaluarachchi, C.~P. Tsokos, and S.~M. Rajasooriya, ``Non-homogeneous
  stochastic model for cyber security predictions,'' \emph{Journal of
  Information Security}, vol.~9, no.~1, pp. 12--24, 2017.

\bibitem{pokhrel2017cybersecurity}
N.~R. Pokhrel, H.~Rodrigo, C.~P. Tsokos \emph{et~al.}, ``Cybersecurity: time
  series predictive modeling of vulnerabilities of desktop operating system
  using linear and non-linear approach,'' \emph{Journal of Information
  Security}, vol.~8, no.~04, p. 362, 2017.

\bibitem{le2022use}
T.~H.~M. Le and M.~A. Babar, ``On the use of fine-grained vulnerable code
  statements for software vulnerability assessment models,'' \emph{arXiv
  preprint arXiv:2203.08417}, 2022.

\bibitem{bugzilla}
\BIBentryALTinterwordspacing
M.~Foundation, ``Bugzilla issue tracking system.'' [Online]. Available:
  \url{https://www.bugzilla.org/}
\BIBentrySTDinterwordspacing

\bibitem{croft2021investigation}
R.~Croft, A.~Babar, and L.~Li, ``An investigation into inconsistency of
  software vulnerability severity across data sources,'' in \emph{2022 29th
  IEEE International Conference on Software Analysis, Evolution and
  Reengineering (SANER)}, 2022.

\bibitem{johnson2013don}
B.~Johnson, Y.~Song, E.~Murphy-Hill, and R.~Bowdidge, ``Why don't software
  developers use static analysis tools to find bugs?'' in \emph{2013 35th
  International Conference on Software Engineering (ICSE)}.\hskip 1em plus
  0.5em minus 0.4em\relax IEEE, 2013, pp. 672--681.

\bibitem{aloraini2019empirical}
B.~Aloraini, M.~Nagappan, D.~M. German, S.~Hayashi, and Y.~Higo, ``An empirical
  study of security warnings from static application security testing tools,''
  \emph{Journal of Systems and Software}, vol. 158, p. 110427, 2019.

\bibitem{zhou2019devign}
Y.~Zhou, S.~Liu, J.~Siow, X.~Du, and Y.~Liu, ``Devign: Effective vulnerability
  identification by learning comprehensive program semantics via graph neural
  networks,'' \emph{arXiv preprint arXiv:1909.03496}, 2019.

\bibitem{zheng2020impact}
W.~Zheng, J.~Gao, X.~Wu, F.~Liu, Y.~Xun, G.~Liu, and X.~Chen, ``The impact
  factors on the performance of machine learning-based vulnerability detection:
  A comparative study,'' \emph{Journal of Systems and Software}, vol. 168, p.
  110659, 2020.

\bibitem{lin2020deep}
G.~Lin, W.~Xiao, J.~Zhang, and Y.~Xiang, ``Deep learning-based vulnerable
  function detection: A benchmark,'' in \emph{International Conference on
  Information and Communications Security}, 2020, pp. 219--232.

\bibitem{li2021vulnerability}
Y.~Li, S.~Wang, and T.~N. Nguyen, ``Vulnerability detection with fine-grained
  interpretations,'' in \emph{the 22nd ACM SIGSOFT International Symposium on
  Foundations of Software Engineering}, 2021.

\bibitem{nguyen2021information}
V.~Nguyen, T.~Le, O.~De~Vel, P.~Montague, J.~Grundy, and D.~Phung,
  ``Information-theoretic source code vulnerability highlighting,'' in
  \emph{2021 International Joint Conference on Neural Networks (IJCNN)}.\hskip
  1em plus 0.5em minus 0.4em\relax IEEE, 2021, pp. 1--8.

\bibitem{li2021vuldeelocator}
Z.~Li, D.~Zou, S.~Xu, Z.~Chen, Y.~Zhu, and H.~Jin, ``Vuldeelocator: a deep
  learning-based fine-grained vulnerability detector,'' \emph{IEEE Transactions
  on Dependable and Secure Computing}, 2021.

\bibitem{wattanakriengkrai2020predicting}
S.~Wattanakriengkrai, P.~Thongtanunam, C.~Tantithamthavorn, H.~Hata, and
  K.~Matsumoto, ``Predicting defective lines using a model-agnostic
  technique,'' \emph{IEEE Transactions on Software Engineering}, 2020.

\bibitem{li2020dlfix}
Y.~Li, S.~Wang, and T.~N. Nguyen, ``Dlfix: Context-based code transformation
  learning for automated program repair,'' in \emph{The ACM/IEEE 42nd
  International Conference on Software Engineering}, 2020, pp. 602--614.

\bibitem{zhou2021finding}
J.~Zhou, M.~Pacheco, Z.~Wan, X.~Xia, D.~Lo, Y.~Wang, and A.~E. Hassan,
  ``Finding a needle in a haystack: Automated mining of silent vulnerability
  fixes,'' in \emph{2021 36th IEEE/ACM International Conference on Automated
  Software Engineering (ASE)}, 2021.

\bibitem{jira}
\BIBentryALTinterwordspacing
Atlassian, ``Jira issue tracking system.'' [Online]. Available:
  \url{https://www.atlassian.com/software/jira}
\BIBentrySTDinterwordspacing

\bibitem{li2021sysevr}
Z.~Li, D.~Zou, S.~Xu, H.~Jin, Y.~Zhu, and Z.~Chen, ``Sysevr: A framework for
  using deep learning to detect software vulnerabilities,'' \emph{IEEE
  Transactions on Dependable and Secure Computing}, 2021.

\bibitem{weiser1984program}
M.~Weiser, ``Program slicing,'' \emph{IEEE Transactions on software
  engineering}, no.~4, pp. 352--357, 1984.

\bibitem{ponta2019manually}
S.~E. Ponta, H.~Plate, A.~Sabetta, M.~Bezzi, and C.~Dangremont, ``A
  manually-curated dataset of fixes to vulnerabilities of open-source
  software,'' in \emph{2019 IEEE/ACM 16th International Conference on Mining
  Software Repositories (MSR)}.\hskip 1em plus 0.5em minus 0.4em\relax IEEE,
  2019, pp. 383--387.

\bibitem{mcintosh2017fix}
S.~McIntosh and Y.~Kamei, ``Are fix-inducing changes a moving target? a
  longitudinal case study of just-in-time defect prediction,'' \emph{IEEE
  Transactions on Software Engineering}, vol.~44, no.~5, pp. 412--428, 2017.

\bibitem{hoang2019deepjit}
T.~Hoang, H.~K. Dam, Y.~Kamei, D.~Lo, and N.~Ubayashi, ``Deepjit: an end-to-end
  deep learning framework for just-in-time defect prediction,'' in \emph{2019
  IEEE/ACM 16th International Conference on Mining Software Repositories
  (MSR)}.\hskip 1em plus 0.5em minus 0.4em\relax IEEE, 2019, pp. 34--45.

\bibitem{alon2019code2vec}
U.~Alon, M.~Zilberstein, O.~Levy, and E.~Yahav, ``code2vec: Learning
  distributed representations of code,'' \emph{the ACM on Programming
  Languages}, vol.~3, no. POPL, pp. 1--29, 2019.

\bibitem{le2021deepcva}
T.~H.~M. Le, D.~Hin, R.~Croft, and M.~A. Babar, ``Deepcva: Automated
  commit-level vulnerability assessment with deep multi-task learning,'' in
  \emph{2021 36th IEEE/ACM International Conference on Automated Software
  Engineering (ASE)}.\hskip 1em plus 0.5em minus 0.4em\relax IEEE, 2021, pp.
  717--729.

\bibitem{rezk2021ghost}
C.~Rezk, Y.~Kamei, and S.~Mcintosh, ``The ghost commit problem when identifying
  fix-inducing changes: An empirical study of apache projects,'' \emph{IEEE
  Transactions on Software Engineering}, 2021.

\bibitem{cochran2007sampling}
W.~G. Cochran, \emph{Sampling techniques}.\hskip 1em plus 0.5em minus
  0.4em\relax John Wiley \& Sons, 2007.

\bibitem{viera2005understanding}
A.~J. Viera, J.~M. Garrett \emph{et~al.}, ``Understanding interobserver
  agreement: the kappa statistic,'' \emph{Fam med}, vol.~37, no.~5, pp.
  360--363, 2005.

\bibitem{tian2020evaluating}
H.~Tian, K.~Liu, A.~K. Kabor{\'e}, A.~Koyuncu, L.~Li, J.~Klein, and T.~F.
  Bissyand{\'e}, ``Evaluating representation learning of code changes for
  predicting patch correctness in program repair,'' in \emph{2020 35th IEEE/ACM
  International Conference on Automated Software Engineering (ASE)}.\hskip 1em
  plus 0.5em minus 0.4em\relax IEEE, 2020, pp. 981--992.

\bibitem{zheng2021vulspg}
W.~Zheng, Y.~Jiang, and X.~Su, ``Vulspg: Vulnerability detection based on slice
  property graph representation learning,'' \emph{arXiv preprint
  arXiv:2109.02527}, 2021.

\bibitem{salimi2020improving}
S.~Salimi, M.~Ebrahimzadeh, and M.~Kharrazi, ``Improving real-world
  vulnerability characterization with vulnerable slices,'' in \emph{The 16th
  ACM International Conference on Predictive Models and Data Analytics in
  Software Engineering}, 2020, pp. 11--20.

\bibitem{ferrante1987program}
J.~Ferrante, K.~J. Ottenstein, and J.~D. Warren, ``The program dependence graph
  and its use in optimization,'' \emph{ACM Transactions on Programming
  Languages and Systems (TOPLAS)}, vol.~9, no.~3, pp. 319--349, 1987.

\bibitem{dashevskyi2018screening}
S.~Dashevskyi, A.~D. Brucker, and F.~Massacci, ``A screening test for disclosed
  vulnerabilities in foss components,'' \emph{IEEE Transactions on Software
  Engineering}, vol.~45, no.~10, pp. 945--966, 2018.

\bibitem{shen2018baseline}
D.~Shen, G.~Wang, W.~Wang, M.~R. Min, Q.~Su, Y.~Zhang, C.~Li, R.~Henao, and
  L.~Carin, ``Baseline needs more love: On simple word-embedding-based models
  and associated pooling mechanisms,'' \emph{arXiv preprint arXiv:1805.09843},
  2018.

\bibitem{feng2020codebert}
Z.~Feng, D.~Guo, D.~Tang, N.~Duan, X.~Feng, M.~Gong, L.~Shou, B.~Qin, T.~Liu,
  D.~Jiang \emph{et~al.}, ``Codebert: A pre-trained model for programming and
  natural languages,'' \emph{arXiv preprint arXiv:2002.08155}, 2020.

\bibitem{sennrich2015neural}
R.~Sennrich, B.~Haddow, and A.~Birch, ``Neural machine translation of rare
  words with subword units,'' \emph{arXiv preprint arXiv:1508.07909}, 2015.

\bibitem{zhou2021assessing}
X.~Zhou, D.~Han, and D.~Lo, ``Assessing generalizability of codebert,'' in
  \emph{2021 IEEE International Conference on Software Maintenance and
  Evolution (ICSME)}.\hskip 1em plus 0.5em minus 0.4em\relax IEEE, 2021, pp.
  425--436.

\bibitem{hoang2020cc2vec}
T.~Hoang, H.~J. Kang, D.~Lo, and J.~Lawall, ``Cc2vec: Distributed
  representations of code changes,'' in \emph{the ACM/IEEE 42nd International
  Conference on Software Engineering}, 2020, pp. 518--529.

\bibitem{altman1992introduction}
N.~S. Altman, ``An introduction to kernel and nearest-neighbor nonparametric
  regression,'' \emph{The American Statistician}, vol.~46, no.~3, pp. 175--185,
  1992.

\bibitem{le2020puminer}
T.~H.~M. Le, D.~Hin, R.~Croft, and M.~A. Babar, ``Puminer: Mining security
  posts from developer question and answer websites with pu learning,'' in
  \emph{the 17th International Conference on Mining Software Repositories},
  2020, pp. 350--361.

\bibitem{hanif2021rise}
H.~Hanif, M.~H. N.~M. Nasir, M.~F. Ab~Razak, A.~Firdaus, and N.~B. Anuar, ``The
  rise of software vulnerability: Taxonomy of software vulnerabilities
  detection and machine learning approaches,'' \emph{Journal of Network and
  Computer Applications}, p. 103009, 2021.

\bibitem{sonnekalb2022deep}
T.~Sonnekalb, T.~S. Heinze, and P.~M{\"a}der, ``Deep security analysis of
  program code,'' \emph{Empirical Software Engineering}, vol.~27, no.~1, pp.
  1--39, 2022.

\bibitem{luque2019impact}
A.~Luque, A.~Carrasco, A.~Mart{\'\i}n, and A.~de~las Heras, ``The impact of
  class imbalance in classification performance metrics based on the binary
  confusion matrix,'' \emph{Pattern Recognition}, vol.~91, pp. 216--231, 2019.

\bibitem{tomczak2014need}
M.~Tomczak and E.~Tomczak, ``The need to report effect size estimates
  revisited. an overview of some recommended measures of effect size,''
  \emph{Trends in Sport Sciences}, vol.~1, no.~21, pp. 19--25, 2014.

\bibitem{field2013discovering}
A.~Field, \emph{Discovering statistics using IBM SPSS statistics}.\hskip 1em
  plus 0.5em minus 0.4em\relax sage, 2013.

\bibitem{cohen2013statistical}
J.~Cohen, \emph{Statistical power analysis for the behavioral sciences}.\hskip
  1em plus 0.5em minus 0.4em\relax Academic press, 2013.

\bibitem{macbeth2011cliff}
G.~Macbeth, E.~Razumiejczyk, and R.~D. Ledesma, ``Cliff's delta calculator: A
  non-parametric effect size program for two groups of observations,''
  \emph{Universitas Psychologica}, vol.~10, no.~2, pp. 545--555, 2011.

\bibitem{treude2019predicting}
C.~Treude and M.~Wagner, ``Predicting good configurations for github and stack
  overflow topic models,'' in \emph{2019 IEEE/ACM 16th International Conference
  on Mining Software Repositories (MSR)}.\hskip 1em plus 0.5em minus
  0.4em\relax IEEE, 2019, pp. 84--95.

\bibitem{tantithamthavorn2018impact}
C.~Tantithamthavorn, A.~E. Hassan, and K.~Matsumoto, ``The impact of class
  rebalancing techniques on the performance and interpretation of defect
  prediction models,'' \emph{IEEE Transactions on Software Engineering},
  vol.~46, no.~11, pp. 1200--1219, 2018.

\bibitem{li2018vuldeepecker}
Z.~Li, D.~Zou, S.~Xu, X.~Ou, H.~Jin, S.~Wang, Z.~Deng, and Y.~Zhong,
  ``Vuldeepecker: A deep learning-based system for vulnerability detection,''
  \emph{arXiv preprint arXiv:1801.01681}, 2018.

\bibitem{le2021large}
T.~H.~M. Le, R.~Croft, D.~Hin, and M.~A. Babar, ``A large-scale study of
  security vulnerability support on developer q\&a websites,'' in
  \emph{Evaluation and Assessment in Software Engineering}, 2021, pp. 109--118.

\bibitem{kim2014survey}
J.~Kim, T.~Kim, and E.~G. Im, ``Survey of dynamic taint analysis,'' in
  \emph{2014 4th IEEE International Conference on Network Infrastructure and
  Digital Content}.\hskip 1em plus 0.5em minus 0.4em\relax IEEE, 2014, pp.
  269--272.

\bibitem{neuhaus2007predicting}
S.~Neuhaus, T.~Zimmermann, C.~Holler, and A.~Zeller, ``Predicting vulnerable
  software components,'' in \emph{The 14th ACM Conference on Computer and
  Communications Security}, 2007, pp. 529--540.

\bibitem{chowdhury2011using}
I.~Chowdhury and M.~Zulkernine, ``Using complexity, coupling, and cohesion
  metrics as early indicators of vulnerabilities,'' \emph{Journal of Systems
  Architecture}, vol.~57, no.~3, pp. 294--313, 2011.

\bibitem{shin2013can}
Y.~Shin and L.~Williams, ``Can traditional fault prediction models be used for
  vulnerability prediction?'' \emph{Empirical Software Engineering}, vol.~18,
  no.~1, pp. 25--59, 2013.

\bibitem{tang2015predicting}
Y.~Tang, F.~Zhao, Y.~Yang, H.~Lu, Y.~Zhou, and B.~Xu, ``Predicting vulnerable
  components via text mining or software metrics? an effort-aware
  perspective,'' in \emph{2015 IEEE International Conference on Software
  Quality, Reliability and Security}.\hskip 1em plus 0.5em minus 0.4em\relax
  IEEE, 2015, pp. 27--36.

\bibitem{lin2018cross}
G.~Lin, J.~Zhang, W.~Luo, L.~Pan, Y.~Xiang, O.~De~Vel, and P.~Montague,
  ``Cross-project transfer representation learning for vulnerable function
  discovery,'' \emph{IEEE Transactions on Industrial Informatics}, vol.~14,
  no.~7, pp. 3289--3297, 2018.

\bibitem{nguyen2019deep}
V.~Nguyen, T.~Le, T.~Le, K.~Nguyen, O.~DeVel, P.~Montague, L.~Qu, and D.~Phung,
  ``Deep domain adaptation for vulnerable code function identification,'' in
  \emph{2019 International Joint Conference on Neural Networks (IJCNN)}.\hskip
  1em plus 0.5em minus 0.4em\relax IEEE, 2019, pp. 1--8.

\bibitem{bilgin2020vulnerability}
Z.~Bilgin, M.~A. Ersoy, E.~U. Soykan, E.~Tomur, P.~{\c{C}}omak, and
  L.~Kara{\c{c}}ay, ``Vulnerability prediction from source code using machine
  learning,'' \emph{IEEE Access}, vol.~8, pp. 150\,672--150\,684, 2020.

\bibitem{wang2020combining}
H.~Wang, G.~Ye, Z.~Tang, S.~H. Tan, S.~Huang, D.~Fang, Y.~Feng, L.~Bian, and
  Z.~Wang, ``Combining graph-based learning with automated data collection for
  code vulnerability detection,'' \emph{IEEE Transactions on Information
  Forensics and Security}, vol.~16, pp. 1943--1958, 2020.

\bibitem{ding2021velvet}
Y.~Ding, S.~Suneja, Y.~Zheng, J.~Laredo, A.~Morari, G.~Kaiser, and B.~Ray,
  ``Velvet: a novel ensemble learning approach to automatically locate
  vulnerable statements,'' in \emph{2022 29th IEEE International Conference on
  Software Analysis, Evolution and Reengineering (SANER)}, 2022.

\bibitem{duan2021automated}
X.~Duan, M.~Ge, T.~H.~M. Le, F.~Ullah, S.~Gao, X.~Lu, and M.~A. Babar,
  ``Automated security assessment for the internet of things,'' in \emph{2021
  IEEE 26th Pacific Rim International Symposium on Dependable Computing
  (PRDC)}.\hskip 1em plus 0.5em minus 0.4em\relax IEEE, 2021, pp. 47--56.

\bibitem{ponta2018beyond}
S.~E. Ponta, H.~Plate, and A.~Sabetta, ``Beyond metadata: Code-centric and
  usage-based analysis of known vulnerabilities in open-source software,'' in
  \emph{2018 IEEE International Conference on Software Maintenance and
  Evolution (ICSME)}.\hskip 1em plus 0.5em minus 0.4em\relax IEEE, 2018, pp.
  449--460.

\bibitem{lamkanfi2010predicting}
A.~Lamkanfi, S.~Demeyer, E.~Giger, and B.~Goethals, ``Predicting the severity
  of a reported bug,'' in \emph{2010 7th IEEE Working Conference on Mining
  Software Repositories (MSR 2010)}.\hskip 1em plus 0.5em minus 0.4em\relax
  IEEE, 2010, pp. 1--10.

\bibitem{perl2015vccfinder}
H.~Perl, S.~Dechand, M.~Smith, D.~Arp, F.~Yamaguchi, K.~Rieck, S.~Fahl, and
  Y.~Acar, ``Vccfinder: Finding potential vulnerabilities in open-source
  projects to assist code audits,'' in \emph{the 22nd SIGSAC ACM Conference on
  Computer and Communications Security}, 2015, pp. 426--437.

\bibitem{yang2017vuldigger}
L.~Yang, X.~Li, and Y.~Yu, ``Vuldigger: A just-in-time and cost-aware tool for
  digging vulnerability-contributing changes,'' in \emph{GLOBECOM 2017-2017
  IEEE Global Communications Conference}.\hskip 1em plus 0.5em minus
  0.4em\relax IEEE, 2017, pp. 1--7.

\bibitem{rodriguez2018analysis}
L.~G.~A. Rodriguez, J.~S. Trazzi, V.~Fossaluza, R.~Campiolo, and D.~M. Batista,
  ``Analysis of vulnerability disclosure delays from the national vulnerability
  database,'' in \emph{Anais do I Workshop de Seguran{\c{c}}a Cibern{\'e}tica
  em Dispositivos Conectados}.\hskip 1em plus 0.5em minus 0.4em\relax SBC,
  2018.

\bibitem{sawadogo2020learning}
A.~D. Sawadogo, T.~F. Bissyand{\'e}, N.~Moha, K.~Allix, J.~Klein, L.~Li, and
  Y.~L. Traon, ``Learning to catch security patches,'' \emph{arXiv preprint
  arXiv:2001.09148}, 2020.

\bibitem{thung2012would}
F.~Thung, D.~Lo, L.~Jiang, F.~Rahman, P.~T. Devanbu \emph{et~al.}, ``When would
  this bug get reported?'' in \emph{2012 28th IEEE International Conference on
  Software Maintenance (ICSM)}.\hskip 1em plus 0.5em minus 0.4em\relax IEEE,
  2012, pp. 420--429.

\bibitem{bosu2012peer}
A.~Bosu and J.~C. Carver, ``Peer code review in open source communities using
  reviewboard,'' in \emph{the ACM 4th Annual Workshop on Evaluation and
  Usability of Programming Languages and Tools}, 2012, pp. 17--24.

\bibitem{thongtanunam2015investigating}
P.~Thongtanunam, S.~McIntosh, A.~E. Hassan, and H.~Iida, ``Investigating code
  review practices in defective files: An empirical study of the qt system,''
  in \emph{2015 IEEE/ACM 12th Working Conference on Mining Software
  Repositories}.\hskip 1em plus 0.5em minus 0.4em\relax IEEE, 2015, pp.
  168--179.

\bibitem{moran2015auto}
K.~Moran, M.~Linares-V{\'a}squez, C.~Bernal-C{\'a}rdenas, and D.~Poshyvanyk,
  ``Auto-completing bug reports for android applications,'' in \emph{the 2015
  10th Joint Meeting on Foundations of Software Engineering}, 2015, pp.
  673--686.

\bibitem{hoang2019patchnet}
T.~Hoang, J.~Lawall, Y.~Tian, R.~J. Oentaryo, and D.~Lo, ``Patchnet:
  Hierarchical deep learning-based stable patch identification for the linux
  kernel,'' \emph{IEEE Transactions on Software Engineering}, 2019.

\bibitem{chowdhuri2019multinet}
S.~Chowdhuri, T.~Pankaj, and K.~Zipser, ``Multinet: Multi-modal multi-task
  learning for autonomous driving,'' in \emph{2019 IEEE Winter Conference on
  Applications of Computer Vision}.\hskip 1em plus 0.5em minus 0.4em\relax
  IEEE, 2019, pp. 1496--1504.

\bibitem{sabetta2018practical}
A.~Sabetta and M.~Bezzi, ``A practical approach to the automatic classification
  of security-relevant commits,'' in \emph{2018 IEEE International Conference
  on Software Maintenance and Evolution (ICSME)}.\hskip 1em plus 0.5em minus
  0.4em\relax IEEE, 2018, pp. 579--582.

\bibitem{sahal2018identifying}
E.~Sahal and A.~Tosun, ``Identifying bug-inducing changes for code additions,''
  in \emph{the 12th ACM/IEEE International Symposium on Empirical Software
  Engineering and Measurement}, 2018, pp. 1--2.

\bibitem{nair2010rectified}
V.~Nair and G.~E. Hinton, ``Rectified linear units improve restricted boltzmann
  machines,'' in \emph{Icml}, 2010.

\bibitem{ruder2016overview}
S.~Ruder, ``An overview of gradient descent optimization algorithms,''
  \emph{arXiv preprint arXiv:1609.04747}, 2016.

\bibitem{rumelhart1986learning}
D.~E. Rumelhart, G.~E. Hinton, and R.~J. Williams, ``Learning representations
  by back-propagating errors,'' \emph{Nature}, vol. 323, no. 6088, pp.
  533--536, 1986.

\bibitem{sliwerski2005changes}
J.~{\'S}liwerski, T.~Zimmermann, and A.~Zeller, ``When do changes induce
  fixes?'' \emph{ACM SIGSOFT Software Engineering Notes}, vol.~30, no.~4, pp.
  1--5, 2005.

\bibitem{mchugh2012interrater}
M.~L. McHugh, ``Interrater reliability: the kappa statistic,'' \emph{Biochemia
  medica}, vol.~22, no.~3, pp. 276--282, 2012.

\bibitem{hata20199}
H.~Hata, C.~Treude, R.~G. Kula, and T.~Ishio, ``9.6 million links in source
  code comments: Purpose, evolution, and decay,'' in \emph{2019 IEEE/ACM 41st
  International Conference on Software Engineering (ICSE)}.\hskip 1em plus
  0.5em minus 0.4em\relax IEEE, 2019, pp. 1211--1221.

\bibitem{fan2019impact}
Y.~Fan, X.~Xia, D.~A. Da~Costa, D.~Lo, A.~E. Hassan, and S.~Li, ``The impact of
  changes mislabeled by szz on just-in-time defect prediction,'' \emph{IEEE
  Transactions on Software Engineering}, 2019.

\bibitem{falessi2020need}
D.~Falessi, J.~Huang, L.~Narayana, J.~F. Thai, and B.~Turhan, ``On the need of
  preserving order of data when validating within-project defect classifiers,''
  \emph{Empirical Software Engineering}, vol.~25, no.~6, pp. 4805--4830, 2020.

\bibitem{jimenez2019importance}
M.~Jimenez, R.~Rwemalika, M.~Papadakis, F.~Sarro, Y.~Le~Traon, and M.~Harman,
  ``The importance of accounting for real-world labelling when predicting
  software vulnerabilities,'' in \emph{2019 27th ACM Joint Meeting on European
  Software Engineering Conference and Symposium on the Foundations of Software
  Engineering}, 2019, pp. 695--705.

\bibitem{gorodkin2004comparing}
J.~Gorodkin, ``Comparing two k-category assignments by a k-category correlation
  coefficient,'' \emph{Computational Biology and Chemistry}, vol.~28, no. 5-6,
  pp. 367--374, 2004.

\bibitem{radford2019language}
A.~Radford, J.~Wu, R.~Child, D.~Luan, D.~Amodei, and I.~Sutskever, ``Language
  models are unsupervised multitask learners,'' \emph{OpenAI blog}, vol.~1,
  no.~8, p.~9, 2019.

\bibitem{pradel2018deepbugs}
M.~Pradel and K.~Sen, ``Deepbugs: A learning approach to name-based bug
  detection,'' \emph{the ACM on Programming Languages}, vol.~2, no. OOPSLA, pp.
  1--25, 2018.

\bibitem{kingma2014adam}
D.~P. Kingma and J.~Ba, ``Adam: A method for stochastic optimization,''
  \emph{arXiv preprint arXiv:1412.6980}, 2014.

\bibitem{srivastava2014dropout}
N.~Srivastava, G.~Hinton, A.~Krizhevsky, I.~Sutskever, and R.~Salakhutdinov,
  ``Dropout: a simple way to prevent neural networks from overfitting,''
  \emph{The Journal of Machine Learning Research}, vol.~15, no.~1, pp.
  1929--1958, 2014.

\bibitem{ioffe2015batch}
S.~Ioffe and C.~Szegedy, ``Batch normalization: Accelerating deep network
  training by reducing internal covariate shift,'' in \emph{International
  Conference on Machine Learning}.\hskip 1em plus 0.5em minus 0.4em\relax PMLR,
  2015, pp. 448--456.

\bibitem{lloyd1982least}
S.~Lloyd, ``Least squares quantization in pcm,'' \emph{IEEE Transactions on
  Information Theory}, vol.~28, no.~2, pp. 129--137, 1982.

\bibitem{zhou2017automated}
Y.~Zhou and A.~Sharma, ``Automated identification of security issues from
  commit messages and bug reports,'' in \emph{the 2017 11th Joint Meeting on
  Foundations of Software Engineering}, 2017, pp. 914--919.

\bibitem{li2019comparative}
Z.~Li, D.~Zou, J.~Tang, Z.~Zhang, M.~Sun, and H.~Jin, ``A comparative study of
  deep learning-based vulnerability detection system,'' \emph{IEEE Access},
  vol.~7, pp. 103\,184--103\,197, 2019.

\bibitem{chawla2002smote}
N.~V. Chawla, K.~W. Bowyer, L.~O. Hall, and W.~P. Kegelmeyer, ``Smote:
  synthetic minority over-sampling technique,'' \emph{Journal of Artificial
  Intelligence Research}, vol.~16, pp. 321--357, 2002.

\bibitem{pascarella2019fine}
L.~Pascarella, F.~Palomba, and A.~Bacchelli, ``Fine-grained just-in-time defect
  prediction,'' \emph{Journal of Systems and Software}, vol. 150, pp. 22--36,
  2019.

\bibitem{gu2018deep}
X.~Gu, H.~Zhang, and S.~Kim, ``Deep code search,'' in \emph{2018 IEEE/ACM 40th
  International Conference on Software Engineering (ICSE)}.\hskip 1em plus
  0.5em minus 0.4em\relax IEEE, 2018, pp. 933--944.

\bibitem{hu2018deep}
X.~Hu, G.~Li, X.~Xia, D.~Lo, and Z.~Jin, ``Deep code comment generation,'' in
  \emph{2018 IEEE/ACM 26th International Conference on Program Comprehension
  (ICPC)}.\hskip 1em plus 0.5em minus 0.4em\relax IEEE, 2018, pp. 200--210.

\bibitem{ponta2020detection}
S.~E. Ponta, H.~Plate, and A.~Sabetta, ``Detection, assessment and mitigation
  of vulnerabilities in open source dependencies,'' \emph{Empirical Software
  Engineering}, vol.~25, no.~5, pp. 3175--3215, 2020.

\bibitem{yang2015deep}
X.~Yang, D.~Lo, X.~Xia, Y.~Zhang, and J.~Sun, ``Deep learning for just-in-time
  defect prediction,'' in \emph{2015 IEEE International Conference on Software
  Quality, Reliability and Security}.\hskip 1em plus 0.5em minus 0.4em\relax
  IEEE, 2015, pp. 17--26.

\bibitem{camilo2015bugs}
F.~Camilo, A.~Meneely, and M.~Nagappan, ``Do bugs foreshadow vulnerabilities? a
  study of the chromium project,'' in \emph{2015 IEEE/ACM 12th Working
  Conference on Mining Software Repositories}.\hskip 1em plus 0.5em minus
  0.4em\relax IEEE, 2015, pp. 269--279.

\bibitem{peters2017text}
F.~Peters, T.~T. Tun, Y.~Yu, and B.~Nuseibeh, ``Text filtering and ranking for
  security bug report prediction,'' \emph{IEEE Transactions on Software
  Engineering}, vol.~45, no.~6, pp. 615--631, 2017.

\bibitem{gegick2010identifying}
M.~Gegick, P.~Rotella, and T.~Xie, ``Identifying security bug reports via text
  mining: An industrial case study,'' in \emph{2010 7th IEEE Working Conference
  on Mining Software Repositories (MSR 2010)}.\hskip 1em plus 0.5em minus
  0.4em\relax IEEE, 2010, pp. 11--20.

\bibitem{bosu2014identifying}
A.~Bosu, J.~C. Carver, M.~Hafiz, P.~Hilley, and D.~Janni, ``Identifying the
  characteristics of vulnerable code changes: An empirical study,'' in
  \emph{the 22nd ACM SIGSOFT International Symposium on Foundations of Software
  Engineering}, 2014, pp. 257--268.

\bibitem{chen2019large}
X.~Chen, Y.~Zhao, Z.~Cui, G.~Meng, Y.~Liu, and Z.~Wang, ``Large-scale empirical
  studies on effort-aware security vulnerability prediction methods,''
  \emph{IEEE Transactions on Reliability}, vol.~69, no.~1, pp. 70--87, 2019.

\bibitem{yang2016security}
X.-L. Yang, D.~Lo, X.~Xia, Z.-Y. Wan, and J.-L. Sun, ``What security questions
  do developers ask? a large-scale study of stack overflow posts,''
  \emph{Journal of Computer Science and Technology}, vol.~31, no.~5, pp.
  910--924, 2016.

\bibitem{bayati2016information}
S.~Bayati and M.~Heidary, ``Information security in software engineering,
  analysis of developers communications about security in social q\&a
  website,'' in \emph{Pacific-Asia Workshop on Intelligence and Security
  Informatics}.\hskip 1em plus 0.5em minus 0.4em\relax Springer, 2016, pp.
  193--202.

\bibitem{lopez2019anatomy}
T.~Lopez, T.~Tun, A.~Bandara, L.~Mark, B.~Nuseibeh, and H.~Sharp, ``An anatomy
  of security conversations in stack overflow,'' in \emph{2019 IEEE/ACM 41st
  International Conference on Software Engineering: Software Engineering in
  Society (ICSE-SEIS)}.\hskip 1em plus 0.5em minus 0.4em\relax IEEE, 2019, pp.
  31--40.

\bibitem{barua2014developers}
A.~Barua, S.~W. Thomas, and A.~E. Hassan, ``What are developers talking about?
  an analysis of topics and trends in stack overflow,'' \emph{Empirical
  Software Engineering}, vol.~19, no.~3, pp. 619--654, 2014.

\bibitem{ahmed2018concurrency}
S.~Ahmed and M.~Bagherzadeh, ``What do concurrency developers ask about? a
  large-scale study using stack overflow,'' in \emph{the 12th ACM/IEEE
  International Symposium on Empirical Software Engineering and Measurement},
  2018, pp. 1--10.

\bibitem{rosen2016mobile}
C.~Rosen and E.~Shihab, ``What are mobile developers asking about? a large
  scale study using stack overflow,'' \emph{Empirical Software Engineering},
  vol.~21, no.~3, pp. 1192--1223, 2016.

\bibitem{bagherzadeh2019going}
M.~Bagherzadeh and R.~Khatchadourian, ``Going big: a large-scale study on what
  big data developers ask,'' in \emph{the 2019 27th ACM Joint Meeting on
  European Software Engineering Conference and Symposium on the Foundations of
  Software Engineering}, 2019, pp. 432--442.

\bibitem{bangash2019developers}
A.~A. Bangash, H.~Sahar, S.~Chowdhury, A.~W. Wong, A.~Hindle, and K.~Ali,
  ``What do developers know about machine learning: a study of ml discussions
  on stackoverflow,'' in \emph{2019 IEEE/ACM 16th International Conference on
  Mining Software Repositories (MSR)}.\hskip 1em plus 0.5em minus 0.4em\relax
  IEEE, 2019, pp. 260--264.

\bibitem{han2020programmers}
J.~Han, E.~Shihab, Z.~Wan, S.~Deng, and X.~Xia, ``What do programmers discuss
  about deep learning frameworks,'' \emph{Empirical Software Engineering},
  vol.~25, no.~4, pp. 2694--2747, 2020.

\bibitem{shahzad2012large}
M.~Shahzad, M.~Z. Shafiq, and A.~X. Liu, ``A large scale exploratory analysis
  of software vulnerability life cycles,'' in \emph{2012 34th International
  Conference on Software Engineering (ICSE)}.\hskip 1em plus 0.5em minus
  0.4em\relax IEEE, 2012, pp. 771--781.

\bibitem{meng2018secure}
N.~Meng, S.~Nagy, D.~Yao, W.~Zhuang, and G.~A. Argoty, ``Secure coding
  practices in java: Challenges and vulnerabilities,'' in \emph{the 40th
  International Conference on Software Engineering}, 2018, pp. 372--383.

\bibitem{rahman2019snakes}
A.~Rahman, E.~Farhana, and N.~Imtiaz, ``Snakes in paradise?: Insecure
  python-related coding practices in stack overflow,'' in \emph{2019 IEEE/ACM
  16th International Conference on Mining Software Repositories (MSR)}.\hskip
  1em plus 0.5em minus 0.4em\relax IEEE, 2019, pp. 200--204.

\bibitem{hanrahan2012modeling}
B.~V. Hanrahan, G.~Convertino, and L.~Nelson, ``Modeling problem difficulty and
  expertise in stackoverflow,'' in \emph{the ACM 2012 conference on Computer
  Supported Cooperative Work Companion}, 2012, pp. 91--94.

\bibitem{dey2020representation}
T.~Dey, A.~Karnauch, and A.~Mockus, ``Representation of developer expertise in
  open source software,'' in \emph{2021 IEEE/ACM 43rd International Conference
  on Software Engineering (ICSE)}.\hskip 1em plus 0.5em minus 0.4em\relax IEEE,
  2021.

\bibitem{seaman1999qualitative}
C.~B. Seaman, ``Qualitative methods in empirical studies of software
  engineering,'' \emph{IEEE Transactions on software engineering}, vol.~25,
  no.~4, pp. 557--572, 1999.

\bibitem{chen2020comprehensive}
Z.~Chen, Y.~Cao, Y.~Liu, H.~Wang, T.~Xie, and X.~Liu, ``A comprehensive study
  on challenges in deploying deep learning based software,'' in \emph{the 28th
  ACM Joint Meeting on European Software Engineering Conference and Symposium
  on the Foundations of Software Engineering}, 2020, pp. 750--762.

\bibitem{treude2011programmers}
C.~Treude, O.~Barzilay, and M.-A. Storey, ``How do programmers ask and answer
  questions on the web?(nier track),'' in \emph{the 33rd international
  conference on software engineering}, 2011, pp. 804--807.

\bibitem{bekker2018learning}
J.~Bekker and J.~Davis, ``Learning from positive and unlabeled data: A
  survey,'' \emph{arXiv preprint arXiv:1811.04820}, 2018.

\bibitem{scholkopf2000support}
B.~Sch{\"o}lkopf, R.~C. Williamson, A.~J. Smola, J.~Shawe-Taylor, and J.~C.
  Platt, ``Support vector method for novelty detection,'' in \emph{Advances in
  neural information processing systems}, 2000, pp. 582--588.

\bibitem{manevitz2001one}
L.~M. Manevitz and M.~Yousef, ``One-class svms for document classification,''
  \emph{Journal of machine Learning research}, vol.~2, no. Dec, pp. 139--154,
  2001.

\bibitem{mendsaikhan2019identification}
O.~Mendsaikhan, H.~Hasegawa, Y.~Yamaguchi, and H.~Shimada, ``Identification of
  cybersecurity specific content using the doc2vec language model,'' in
  \emph{2019 IEEE 43rd Annual Computer Software and Applications Conference
  (COMPSAC)}, vol.~1.\hskip 1em plus 0.5em minus 0.4em\relax IEEE, 2019, pp.
  396--401.

\bibitem{haque2020challenges}
M.~U. Haque, L.~H. Iwaya, and M.~A. Babar, ``Challenges in docker development:
  A large-scale study using stack overflow,'' in \emph{the 14th ACM/IEEE
  International Symposium on Empirical Software Engineering and Measurement
  (ESEM)}, 2020, pp. 1--11.

\bibitem{roder2015exploring}
M.~R{\"o}der, A.~Both, and A.~Hinneburg, ``Exploring the space of topic
  coherence measures,'' in \emph{the eighth ACM international conference on Web
  search and data mining}, 2015, pp. 399--408.

\bibitem{abdellatif2020challenges}
A.~Abdellatif, D.~Costa, K.~Badran, R.~Abdalkareem, and E.~Shihab, ``Challenges
  in chatbot development: A study of stack overflow posts,'' in \emph{the 17th
  International Conference on Mining Software Repositories}, 2020, pp.
  174--185.

\bibitem{zahedi2020mining}
M.~Zahedi, R.~N. Rajapakse, and M.~A. Babar, ``Mining questions asked about
  continuous software engineering: A case study of stack overflow,'' in
  \emph{the Evaluation and Assessment in Software Engineering}, 2020, pp.
  41--50.

\bibitem{mann1945nonparametric}
H.~B. Mann, ``Nonparametric tests against trend,'' \emph{Econometrica: Journal
  of the Econometric Society}, pp. 245--259, 1945.

\bibitem{knight1966computer}
W.~R. Knight, ``A computer method for calculating kendall's tau with ungrouped
  data,'' \emph{Journal of the American Statistical Association}, vol.~61, no.
  314, pp. 436--439, 1966.

\bibitem{mann1947test}
H.~B. Mann and D.~R. Whitney, ``On a test of whether one of two random
  variables is stochastically larger than the other,'' \emph{The annals of
  mathematical statistics}, pp. 50--60, 1947.

\bibitem{exploit_db}
\BIBentryALTinterwordspacing
E.~Database, ``Exploit database.'' [Online]. Available:
  \url{www.exploit-db.com}
\BIBentrySTDinterwordspacing

\bibitem{zahedi2018empirical}
M.~Zahedi, M.~Ali~Babar, and C.~Treude, ``An empirical study of security issues
  posted in open source projects,'' in \emph{the 51st Hawaii International
  Conference on System Sciences}, 2018.

\bibitem{white2015toward}
M.~White, C.~Vendome, M.~Linares-V{\'a}squez, and D.~Poshyvanyk, ``Toward deep
  learning software repositories,'' in \emph{2015 IEEE/ACM 12th Working
  Conference on Mining Software Repositories}.\hskip 1em plus 0.5em minus
  0.4em\relax IEEE, 2015, pp. 334--345.

\bibitem{elkan2008learning}
C.~Elkan and K.~Noto, ``Learning classifiers from only positive and unlabeled
  data,'' in \emph{the 14th ACM SIGKDD international conference on Knowledge
  discovery and data mining}, 2008, pp. 213--220.

\bibitem{mordelet2014bagging}
F.~Mordelet and J.-P. Vert, ``A bagging svm to learn from positive and
  unlabeled examples,'' \emph{Pattern Recognition Letters}, vol.~37, pp.
  201--209, 2014.

\bibitem{li2003learning}
X.~Li and B.~Liu, ``Learning to classify texts using positive and unlabeled
  data,'' in \emph{IJCAI}, vol.~3, no. 2003, 2003, pp. 587--592.

\bibitem{fusilier2015detecting}
D.~H. Fusilier, M.~Montes-y G{\'o}mez, P.~Rosso, and R.~G. Cabrera, ``Detecting
  positive and negative deceptive opinions using pu-learning,''
  \emph{Information processing \& management}, vol.~51, no.~4, pp. 433--443,
  2015.

\bibitem{zimmermann2010makes}
T.~Zimmermann, R.~Premraj, N.~Bettenburg, S.~Just, A.~Schroter, and C.~Weiss,
  ``What makes a good bug report?'' \emph{IEEE Transactions on Software
  Engineering}, vol.~36, no.~5, pp. 618--643, 2010.

\bibitem{bhuiyan2021security}
F.~A. Bhuiyan, M.~B. Sharif, and A.~Rahman, ``Security bug report usage for
  software vulnerability research: A systematic mapping study,'' \emph{IEEE
  Access}, vol.~9, pp. 28\,471--28\,495, 2021.

\bibitem{wu2021data}
X.~Wu, W.~Zheng, X.~Xia, and D.~Lo, ``Data quality matters: A case study on
  data label correctness for security bug report prediction,'' \emph{IEEE
  Transactions on Software Engineering}, 2021.

\bibitem{ohira2015dataset}
M.~Ohira, Y.~Kashiwa, Y.~Yamatani, H.~Yoshiyuki, Y.~Maeda, N.~Limsettho,
  K.~Fujino, H.~Hata, A.~Ihara, and K.~Matsumoto, ``A dataset of high impact
  bugs: Manually-classified issue reports,'' in \emph{2015 IEEE/ACM 12th
  Working Conference on Mining Software Repositories}.\hskip 1em plus 0.5em
  minus 0.4em\relax IEEE, 2015, pp. 518--521.

\bibitem{chen2020machine}
Y.~Chen, A.~E. Santosa, A.~M. Yi, A.~Sharma, A.~Sharma, and D.~Lo, ``A machine
  learning approach for vulnerability curation,'' in \emph{the 17th
  International Conference on Mining Software Repositories}, 2020, pp. 32--42.

\bibitem{russell2018automated}
R.~Russell, L.~Kim, L.~Hamilton, T.~Lazovich, J.~Harer, O.~Ozdemir,
  P.~Ellingwood, and M.~McConley, ``Automated vulnerability detection in source
  code using deep representation learning,'' in \emph{2018 17th IEEE
  international conference on machine learning and applications (ICMLA)}.\hskip
  1em plus 0.5em minus 0.4em\relax IEEE, 2018, pp. 757--762.

\bibitem{duan2019vulsniper}
X.~Duan, J.~Wu, S.~Ji, Z.~Rui, T.~Luo, M.~Yang, and Y.~Wu, ``Vulsniper: Focus
  your attention to shoot fine-grained vulnerabilities,'' in \emph{IJCAI},
  2019, pp. 4665--4671.

\bibitem{warnecke2020evaluating}
A.~Warnecke, D.~Arp, C.~Wressnegger, and K.~Rieck, ``Evaluating explanation
  methods for deep learning in security,'' in \emph{2020 IEEE European
  Symposium on Security and Privacy (EuroS\&P)}.\hskip 1em plus 0.5em minus
  0.4em\relax IEEE, 2020, pp. 158--174.

\bibitem{zou2021interpreting}
D.~Zou, Y.~Zhu, S.~Xu, Z.~Li, H.~Jin, and H.~Ye, ``Interpreting deep
  learning-based vulnerability detector predictions based on heuristic
  searching,'' \emph{ACM Transactions on Software Engineering and Methodology
  (TOSEM)}, vol.~30, no.~2, pp. 1--31, 2021.

\bibitem{chandrashekar2014survey}
G.~Chandrashekar and F.~Sahin, ``A survey on feature selection methods,''
  \emph{Computers \& Electrical Engineering}, vol.~40, no.~1, pp. 16--28, 2014.

\bibitem{rosenberg2021adversarial}
I.~Rosenberg, A.~Shabtai, Y.~Elovici, and L.~Rokach, ``Adversarial machine
  learning attacks and defense methods in the cyber security domain,''
  \emph{ACM Computing Surveys (CSUR)}, vol.~54, no.~5, pp. 1--36, 2021.

\bibitem{wan2019does}
Z.~Wan, X.~Xia, D.~Lo, and G.~C. Murphy, ``How does machine learning change
  software development practices?'' \emph{IEEE Transactions on Software
  Engineering}, 2019.

\bibitem{liu2020using}
J.~Liu, Q.~Huang, X.~Xia, E.~Shihab, D.~Lo, and S.~Li, ``Is using deep learning
  frameworks free? characterizing technical debt in deep learning frameworks,''
  in \emph{the ACM/IEEE 42nd International Conference on Software Engineering:
  Software Engineering in Society}, 2020, pp. 1--10.

\end{thebibliography}
